\documentclass[12pt]{article}
\usepackage{amsmath,amssymb,amsfonts,color,graphicx,cite,color,feynarts,soul}
\input paperdef

\graphicspath{{figs/}}

\evensidemargin \oddsidemargin
\allowdisplaybreaks

\begin{document}
\thispagestyle{empty}

\def\thefootnote{\fnsymbol{footnote}}

\begin{flushright}
KA-TP--40--2011 
\end{flushright}

\vspace{0.5cm}

\begin{center}

{\large\sc {\bf Gluino Decays in the Complex MSSM:}}

\vspace{0.4cm}

{\large\sc {\bf A Full One-Loop Analysis}}

\vspace{1cm}

{\sc
S.~Heinemeyer$^{1}$%
\footnote{email: Sven.Heinemeyer@cern.ch}%
~and C.~Schappacher$^{2}$%
\footnote{email: cs@particle.uni-karlsruhe.de}%
}

\vspace*{.7cm}

{\sl
$^1$Instituto de F\'isica de Cantabria (CSIC-UC), Santander,  Spain

\vspace*{0.1cm}

$^2$Institut f\"ur Theoretische Physik,
Karlsruhe Institute of Technology, \\
D--76128 Karlsruhe, Germany
}

\end{center}

\vspace*{0.1cm}

\begin{abstract}
\noindent
We evaluate all two-body decay modes of the gluino, in the 
Minimal Supersymmetric Standard Model with complex parameters (cMSSM). 
This constitutes an important step in the cascade decays of SUSY 
particles at the LHC.
The evaluation is based on a full one-loop calculation of all two-body 
decay channels, also including hard QED and QCD radiation. 
The dependence of the gluino decay to a scalar quark and a 
quark on the relevant cMSSM parameters is analyzed numerically. 
We find sizable contributions to the decay widths and branching
ratios. They are, roughly of \order{\pm 5\%}, but can go up to 
$\pm 10\%$ or higher, where the pure SUSY QCD contributions alone
can give an insufficient approximation to the full one-loop result.
Therefore the  full corrections are important for the correct 
interpretation of gluino decays at the LHC.
The results will be implemented into the Fortran code {\tt FeynHiggs}.
\end{abstract}


\def\thefootnote{\arabic{footnote}}
\setcounter{page}{0}
\setcounter{footnote}{0}

\newpage

\newcommand{\decaySqi}{\gl \to \tilde{q}_i\, q}
\newcommand{\decaySqe}{\gl \to \tilde{q}_1\, q}
\newcommand{\decaySqz}{\gl \to \tilde{q}_2\, q}
\newcommand{\decayaSqi}{\gl \to \tilde{q}^{\dagger}_i\, q}
\newcommand{\decaySqai}{\gl \to \tilde{q}_i\, \bar{q}}
\newcommand{\decayaSqe}{\gl \to \tilde{q}^{\dagger}_1\, q}
\newcommand{\decayaSqz}{\gl \to \tilde{q}^{\dagger}_2\, q}
\newcommand{\decaySqae}{\gl \to \tilde{q}_1\, \bar q}
\newcommand{\decaySqaz}{\gl \to \tilde{q}_2\, \bar q}
\newcommand{\decaySti}{\gl \to \tilde{t}_i\, t}
\newcommand{\decaySbi}{\gl \to \tilde{b}_i\, b}
\newcommand{\decaySci}{\gl \to \tilde{c}_i\, c}
\newcommand{\decaySsi}{\gl \to \tilde{s}_i\, s}
\newcommand{\decaySui}{\gl \to \tilde{u}_i\, u}
\newcommand{\decaySdi}{\gl \to \tilde{d}_i\, d}
\newcommand{\decaySte}{\gl \to \Stope\, t}
\newcommand{\decayStec}{\gl \to \Stope\, c}
\newcommand{\decayStz}{\gl \to \Stopz\, t}
\newcommand{\decaySbe}{\gl \to \Sbote\, b}
\newcommand{\decaySbz}{\gl \to \Sbotz\, b}
\newcommand{\decaySce}{\gl \to \Schae\, c}
\newcommand{\decayScz}{\gl \to \Schaz\, c}
\newcommand{\decaySse}{\gl \to \Sstre\, s}
\newcommand{\decaySsz}{\gl \to \Sstrz\, s}
\newcommand{\decaySue}{\gl \to \Supe\, u}
\newcommand{\decaySuz}{\gl \to \Supz\, u}
\newcommand{\decaySde}{\gl \to \Sdowne\, d}
\newcommand{\decaySdz}{\gl \to \Sdownz\, d}

\newcommand{\decayaSte}{\gl \to \Stop^{\dagger}_1\, t}
\newcommand{\decayaStz}{\gl \to \Stop^{\dagger}_2\, t}
\newcommand{\decayaSbe}{\gl \to \Sbot^{\dagger}_1\, b}
\newcommand{\decayaSbz}{\gl \to \Sbot^{\dagger}_2\, b}
\newcommand{\decayaSce}{\gl \to \Scha^{\dagger}_1\, c}
\newcommand{\decayaScz}{\gl \to \Scha^{\dagger}_2\, c}
\newcommand{\decayaSse}{\gl \to \Sstr^{\dagger}_1\, s}
\newcommand{\decayaSsz}{\gl \to \Sstr^{\dagger}_2\, s}
\newcommand{\decayaSue}{\gl \to \Sup^{\dagger}_1\, u}
\newcommand{\decayaSuz}{\gl \to \Sup^{\dagger}_2\, u}
\newcommand{\decayaSde}{\gl \to \Sdown^{\dagger}_1\, d}
\newcommand{\decayaSdz}{\gl \to \Sdown^{\dagger}_2\, d}

\newcommand{\decayStae}{\gl \to \Stope\, \bar t}
\newcommand{\decayStaz}{\gl \to \Stopz\, \bar t}
\newcommand{\decaySbae}{\gl \to \Sbote\, \bar b}
\newcommand{\decaySbaz}{\gl \to \Sbotz\, \bar b}
\newcommand{\decayScae}{\gl \to \Schae\, \bar c}
\newcommand{\decayScaz}{\gl \to \Schaz\, \bar c}
\newcommand{\decaySsae}{\gl \to \Sstre\, \bar s}
\newcommand{\decaySsaz}{\gl \to \Sstrz\, \bar s}
\newcommand{\decaySuae}{\gl \to \Supe\, \bar u}
\newcommand{\decaySuaz}{\gl \to \Supz\, \bar u}
\newcommand{\decaySdae}{\gl \to \Sdowne\, \bar d}
\newcommand{\decaySdaz}{\gl \to \Sdownz\, \bar d}

\newcommand{\decayNeug}[1]{\gl \to \neu{#1}\, g}
\newcommand{\decayNeg}{\decayNeug{1}}
\newcommand{\decayNzg}{\decayNeug{2}}
\newcommand{\decayNdg}{\decayNeug{3}}
\newcommand{\decayNvg}{\decayNeug{4}}
\newcommand{\decayNkg}{\decayNeug{k}}
\newcommand{\decayxy}{\gl \to {\rm xy}}

\newcommand{\SE}{${\cal S}$}


\section{Introduction}

One of the most important tasks at the LHC is to search for physics
effects beyond the Standard Model (SM), where the Minimal Supersymmetric
Standard Model (MSSM)~\cite{mssm} is one of the leading candidates. 
Supersymmetry (SUSY) predicts two scalar partners for all SM fermions as well
as fermionic partners to all SM bosons.
Especially it predicts the fermionic gluino as a superpartner of the
bosonic gluon.

If SUSY is realized in nature and the gluino and/or scalar quarks 
are in the kinematic reach of the LHC, it is expected that these
strongly interacting particles are copiously produced~\citeres{atlas,cms}.
After the (pair) production of these particles they are expected to
decay via cascades to lighter SUSY particles and quarks and/or leptons. 
Many models predict that the heaviest SUSY particle is indeed the gluino
(see, for instance, \citere{mc3} and references therein), and
consequently, the decay to a scalar quark and a quark will one of the
most relevant decays in such a cascade. Only if these two-body decays
are kinematically forbidden other decay channels can become numerically
relevant. These are either two-body decays that are purely loop-induced,
or three-body decays via a virtual squark. 
In this paper we will concentrate on the two-body decays that are
allowed at tree-level already. We will also, for the sake of
completeness, evaluate the two-body decays that appear 
only at the loop-level, the decays of the gluino to a gluon and a 
neutralino, 
which can potentially serve as a production source of the cMSSM 
cold dark matter (CDM) candidate.

In order to yield a sufficient accuracy one-loop corrections to
the various gluino decay modes have to be considered.
As outlined above, we take into account all two-body decay modes of the
gluino in the MSSM with complex parameters (cMSSM).
More specifically we calculate the full one-loop corrections to
\begin{align}
\label{glsqq}
&\Ga(\decayaSqi) \qquad (i = 1,2; \; q = t,b,c,s,u,d)~, \\
&\Ga(\decaySqai) \qquad (i = 1,2; \; q = t,b,c,s,u,d)~, \\
\label{glneg}
&\Ga(\decayNkg) \qquad (k = 1,2,3,4)~.
\end{align}
A difference between $\Ga(\decayaSqi)$ and $\Ga(\decaySqai)$ can
arise from complex parameters. In the case when all parameters are
real (and sometimes to simplify the notation) we also use the notation 
$\Ga(\decaySqi)$ referring to {\em both} channels.
The total width is defined as the sum of the channels (\ref{glsqq}) --
(\ref{glneg}), 
where for a given parameter point several of the~28 channels may be 
kinematically forbidden. As it is expected, the channel
(\ref{glneg}) is numerically irrelevant for the total width as long
as one or more of the two-body tree-level decays is kinematically
allowed. 

One-loop QCD corrections to gluino decays have been evaluated in
\citere{glsqq_als}. The calculation was done in the MSSM with real
parameters (rMSSM).
Those \order{\als} corrections have been implemented into the code
{\tt SDECAY}~\cite{sdecay}.
To our knowledge, no evaluation of electroweak (EW)
corrections to gluino decays, nor any higher-order corrections
  involving complex phases have been performed so far.
Gluino decays at the tree-level were employed, for instance, in
\citere{majoranagluinos} to determine their Majorana/Dirac character. 
Gluino polarizations were analyzed in \citere{gluinopol}. Again, both
analyses were performed in the rMSSM.

Several methods have been discussed in the literature to
extract the complex parameters of the model from experimental
measurements. 
However, no such an analysis, to our knowledge, of the phase of the
gluino mass parameter, $M_3$, has been performed so far.
We analyze the effects of this phase on the branching ratios (BR's)
of the gluino, also as a motivation to devise experimental strategies to
determine this parameter at the LHC (or other future
colliders)~\cite{lhc2fc}. 

In this paper we present
for the first time a full one-loop calculation, including
electroweak effects, for all two-body decay
channels of the gluino in the cMSSM. The calculation includes
soft and hard QED and QCD radiation.
In \refse{sec:cMSSM} we review the relevant sectors of the cMSSM. 
Details about the calculation can be
found in \refse{sec:calc}, and the numerical results for all decay
channels are presented in \refse{sec:numeval}, were we discuss
especially the size of the EW corrections and the effects
from the complex gluino phase. The conclusions can be
found in \refse{sec:conclusions}.
The evaluation of the branching ratios of the gluino will be 
implemented into the Fortran code 
{\tt FeynHiggs}~\cite{feynhiggs,mhiggslong,mhiggsAEC,mhcMSSMlong}.


\section{The relevant sectors of the complex MSSM}
\label{sec:cMSSM}

All the channels (\ref{glsqq}) -- (\ref{glneg}) are calculated at the
one-loop level, including hard QED and QCD radiation (for the decays
that exist already at the tree-level). This requires the simultaneous 
renormalization of several sectors of the cMSSM, where details can be
found in \citeres{dissTF,SbotRen,Stop2decay}. In the following 
subsections we briefly review these relevant sectors.


\subsection{The squark sector of the cMSSM}
\label{sec:squark}

For the evaluation of the one-loop contributions to the decay 
of the $\gl$ to a squark and quark, a renormalization of the 
scalar quark sector is needed.
The squark mass matrix $\matr{M}_{\tilde{q}}$ read
\begin{align}\label{Sfermionmassenmatrix}
\matr{M}_{\tilde{q}} &= \begin{pmatrix} 
M_{\tilde q_L}^2 + m_q^2 + \MZ^2\, (I_q^3 - Q_q \sw^2)\, \CZb & 
 m_q \Xq^* \\[.2em]
 m_q \Xq &
 M_{\tilde{q}_R}^2 + m_q^2 + \MZ^2\, Q_q\, \sw^2\, \CZb
\end{pmatrix}
\end{align}
with
\begin{align}\label{kappa}
\Xq &= \Aq - \mu^*\kappa~, \qquad \kappa = \{\cot\beta, \tan\beta\} 
        \quad {\rm for} \quad q = \{u{\rm -type}, d{\rm -type} \}~.
\end{align}
$M_{\tilde q_L}$ and $M_{\tilde{q}_R}$ are the soft SUSY-breaking mass
parameters, where $M_{\tilde q_L}$ is equal for all members of an
$SU(2)_L$ doublet and $M_{\tilde q_R}$ depends on the scalar quark flavor.
$m_q$ is the mass of the corresponding quark.
$Q_{{q}}$ and $I_q^3$ denote the charge and isospin of $q$, and
$A_q$ is the trilinear soft-breaking parameter.
$\MZ$ and $\MW$ are the masses of the $Z$~and $W$~boson, 
$\cw = \MW/\MZ$, and $\sw = \sqrt{1 - \cw^2}$.
The mass matrix can be diagonalized with the help of a unitary
 transformation ${\matr{U}}_{\tilde{q}}$,
\begin{align}\label{transformationkompl}
\matr{D}_{\tilde{q}} &= 
\matr{U}_{\tilde{q}}\, \matr{M}_{\tilde{q}} \, 
{\matr{U}}_{\tilde{q}}^\dagger = 
\begin{pmatrix} \msqe^2 & 0 \\ 0 & \msqz^2 \end{pmatrix}~, \qquad
{\matr{U}}_{\tilde{q}}= 
\begin{pmatrix} U_{\tilde{q}_{11}}  & U_{\tilde{q}_{12}} \\  
                U_{\tilde{q}_{21}} & U_{\tilde{q}_{22}}  \end{pmatrix}~.
\end{align}
The mass eigenvalues depend only on $|\Xq|$.
The scalar quark masses will always be mass ordered, i.e.\
$m_{\tilde{q}_1} \le m_{\tilde{q}_2}$:
\begin{align}
m_{\tilde{q}_{1,2}}^2 &= \edz \KL M_{\tilde{q}_L}^2 + M_{\tilde{q}_R}^2 \KR
       + m_q^2 + \edz I_q^3\, \MZ^2\, \CZb \non \\
&\quad \mp \edz \sqrt{\KKL M_{\tilde{q}_L}^2 - M_{{\tilde{q}_R}}^2
       + \MZ^2\, (I_q^3 - 2 Q_q \sw^2)\, \CZb \KKR^2 + 4 m_q^2 |\Xq|^2}~.
\label{MSbot}
\end{align}

Details about the renormalization of the scalar quark sector of the
cMSSM can be found in \citeres{SbotRen,Stop2decay}. The employed
renormalization preserves the $SU(2)_L$ symmetry of the model. It is
ensured that all external particles fulfill the required on-shell
(OS) properties for masses and $Z$~factors. 
The field renormalization requires the renormalization of the
off-diagonal entry in the squark mass matrix, which leads to a
renormalization of the trilinear couplings $A_q$, $\mu$ and $\tb$, 
see below.
It was shown in \citeres{SbotRen,Stop2decay} that the renormalization 
produces stable and well-behaved results for nearly the whole cMSSM 
parameter space.

Finally it should be noted that we take into account the absorptive part 
of the self-energy type contributions on the external legs via combined 
$Z$ factors which are different for incoming squarks/outgoing antisquarks 
(unbarred) and outgoing squarks/incoming antisquarks
(barred)~\cite{Stop2decay}: 

\begin{itemize}
\item[(i)]
The diagonal $Z$~factors read
\begin{align}
\bigl[\de{\cZ}_{\sq}\bigr]_{ii} &= 
\bigl[\de{\bar{\cZ}}_{\sq}\bigr]_{ii} = - \Si_{\sq_{ii}}'(\msqi^2)
\qquad (i = 1,2)~.
\end{align}
\item[(ii)]
The off-diagonal $Z$~factors read
\begin{align}
\bigl[\dcZ{\sq}\bigr]_{12} &= 
+ 2 \frac{\Si_{\sq_{12}}(\msqz^2) - \de Y_q}{(\msqe^2 - \msqz^2)}~,
& \bigl[\dcZ{\sq}\bigr]_{21} &= 
- 2 \frac{\Si_{\sq_{21}}(\msqe^2) - \de Y_q^*}{(\msqe^2 - \msqz^2)}~, \\ 
\bigl[\de\bar{\cZ}_{\sq}\bigr]_{12} &= 
+ 2 \frac{\Si_{\sq_{21}}(\msqz^2) - \de Y_q^*}{(\msqe^2 - \msqz^2)}~,
& \bigl[\de\bar{\cZ}_{\sq}\bigr]_{21} &= 
- 2 \frac{\Si_{\sq_{12}}(\msqe^2) - \de Y_q}{(\msqe^2 - \msqz^2)}~.
\end{align}
\end{itemize}
$\Si_{\sq_{ij}}$ denotes the scalar quark self-energies, and 
$\Si'(m^2) \equiv 
\frac{\partial \Si(p^2)}{\partial p^2}\big|_{p^2 = m^2}$.
See \citere{Stop2decay} for the other renormalization constants 
and further details.


\subsection{The quark sector of the cMSSM}
\label{sec:quark}

In this section we briefly describe the quark sector of the cMSSM
and its renormalization, extending the corresponding discussions in
\citeres{SbotRen,Stop2decay,imim}.
The quark mass, $m_q$, and the quark fields $q_L$, $q_R$ are 
renormalized in the following way:
\begin{align}
m_q &\to m_q + \de m_q~,\\
q_{L/R} &\to (1 + \tedz \dcZ{q}^{L/R})\, q_{L/R}~, \\
\bar{q}_{L/R} &\to (1 + \tedz \dbcZ{q}^{L/R})\, \bar{q}_{L/R}~,
\end{align}
with $\de m_q$ being the quark mass counterterm and $\dcZ{q}^{L/R}$
being the combined $Z$~factors of the left/right-handed
quark fields, respectively. 
They are determined separately to include effects from the
absorptive parts of the self-energy type contribution on an 
external quark leg~\cite{Stop2decay}.
The unbarred quantities denote incoming quarks/outgoing antiquarks and 
the barred denote outgoing quarks/incoming antiquarks.
The renormalized self energy, $\hSi_{q}$, can be decomposed
into left/right-handed and scalar left/right-handed parts, 
${\Si}_q^{L/R}$ and ${\Si}_q^{SL/SR}$, respectively,
\begin{align}\label{decomposition}
\hSi_{q} (p) &= \not\! p\, {\omega}_{-} \hSi_q^L (p^2)
                   + \not\! p\, {\omega}_{+} \hSi_q^R (p^2)
                   + {\omega}_{-} \hSi_q^{SL} (p^2) 
                   + {\omega}_{+} \hSi_q^{SR} (p^2)~,
\end{align}
where the components are given by
\begin{align}
\hSi_q^{L/R} (p^2) &= {\Si}_q^{L/R} (p^2) 
   + \frac{1}{2} (\dcZ{q}^{L/R} + \dbcZ{q}^{L/R})~, \\
\hSi_q^{SL} (p^2) &=  {\Si}_q^{SL} (p^2) 
   - \frac{m_q}{2} (\dcZ{q}^L + \dbcZ{q}^R) - \de m_q~,  \\
\hSi_q^{SR} (p^2) &=  {\Si}_q^{SR} (p^2) 
   - \frac{m_q}{2} (\dcZ{q}^R + \dbcZ{q}^L) - \de m_q~, 
\end{align}
and ${\omega}_{\pm} = \frac{1}{2}(1 \pm \gamma_5)$ 
are the right- and left-handed projectors, respectively.

The quark mass is defined on-shell~\cite{denner}, yielding the one-loop 
counterterm $\de \mq$:
\begin{align}\label{dmq}
\de \mq^{\OS} &= \tedz \wtre \KKKL 
    \mq \KKL\Si_q^L (\mq^2) + \Si_q^R (\mq^2) \KKR  
  + \KKL \Si_q^{SL} (\mq^2) + \Si_q^{SR} (\mq^2) \KKR \KKKR~,
\end{align}
referring to the Lorentz decomposition of the self energy 
${\hSi}_{q}(p)$, see \refeq{decomposition}.
$\widetilde{\text{Re}}$ denotes the real part with respect to
contributions from the loop integral, but leaves the complex
couplings unaffected.

Special care is needed for the bottom sector due to potentially
large effects at large $\tb$ (denoting the ratio of the two vacuum
expectation values). If (and only if) there are no external bottom
quarks, the bottom-quark mass is defined \DRbar\ 
(see \citere{Stop2decay} for more details),
yielding the one-loop counterterm $\de\mb^{\DRbar}$:
\begin{align}\label{dmb}
\de\mb^{\DRbar} = \tedz \wtre \KKKL
  \mb \KKL \Si_b^L (\mb^2) + \Si_b^R (\mb^2) \KKR_{\rm div}
+ \KKL \Si_b^{SL} (\mb^2) + \Si_b^{SR} (\mb^2) \KKR_{\rm div} \KKKR~.
\end{align}

The new (diagonal) field renormalization constants, taking into 
account the absorptive part of the self-energy type contribution on an 
external quark leg are different for incoming quarks/outgoing antiquarks 
(unbarred) and outgoing quarks/incoming antiquarks (barred),
\begin{align}
\dcZ{q}^{L/R} &= - \Big[ \Si_q^{L/R} (m_q^2)
    + m_q^2 \KL {{\Si}_q^{L}}'(m_q^2) + {{\Si}_q^{R}}'(m_q^2) \KR
    + m_q \KL {{\Si}_q^{SL}}'(m_q^2) + {{\Si}_q^{SR}}'(m_q^2) \KR
                  \Big] \non \\
&\qquad \pm \frac{1}{2\, \mq} 
        \KKL {\Si}_q^{SL}(\mq^2) - {\Si}_q^{SR}(\mq^2) \KKR~, \\
\dbcZ{q}^{L/R} &= - \Big[ \Si_q^{L/R} (m_q^2)
    + m_q^2 \KL {{\Si}_q^{L}}'(m_q^2) + {{\Si}_q^{R}}'(m_q^2) \KR
    + m_q \KL {{\Si}_q^{SL}}'(m_q^2) + {{\Si}_q^{SR}}'(m_q^2) \KR
                  \Big] \non \\
&\qquad \mp \frac{1}{2\, \mq} 
        \KKL {\Si}_q^{SL}(\mq^2) - {\Si}_q^{SR}(\mq^2) \KKR~.
\end{align}
For further details see \citere{Stop2decay}.

\medskip

The input parameters in the $b$ sector have to correspond to the
chosen renormalization. We start by defining the bottom mass, where the
experimental input is the SM \MSbar\ mass \cite{pdg}.
The value of $\mb^{\MSbar}(\mu_R)$ 
(at the renormalization scale $\mu_R$) is calculated from 
$\mb^{\MSbar}(\mb)$ at the three loop level following the prescription 
given in~\citere{RunDec}.
The input parameters can be found in \refse{input}.
The ``on-shell'' mass is connected to the \MSbar\ mass via
\begin{align}
\mb^{\os} &= \mb^{\MSbar}(\mu_R) \; 
   \KKL 1 + \frac{\als^{\MSbar}(\mu_R)}{\pi} 
        \KL \frac{4}{3} + 2\, \ln \frac{\mu_R}{\mb^{\MSbar}(\mu_R)} \KR 
   \KKR~.
\end{align}
The $\DRbar$ bottom quark mass is calculated iteratively from
\begin{align}\label{eq:mbDR}
\mb^{\DRbar} &= \frac{\mb^{\os} |1 + \db| + \de\mb^{\OS} - \de\mb^{\DRbar}}
            {|1 + \db|}
\end{align}
with an accuracy of 
$|1 - (\mb^{\DRbar})^{(n)}/(\mb^{\DRbar})^{(n-1)}| < 10^{-5}$
reached in the $n$th step of the iteration 
and $\de\mb^{\OS}$ as given in \refeq{dmq} with $q = b$.
The quantity $\db$~\cite{deltab1,deltab2} resums the \order{(\als\tb)^n}
and \order{(\alt\tb)^n} terms and is given 
in Eq. (66) of \citere{Stop2decay}.

In the case of external bottom quarks we use an OS renormalization 
scheme for the bottom sector. The bottom quark mass is then obtained 
from 
\begin{align}\label{eq:mbcorr}
\mb^{\OS} &= \mb^{\DRbar} + \de\mb^{\DRbar} - \de\mb^{\OS}~,
\end{align}
again with $\de\mb^{\OS}$ from \refeq{dmq} with $q = b$.


\subsection{The gluino and the strong coupling constant}
\label{sec:gluino}

The soft-breaking gluino mass parameter $M_3$ is in general complex, 
\begin{align}
M_3 = |M_3| e^{i \phigl} \qquad (\text{with the gluino mass}~\mgl = |M_3|)~.
\end{align}
We choose an OS renormalization for the gluino, where the renormalization 
constants can be found in \citere{Stop2decay}.
It properly takes into account the complex phase of $M_3$ and yields an 
on-shell gluino, as it is required for an external particle in our decays.%
\footnote{
  The general renormalization procedure is described in 
  \citeres{dissTF,SbotRen,Stop2decay}.}
It is possible to modify the field renormalization constants to take 
into account the absorptive part of the self-energy type contribution on 
the external gluino leg.
The renormalization constants then read~\cite{Stop2decay}
(unbarred (barred) for an incoming (outgoing) gluino), 
\begin{align}
\de M_3 &= \edz \wtre \KKKL
      \mgl \KKL \Si_{\gl}^L (\mgl^2) + \Si_{\gl}^R (\mgl^2) \KKR
    + \KKL \Si_{\gl}^{SL} (\mgl^2) + \Si_{\gl}^{SR} (\mgl^2) \KKR 
      \KKKR e^{i \phigl}~, \\
\dcZ{\gl}^{L/R} &= - \Big[ \Si_{\gl}^{L/R}(\mgl^2)
    + \mgl^2 \KL \Si_{\gl}^{L'}(\mgl^2) + \Si_{\gl}^{R'}(\mgl^2) \KR
    + \mgl \KL \Si_{\gl}^{SL'}(\mgl^2) + \Si_{\gl}^{SR'}(\mgl^2) \KR
                   \Big] \non \\
&\qquad \pm \ed{2 \mgl} \KKL 
            \Si_{\gl}^{SL}(\mgl^2) - \Si_{\gl}^{SR}(\mgl^2) \KKR~, \\
\dbcZ{\gl}^{L/R} &= \dcZ{\gl}^{R/L}~.
\end{align}
The last formula holds due to the Majorana character of the gluino 
and we have chosen $\de\phigl = 0$.
This choice is possible, since the imaginary part of $M_3$ does not
contain any divergence.%
\footnote{
    Formulas including the $\de\phigl$ contributions can be found in 
    Sect. 2.1.2 in \citere{Stop2decay}.}

\medskip

Our renormalization of the strong coupling constant, $\als$, is
described in \citeres{dissTF,SbotRen,Stop2decay} (and references
therein). 
The decoupling of the heavy particles is taken into account in 
the definition of $\als$: Starting point is 
$\als^{\MSbar}(\MZ)$~\cite{pdg},
where the running can also be found in \citere{pdg}.
From the \MSbar\ value the \DRbar\ value is obtained at the two-loop 
level (with $n_f = 6$ for $\mu_R > \mt$) via~\cite{alsDRbar}
\begin{align}
\als^{\DRbar,(6)}(\mu_R) &= \als^{\MSbar,(6)}(\mu_R)\, \KKL 1 + 
      \frac{\als^{\MSbar,(6)}(\mu_R)}{4\,\pi} + \frac{7}{8}
      \frac{\bigl( \als^{\MSbar,(6)}(\mu_R) \bigr)^2}{\pi^2} \KKR
\end{align}
Within the MSSM $\als$ to one-loop reads
\begin{align}
\als^{\rm MSSM}(\mu_R) &= \als^{\DRbar,(6)}(\mu_R) \, \KKL 1 + 
                         \frac{\als^{\DRbar,(6)}(\mu_R)}{\pi}
                         \KL \ln \frac{\mu_R}{\mgl} + 
                         \ln \frac{\mu_R}{M_{\tilde{q}}} \KR \KKR~,
\end{align}
with $M_{\sq} = \Pi_{\sq}(\msqe\msqz)^{\frac{1}{12}}$.
The log~terms origin from the decoupling of the SUSY QCD (SQCD) 
particles from the running of $\als$ at lower scales 
$\mu_R \leq \mu_{\rm dec.} = \mgl$.


\subsection{The Higgs and gauge boson sector of the cMSSM}
\label{sec:higgs}

The two Higgs doublets of the cMSSM are decomposed in the following way,
\begin{align}
\label{eq:higgsdoublets}
\cHe = \begin{pmatrix} H_{11} \\ H_{12} \end{pmatrix} &=
\begin{pmatrix} v_1 + \tfrac{1}{\sqrt{2}} (\phi_1-i \chi_1) \\
  -\phi^-_1 \end{pmatrix}, \notag \\ 
\cHz = \begin{pmatrix} H_{21} \\ H_{22} \end{pmatrix} &= e^{i \xi}
\begin{pmatrix} \phi^+_2 \\ v_2 + \tfrac{1}{\sqrt{2}} (\phi_2+i
  \chi_2) \end{pmatrix}. 
\end{align}
Besides the vacuum expectation values $v_1$ and $v_2$, in 
\refeq{eq:higgsdoublets} a possible new phase $\xi$ between the two
Higgs doublets is introduced. 
The Higgs potential $\VHiggs$ can be written in powers of the Higgs fields%
\footnote{
  Corresponding to the convention used in \fa/\fc, we exchanged in the 
  charged part the positive Higgs fields with the negative ones, which 
  is in contrast to \citere{mhcMSSMlong}. 
  As we keep the definition of the matrix $\matr{M}_{\phi^\pm\phi^\pm}$ 
  used in \citere{mhcMSSMlong} the transposed matrix will appear in the 
  expression for $\matr{M}_{H^\pm G^\pm}^{\rm diag}$.
},
\begin{align}
\VHiggs &=  \ldots + T_{\phi_1}\, \phi_1  +T_{\phi_2}\, \phi_2 +
        T_{\chi_1}\, \chi_1 + T_{\chi_2}\, \chi_2 \non \\ 
&\quad - \edz \begin{pmatrix} \phi_1,\phi_2,\chi_1,\chi_2
        \end{pmatrix} 
\matr{M}_{\phi\phi\chi\chi}
\begin{pmatrix} \phi_1 \\ \phi_2 \\ \chi_1 \\ \chi_2 \end{pmatrix} -
\begin{pmatrix} \phi^{+}_1,\phi^{+}_2  \end{pmatrix}
\matr{M}^{\top}_{\phi^\pm\phi^\pm}
\begin{pmatrix} \phi^{-}_1 \\ \phi^{-}_2  \end{pmatrix} + \ldots~,
\end{align}
where the coefficients of the linear terms are called tadpoles and
those of the bilinear terms are the mass matrices
$\matr{M}_{\phi\phi\chi\chi}$ and $\matr{M}_{\phi^\pm\phi^\pm}$. 
After a rotation to the physical fields 
one obtains
\begin{align}
\label{VHiggs}
\VHiggs &=  \ldots + T_{h}\, h + T_{H}\, H + T_{A}\, A \non \\ 
&\quad  - \edz \begin{pmatrix} h, H, A, G 
        \end{pmatrix} 
\matr{M}_{hHAG}^{\rm diag}
\begin{pmatrix} h \\ H \\ A \\ G  \end{pmatrix} -
\begin{pmatrix} H^{+}, G^{+}  \end{pmatrix}
\matr{M}_{H^\pm G^\pm}^{\rm diag}
\begin{pmatrix} H^{-} \\ G^{-} \end{pmatrix} + \ldots~,
\end{align}
where the tree-level masses are denoted as
$\mh$, $\mH$, $\mA$, $\mG$, $\MHp$, $\mGp$.
With the help of a Peccei-Quinn transformation~\cite{Peccei} $\mu$ and 
the complex soft SUSY-breaking parameters in the Higgs sector can be 
redefined~\cite{MSSMcomplphasen} such that the complex phases vanish 
at tree-level.
As input parameter we choose the mass of the charged Higgs boson, $\MHp$.
All details can be found in \citeres{Stop2decay,mhcMSSMlong}.

Higgs bosons, $h^0$, $H^0$, $H^\pm$, $A^0$,
and the electroweak gauge bosons, $Z$, $W^\pm$, $\ga$, 
appear only as internal particles, hence no renormalization is required. 
Furthermore we use tree-level masses and couplings for the Higgs bosons
that appear internally in the loops, see \refse{sec:calc}. 
The only exception is the renormalization of $\tb$ that enters the
squark field renormalization. Here we choose a \DRbar\ scheme as
defined, for instance, in Eq.~(120c) in \citere{Stop2decay}.


\subsection{The chargino/neutralino sector of the cMSSM}
\label{sec:chaneu}

The mass eigenstates of the charginos can be determined from the matrix
\begin{align}
  \matr{X} =
  \begin{pmatrix}
    \MTwo & \sqrt{2} \sinb \MW \\
    \sqrt{2} \cosb \MW & \mu
  \end{pmatrix}.
\end{align}
In addition to the higgsino mass parameter $\mu$ it contains the 
soft breaking term $\MTwo$, which can also be complex in the cMSSM.
The rotation to the chargino mass eigenstates is done by transforming
the original wino and higgsino fields with the help of two unitary 
2$\times$2 matrices $\matr{U}$ and $\matr{V}$,
\begin{align}
\label{eq:charginotransform}
\tilde{\chi}^-_i = 
\begin{pmatrix} \psi^L_i \\[.2em] \overline{\psi^R_i} \end{pmatrix}
\quad \text{with} \quad \psi^L_{i} = U_{ij} \begin{pmatrix} \tilde{W}^-
  \\ \tilde{H}^-_1 \end{pmatrix}_{j} \quad \text{and} \quad
 \psi^R_{i} = V_{ij} \begin{pmatrix} \tilde{W}^+
  \\ \tilde{H}^+_2 \end{pmatrix}_{j}~,
\end{align}
where the $i$th mass eigenstate can be expressed in terms of either the Weyl
spinors $\psi^L_i$ and $\psi^R_i$  or the Dirac spinor $\tilde{\chi}^-_i$
with $i,j = 1,2$.

These rotations lead to the diagonal mass matrix
\begin{align}
\matr{M}_{\cham{}} = 
\matr{V}^* \, \matr{X}^{\top} \, \matr{U}^{\dagger} =
\begin{pmatrix} m_{\cha{1}} & 0 \\ 0 & m_{\cha{2}} \end{pmatrix}~.
\end{align}
From this relation, it becomes clear that the chargino masses $\mcha{1,2}$
can be determined as the (real and positive) singular values of $\matr{X}$.
The singular value decomposition of $\matr{X}$ also yields results for
$\matr{U}$ and~$\matr{V}$.

A similar procedure is used for the determination of the neutralino masses and
mixing matrix, which can both be calculated from the mass matrix
\begin{align}
  \matr{Y} =
  \begin{pmatrix}
    \MOne                  & 0                & -\MZ \, \sw \cosb
    & \MZ \, \sw \sinb \\ 
    0                      & \MTwo            & \quad \MZ \, \cw \cosb
    & -\MZ \, \cw \sinb \\ 
    -\MZ \, \sw \cosb      & \MZ \, \cw \cosb & 0
    & -\mu             \\ 
    \quad \MZ \, \sw \sinb & -\MZ \, \cw \sinb & -\mu                   & 0
  \end{pmatrix}.
\end{align}
This symmetric matrix contains the additional complex soft-breaking
parameter $\MOne$. 
The diagonalization of the matrix
is achieved by a transformation starting from the original
bino/wino/higgsino basis,
\begin{align}
\tilde{\chi}^0_{k} = \begin{pmatrix} \psi^0_{k} \\[.2em] \overline{\psi^0_{k}} 
                   \end{pmatrix}~, \quad
\psi^0_{k} = N_{{kl}} \begin{pmatrix} \tilde{B}^0 \\ \tilde{W}^0 \\ \tilde{H}^0_1
  \\ \tilde{H}^0_2 \end{pmatrix}_{l}~, \quad
\matr{M}_{\neu{}} = \matr{N}^* \, \matr{Y} \, \matr{N}^{\dagger} =
  \begin{pmatrix}
    \mneu{1} & 0 & 0 & 0 \\
    0 & \mneu{2} & 0 & 0 \\
    0 & 0 & \mneu{3} & 0 \\
    0 & 0 & 0 & \mneu{4}
  \end{pmatrix}~,
\end{align}
where $\psi^0_{k}$ denotes the two component Weyl spinor and 
$\tilde{\chi}^0_{k}$
the four component Majorana spinor of the $k$th neutralino field
with $k,l = 1,2,3,4$.
The unitary 4$\times$4 matrix $\matr{N}$ and the physical neutralino
masses again result from a numerical singular value decomposition of 
$\matr{Y}$.
The symmetry of $\matr{Y}$ permits the non-trivial condition of
using only one matrix $\matr{N}$ for its diagonalization, 
in contrast to the chargino case shown above. 

A renormalization of this sector, as described in detail in
\citeres{dissTF,Stop2decay}, is not required.
Neutralinos and charginos appear only as internal particles 
in the loop corrections of the two-body decays, see \refse{sec:calc}.
In addition, neutralinos appear as external particles only in purely 
loop-induced processes, see \refeqs{glsqq} -- (\ref{glneg}).
Consequently, we obtained our results by using tree-level masses and
couplings in the chargino and neutralino sector.
Again, the only exception is the renormalization of $\mu$ entering 
the squark field renormalization. We follow the on-shell prescription
given in \citere{Stop2decay}, where $\de\mu$ is defined in Eq.~(181).


\section{Calculation of loop diagrams}
\label{sec:calc}

\begin{figure}[htb!]
\begin{center}
\includegraphics[width=0.90\textwidth]{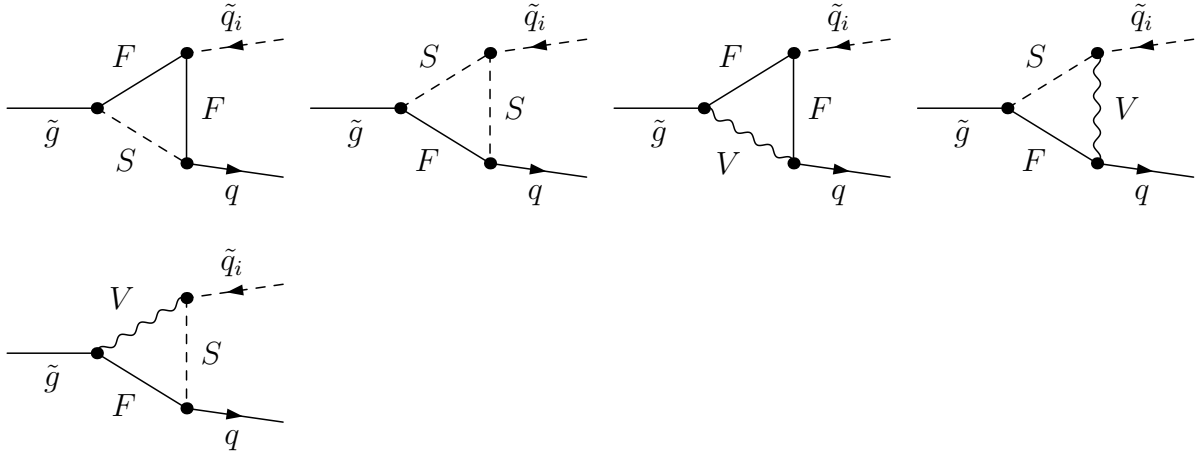}
\caption{
  Generic Feynman diagrams for the decay $\decaySqi$ ($i = 1, 2$). 
  $F$ can be a quark, chargino, neutralino or $\gl$, $S$ can be a
  squark or a Higgs boson, $V$ can be a $\ga$, $Z$, $W^\pm$ or $g$. 
  The tree-level diagram is not shown here.
}
\label{fig:glsqq}
\end{center}
\end{figure}

\begin{figure}[htb!]
\begin{center}
\includegraphics[width=0.40\textwidth]{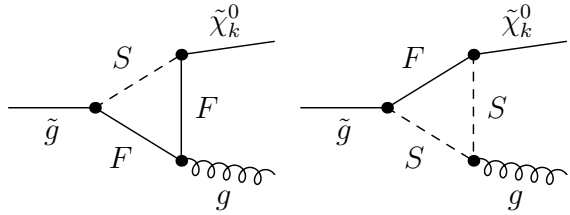}
\caption{
  Generic Feynman diagrams for the decay $\decayNkg$ ($k = 1,2,3,4$). 
  Here $F$ denotes a quark and $S$ a squark. 
  It should be noted that this process exists only at the loop-level.
}
\label{fig:glneg}
\end{center}
\end{figure}

In this section we give some details about the calculation of the
higher-order corrections to the gluino decays. Sample diagrams are
shown in \reffis{fig:glsqq}, \ref{fig:glneg}. 
Not shown are the diagrams for real (hard and soft) photon and gluon
radiation. 
They are included via analytical formulas following the description 
given in \citere{denner}.

The internal generically depicted particles in
\reffis{fig:glsqq}, \ref{fig:glneg} are labeled as follows:
$F$ can be a quark, chargino, neutralino or $\gl$, $S$ can be a 
squark or a Higgs boson, $V$ can be a $\ga$, $Z$, $W^\pm$ or $g$. 
Internally appearing Higgs bosons do not receive higher-order
corrections in their masses or couplings, which would correspond to
effects beyond one-loop.%
\footnote{
  We found that using loop corrected Higgs boson masses for the 
  internal Higgs bosons also result in finite results, where the 
  numerical difference for the various decay widths is negligible.}

In \reffi{fig:glsqq} we have furthermore omitted in general diagrams of
self-energy type 
of external (on-shell) particles. While the real part 
of such a loop
does not contribute to the decay width due to the on-shell
renormalization, the imaginary part, in product with an imaginary part
of a complex coupling (in our case coming from $\phigl$) can give a real
contribution to the decay width. 
While these diagrams are not shown explicitly, they have been taken into 
account in the analytical and numerical evaluation 
via renormalization constants, see \refse{sec:cMSSM}.
The impact of those absorptive contributions will be discussed in 
\refse{sec:full1Lphigl}.

The diagrams and corresponding amplitudes have been obtained with 
\fa~\cite{feynarts}. The model file, including the MSSM counter
terms, is based on \citere{dissTF,SbotRen,Stop2decay}, i.e.\ to match 
exactly the renormalization prescription described in those articles
(see also \citere{LHCxC} for other sectors).
The further evaluation has been performed with \fc\ 
(and \lt)~\cite{formcalc}. 
As regularization scheme for the UV-divergences we
have used constrained differential renormalization~\cite{cdr}, 
which has been shown to be equivalent to 
dimensional reduction~\cite{dred} at the \onel\ level~\cite{formcalc}. 
Thus the employed regularization preserves SUSY~\cite{dredDS,dredDS2}. 
All UV-divergences cancel in the final result.

The IR-divergences from diagrams with an internal photon or gluon have
to cancel with the ones from the corresponding real soft radiation. 
In the case of QED we have included the soft and hard photon 
contribution following the description given in \citere{denner}. 
In the case of QCD we have modified this prescription by replacing the
product of electric charges by the appropriate combination of color
charges (linear combination of $C_A$ and $C_F$ times $\als$).
The IR-divergences arising from the diagrams involving a $\ga$ or a $g$
are regularized by introducing a finite photon mass, or gluon mass, 
$\la$. While for the QED part this procedure
always works, in the QCD part due to its non-abelian character this
method can fail. However, since no triple or quartic gluon vertices
appear, $\la$ can indeed be used as a regulator. 
All IR-divergences, i.e.\ all divergences in the limit
$\la \to 0$, cancel once virtual and real radiation for one 
decay channel are added.%
\footnote{
  Using tree-level, as well as one-loop down-type squark masses 
  (see \refse{sec:squark}), yields a cancellation of IR divergences 
  to all orders for all $\gl$ decays.}

\medskip

For completeness we show here also the formulas for the 
tree-level decay widths:
\begin{align}
\Gamma^{\rm tree}(\decaySqi) &= \Big[ \KL |C(\gl, \tilde{q}_i, q)_L|^2 
       + |C(\gl, \tilde{q}_i, q)_R|^2 \KR (\mgl^2 - \msqi^2 + \mq^2) \non \\
&\qquad + 4\, \re \{C(\gl, \tilde{q}_i, q)_L^*\, C(\gl, \tilde{q}_i, q)_R \}\,
       \mgl\, \mq \Big]\,
\frac{\lambda^{1/2}(\mgl^2, \msqi^2,\mq^2)}{64\, \pi\, \mgl^3}~,
\end{align}
where $\lambda(x,y,z) = (x - y - z)^2 - 4yz$ and the couplings 
$C(a, b, c)$ can be found in the \fa~model files~\cite{feynarts-mf}.
$C(a, b, c)_{L,R}$ denote the part of the coupling which
is proportional to $\edz(\id \mp \ga_5)$.


\section{Numerical analysis}
\label{sec:numeval}

In this section we present a numerical analysis of all 28 decay
channels (``xy''). In the various figures we show the decay width and its
relative correction at the tree-level (``tree'') and at the one-loop
level (``full''), 
\begin{align}
\Ga^{\rm tree} \equiv \Ga^{\rm tree}(\decayxy)~, \quad
\Ga^{\rm full} \equiv \Ga^{\rm full}(\decayxy)~, \quad
\de\Ga/\Ga^{\rm tree} \equiv \frac{\Ga^{\rm full} - \Ga^{\rm tree}}
                                 {\Ga^{\rm tree}}~.
\end{align}
The total decay width is defined as the sum of the 24 decay widths%
\footnote{
  We neglected the $\decayNkg$ decays in the total decay width due to their 
  negligible contribution, see \refse{sec:glneug} and \refse{sec:tot}.}~,
\begin{align}\label{tot}
\Ga_{\rm tot}^{\rm tree} \equiv \sum_{{\rm xy}} \Ga^{\rm tree}(\decayxy)~, \quad
\Ga_{\rm tot}^{\rm full} \equiv \sum_{{\rm xy}} \Ga^{\rm full}(\decayxy)~, \quad
\de\Ga_{\rm tot}/\Ga_{\rm tot}^{\rm tree} \equiv 
   \frac{\Ga_{\rm tot}^{\rm full} - \Ga_{\rm tot}^{\rm tree}}
        {\Ga_{\rm tot}^{\rm tree}}~.
\end{align}
We also show the absolute and relative changes of the branching ratios,
\begin{align}
\br^{\rm tree} \equiv \frac{\Ga^{\rm tree}(\decayxy)}
                          {\Ga_{\rm tot}^{\rm tree}}~, \quad
\br^{\rm full} \equiv \frac{\Ga^{\rm full}(\decayxy)}
                          {\Ga_{\rm tot}^{\rm full}}~, \quad
\de\br/\br    \equiv \frac{\br^{\rm full} - \br^{\rm tree}}
                          {\br^{\rm full}}~. 
\label{brrel}
\end{align}
The last quantity is crucial to analyze the impact of the one-loop
corrections on the phenomenology at the LHC.


\subsection{Parameter settings}
\label{input}

The renormalization scale, $\mu_R$, has been set to the mass of the
decaying particle, i.e.\ $\mu_R = \mgl$.
The SM parameters are chosen as follows, see also \cite{pdg}:
\begin{itemize}
\item Fermion masses\index{leptonmasses}:
\begin{align}
m_e    &= 0.51099891\mev~, & m_{\nu_e}     &= 0\mev~, \non \\
m_\mu  &= 105.658367\mev~, & m_{\nu_{\mu}}  &= 0\mev~, \non \\
m_\tau &= 1776.82\mev~,    & m_{\nu_{\tau}} &= 0\mev~, \non \\
m_u &= 53.8\mev~,          & m_d &= 53.8\mev~, \non \\ 
m_c &= 1.27\gev~,          & m_s &= 101\mev~, \non \\
m_t &= 172.0\gev~,         & \mb^{\MSbar}(\mb) &= 4.25\gev~.
\end{align}
$m_u$ and $m_d$ are effective parameters, calculated through the hadronic
contributions to:
\begin{align}
\Delta\alpha_{\text{had}}^{(5)}(\MZ) = 
      \frac{\alpha}{\pi}\sum_{f = u,c,d,s,b}
      Q_f^2 \Bigl(\ln\frac{\MZ^2}{m_f^2} - \frac 53\Bigr)~.
\end{align}

\item The CKM matrix has been set to unity.

\item Gauge boson masses:
\begin{align}  
  \MZ = 91.1876 \gev~, \quad \MW =  80.399 \gev~,
\end{align}  

\item Coupling constants:
\begin{align}  
\al = 1/137.035999679~, \qquad \als^{\MSbar}(\MZ) = 0.1184~.
\end{align}  
\end{itemize}
The Higgs sector quantities (masses, etc.\ at the tree-level) 
have been evaluated using the \fc\ implementation~\cite{formcalc}.

We will show the results for one representative numerical example. 
The parameters are chosen according to the scenario \SE\, shown in 
\refta{tab:para}, but with one of the parameters varied.
The scenarios are defined such that {\em all} decay modes are open
simultaneously to permit an analysis of all channels, i.e.\ not picking
specific parameters for each decay.
We will start with a variation of $\mgl$, and show later the
results for varying $\phigl$.
The scenarios are in agreement with the 
MSSM Higgs boson searches at LEP~\cite{LEPHiggsSM,LEPHiggsMSSM}. 
The choice of relatively large gluino and squark masses also avoids all
LHC bounds~\cite{LHCsusy}.
The slepton/lepton sector of the cMSSM does not enter into our
calculation, hence the parameters are not specified.
Furthermore, also the following exclusion limits \cite{pdg} hold in our
scenario:
\begin{align}
\mneu{1} &> 46 \gev, \;
\mneu{2} > 62 \gev, \;
\mneu{3} > 100 \gev, \;
\mneu{4} > 116 \gev, \;
\mcha{1} > 94 \gev~.
\end{align}

The scalar quark masses of the parameter set \SE\ are shown in 
\refta{tab:squark}. The values of $\mgl$ allow copious production of 
the gluino at the LHC, once $\sqrt{s} = 14 \tev$ is reached. 

The numerical results we will show in the next subsections are of
course dependent on choice of the SUSY parameters. Nevertheless, they
give an idea of the relevance of the full one-loop corrections.
As an example, the various decay widths $\Ga(\decaySqi)$ are
all of similar size, contributing similarly to $\Ga_{\rm tot}$. The same
holds for the various branching ratios. 
If the (artificial) degeneracy of the soft SUSY-breaking parameters in
the squark sector were lifted, these results could change
significantly. However, the size of the loop corrections shown in our
numerical example still gives a good indication about the size of the
expected corrections.
In the special case of $\mgl < \msqi$ (for all $\sq$ and $i = 1,2$) the
loop-induced decays $\decayNkg$ ($k = 1,2,3,4$) 
as well as three-body decays (which are not investigated here) 
could become dominant.

\begin{table}[tb!]
\renewcommand{\arraystretch}{1.5}
\BC
\begin{tabular}{|c||c|c|c|c|c|c|c|c|c|c|c|}
\hline
Scen.\ & $\tb$ & $\MHp$ & $M_{\tilde q_L}$ & $M_{\tilde{q}_R}$ & $\mu$ & 
$\At$ & $\Ab$ & $M_1$ & $M_2$ & $M_3$ 
\\ \hline\hline
\SE & 20 & 200 & 700 & 800 & 200 & 1000 & 800 & 200 & 300 & 1200 
\\ \hline
\end{tabular}
\caption{MSSM parameters for the initial numerical investigation; 
  all masses are in GeV. 
  For the $\sq$~sector the shifts in $M_{\tilde{q}_{L,R}}$ as defined in 
  Eqs.\,(49) and~(50) in \citere{Stop2decay} 
  are taken into account.
  The values for $\At\, (= A_c = A_u)$ and $\Ab\, (= A_s = A_d)$ are 
  chosen such that charge- or color-breaking minima are avoided~\cite{ccb}.
}
\label{tab:para}
\EC
\renewcommand{\arraystretch}{1.0}
\end{table}

\begin{table}[tb!]
\renewcommand{\arraystretch}{1.5}
\BC
\begin{tabular}{|c||c|c|c|c|c|c|} 
\hline
$\sq$-type &   $\tilde u$   &   $\tilde c$   &   $\tilde t$   
           &   $\tilde d$   &   $\tilde s$   &   $\tilde b$   \\
\hline\hline
$\msqe$  & 697.919 & 697.913 & 637.268 & 702.402 & 702.402 & 705.879 \\
\hline
$\msqz$  & 799.232 & 799.240 & 882.565 & 800.384 & 800.384 & 800.848 \\
\hline
\end{tabular}
\caption{The scalar quark masses in \SE\ for the numerical 
  investigation; all masses are in GeV and rounded to one MeV.
}
\label{tab:squark}
\EC
\renewcommand{\arraystretch}{1.0}
\end{table}


\subsection{Full one-loop results for varying \boldmath{$\mgl$}}
\label{sec:full1L}

The results shown in this and the following subsections consist of 
``tree'', which denotes the tree-level 
value and of ``full'', which is the decay width including {\em all} one-loop 
corrections as described in \refse{sec:calc}.
Additionally in this section we also investigate the accuracy of
the pure SUSY QCD corrections for the decay widths as evaluated for real
parameters in \citere{glsqq_als}.%
\footnote{
  We have checked that we are in good agreement with \citere{glsqq_als} 
  using their input parameters, where a small difference remains due to 
  the different renormalization schemes.}
The corresponding curves are labelled ``SQCD''.
We start the numerical analysis with $\gl$~decay widths evaluated as
a function of $\mgl$ starting at $\mgl = 700 \gev$ up to 
$\mgl = 2 \tev$, 
which roughly coincides with the reach of the LHC for high-luminosity running.
The upper panels contain the results for the absolute
value of the various decay widths, $\Ga(\decayxy)$ (left) and
the relative correction from the full one-loop contributions
(right), where we compare ``tree'', ``full'' and ``SQCD''. 
The lower panels show the same results (but leaving out ``SQCD'') 
for $\br(\decayxy)$.

Since all parameters are chosen real no difference between the two
decay modes $\decayaSqi$ and $\decaySqai$ arises, neither in the decay
widths, nor in the branching ratios (where of course both channels are
taken into account in the total decay width).
Consequently, we only show results for the channel ``$\decaySqi$'', 
and the results for $\br(\decayaSqi + \decaySqai)$ are
simply the $\br(\decaySqi)$ multiplied by two.
In \refse{sec:full1Lphigl} both 
channels will be shown separately. Furthermore, due to the absense 
of complex parameters no contributions from absorptive parts of 
self-energy type corrections on external legs can contribute. 
Again, this will be different in \refse{sec:full1Lphigl}.

In our numerical scenario \SE\ we have chosen a small splitting between 
the left- and the right-handed soft SUSY-breaking parameter within 
all squark flavors, see \refta{tab:para}. 
If the effects of mass differences of the quark and squarks in the final
state were neglected one would expect 
$\Sigma_{i=1,2}\, \br(\decaySqi) \approx 1/12$
(or $\Sigma_{i=1,2}\, \br(\decayaSqi + \decaySqai) \approx 1/6$
if both combinations are taken into account). Deviations from this
number are expected due to kinematical effects, and differences in the
loop corrections (including possible effects related to the scalar quark
mixing). For large values of $\mgl$ where mass effects should be small
it is expected to reach the value of $\sim 1/12$.

\begin{figure}[htb!]
\begin{center}
\begin{tabular}{c}
\includegraphics[width=0.49\textwidth,height=8.0cm]{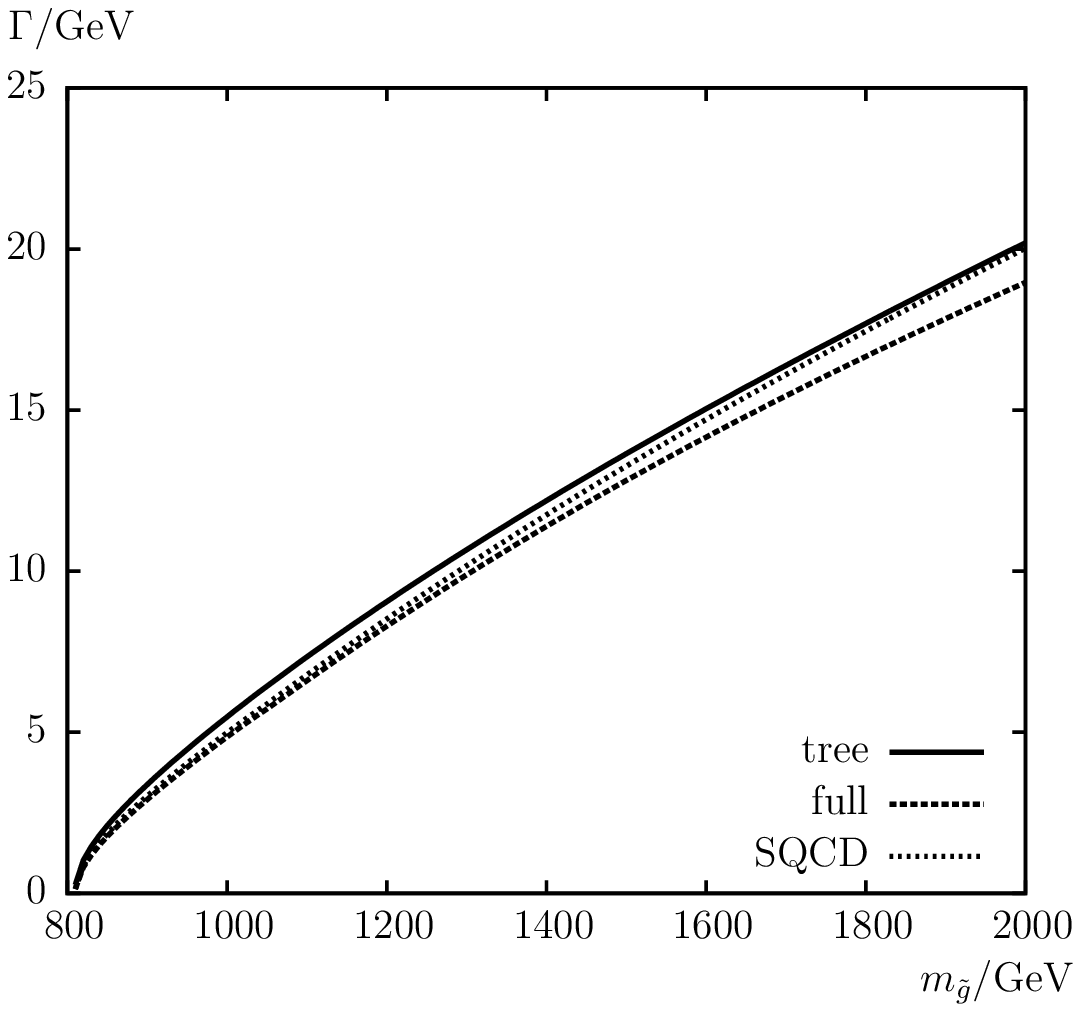}
\hspace{-4mm}
\includegraphics[width=0.49\textwidth,height=8.0cm]{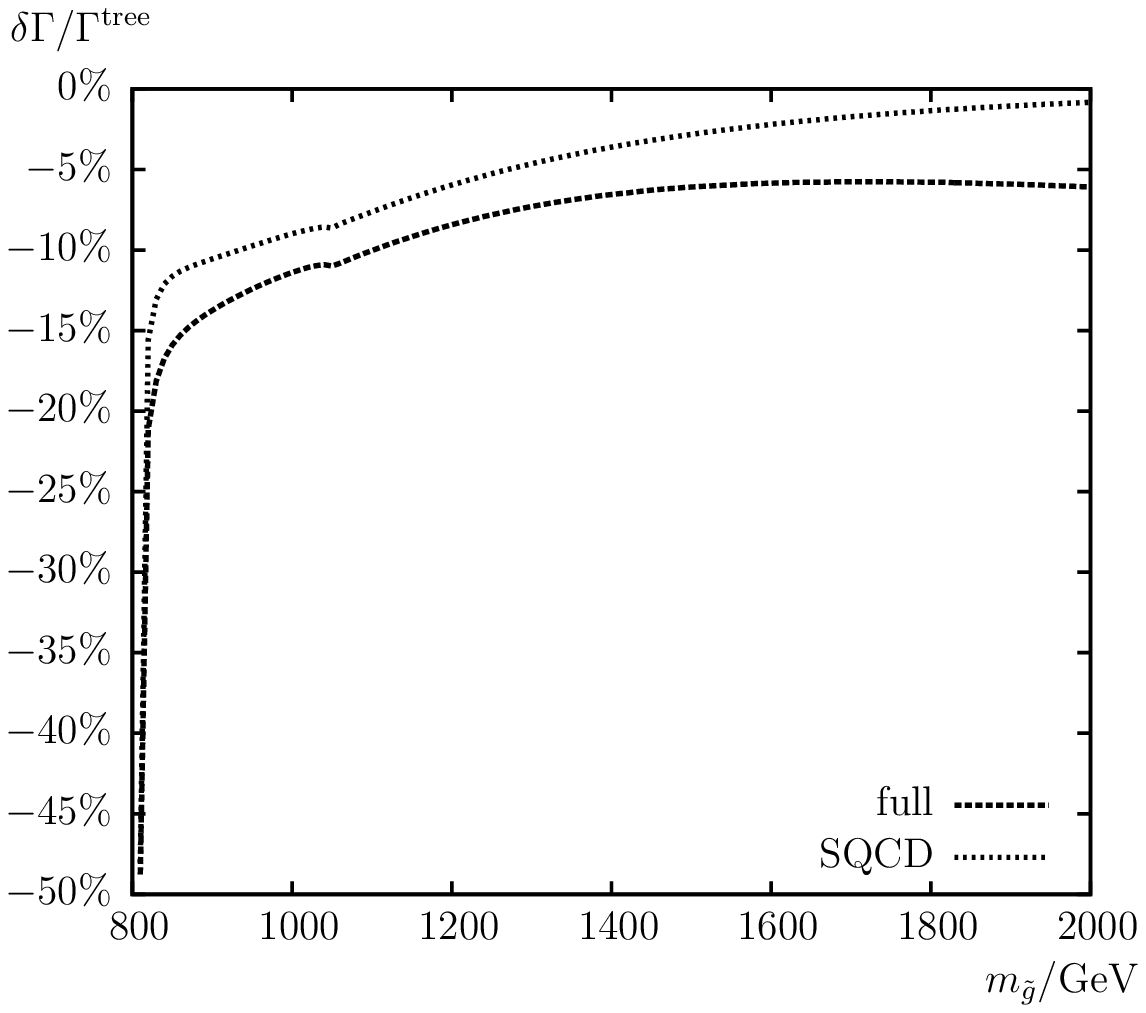}
\\[4em]
\includegraphics[width=0.49\textwidth,height=8.0cm]{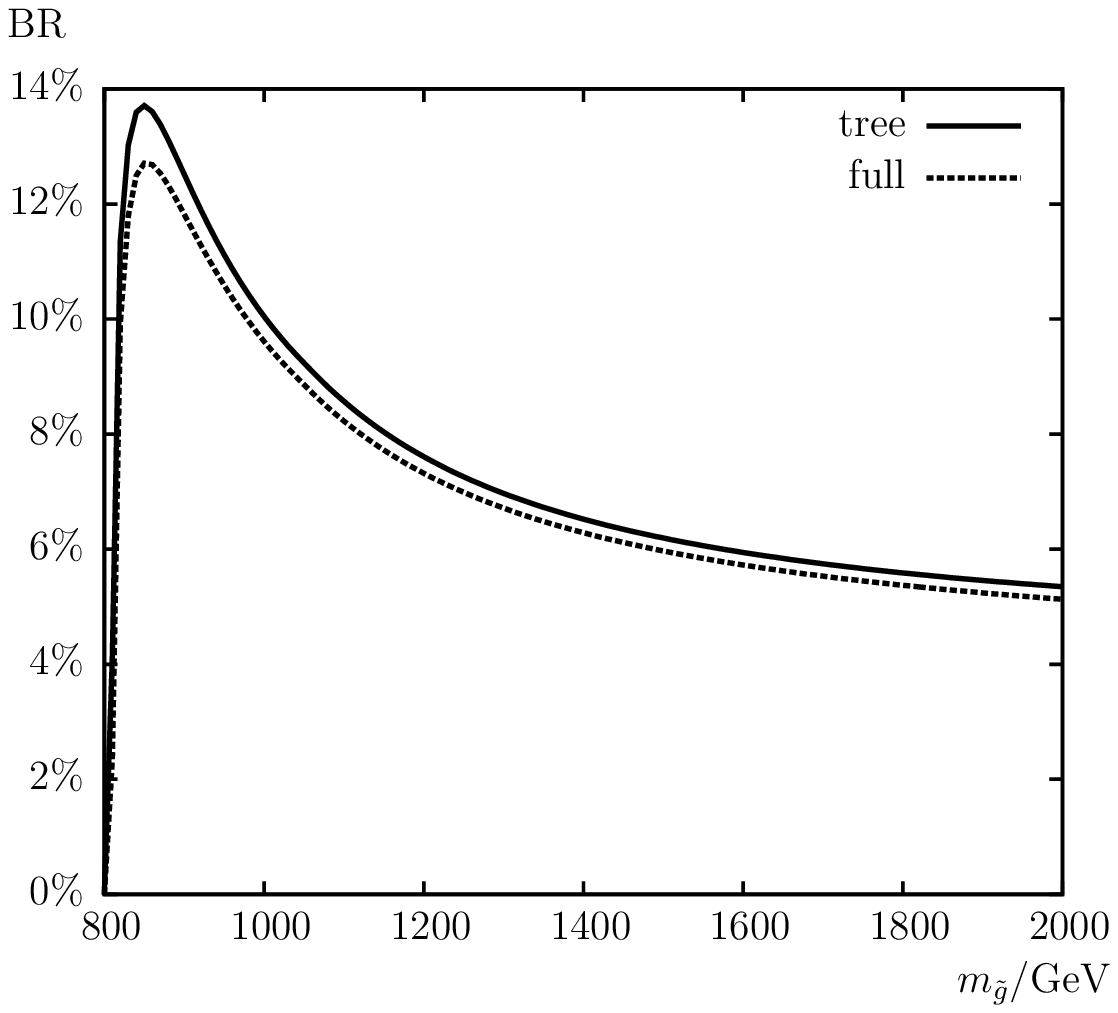}
\hspace{-4mm}
\includegraphics[width=0.49\textwidth,height=8.0cm]{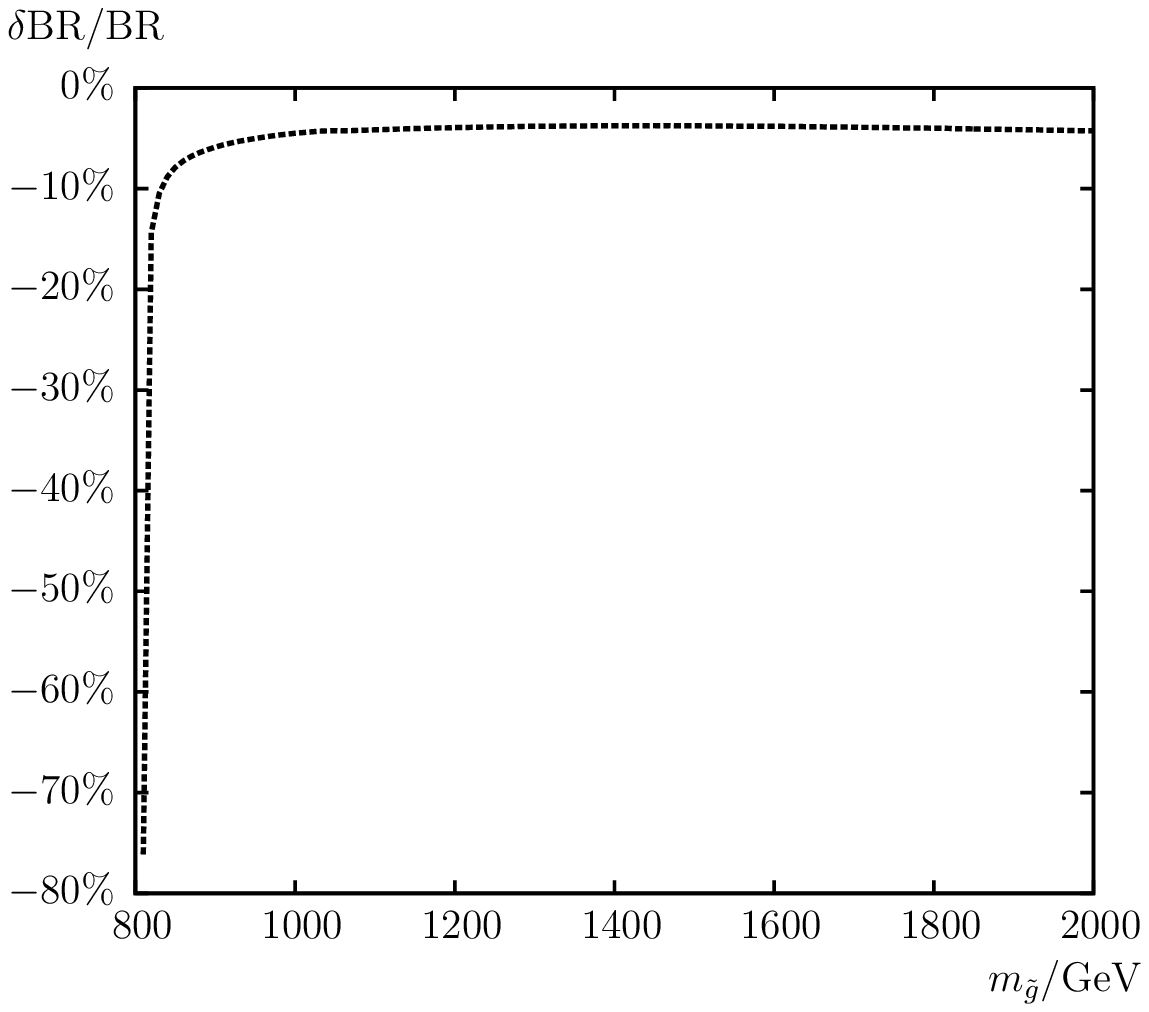}
\end{tabular}
\vspace{2em}
\caption{
  $\Ga(\decaySte)$. Tree-level (``tree'') and full one-loop 
  (``full'') corrected decay widths are shown with the parameters 
  chosen according to \SE\ (see \refta{tab:para}), with $\mgl$ varied.
  The upper left plot shows the decay width, the upper right plot shows 
  the relative size of the corrections.
  Also shown are the pure SQCD corrections (``SQCD'').
  The lower left plot shows the BR, the lower right plot shows 
  the relative size of the BR.
}
\label{fig:AbsM3.glst1t}
\end{center}
\end{figure}

We start with the decay $\decaySte$. The results for this channel are
shown in \reffi{fig:AbsM3.glst1t} as a function of $\mgl$. 
The tiny dip in \SE\ at $\mgl \approx 1054.6 \gev$ is the threshold
$\mgl = \mstz + \mt$ of the self energy $\Si_{\gl}(\mgl^2)$ in the 
renormalization constants $\dZ{\gl}$ and $\de M_3$.
One can see that the size of the corrections of the decay width
is especially large very close to the production threshold.%
\footnote{
  It should be noted that a calculation very close to threshold requires 
  the inclusion of additional (non-relativistic) contributions, which is 
  beyond the scope of this paper. Consequently, very close to threshold 
  our calculation (at the tree- or loop-level) does not provide a very 
  accurate description of the decay width.}
Away from the production threshold relative corrections of 
$\sim -6\%$ are found. The difference between the full and the SQCD
corrections is roughly $5\%$; at low masses the EW corrections are
about one third of the full corrections, while at large $\mgl$ they are
dominating over the SQCD corrections.
The branching ratio in our numerical scenario \SE\ (where we have scalar
quarks very close in mass) can reach more than
$12\%$ close to threshold and stays above $5\%$ for large $\mgl$. 
The correction to the branching ratio is then found to be $\sim -4\%$.

\begin{figure}[htb!]
\begin{center}
\begin{tabular}{c}
\includegraphics[width=0.49\textwidth,height=8.0cm]{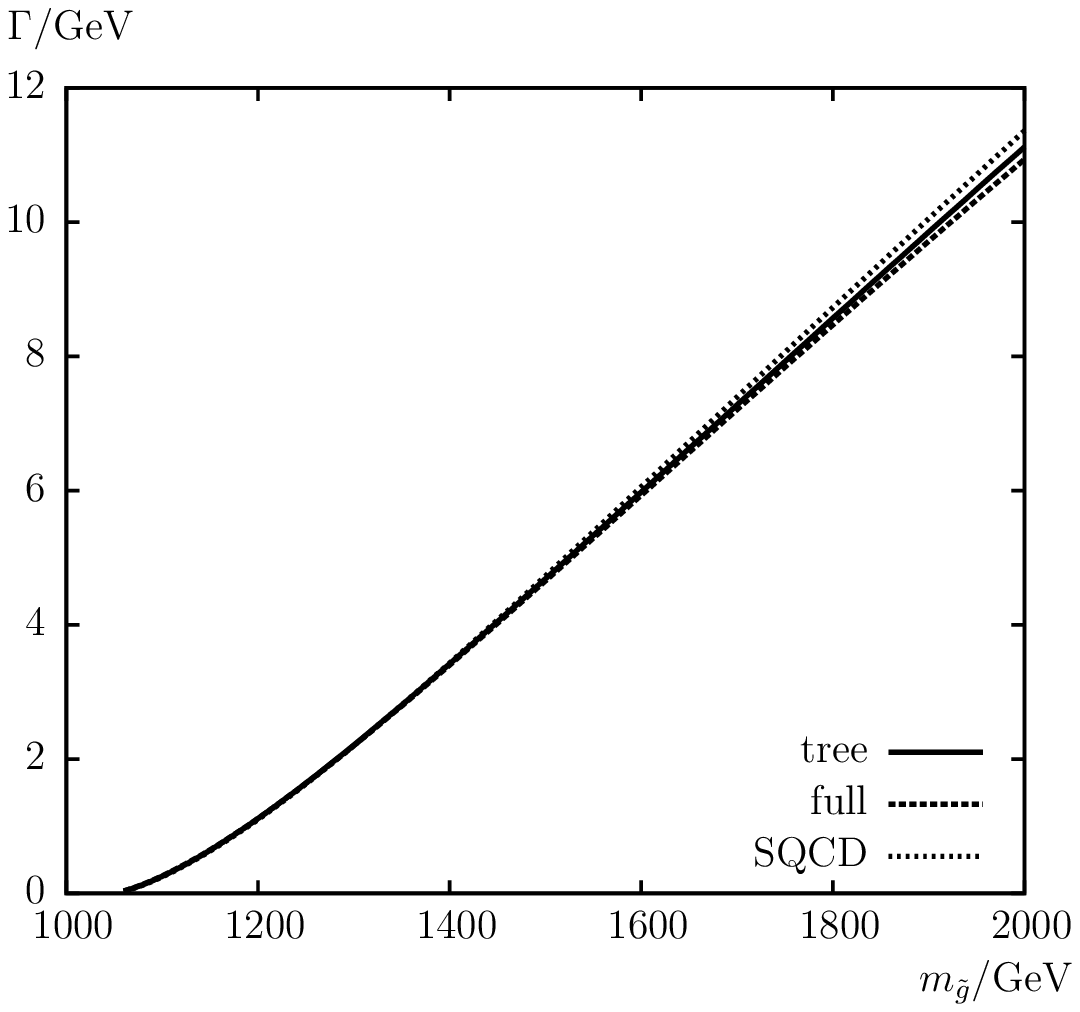}
\hspace{-4mm}
\includegraphics[width=0.49\textwidth,height=8.0cm]{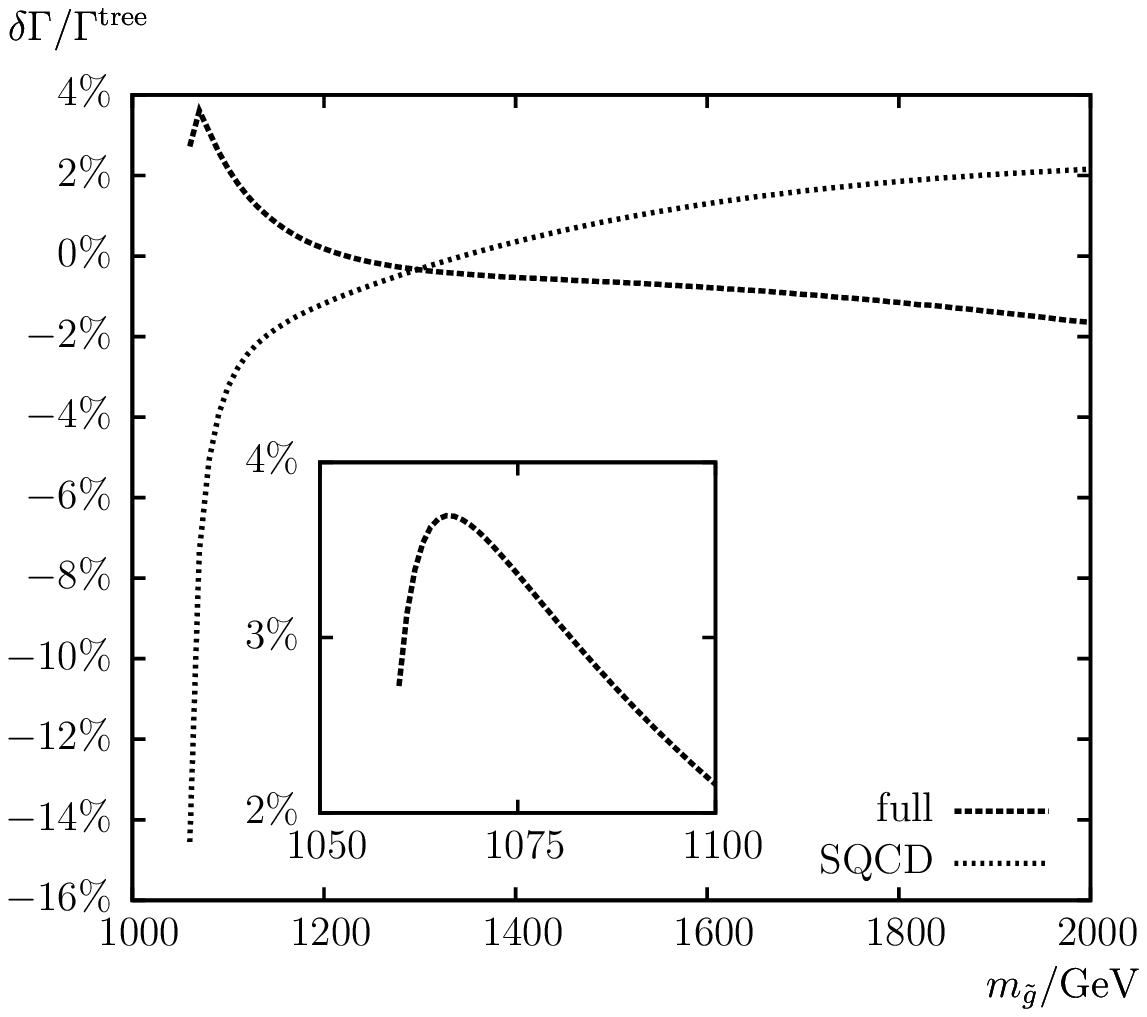}
\\[4em]
\includegraphics[width=0.49\textwidth,height=8.0cm]{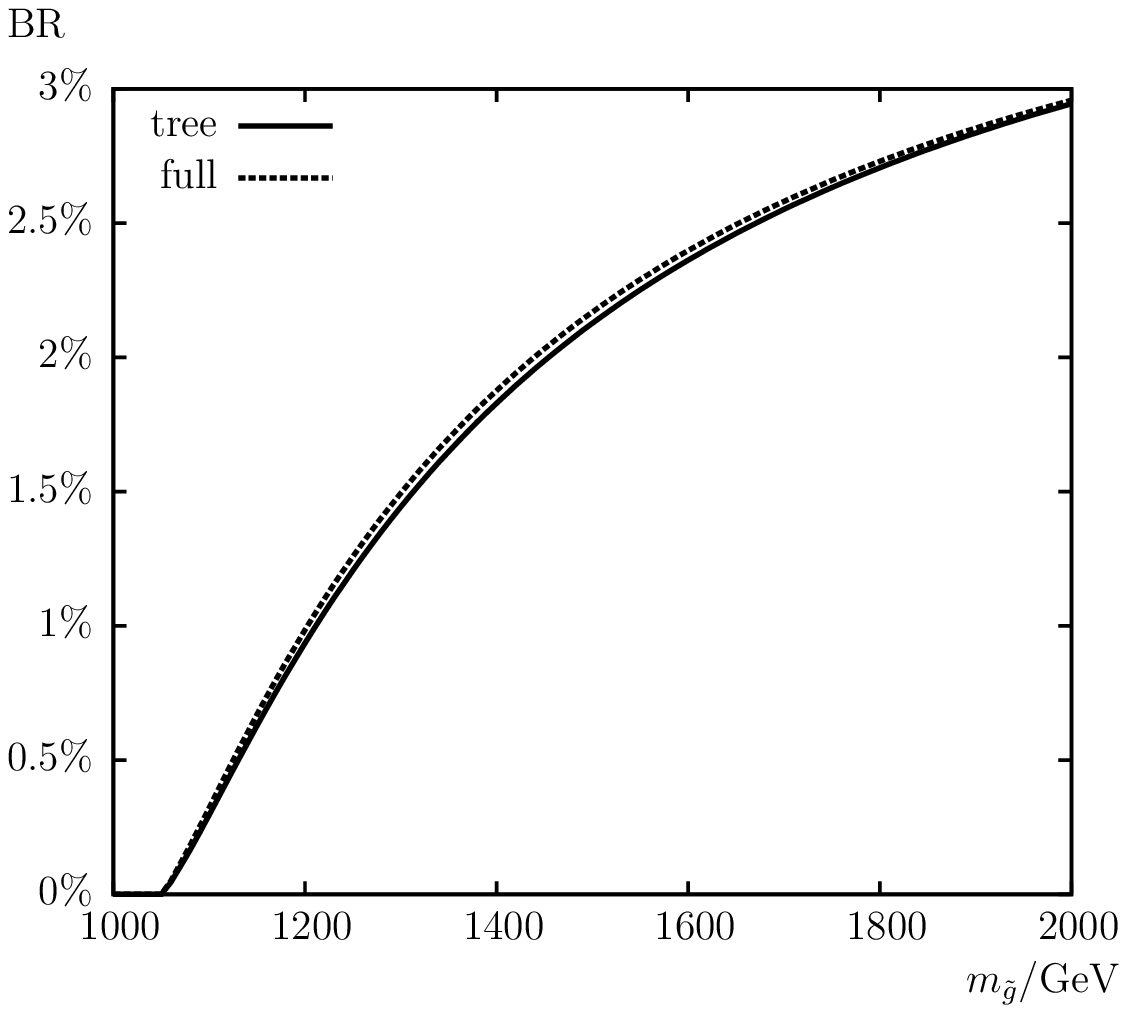}
\hspace{-4mm}
\includegraphics[width=0.49\textwidth,height=8.0cm]{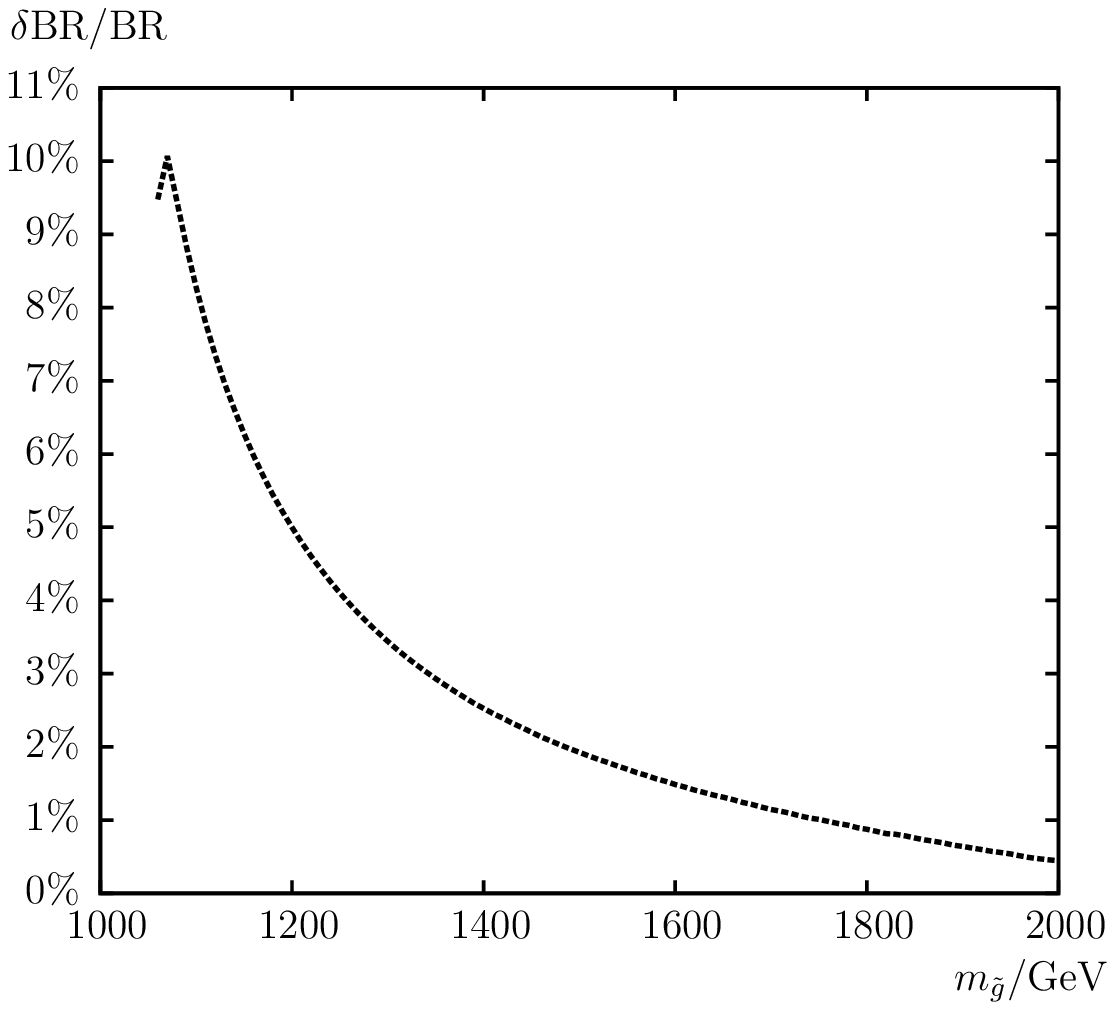}
\end{tabular}
\vspace{2em}
\caption{
  $\Ga(\decayStz)$. Tree-level (``tree'') and full one-loop 
  (``full'') corrected decay widths are shown with the parameters 
  chosen according to \SE\ (see \refta{tab:para}), with $\mgl$ varied.
  The upper left plot shows the decay width, the upper right plot shows 
  the relative size of the corrections.
  Also shown are the pure SQCD corrections (``SQCD'').
  The lower left plot shows the BR, the lower right plot shows 
  the relative size of the BR.
}
\label{fig:AbsM3.glst2t}
\end{center}
\end{figure}

Next, in \reffi{fig:AbsM3.glst2t} we show the decay $\Ga(\decayStz)$.
At low $\mgl$ the decay is kinematically forbidden and reaches $\sim 11 \gev$ 
at $\mgl = 2 \tev$, with small positive corrections to the decay width
at low $\mgl$ and very small negative corrections at large $\mgl$.
The behavior of the SQCD corrections alone is quite different for
this decay. They are large and negative, 
$\sim -14\%$ close to threshold and turn positive up to $+2\%$ for large
$\mgl$. The SQCD contributions alone constitute a very weak
approximation to the full result in this channel.
The BR reaches nearly $3\%$ at large $\mgl$, so that 
$\br(\decaySte) + \br(\decayStz) \approx 8.5\% \sim 1/12$ is reached.
The relative corrections reach from $10\%$ where the width is small
going down below $1\%$ where the width and BR are largest.

\begin{figure}[htb!]
\begin{center}
\begin{tabular}{c}
\includegraphics[width=0.49\textwidth,height=8.0cm]{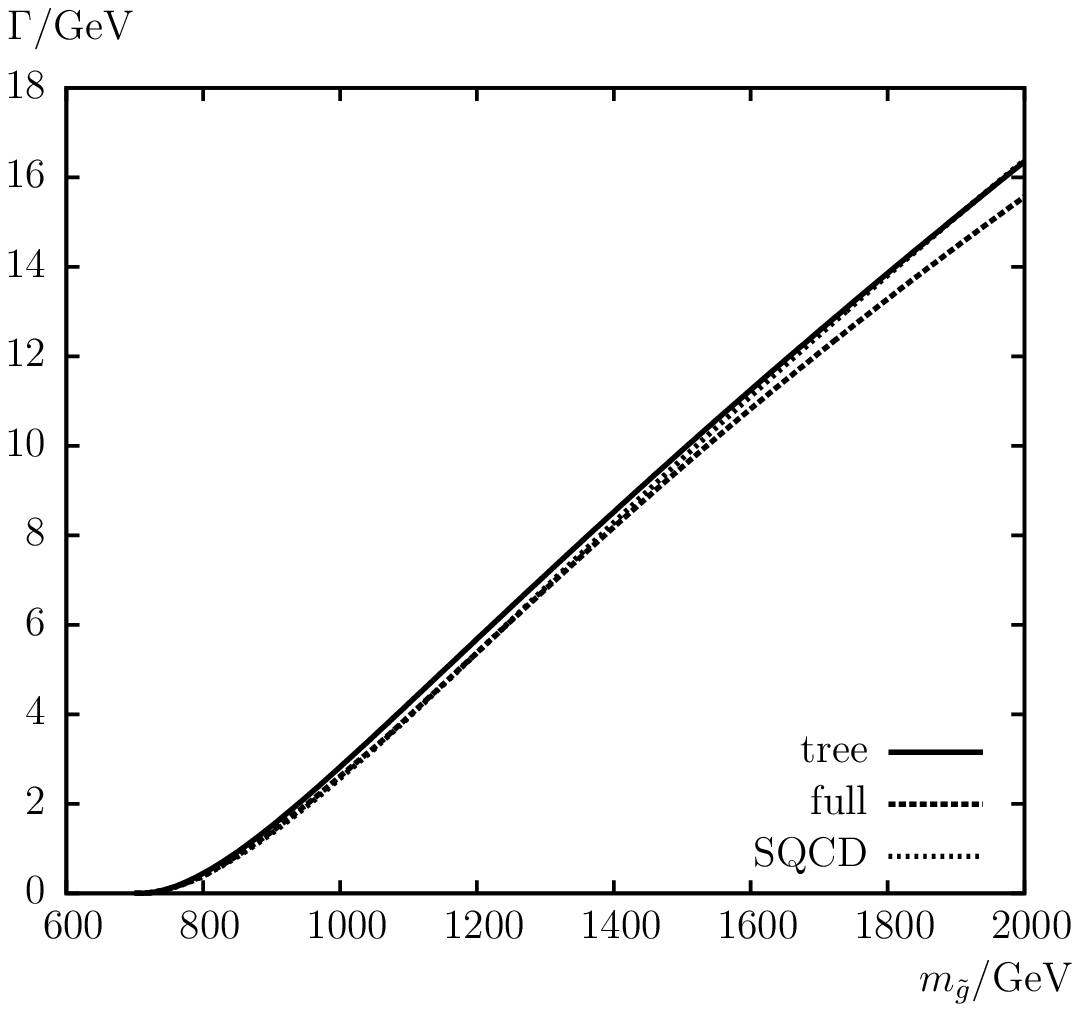}
\hspace{-4mm}
\includegraphics[width=0.49\textwidth,height=8.0cm]{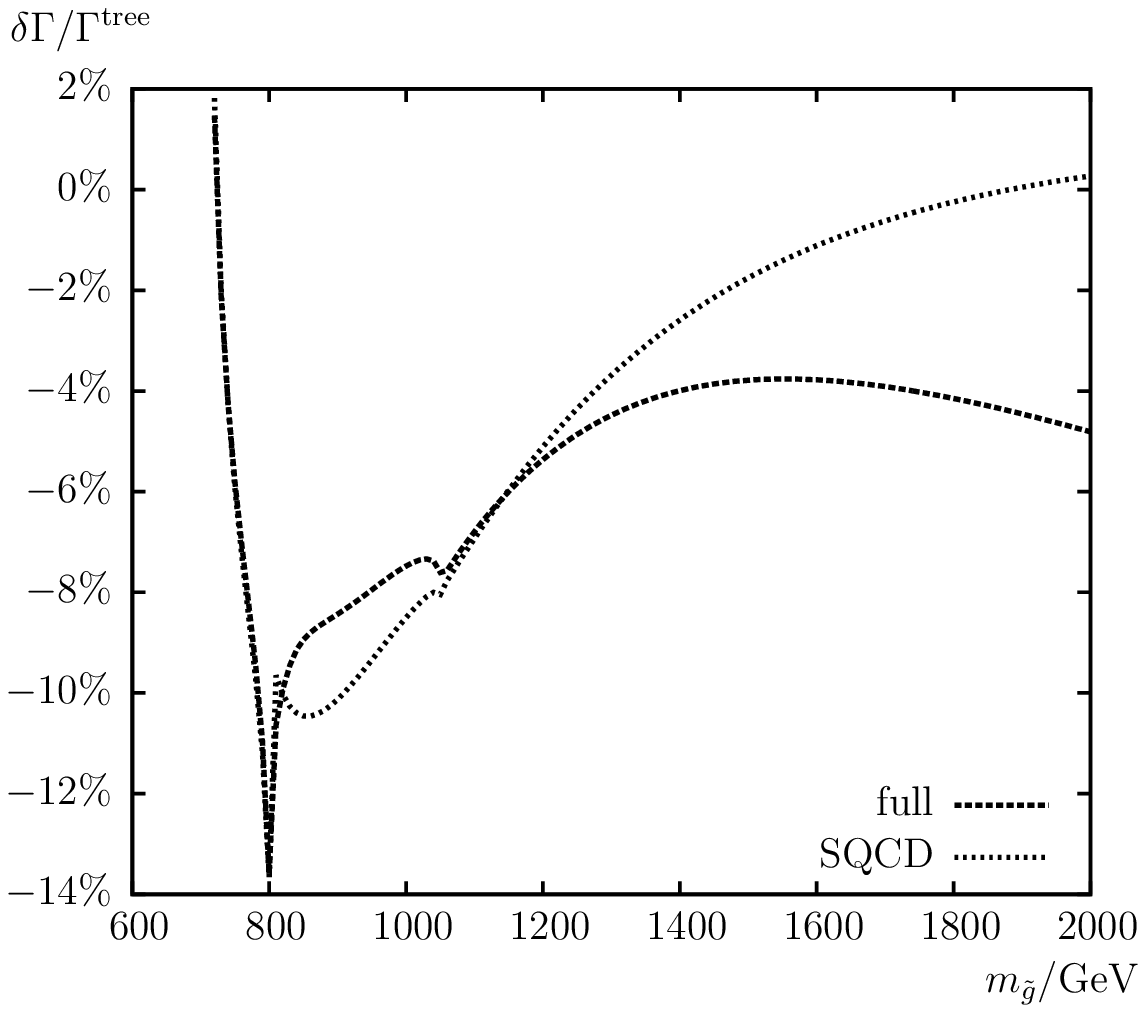}
\\[4em]
\includegraphics[width=0.49\textwidth,height=8.0cm]{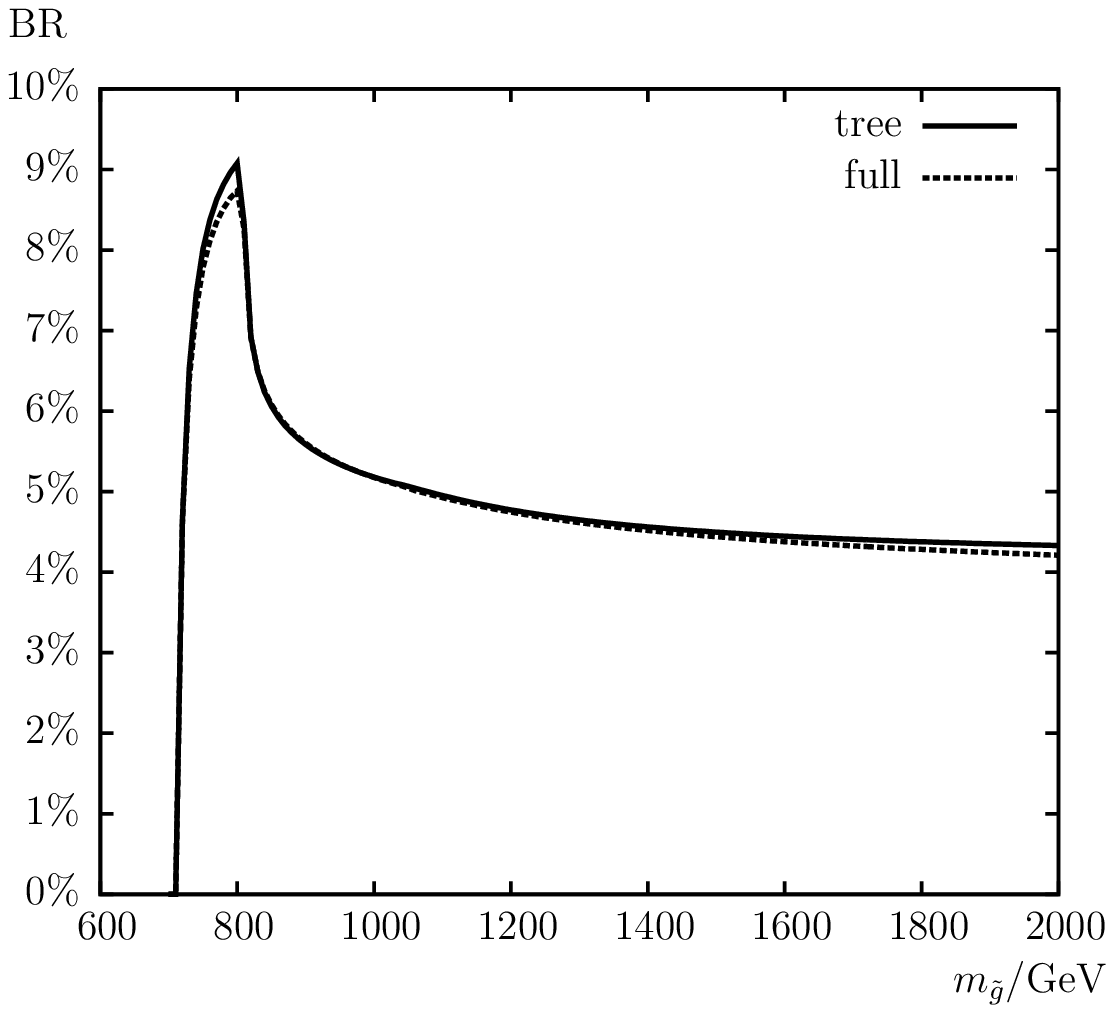}
\hspace{-4mm}
\includegraphics[width=0.49\textwidth,height=8.0cm]{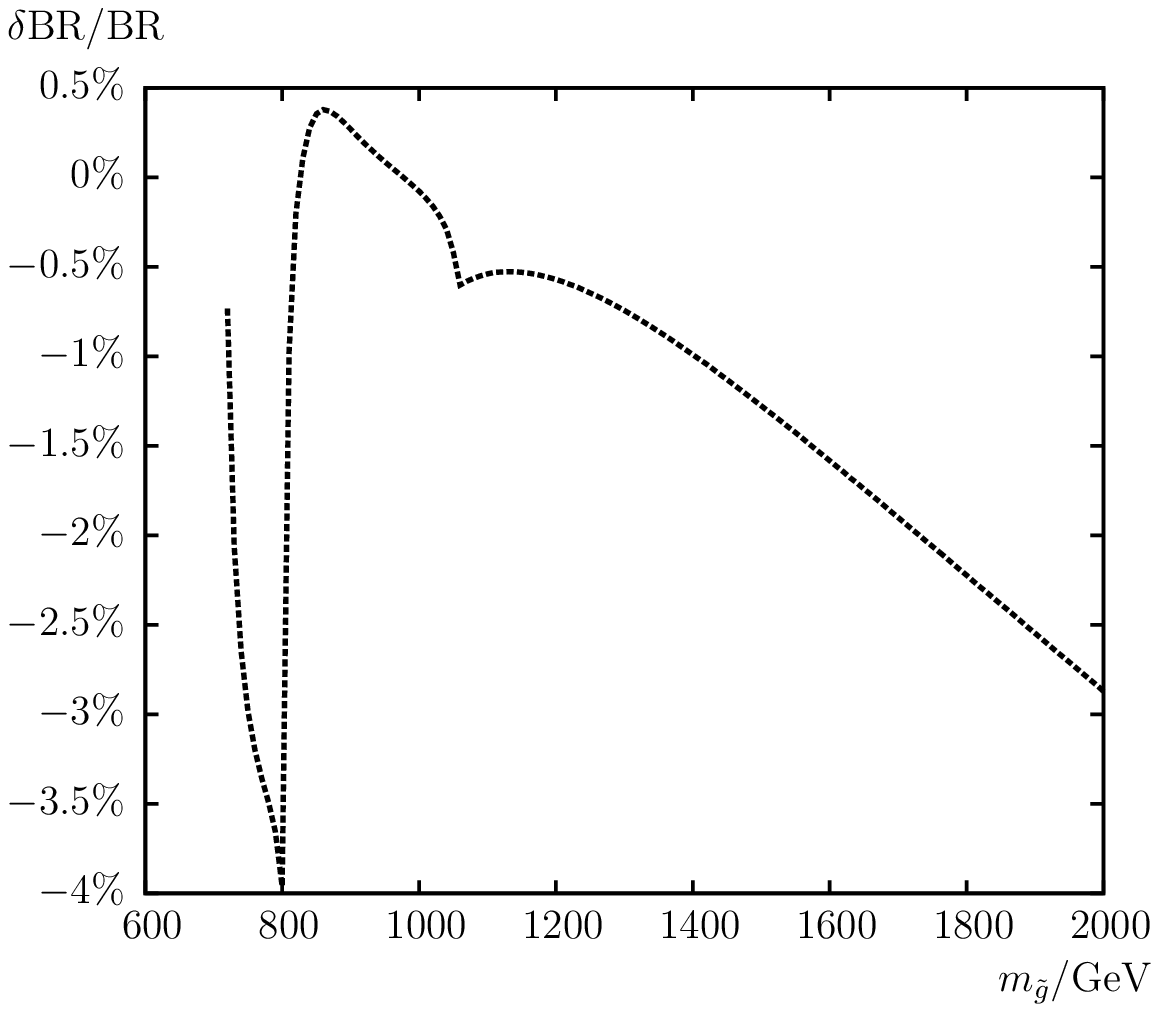}
\end{tabular}
\vspace{2em}
\caption{
  $\Ga(\decaySbe)$. Tree-level (``tree'') and full one-loop 
  (``full'') corrected decay widths are shown with the parameters 
  chosen according to \SE\ (see \refta{tab:para}), with $\mgl$ varied.
  The upper left plot shows the decay width, the upper right plot shows 
  the relative size of the corrections.
  Also shown are the pure SQCD corrections (``SQCD'').
  The lower left plot shows the BR, the lower right plot shows 
  the relative size of the BR.
}
\label{fig:AbsM3.glsb1b}
\end{center}
\end{figure}

Now we turn to the decays to $\Sbot/b$. 
The results for the decay $\decaySbe$ are presented in
\reffi{fig:AbsM3.glsb1b}.
Within \SE\ the first dip at $\mgl \approx 800.51 \gev$ stems from 
the threshold $\mgl = \mscz + m_c$ in the self energy 
$\Si_{\gl}(\mgl^2)$ entering the renormalization constants 
$\dZ{\gl}$ and $\de M_3$.
The second dip at $\mgl \approx 1054.6 \gev$ comes from the
threshold $\mgl = \mstz + \mt$.
As expected the decay width rises from zero at threshold to
$\sim 15.5 \gev$ at $\mgl = 2 \tev$, including the full one-loop
corrections. The first threshold leads to very large corrections of
nearly $-14\%$, which then decreases to $\sim -5\%$ at larger $\mgl$
values. The SQCD corrections approximate the full result quite well
up to $\mgl = 1200 \gev$, but then tend to zero. Again for large $\mgl$
the full corrections are dominated by the EW contributions.
The first threshold is also clearly visible in the BR, where
values around $9\%$ can be reached, going down to $\sim 4\%$ at 
$\mgl = 2 \tev$. The relative corrections to the BR vary between
$-4\%$ and $+0.5\%$ around the first threshold and go to $-3\%$ 
at high $\mgl$ values.

\begin{figure}[htb!]
\begin{center}
\begin{tabular}{c}
\includegraphics[width=0.49\textwidth,height=8.0cm]{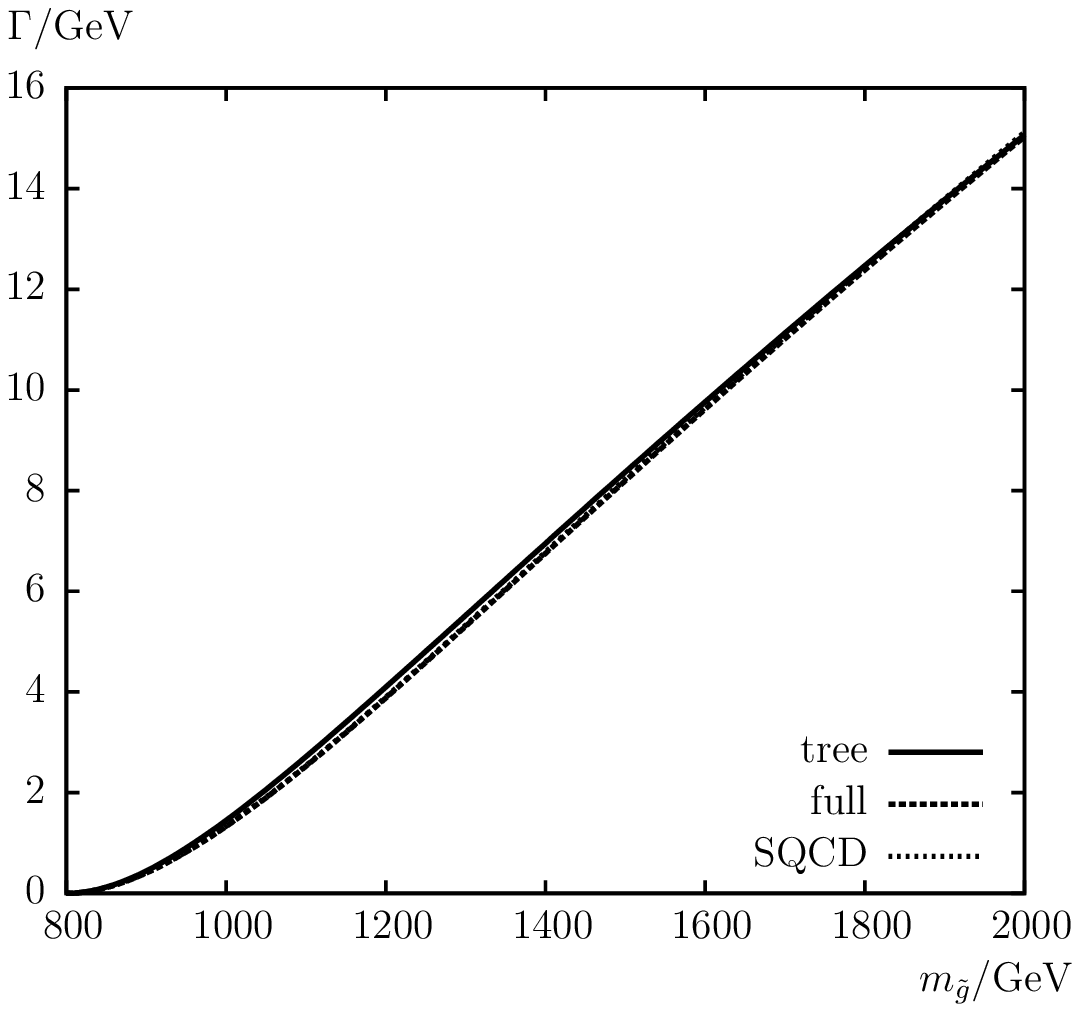}
\hspace{-4mm}
\includegraphics[width=0.49\textwidth,height=8.0cm]{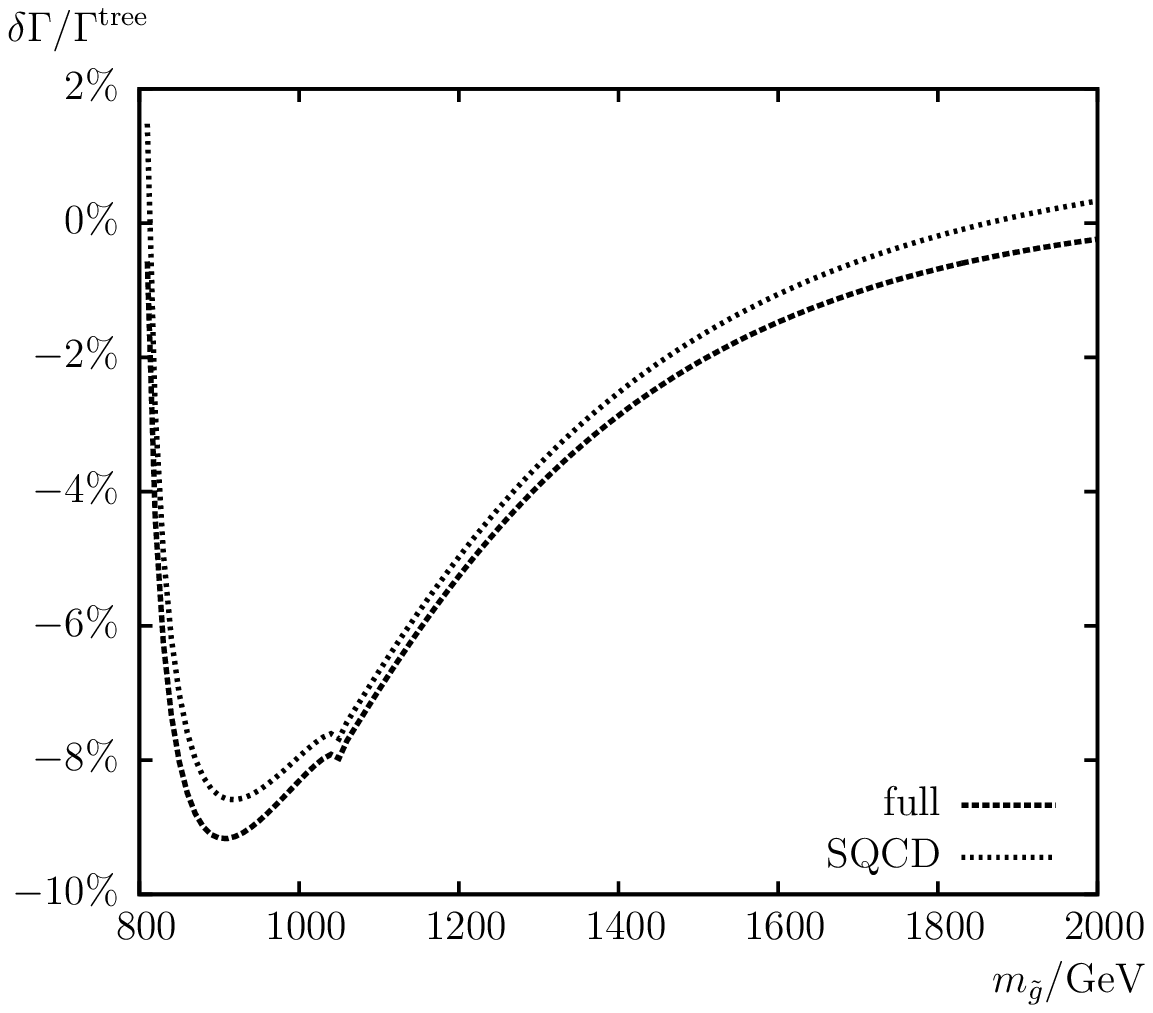}
\\[4em]
\includegraphics[width=0.49\textwidth,height=8.0cm]{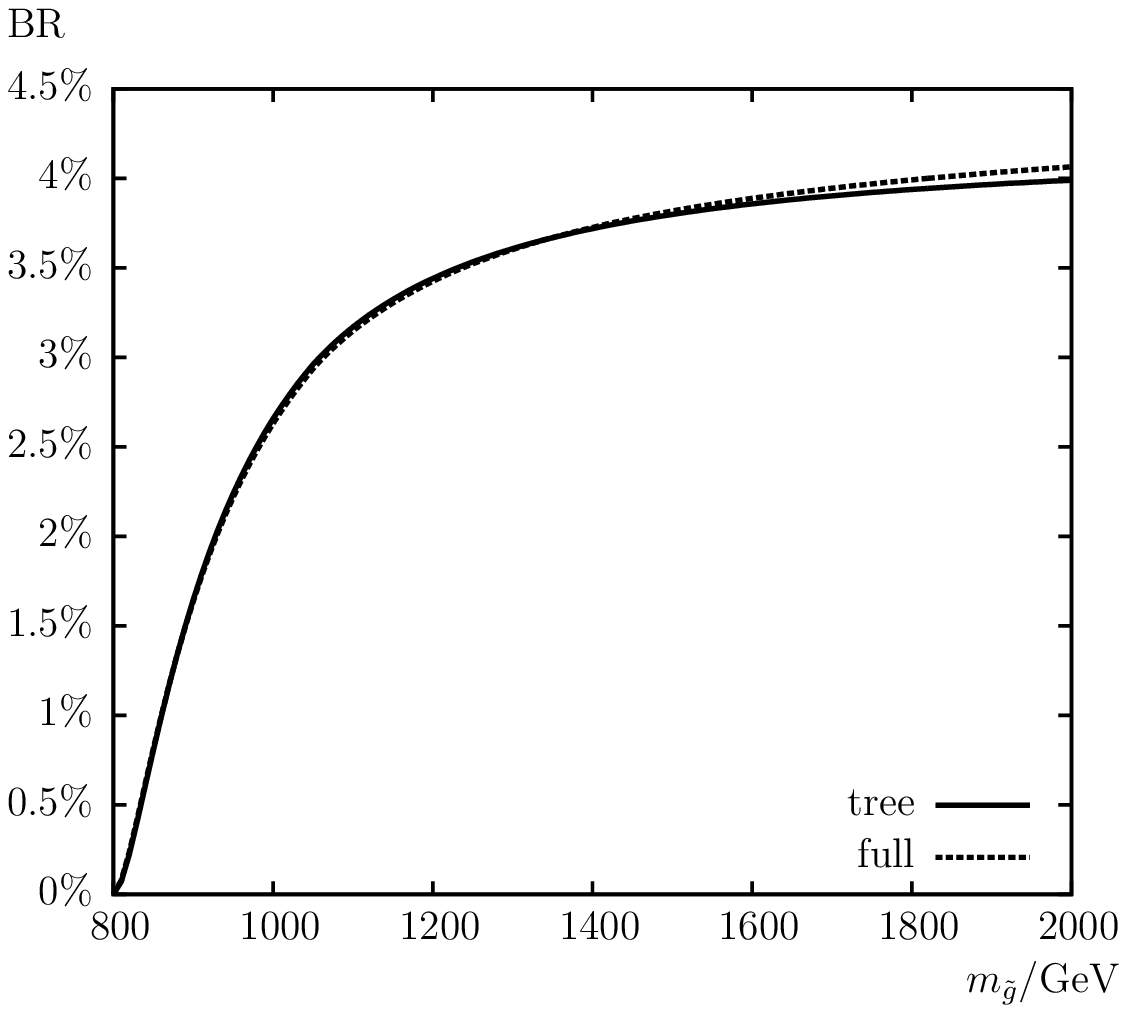}
\hspace{-4mm}
\includegraphics[width=0.49\textwidth,height=8.0cm]{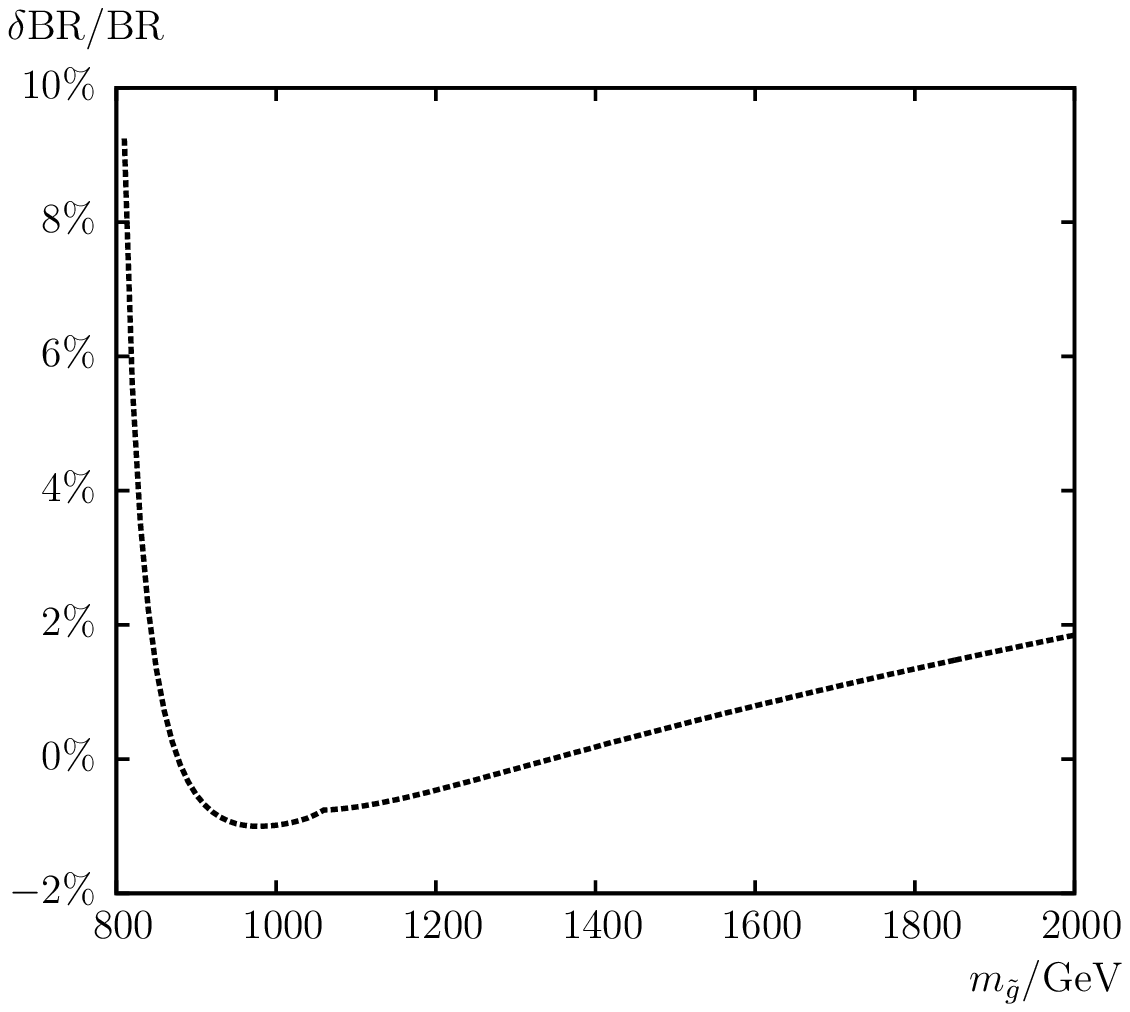}
\end{tabular}
\vspace{2em}
\caption{
  $\Ga(\decaySbz)$. Tree-level (``tree'') and full one-loop 
  (``full'') corrected decay widths are shown with the parameters 
  chosen according to \SE\ (see \refta{tab:para}), with $\mgl$ varied.
  The upper left plot shows the decay width, the upper right plot shows 
  the relative size of the corrections.
  Also shown are the pure SQCD corrections (``SQCD'').
  The lower left plot shows the BR, the lower right plot shows 
  the relative size of the BR.
}
\label{fig:AbsM3.glsb2b}
\end{center}
\end{figure}

The results for $\decaySbz$ are shown in \reffi{fig:AbsM3.glsb2b}.
Again the dip at $\mgl \approx 1054.6 \gev$ comes from the threshold 
$\mgl = \mstz + \mt$.
The threshold $\mgl = \mscz + m_c$ and the corresponding large effects
observed for $\decaySbe$ are absent here due to the fact that
$\msbz = 800.848 \gev$. Apart from this 
we find for the decay widths the expected values with 
$\Ga(\decaySbz) \sim 15 \gev$ at $\mgl = 2 \tev$ (with the SQCD
corrections giving a very good approximation to the full result)
and a corresponding BR of $\sim 4\%$. Again the sum of the two BR's
reaches $\sim 8.5\% \sim 1/12$. The relative corrections can reach
nearly $-9\%$ at $\mgl \approx 900 \gev$ for the decay width and values
between $+9\%$ and $+2\%$ for small and large BR's, respectively.

\begin{figure}[htb!]
\begin{center}
\begin{tabular}{c}
\includegraphics[width=0.49\textwidth,height=8.0cm]{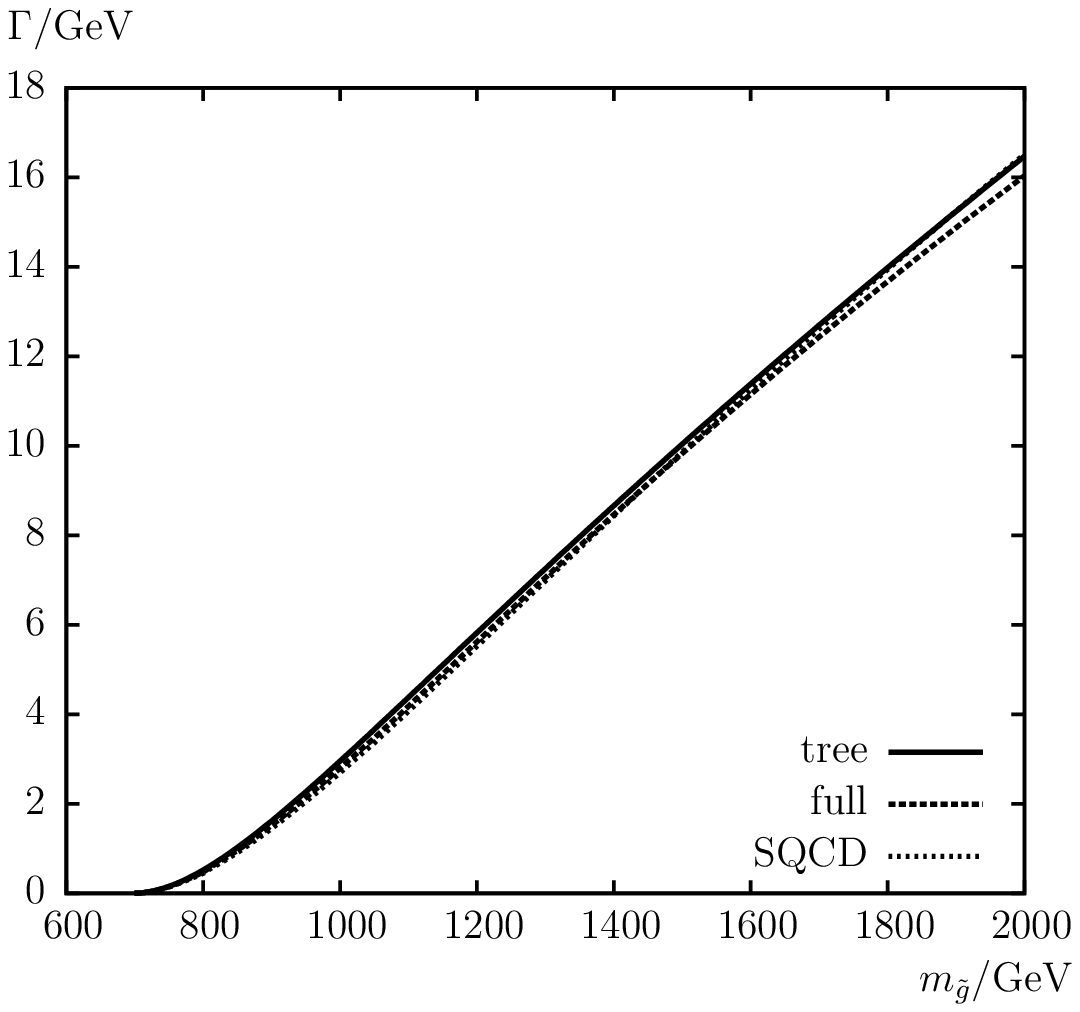}
\hspace{-4mm}
\includegraphics[width=0.49\textwidth,height=8.0cm]{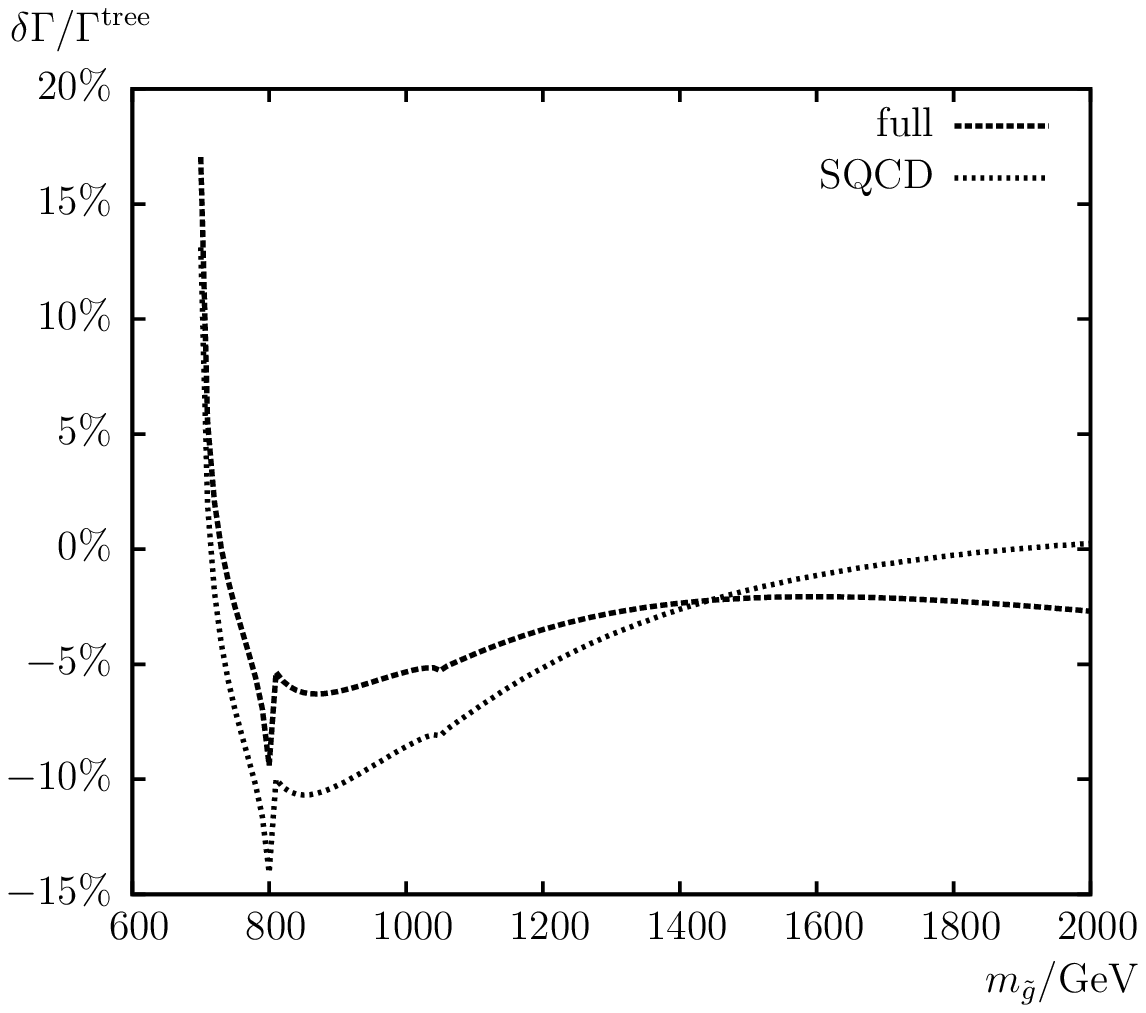}
\\[4em]
\includegraphics[width=0.49\textwidth,height=8.0cm]{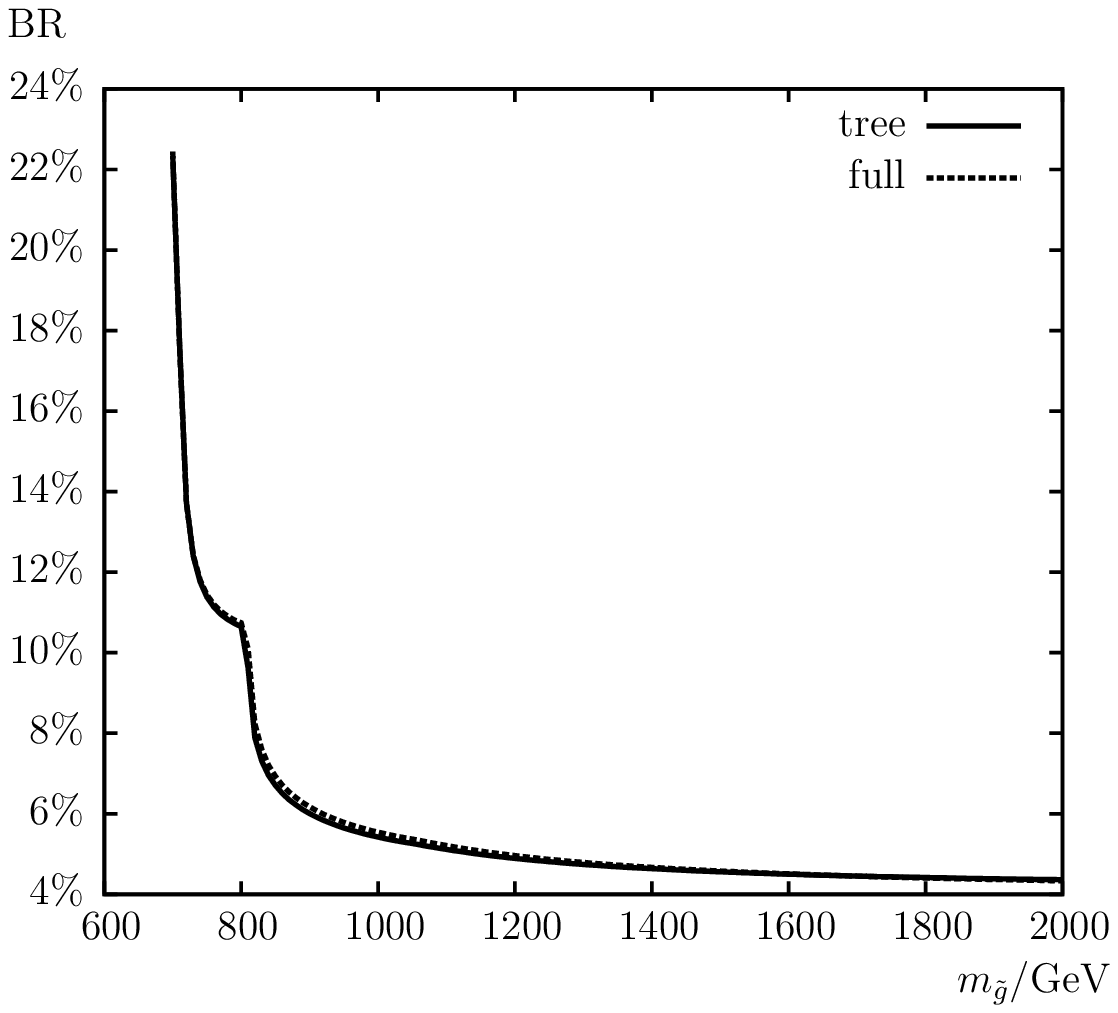}
\hspace{-4mm}
\includegraphics[width=0.49\textwidth,height=8.0cm]{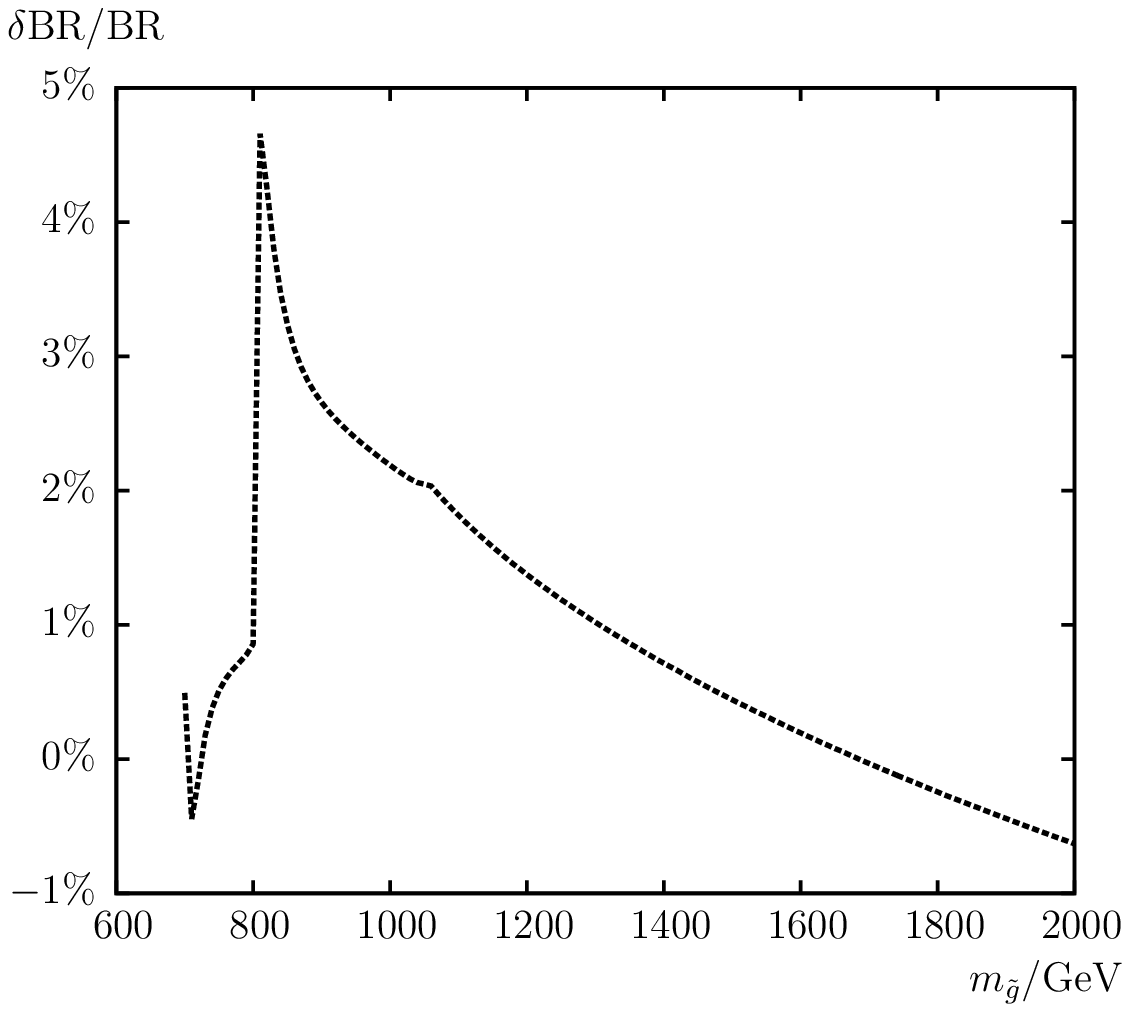}
\end{tabular}
\vspace{2em}
\caption{
  $\Ga(\decaySce)$. Tree-level (``tree'') and full one-loop 
  (``full'') corrected decay widths are shown with the parameters 
  chosen according to \SE\ (see \refta{tab:para}), with $\mgl$ varied.
  The upper left plot shows the decay width, the upper right plot shows 
  the relative size of the corrections.
  Also shown are the pure SQCD corrections (``SQCD'').
  The lower left plot shows the BR, the lower right plot shows 
  the relative size of the BR.
}
\label{fig:AbsM3.glsc1c}
\end{center}
\end{figure}

Next we analyze the effects in the decays involving second generation
(s)quarks. We start with the decay $\decaySce$ shown in
\reffi{fig:AbsM3.glsc1c}. Two dips are visible: 
within \SE\ the first dip appears at $\mgl \approx 809.3 \gev$, 
due to the threshold $\mgl = \mste + \mt$.
The second dip at $\mgl \approx 1054.6\gev$ comes from 
the threshold $\mgl = \mstz + \mt$.
The decay width behaves ``as expected'', rising up to $\sim 16 \gev$ at 
$\mgl = 2 \tev$. The relative corrections are again largest around the
first threshold, reaching nearly $-10\%$, flattening out to $\sim -3\%$
at larger $\mgl$ values. The SQCD corrections alone yield larger
corrections by $\sim -5\%$, going to zero for large $\mgl$, where the EW
contributions are dominating.
The branching ratio is very large at threshold
due to the relative smallness of $\msce + m_c$ in \SE, exceeding
$20\%$. At large $\mgl$ the ``expected'' $\sim 4\%$ are reached. The
relative correction to the BR exhibits a maximum around 
$\mgl = 800 \gev$ due to $\mgl = \mste + \mt$, see above. At large
$\mgl$ the corrections become very small, but potentially reach larger
negative values at $\mgl > 2 \tev$.

\begin{figure}[htb!]
\begin{center}
\begin{tabular}{c}
\includegraphics[width=0.49\textwidth,height=8.0cm]{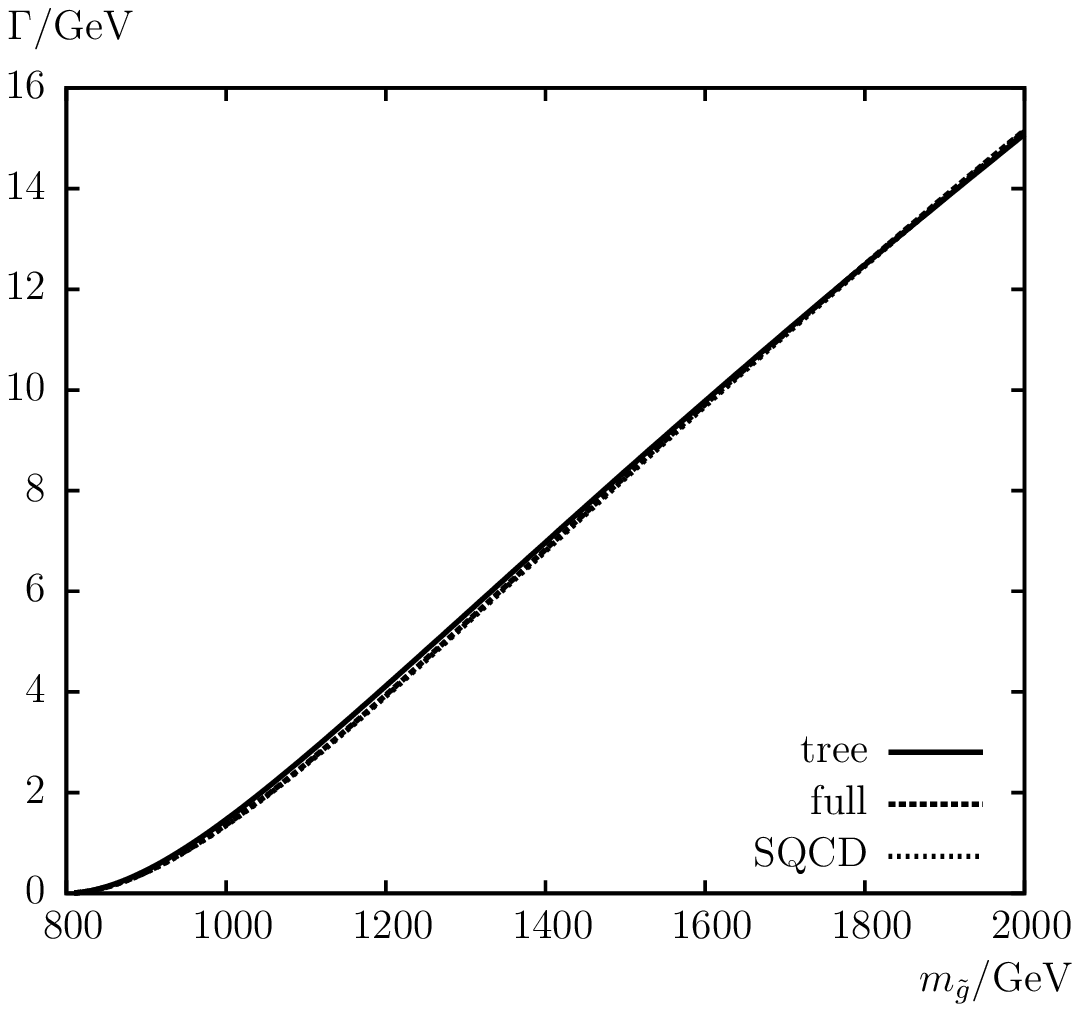}
\hspace{-4mm}
\includegraphics[width=0.49\textwidth,height=8.0cm]{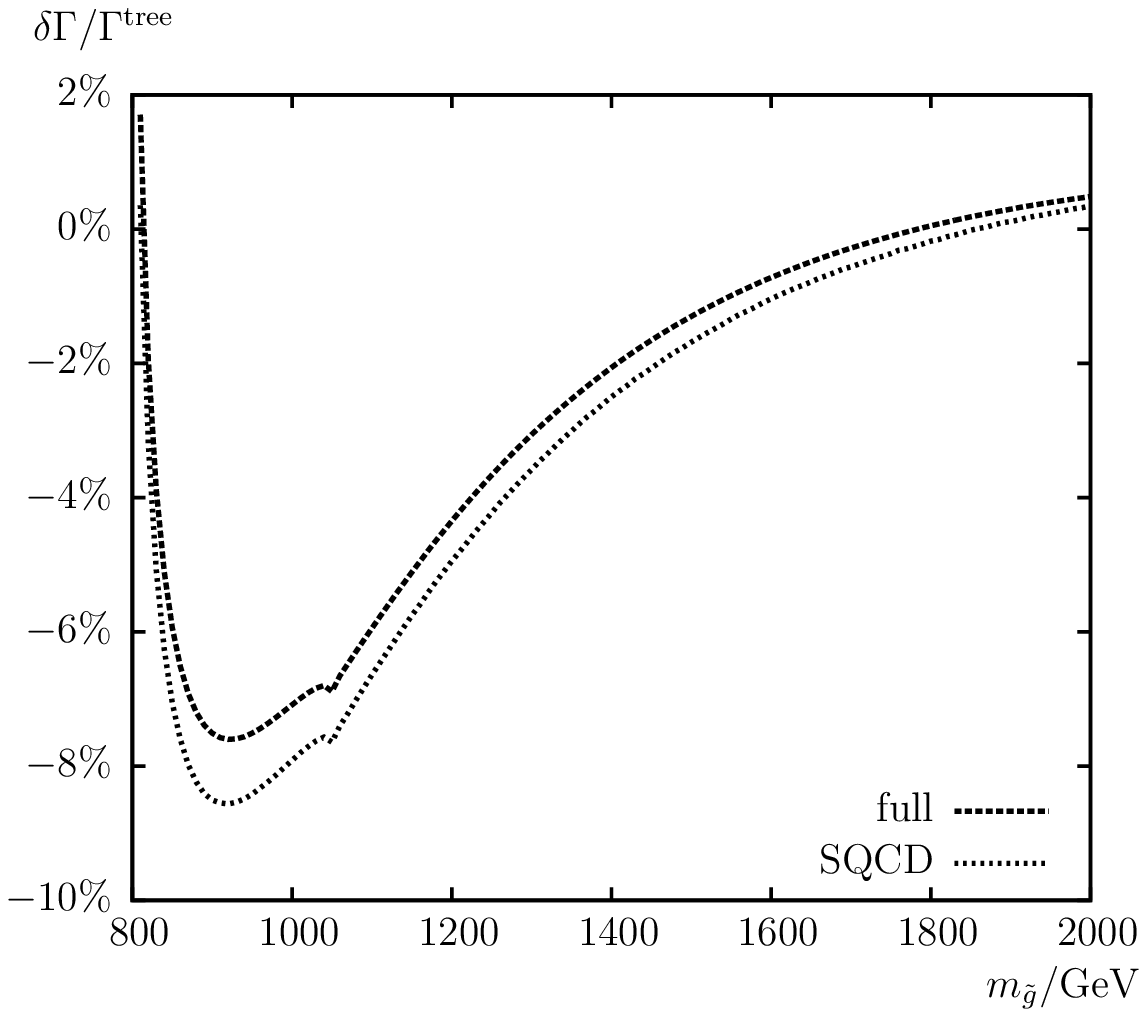}
\\[4em]
\includegraphics[width=0.49\textwidth,height=8.0cm]{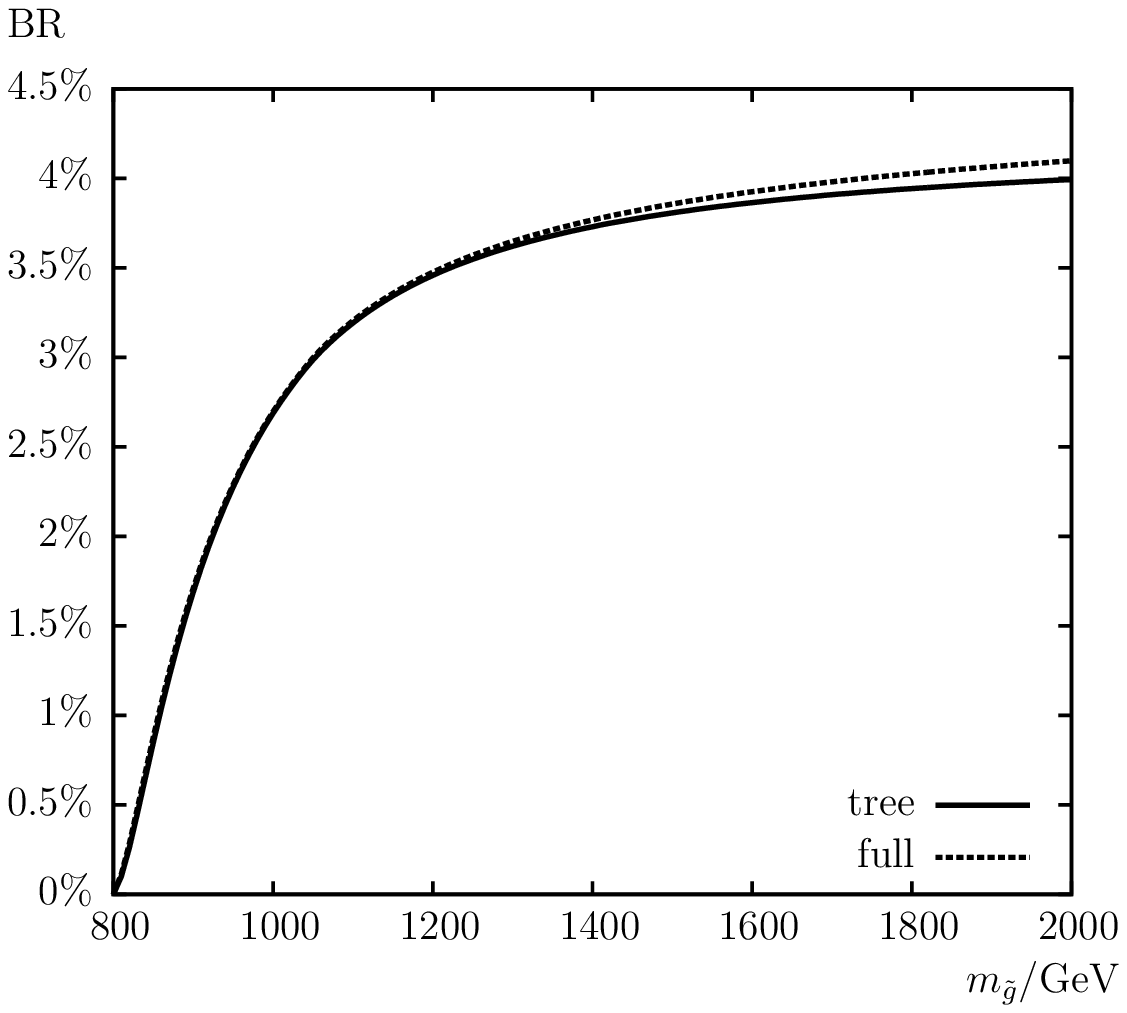}
\hspace{-4mm}
\includegraphics[width=0.49\textwidth,height=8.0cm]{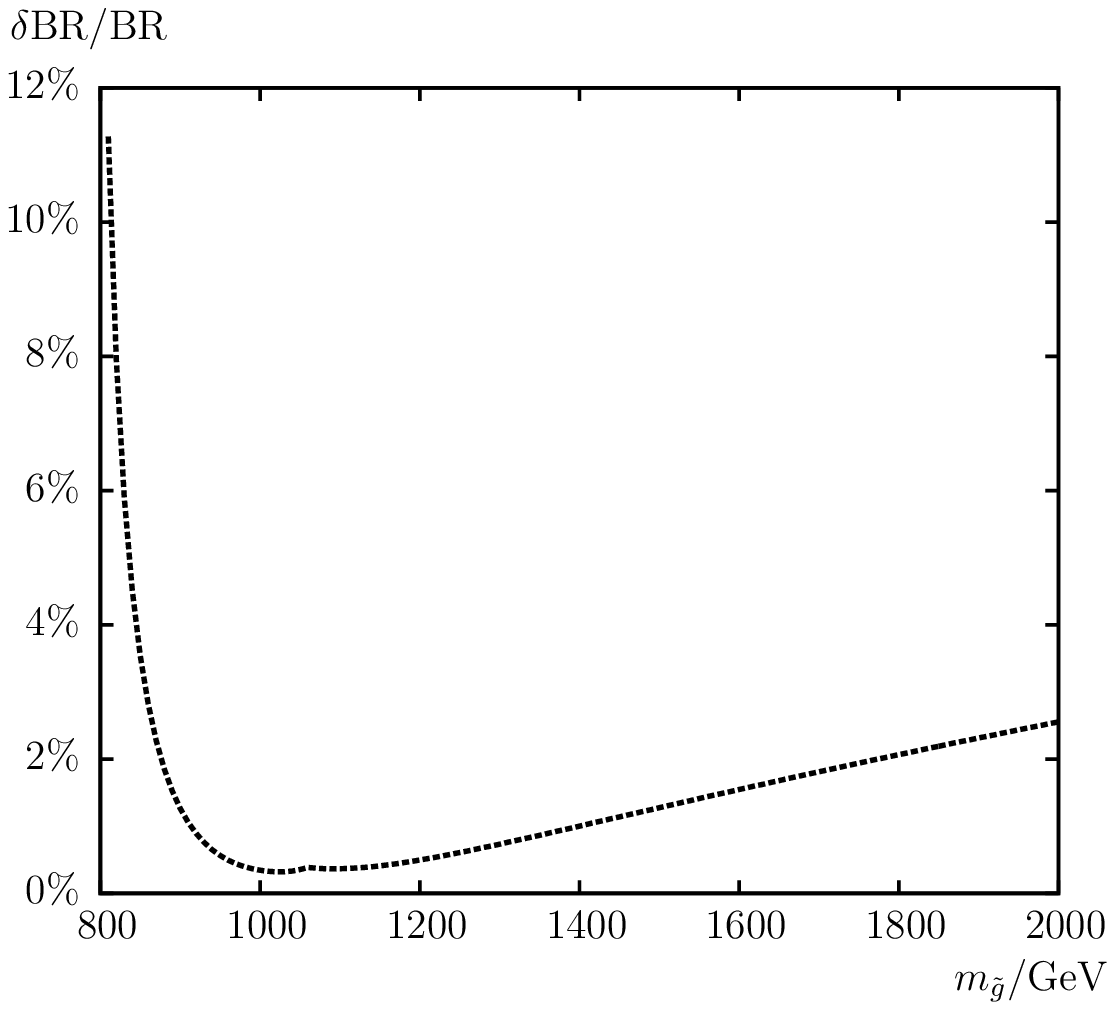}
\end{tabular}
\vspace{2em}
\caption{
  $\Ga(\decayScz)$. Tree-level (``tree'') and full one-loop 
  (``full'') corrected decay widths are shown with the parameters 
  chosen according to \SE\ (see \refta{tab:para}), with $\mgl$ varied.
  The upper left plot shows the decay width, the upper right plot shows 
  the relative size of the corrections.
  Also shown are the pure SQCD corrections (``SQCD'').
  The lower left plot shows the BR, the lower right plot shows 
  the relative size of the BR.
}
\label{fig:AbsM3.glsc2c}
\end{center}
\end{figure}

The results for $\decayScz$ are shown in \reffi{fig:AbsM3.glsc2c}.
Again, the dip at $\mgl \approx 1054.6 \gev$ stems from the 
threshold $\mgl = \mstz + \mt$. Decay width and branching ratio show the
expected behavior. The BR reaches $\sim 4\%$ at large $\mgl$, so that
the sum of the four ``charm BR's'' goes to $\sim 1/6$. 
The corrections to the width are maximal with $\sim -7.5\%$ around 
$\mgl = 900 \gev$ and rise to small positive values for large
$\mgl$, with the SQCD corrections deviating from this by up to
$-1\%$ at most.
Concerning the branching ratio the effects are largest where the
BR is small and reach the level of $\sim 3\%$ at $\mgl = 2 \tev$. 

\begin{figure}[htb!]
\begin{center}
\begin{tabular}{c}
\includegraphics[width=0.49\textwidth,height=8.0cm]{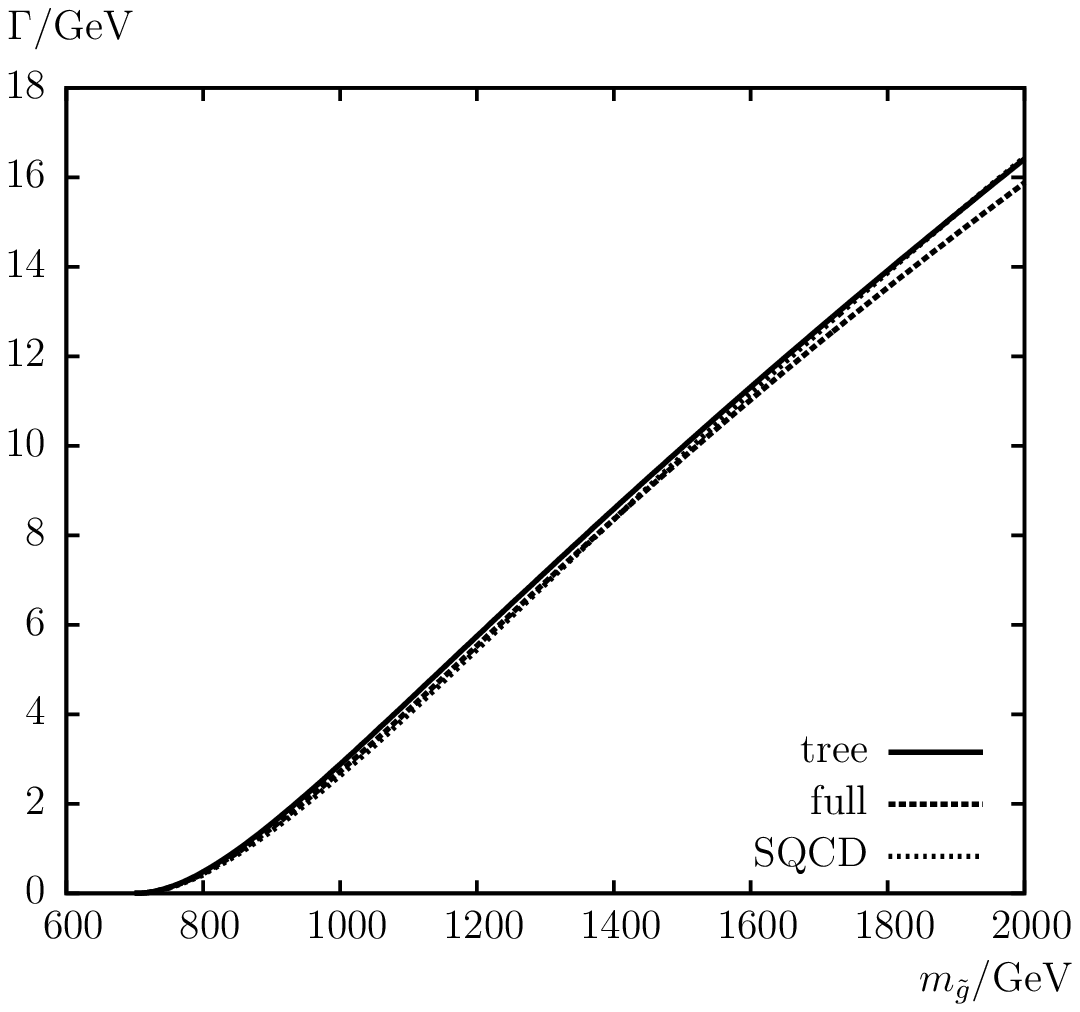}
\hspace{-4mm}
\includegraphics[width=0.49\textwidth,height=8.0cm]{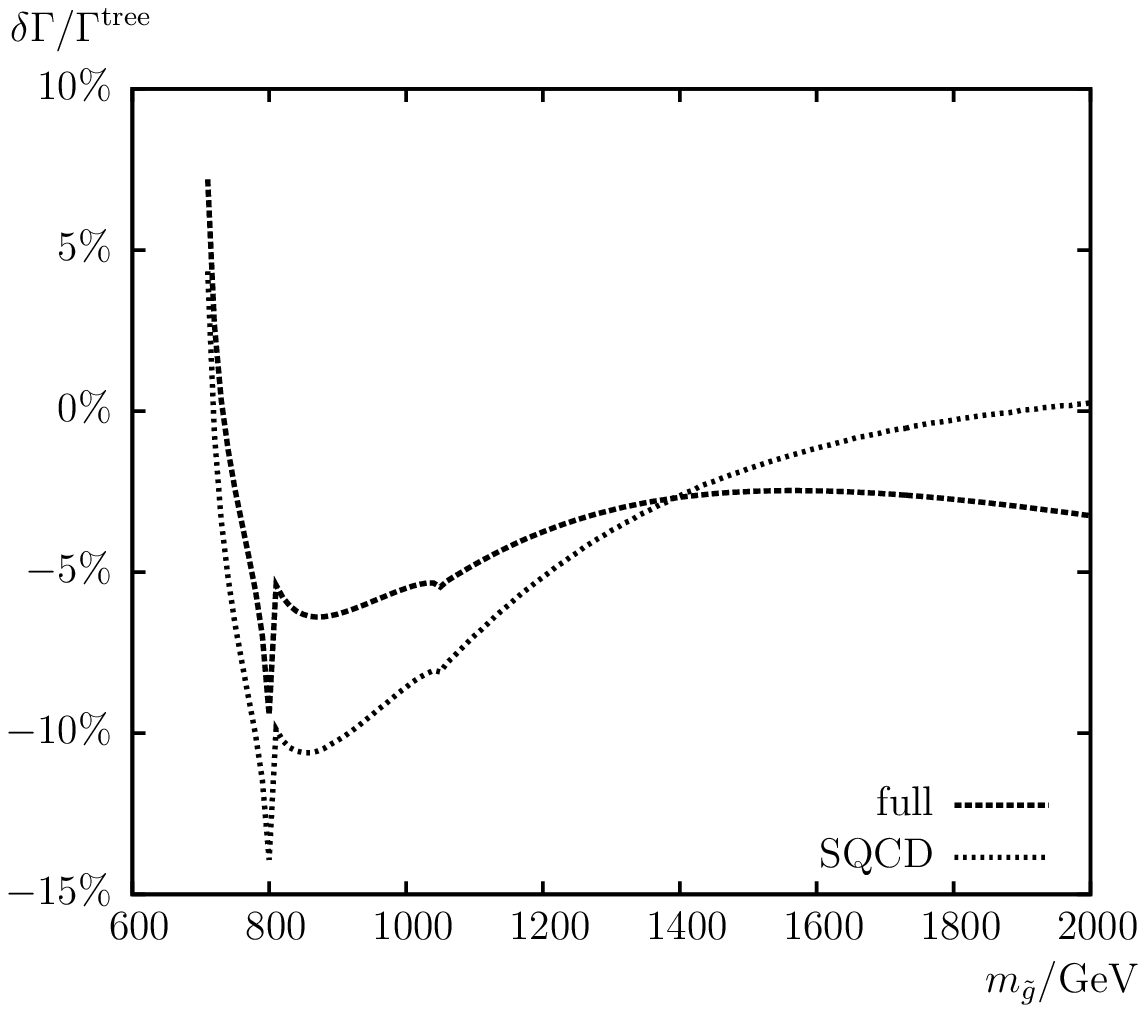}
\\[4em]
\includegraphics[width=0.49\textwidth,height=8.0cm]{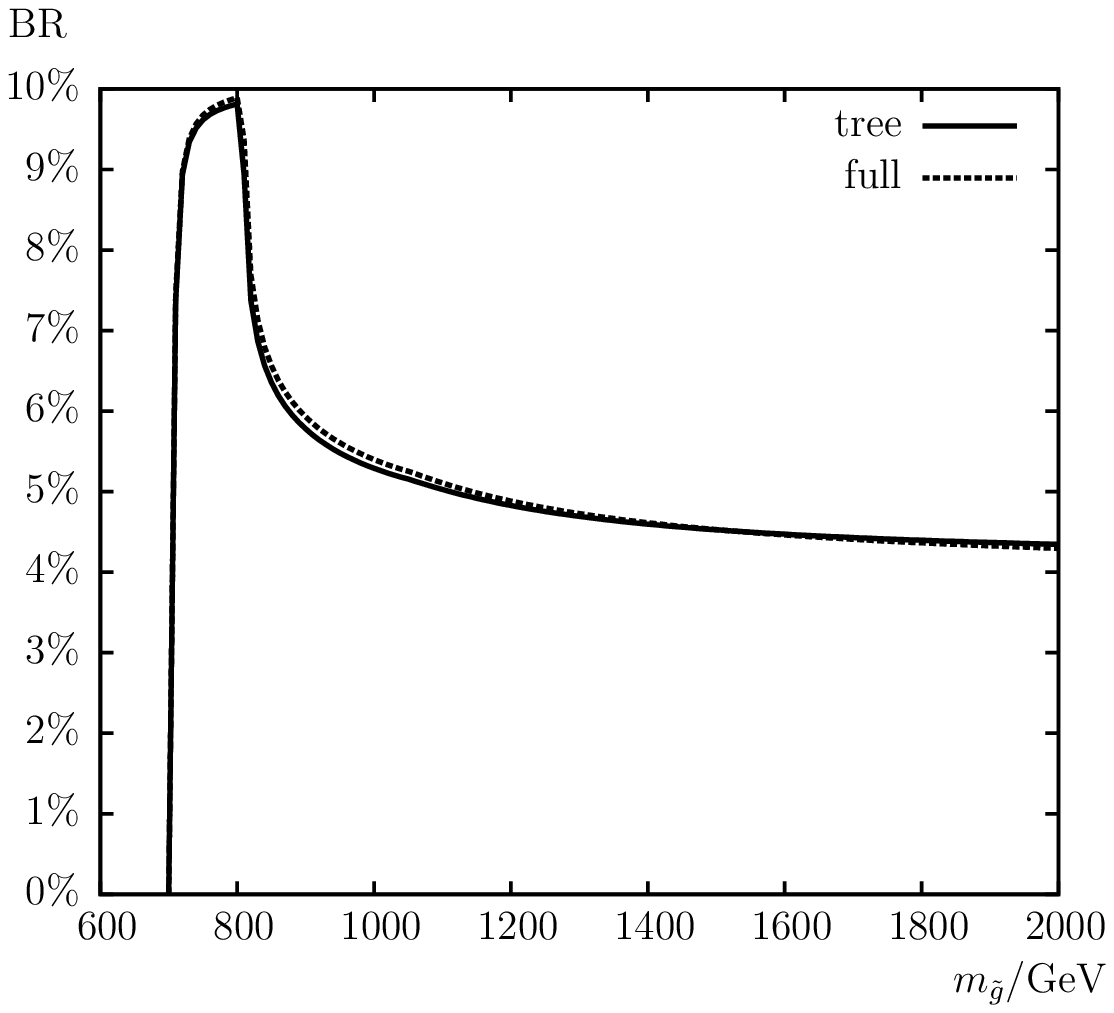}
\hspace{-4mm}
\includegraphics[width=0.49\textwidth,height=8.0cm]{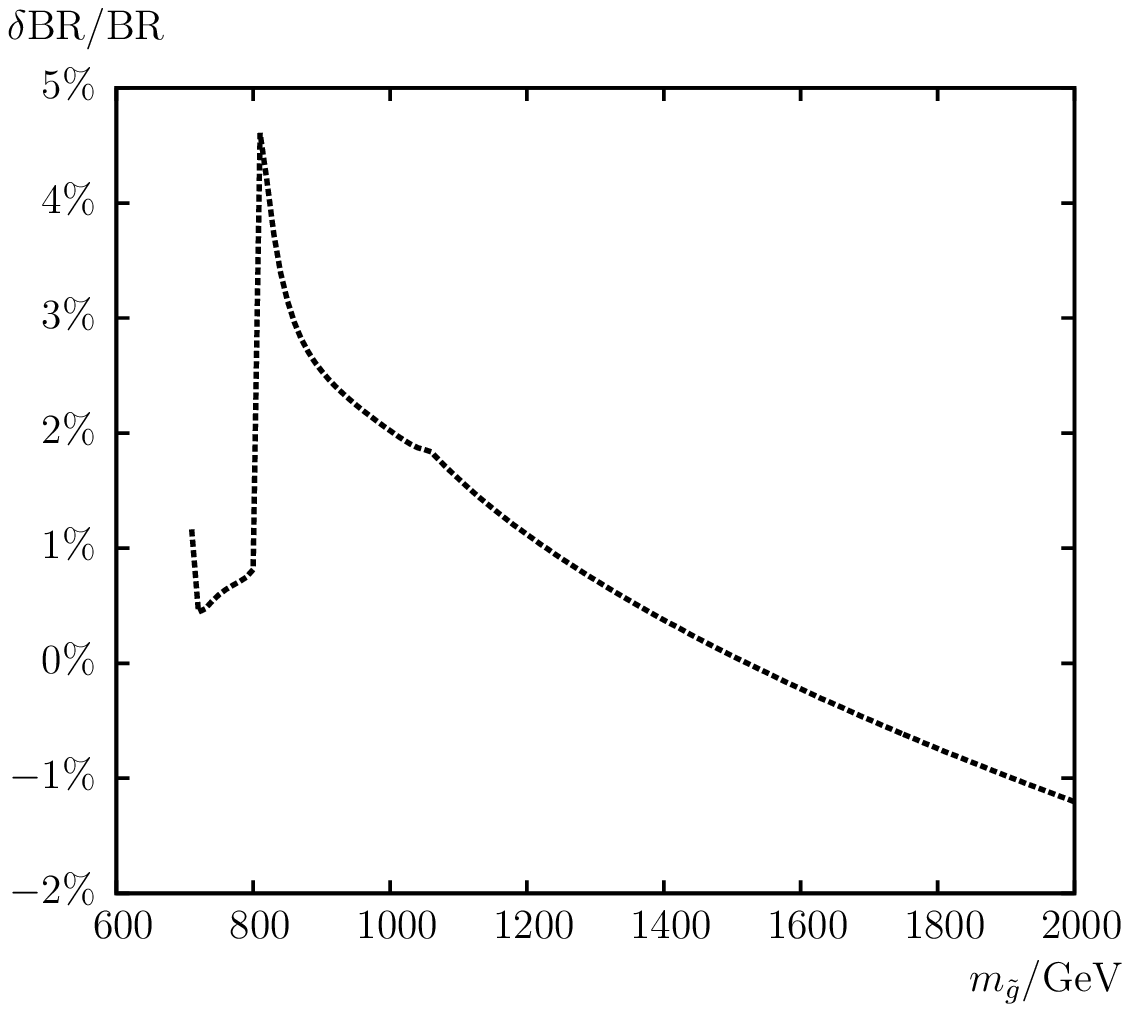}
\end{tabular}
\vspace{2em}
\caption{
  $\Ga(\decaySse)$. Tree-level (``tree'') and full one-loop 
  (``full'') corrected decay widths are shown with the parameters 
  chosen according to \SE\ (see \refta{tab:para}), with $\mgl$ varied.
  The upper left plot shows the decay width, the upper right plot shows 
  the relative size of the corrections.
  Also shown are the pure SQCD corrections (``SQCD'').
  The lower left plot shows the BR, the lower right plot shows 
  the relative size of the BR.
}
\label{fig:AbsM3.glss1s}
\end{center}
\end{figure}

\begin{figure}[htb!]
\begin{center}
\begin{tabular}{c}
\includegraphics[width=0.49\textwidth,height=8.0cm]{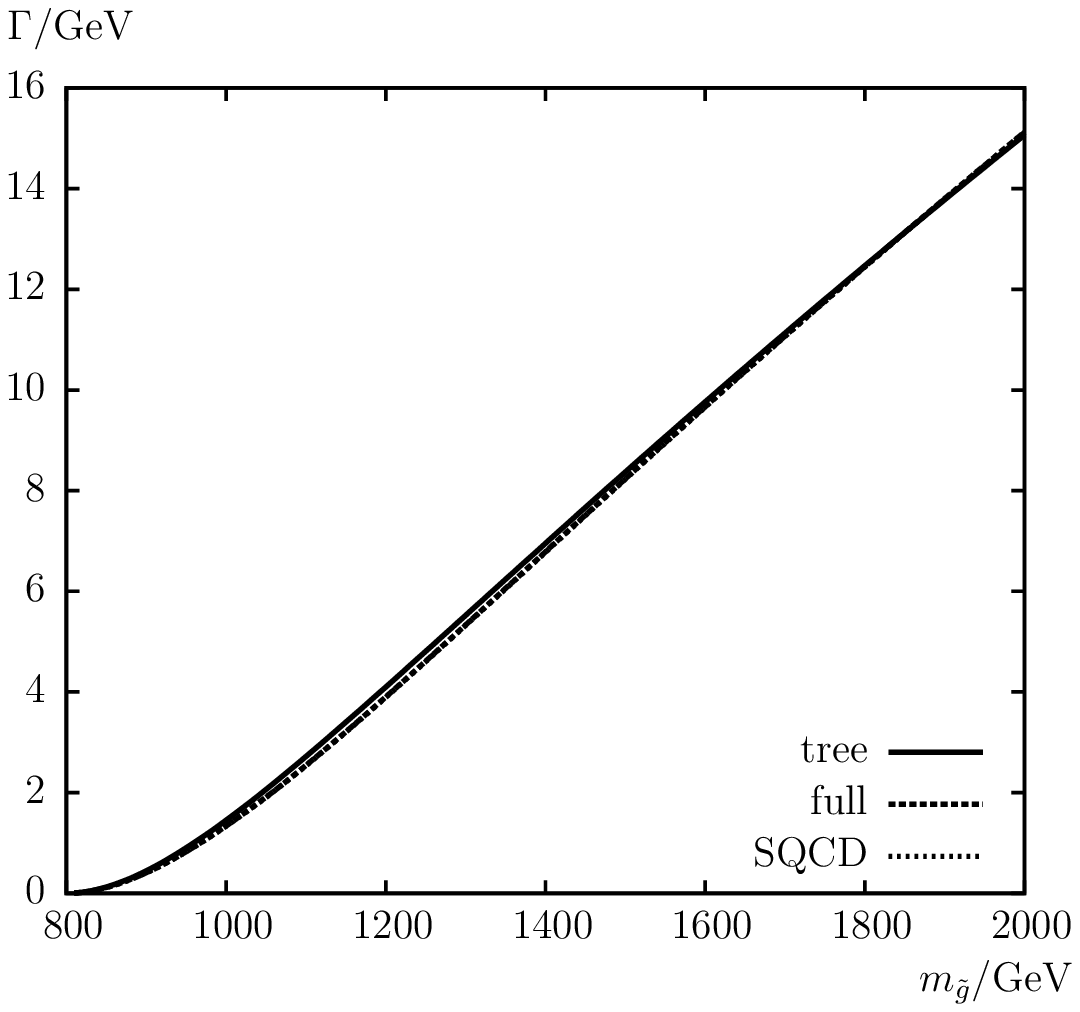}
\hspace{-4mm}
\includegraphics[width=0.49\textwidth,height=8.0cm]{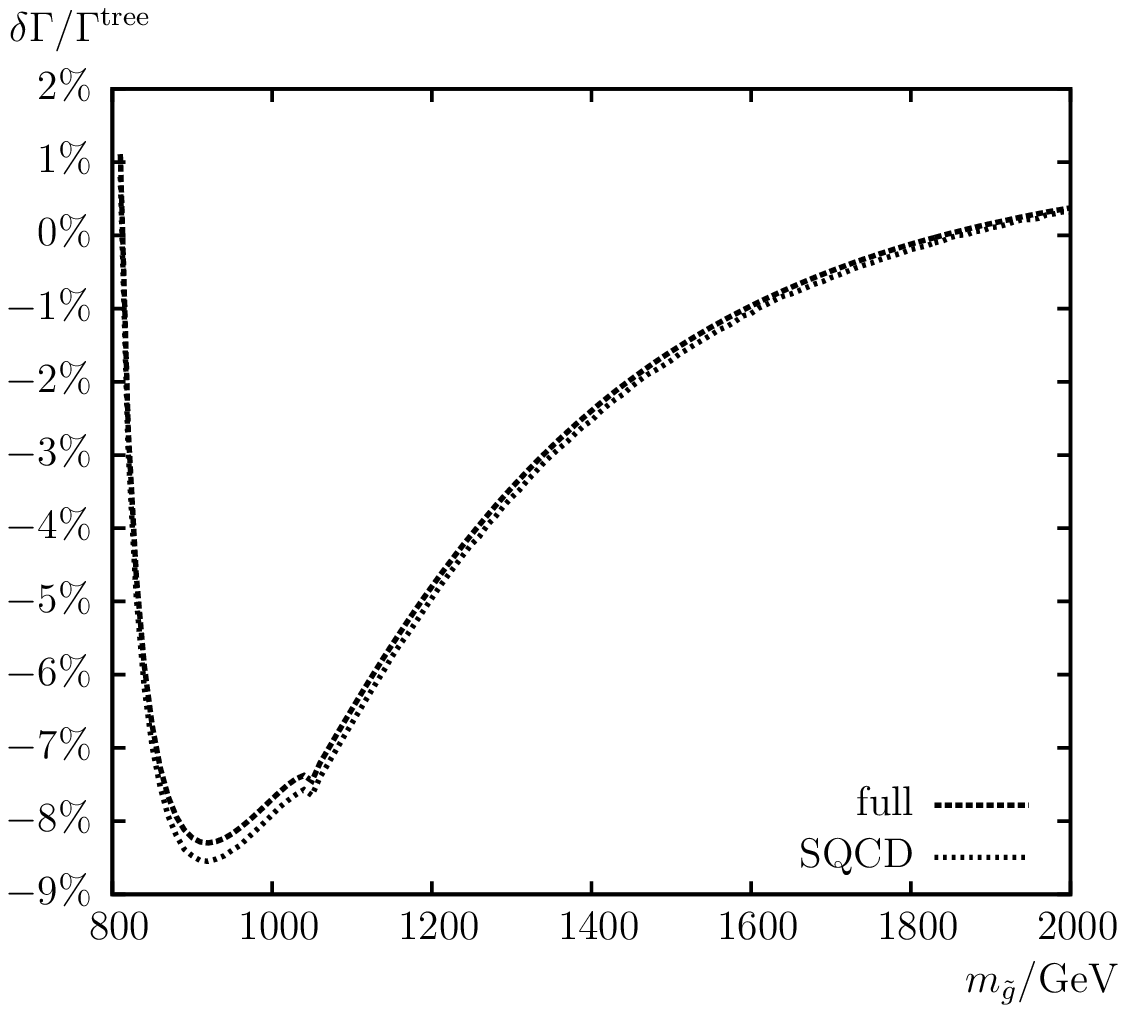}
\\[4em]
\includegraphics[width=0.49\textwidth,height=8.0cm]{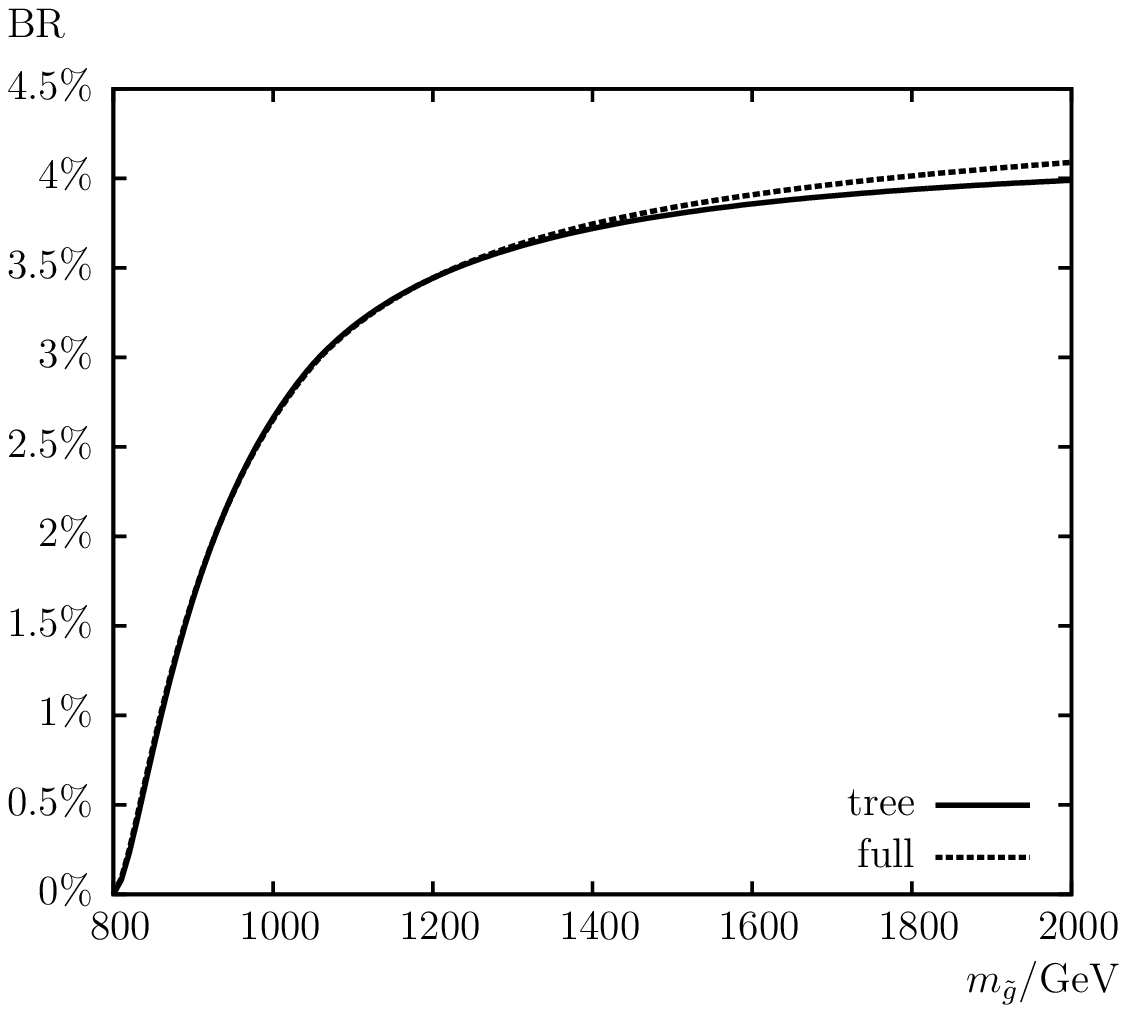}
\hspace{-4mm}
\includegraphics[width=0.49\textwidth,height=8.0cm]{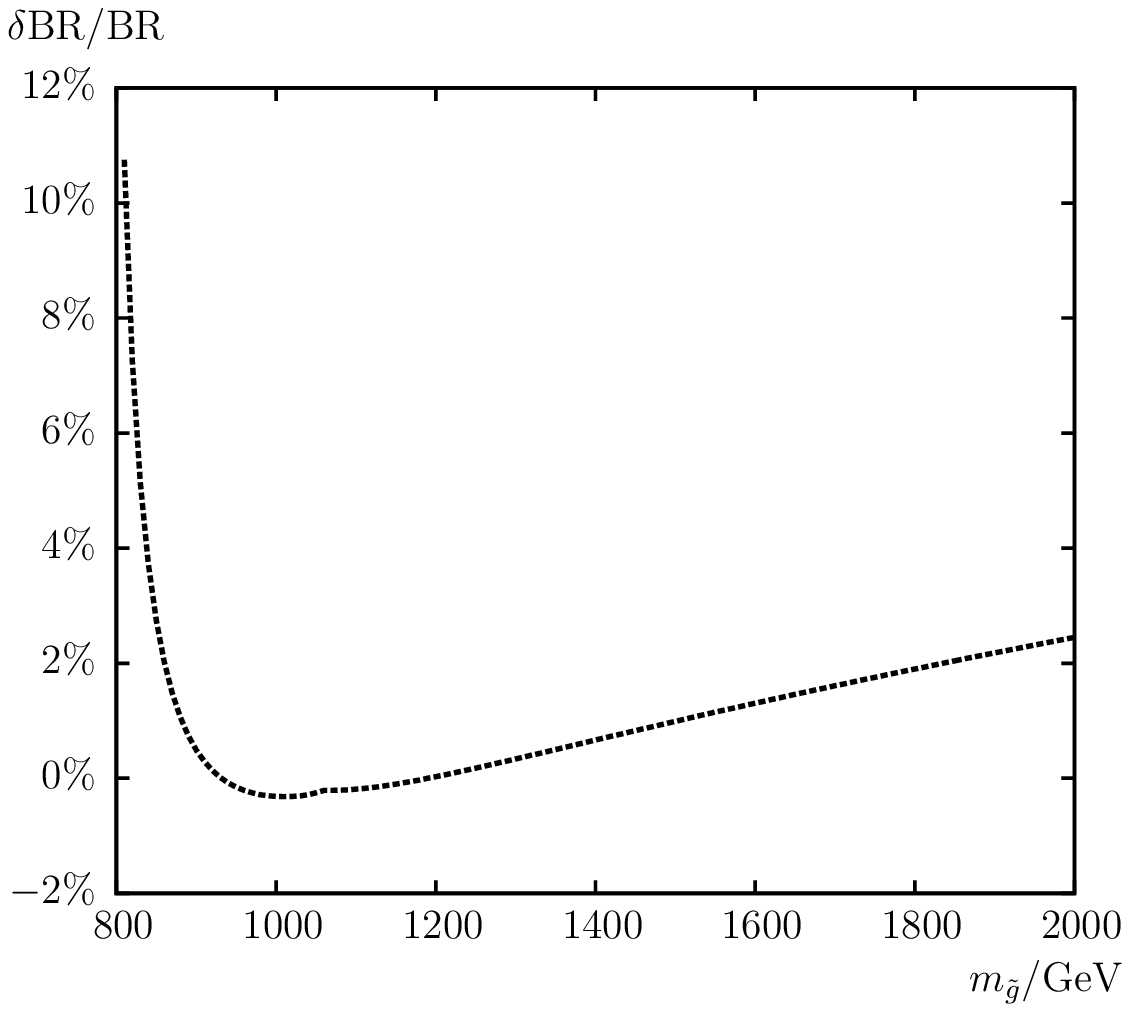}
\end{tabular}
\vspace{2em}
\caption{
  $\Ga(\decaySsz)$. Tree-level (``tree'') and full one-loop 
  (``full'') corrected decay widths are shown with the parameters 
  chosen according to \SE\ (see \refta{tab:para}), with $\mgl$ varied.
  The upper left plot shows the decay width, the upper right plot shows 
  the relative size of the corrections.
  Also shown are the pure SQCD corrections (``SQCD'').
  The lower left plot shows the BR, the lower right plot shows 
  the relative size of the BR.
}
\label{fig:AbsM3.glss2s}
\end{center}
\end{figure}

The decays involving scalar strange quarks, $\decaySsi$ 
($i = 1,2$), are shown in \reffis{fig:AbsM3.glss1s}, \ref{fig:AbsM3.glss2s}.
Again several dips are visible, 
the first dip at $\mgl \approx 809.3 \gev$ comes from the threshold 
$\mgl = \mste + \mt$. Once more the second dip at 
$\mgl \approx 1054.6 \gev$ is the threshold $\mgl = \mstz + \mt$.
The results are very similar to the ones for the decays $\decaySci$,
where differences can only be observed close to threshold due to
slightly different masses of the scalar quarks. The most prominent
example in this respect is the $\br(\decaySse)$ that reaches ``only''
about $10\%$ for $\mgl = 700 \ldots 800 \gev$. Otherwise the
``expected'' behavior is found at the tree and at the loop level,
including also the size of the pure SQCD corrections.

The results for decays involving first generation (s)quarks are very
similar to the ones for the second generation, again differences appear
only due to small deviations in the scalar quark masses, see
\refta{tab:squark}. 
Otherwise again the ``expected'' behavior is found with loop corrections
yielding maximum values of $-10\%$ for the decay widths, again with
the same size of the SQCD corrections, and $+5\%$ for the branching ratios. 
It should be kept in mind that the results found for the branching
ratios are highly model dependent, and much larger/smaller values can be
found for different kinematical situations.

\begin{figure}[htb!]
\begin{center}
\begin{tabular}{c}
\includegraphics[width=0.49\textwidth,height=8.0cm]{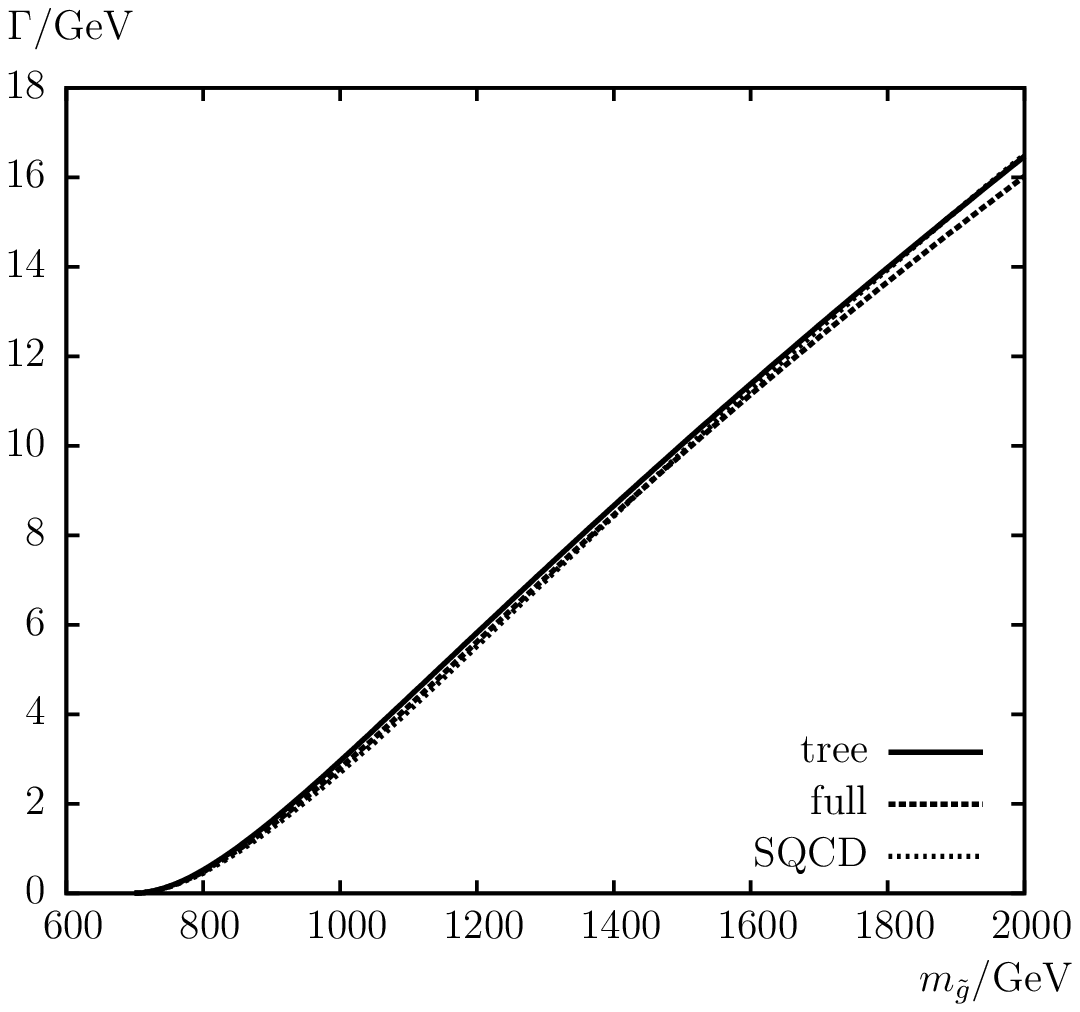}
\hspace{-4mm}
\includegraphics[width=0.49\textwidth,height=8.0cm]{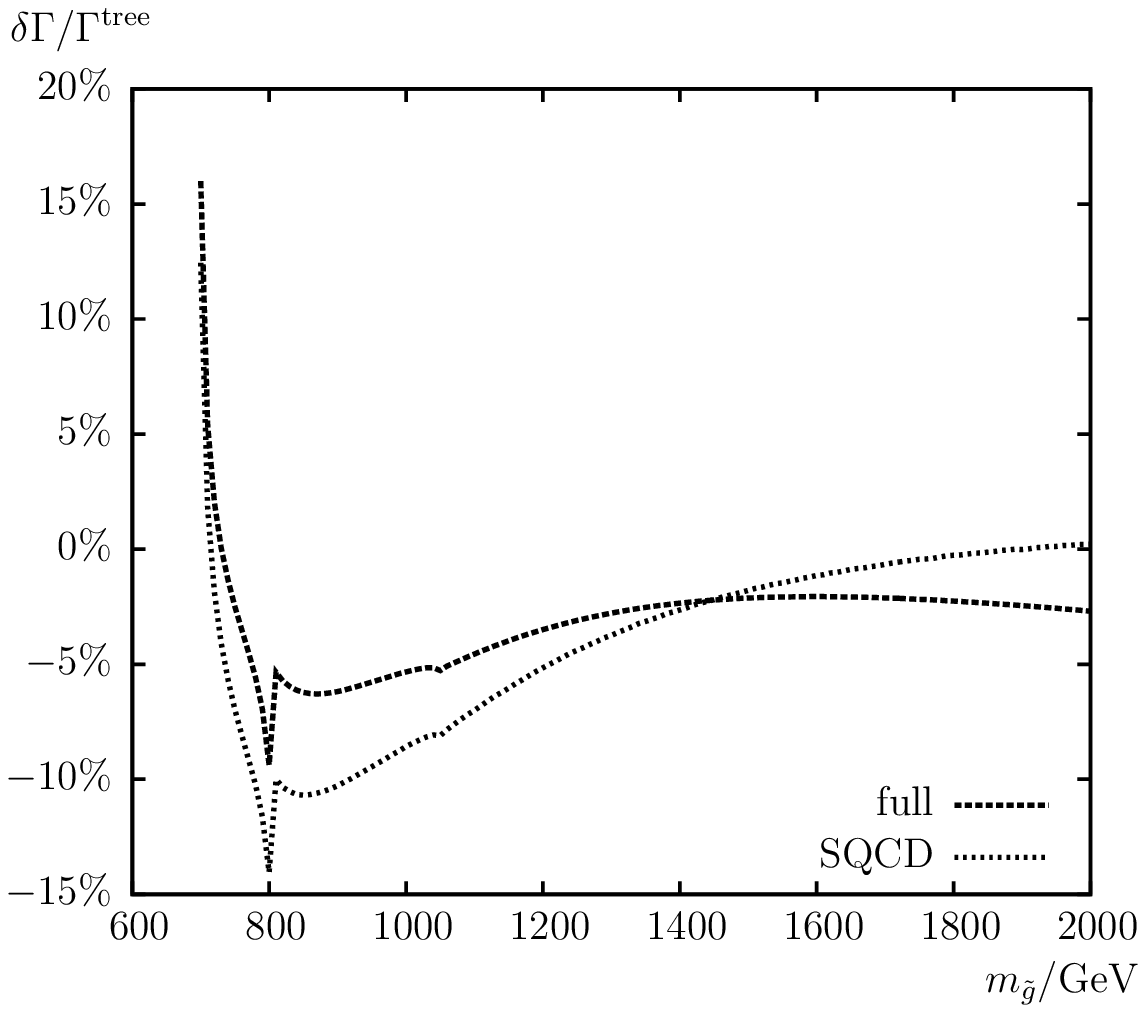}
\\[4em]
\includegraphics[width=0.49\textwidth,height=8.0cm]{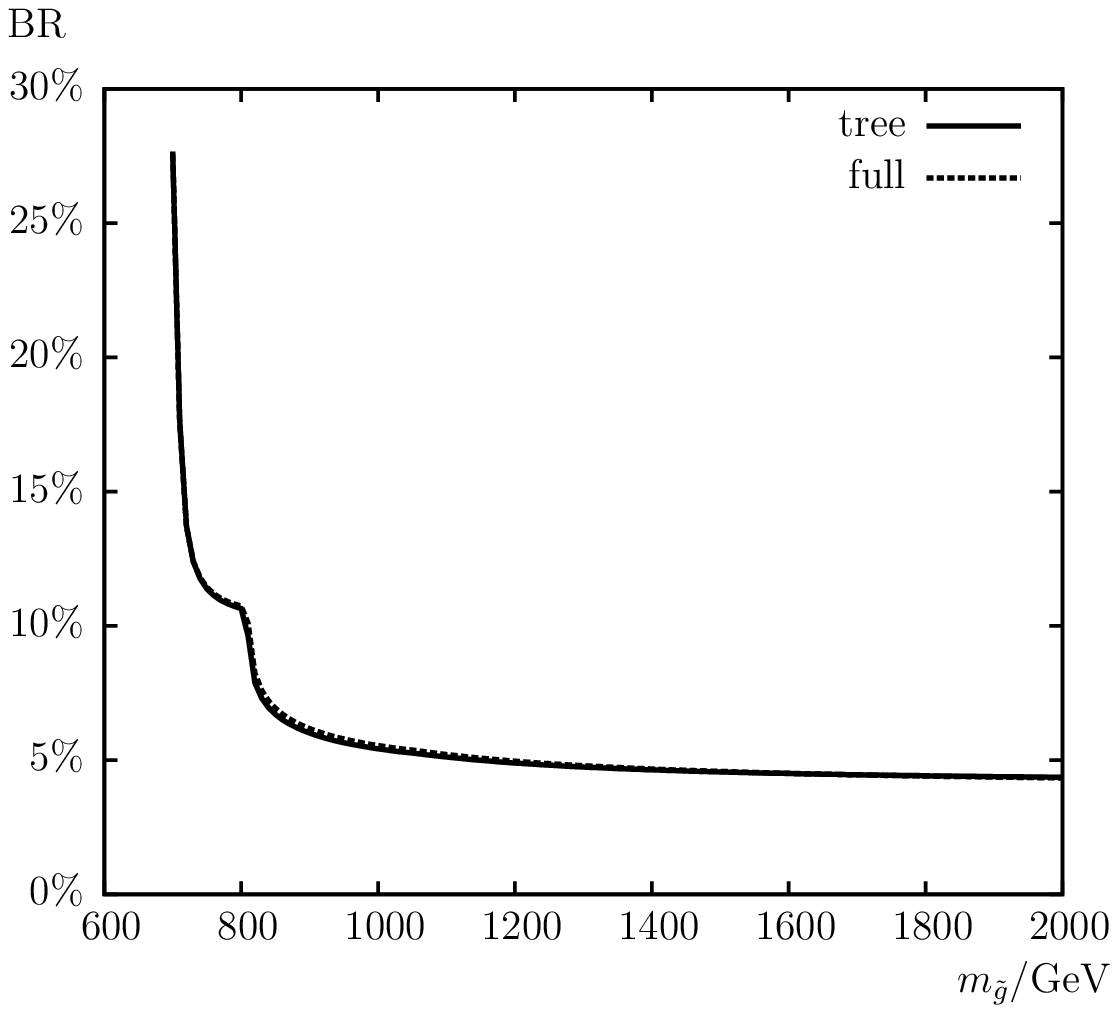}
\hspace{-4mm}
\includegraphics[width=0.49\textwidth,height=8.0cm]{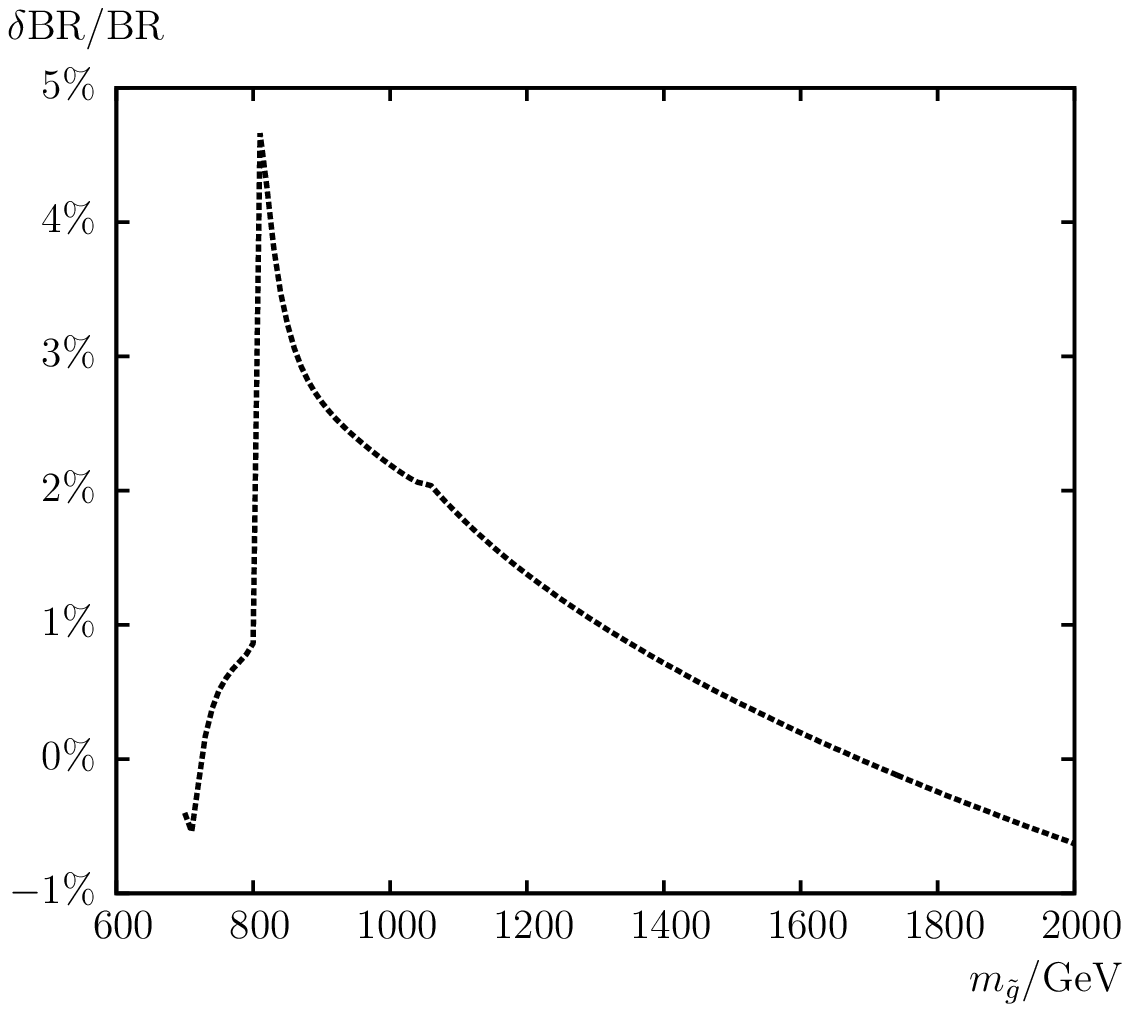}
\end{tabular}
\vspace{2em}
\caption{
  $\Ga(\decaySue)$. Tree-level (``tree'') and full one-loop 
  (``full'') corrected decay widths are shown with the parameters 
  chosen according to \SE\ (see \refta{tab:para}), with $\mgl$ varied.
  The upper left plot shows the decay width, the upper right plot shows 
  the relative size of the corrections.
  Also shown are the pure SQCD corrections (``SQCD'').
  The lower left plot shows the BR, the lower right plot shows 
  the relative size of the BR.
}
\label{fig:AbsM3.glsu1u}
\end{center}
\end{figure}

\begin{figure}[htb!]
\begin{center}
\begin{tabular}{c}
\includegraphics[width=0.49\textwidth,height=8.0cm]{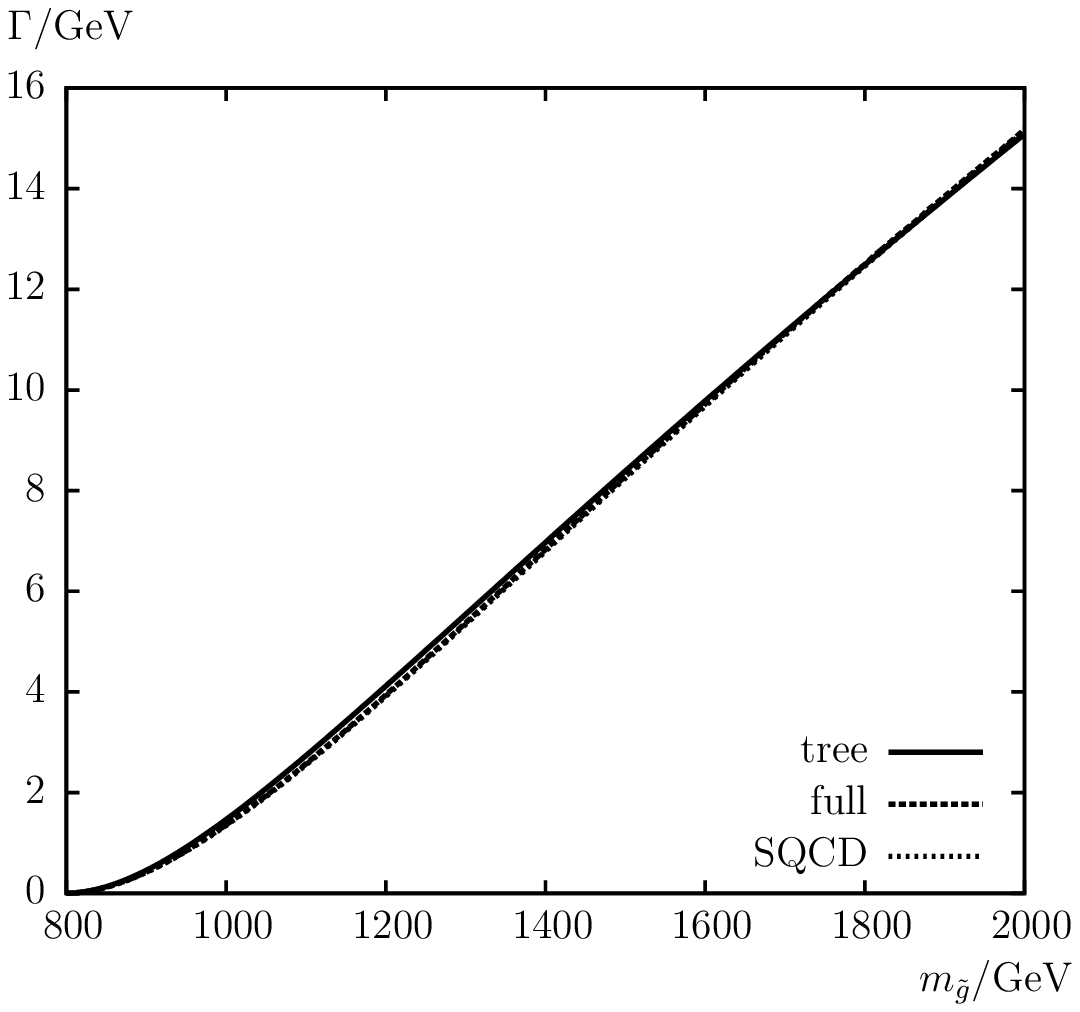}
\hspace{-4mm}
\includegraphics[width=0.49\textwidth,height=8.0cm]{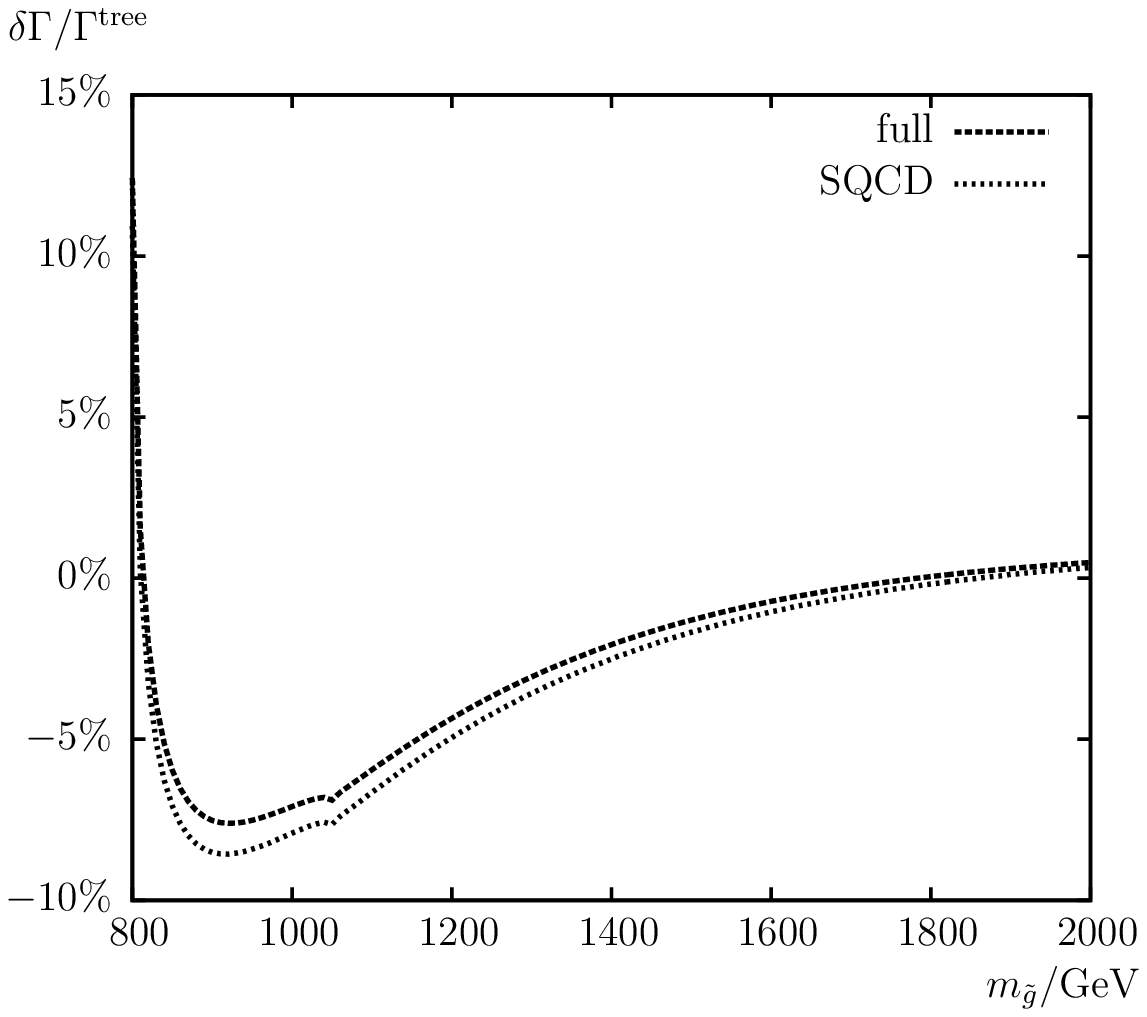}
\\[4em]
\includegraphics[width=0.49\textwidth,height=8.0cm]{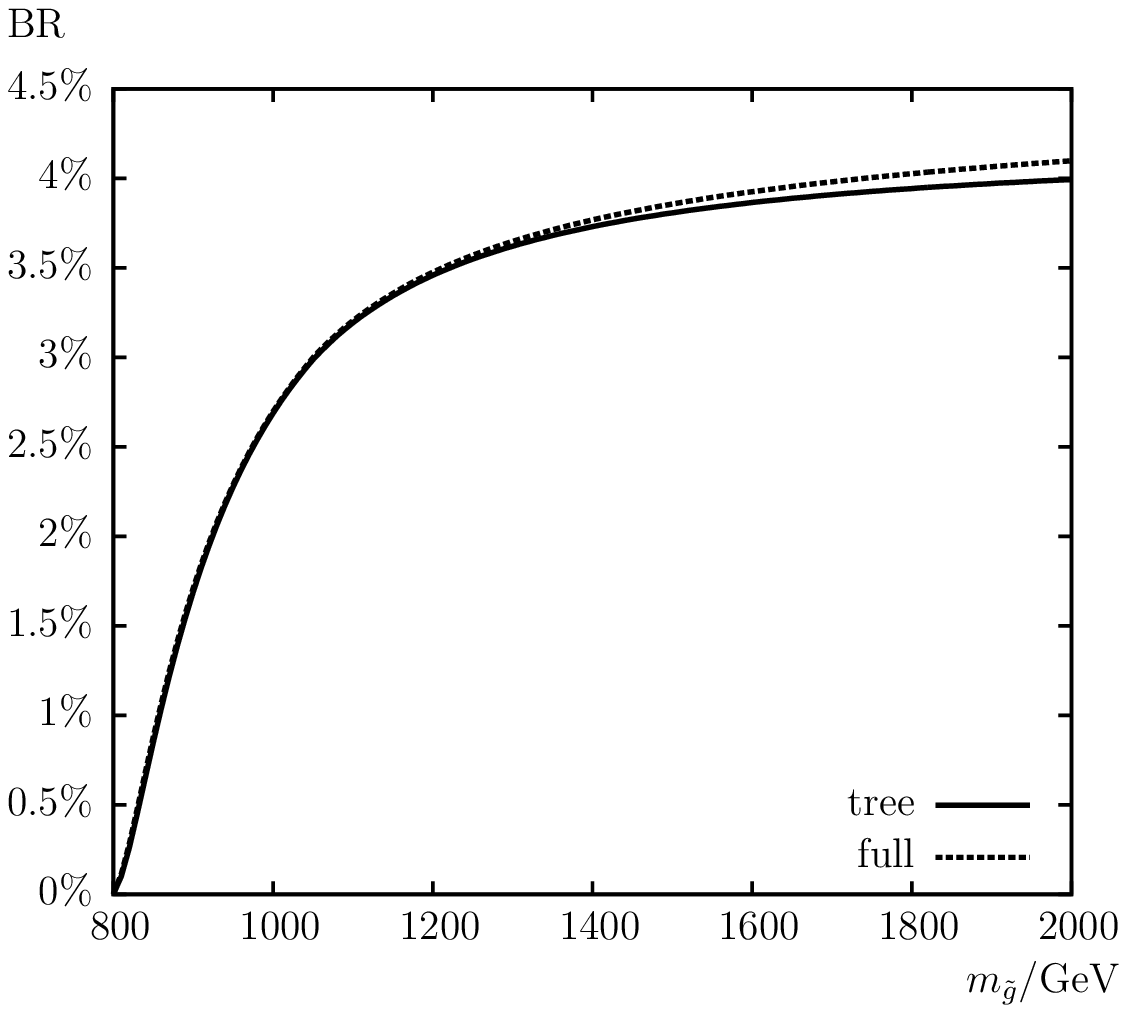}
\hspace{-4mm}
\includegraphics[width=0.49\textwidth,height=8.0cm]{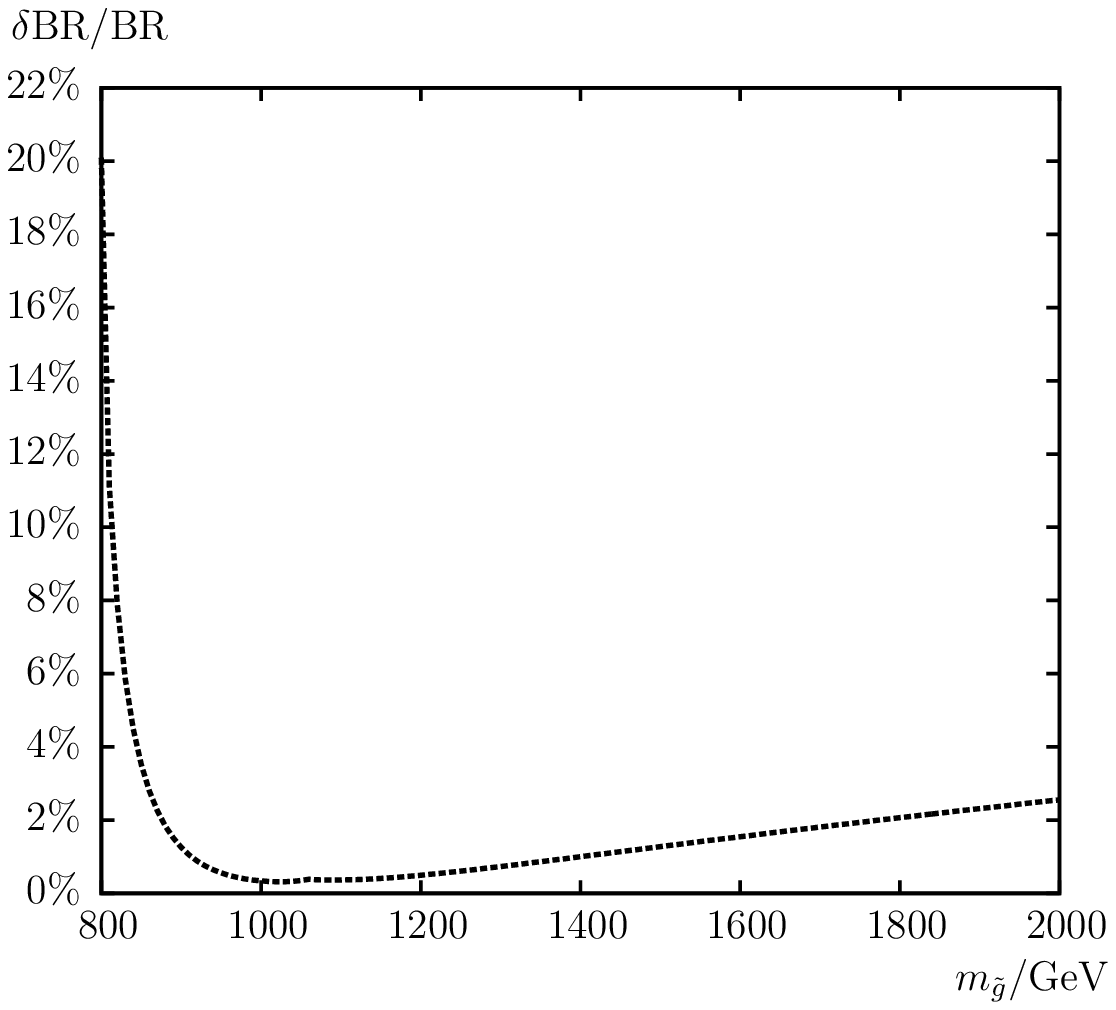}
\end{tabular}
\vspace{2em}
\caption{
  $\Ga(\decaySuz)$. Tree-level (``tree'') and full one-loop 
  (``full'') corrected decay widths are shown with the parameters 
  chosen according to \SE\ (see \refta{tab:para}), with $\mgl$ varied.
  The upper left plot shows the decay width, the upper right plot shows 
  the relative size of the corrections.
  Also shown are the pure SQCD corrections (``SQCD'').
  The lower left plot shows the BR, the lower right plot shows 
  the relative size of the BR.
}
\label{fig:AbsM3.glsu2u}
\end{center}
\end{figure}

\begin{figure}[htb!]
\begin{center}
\begin{tabular}{c}
\includegraphics[width=0.49\textwidth,height=8.0cm]{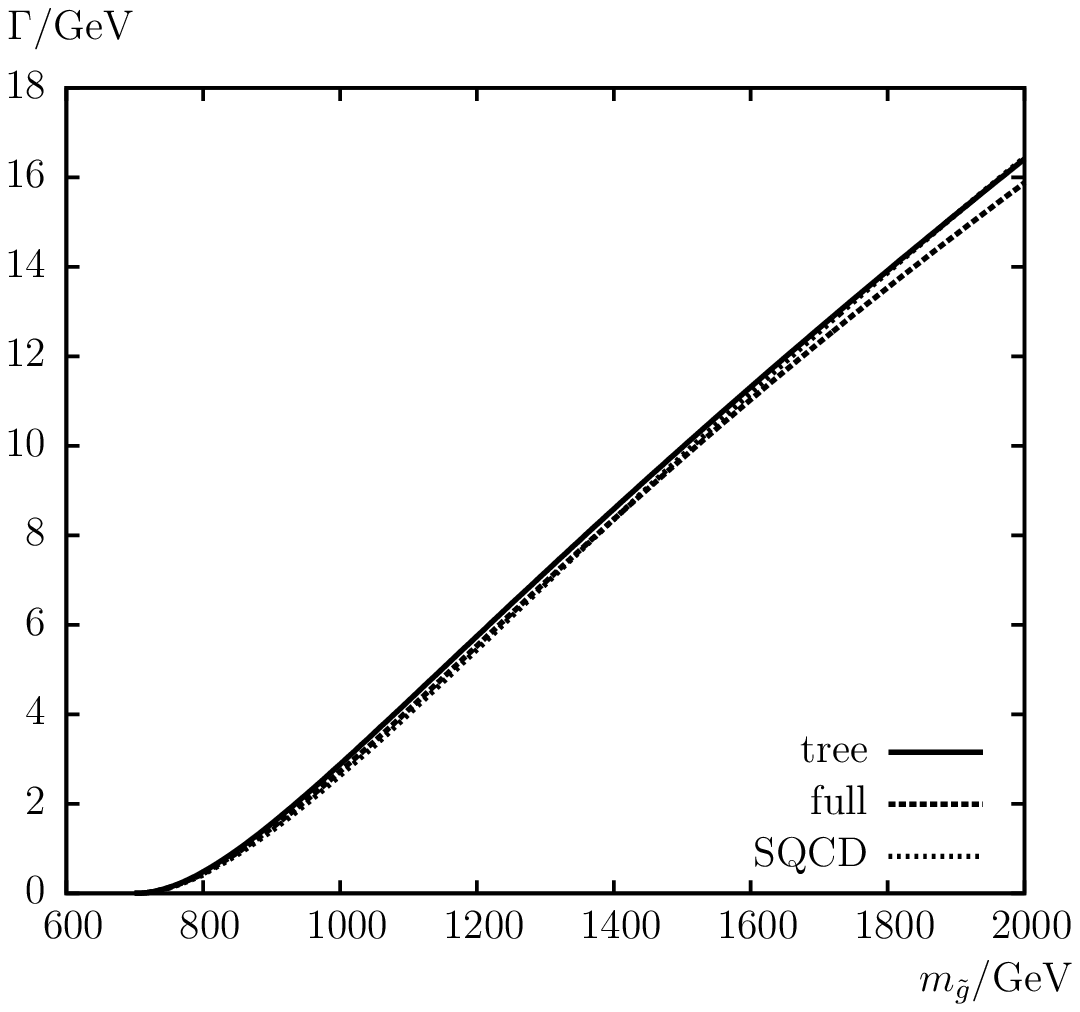}
\hspace{-4mm}
\includegraphics[width=0.49\textwidth,height=8.0cm]{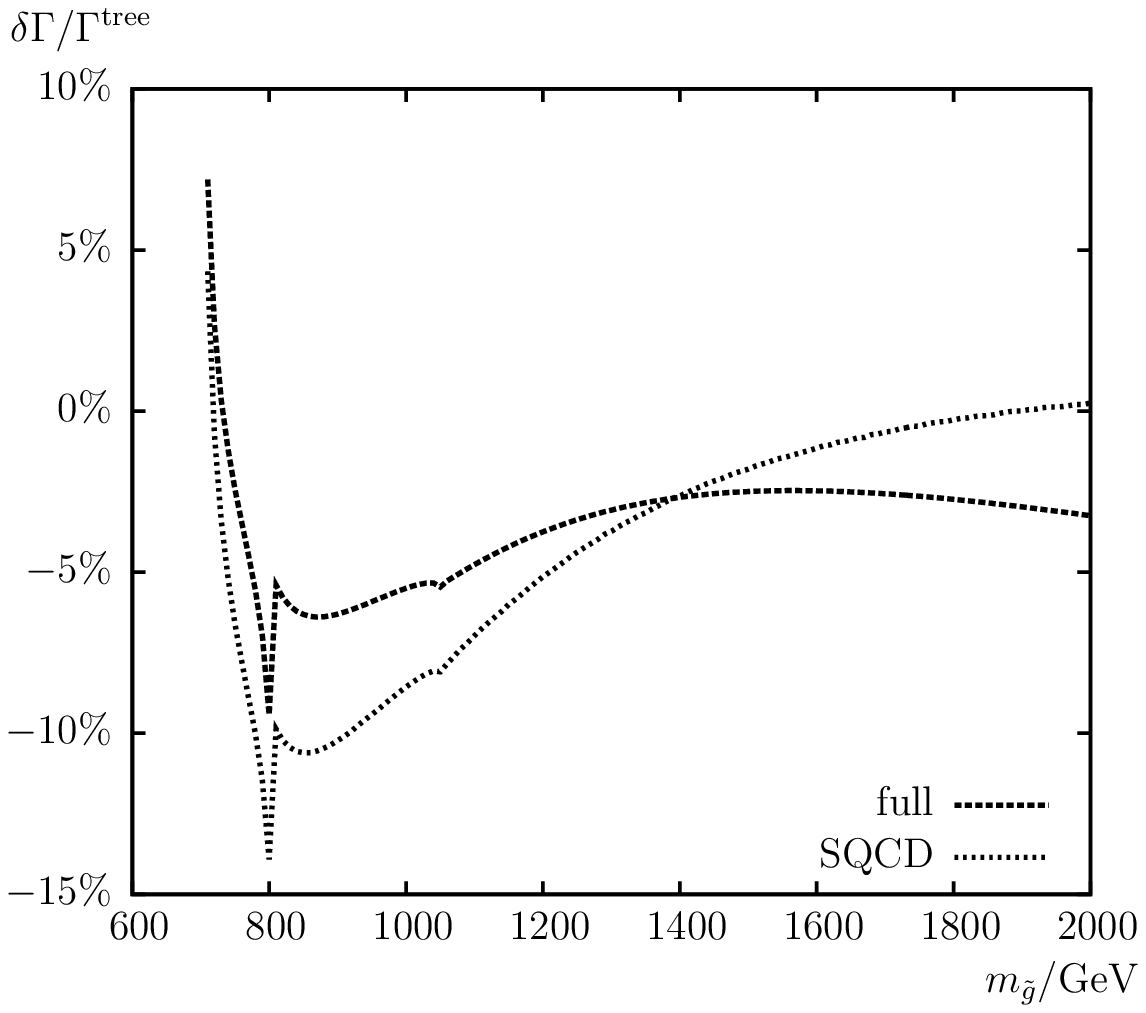}
\\[4em]
\includegraphics[width=0.49\textwidth,height=8.0cm]{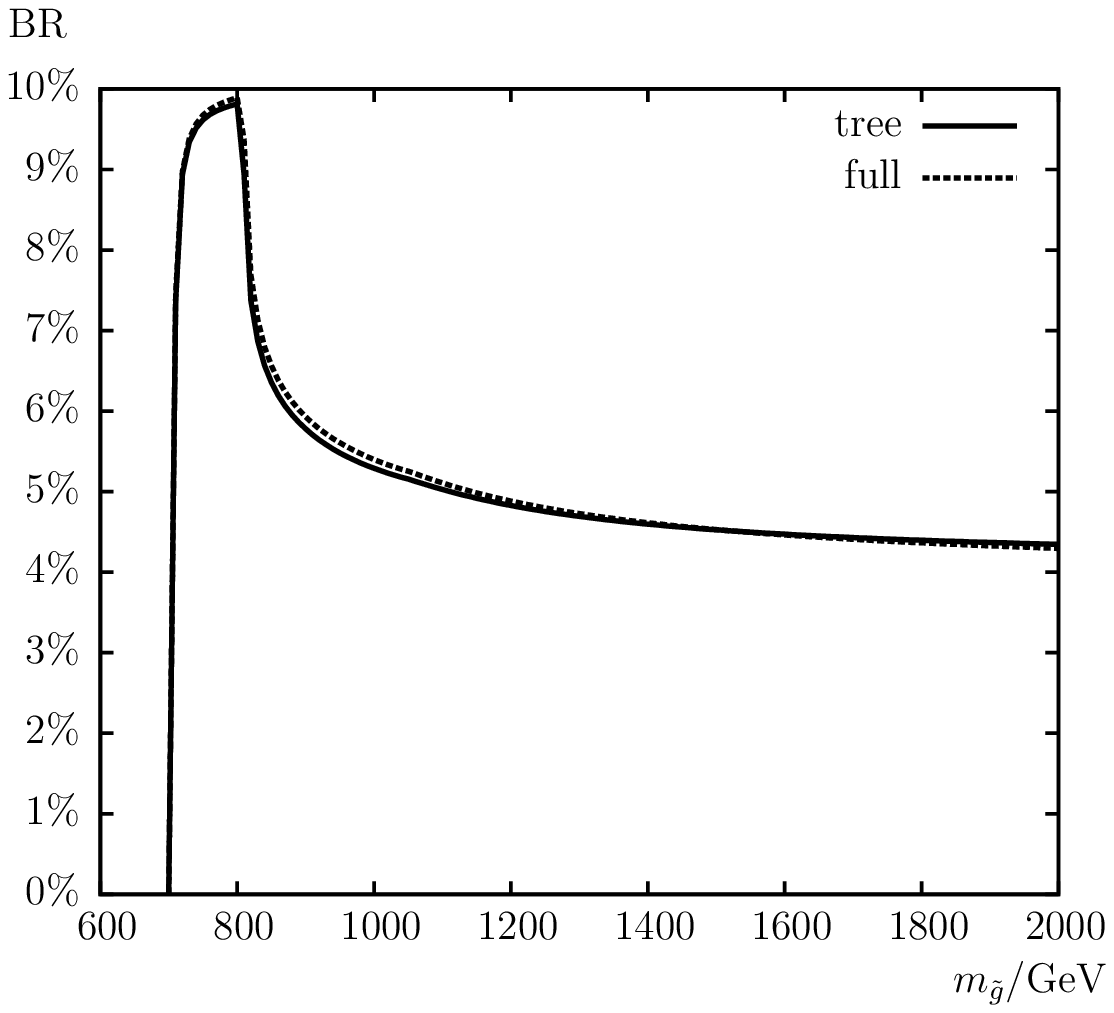}
\hspace{-4mm}
\includegraphics[width=0.49\textwidth,height=8.0cm]{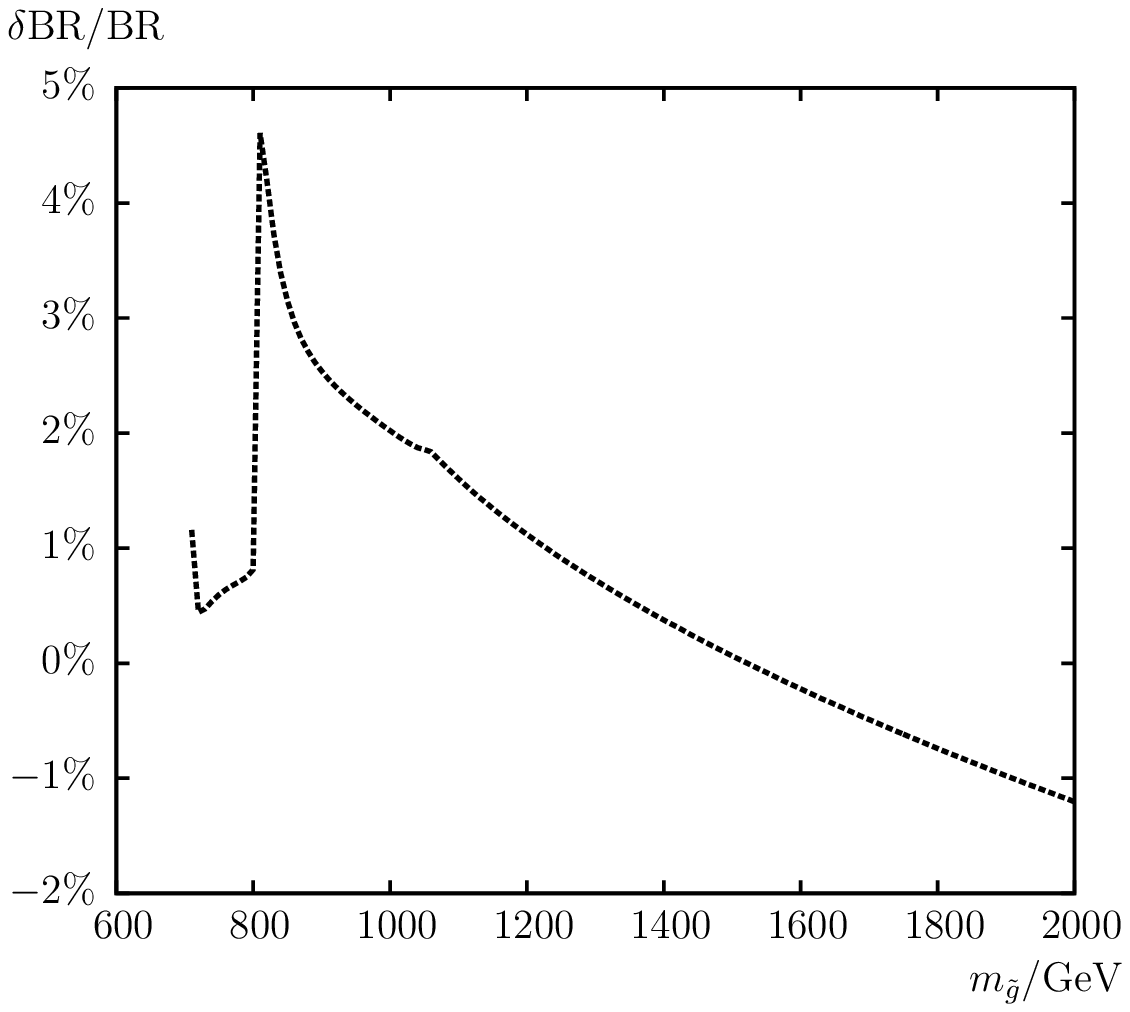}
\end{tabular}
\vspace{2em}
\caption{
  $\Ga(\decaySde)$. Tree-level (``tree'') and full one-loop 
  (``full'') corrected decay widths are shown with the parameters 
  chosen according to \SE\ (see \refta{tab:para}), with $\mgl$ varied.
  The upper left plot shows the decay width, the upper right plot shows 
  the relative size of the corrections.
  Also shown are the pure SQCD corrections (``SQCD'').
  The lower left plot shows the BR, the lower right plot shows 
  the relative size of the BR.
}
\label{fig:AbsM3.glsd1d}
\end{center}
\end{figure}

\begin{figure}[htb!]
\begin{center}
\begin{tabular}{c}
\includegraphics[width=0.49\textwidth,height=8.0cm]{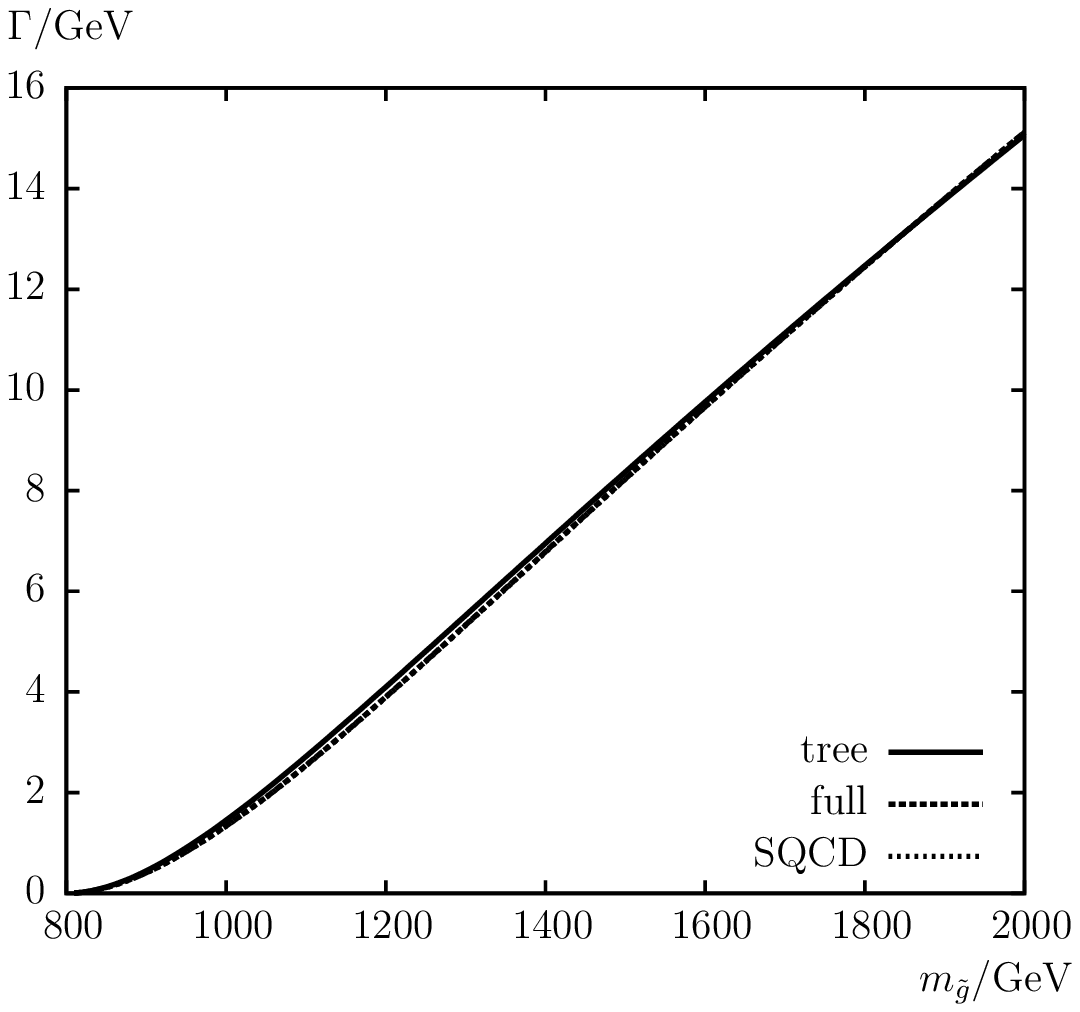}
\hspace{-4mm}
\includegraphics[width=0.49\textwidth,height=8.0cm]{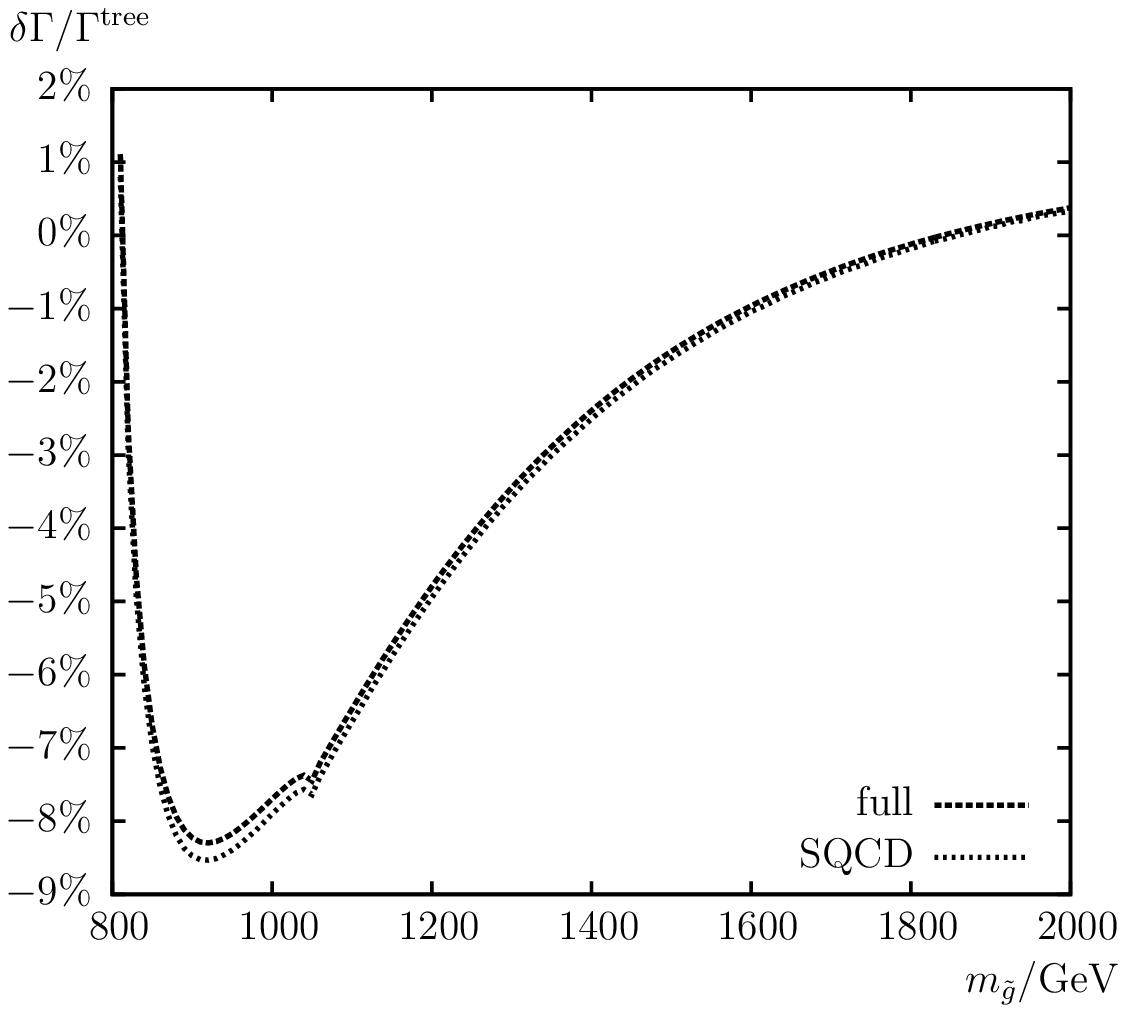}
\\[4em]
\includegraphics[width=0.49\textwidth,height=8.0cm]{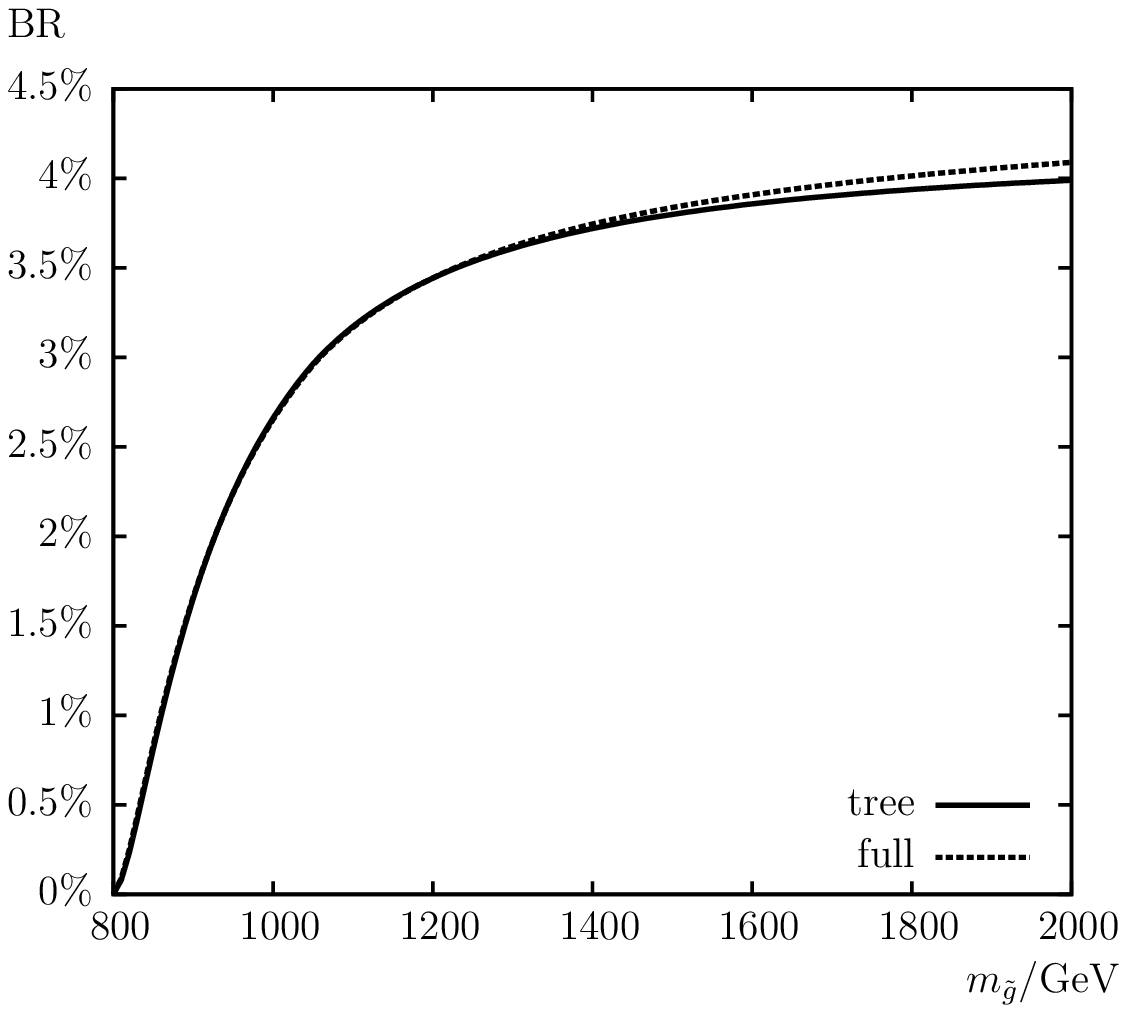}
\hspace{-4mm}
\includegraphics[width=0.49\textwidth,height=8.0cm]{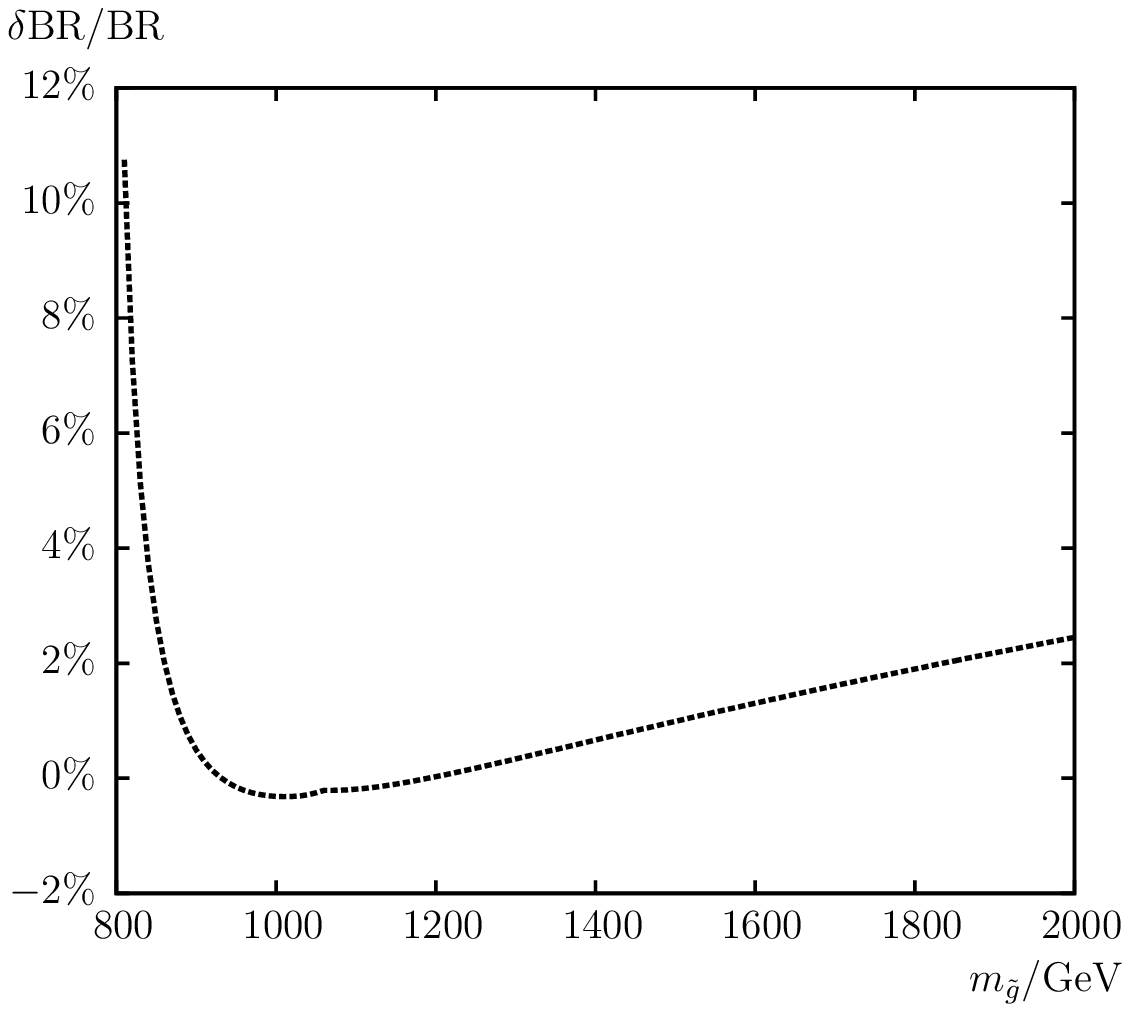}
\end{tabular}
\vspace{2em}
\caption{
  $\Ga(\decaySdz)$. Tree-level (``tree'') and full one-loop 
  (``full'') corrected decay widths are shown with the parameters 
  chosen according to \SE\ (see \refta{tab:para}), with $\mgl$ varied.
  The upper left plot shows the decay width, the upper right plot shows 
  the relative size of the corrections.
  Also shown are the pure SQCD corrections (``SQCD'').
  The lower left plot shows the BR, the lower right plot shows 
  the relative size of the BR.
}
\label{fig:AbsM3.glsd2d}
\end{center}
\end{figure}

\clearpage
\newpage


\subsection{Full one-loop results for varying \boldmath{$\phigl$}}
\label{sec:full1Lphigl}

In this subsection we analyze the various decay widths and branching
ratios as a function of $\phigl$. 
The other parameters are chosen according to \refta{tab:para}. 

When performing an analysis involving complex parameters
it should be noted that the results for physical observables are
affected only 
by certain combinations of the complex phases of the 
parameters $\mu$, the trilinear couplings $\At$, $\Ab$, \ldots, and the
gaugino mass parameters $M_1$, $M_2$,
$M_3$~\cite{MSSMcomplphasen,SUSYphases}.
It is possible, for instance, to rotate the phase $\phiMz$ away,
and we remain with $\phiMe$ and $\phiMd \equiv \phigl$ as independent
physical phases. 
Experimental constraints on the (combinations of) complex phases 
arise in particular from their contributions to electric dipole moments of
heavy quarks~\cite{EDMDoink}, of the electron and 
the neutron (see \citeres{EDMrev2,EDMPilaftsis} and references therein), 
and of deuteron~\cite{EDMRitz}, \citeres{EDMrev1,EDMrev2,EDMrev3} for
reviews. The phase of the gluino mass parameter remains relatively
weakly constrained. Since our decaying particle is the gluino
we focus here on the dependence on $\phigl$ and set all other phases to
zero. Analyses of the full one-loop effects of SUSY decays on $\phiab$,
$\phiat$ and $\phiMe$ can be found in
\citeres{SbotRen,Stop2decay,LHCxC}, respectively.

Since now the complex phase of $M_3$ can appear in the couplings,
two effects arise: first a difference between $\Ga(\decayaSqi)$ and
$\Ga(\decaySqai)$ is observed, and consequently we now show the results
separately for these two decay channels (but stick to the notation
``$\decaySqi$'' when referring in general to a plot). These differences 
arise naturally from the combination of an imaginary (absorptive) part
of a decay diagram and a complex coupling. As required, we always find
$\Ga(\decayaSqi)|_{\phigl} = \Ga(\decaySqai)|_{-\phigl}$. 
A second source of this type of contributions comes 
from absorptive parts of self-energy type corrections on
external legs (called ``absorptive self-energy contributions''
from now on), and they also have been included according to the 
formulas given in \refse{sec:gluino}. 
For $\phigl = 0, \pi, 2\pi$ these ``absorptive effects'' vanish 
(by construction).

\begin{figure}[htb!]
\begin{center}
\begin{tabular}{c}
\includegraphics[width=0.49\textwidth,height=8.0cm]{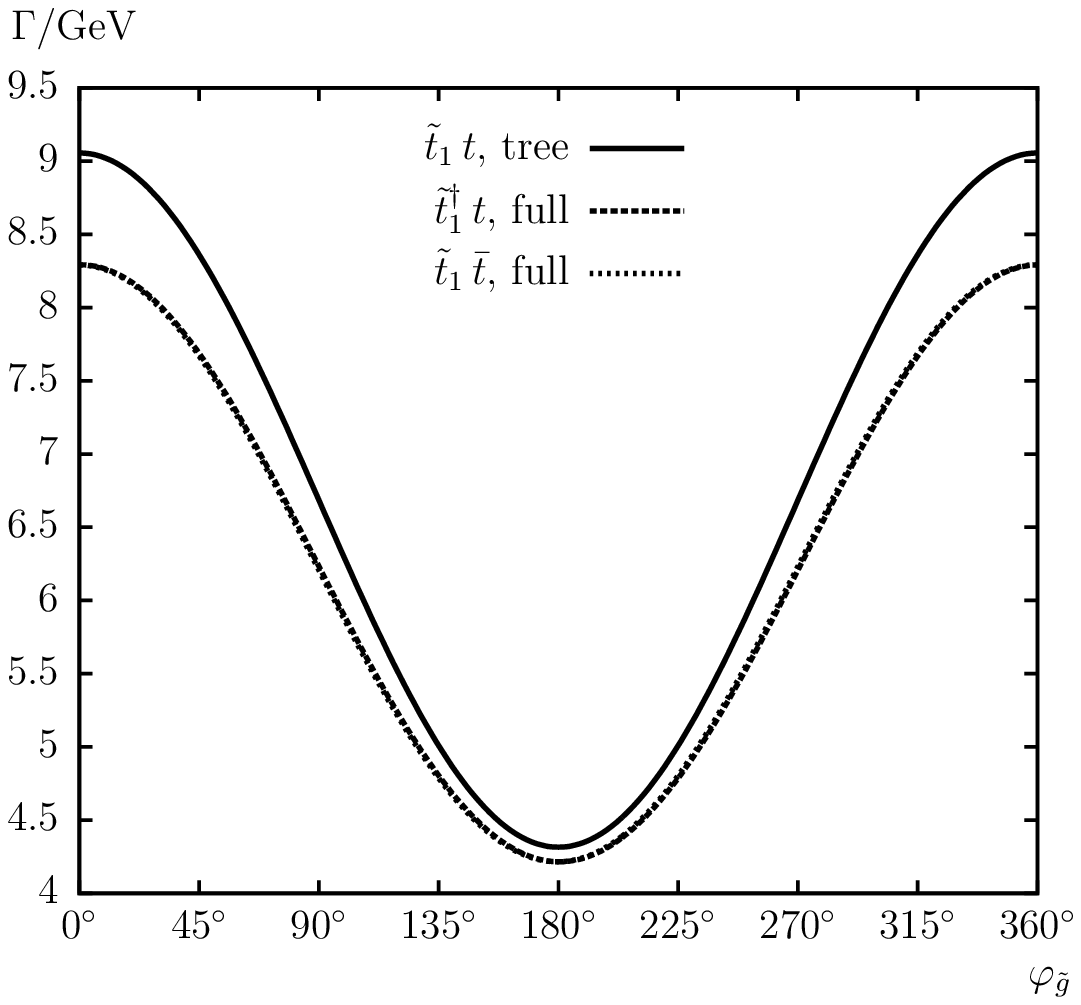}
\hspace{-4mm}
\includegraphics[width=0.49\textwidth,height=8.0cm]{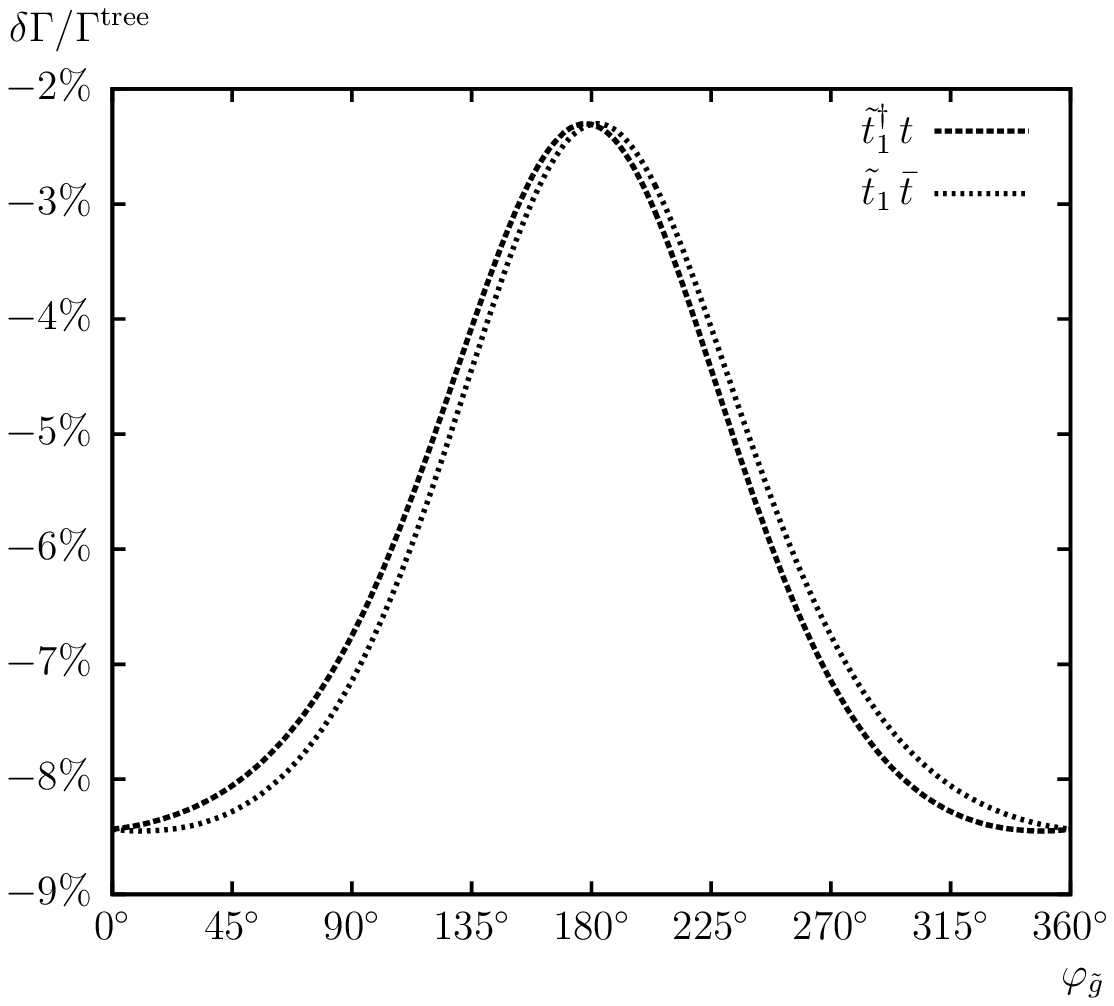}
\\[4em]
\includegraphics[width=0.49\textwidth,height=8.0cm]{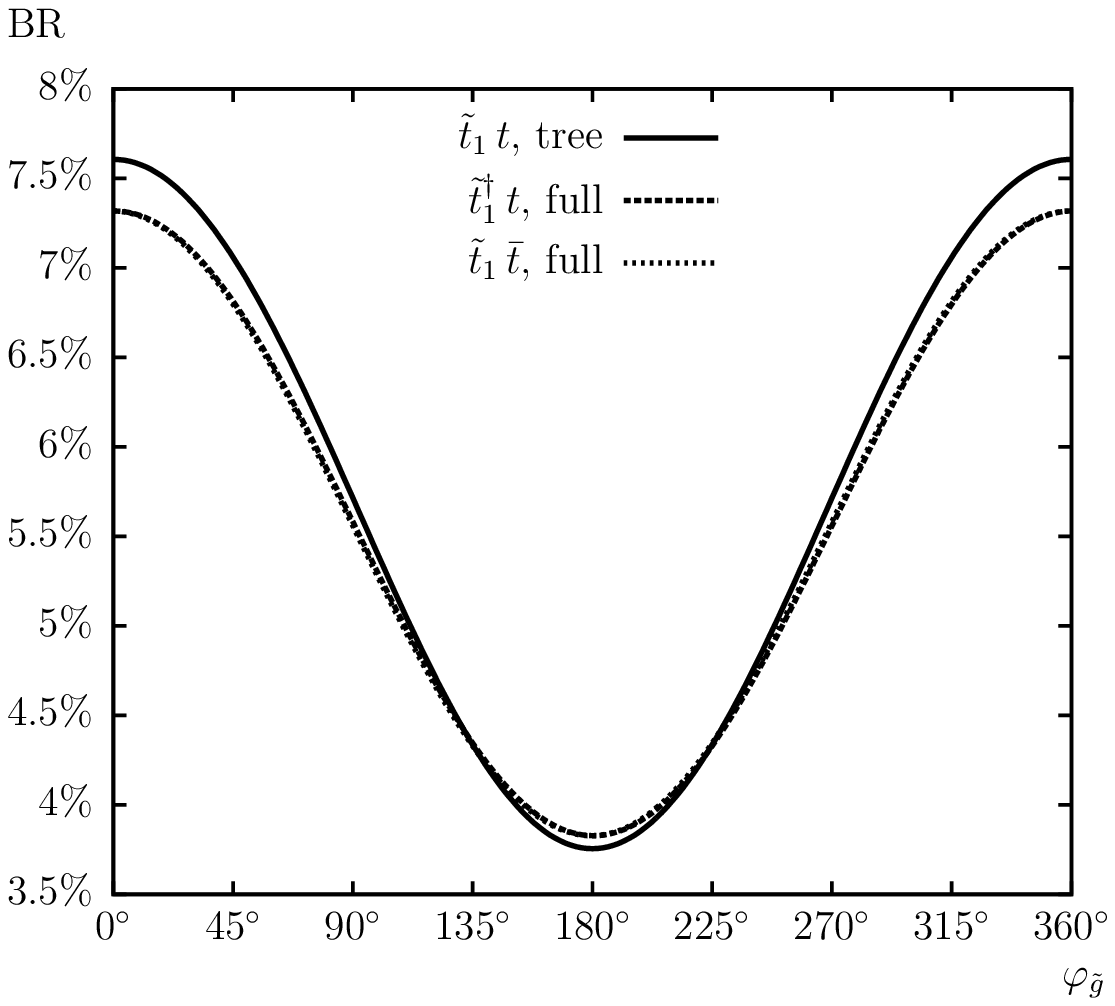}
\hspace{-4mm}
\includegraphics[width=0.49\textwidth,height=8.0cm]{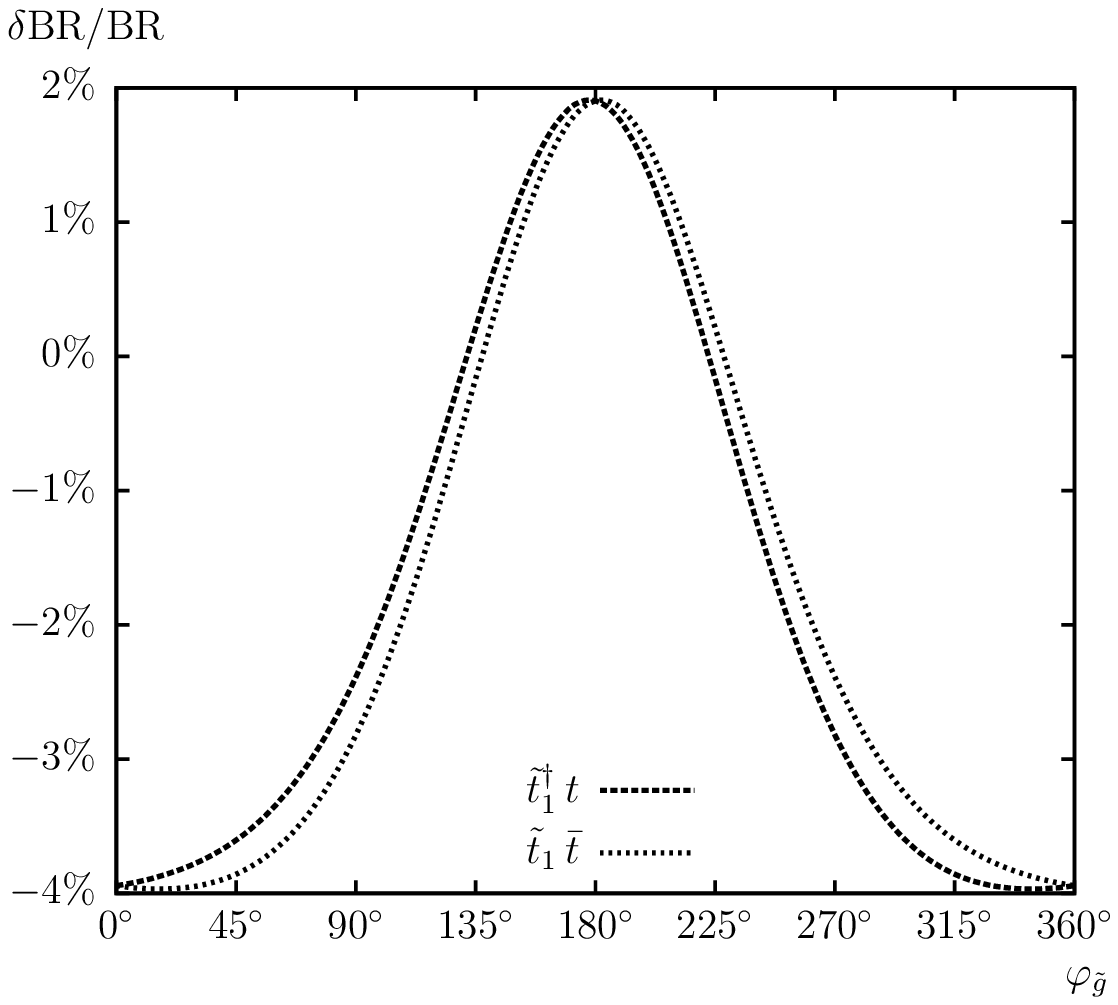}
\end{tabular}
\vspace{2em}
\caption{$\Ga(\decaySte)$.
  Tree-level (``tree'') and full one-loop (``full'') corrected 
  decay widths (including absorptive self-energy contributions) are shown.
  The parameters are chosen according to \SE\ (see \refta{tab:para}), 
  with $\phigl$ varied.
  The upper left plot shows the decay width, the upper right plot shows 
  the relative size of the corrections. 
  The lower left plot shows the BR, the lower right plot shows 
  the relative size of the BR.
}
\label{fig:PhiM3.glst1t}
\end{center}
\end{figure}

\begin{figure}[htb!]
\begin{center}
\begin{tabular}{c}
\includegraphics[width=0.49\textwidth,height=8.0cm]{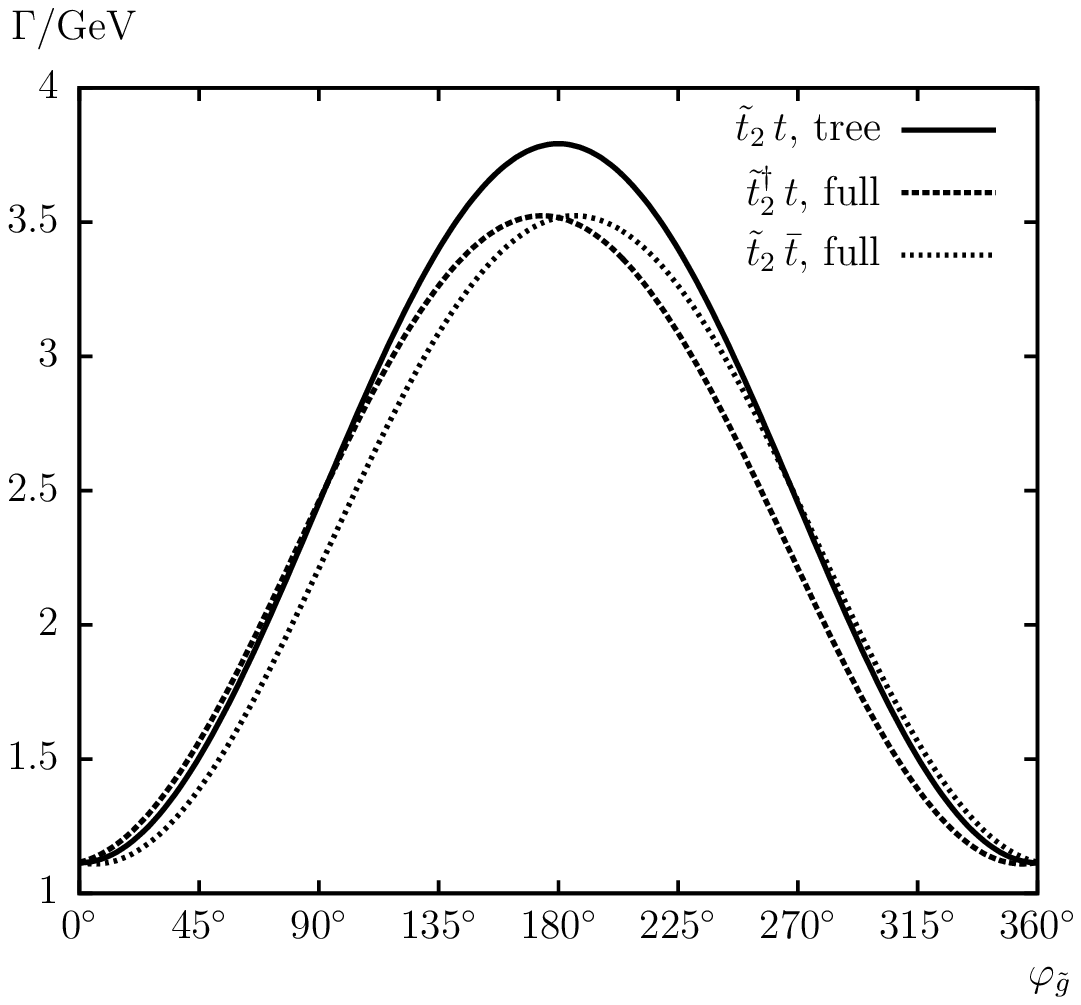}
\hspace{-4mm}
\includegraphics[width=0.49\textwidth,height=8.0cm]{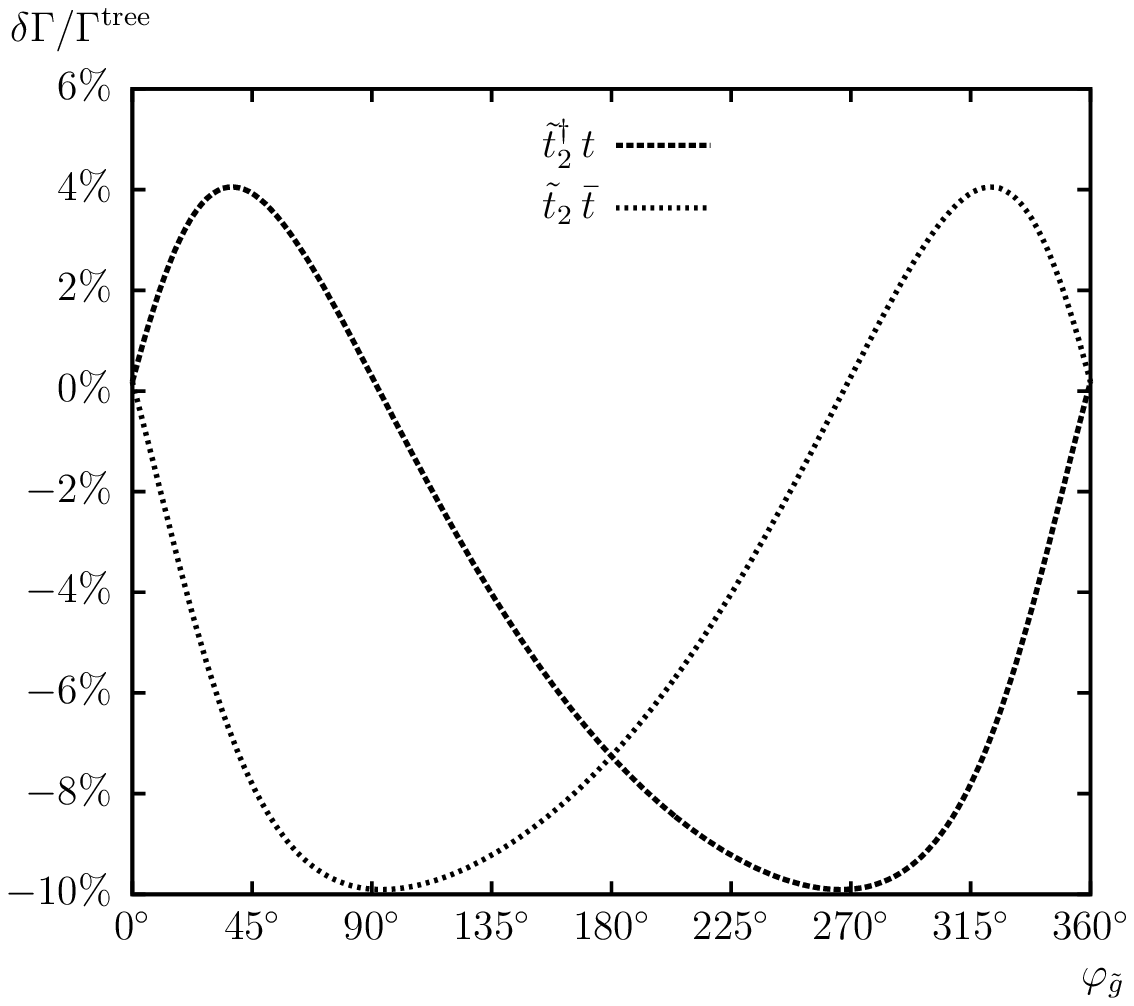}
\\[4em]
\includegraphics[width=0.49\textwidth,height=8.0cm]{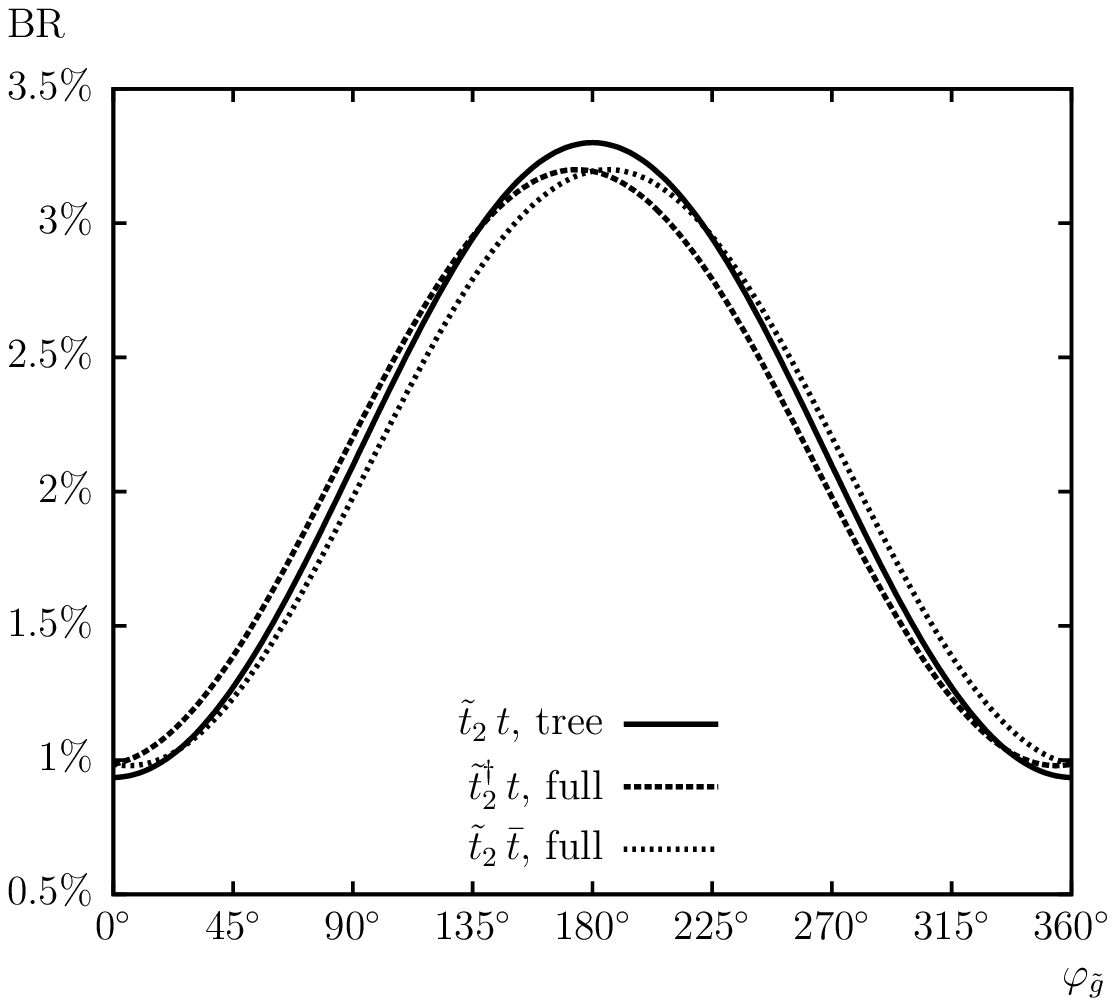}
\hspace{-4mm}
\includegraphics[width=0.49\textwidth,height=8.0cm]{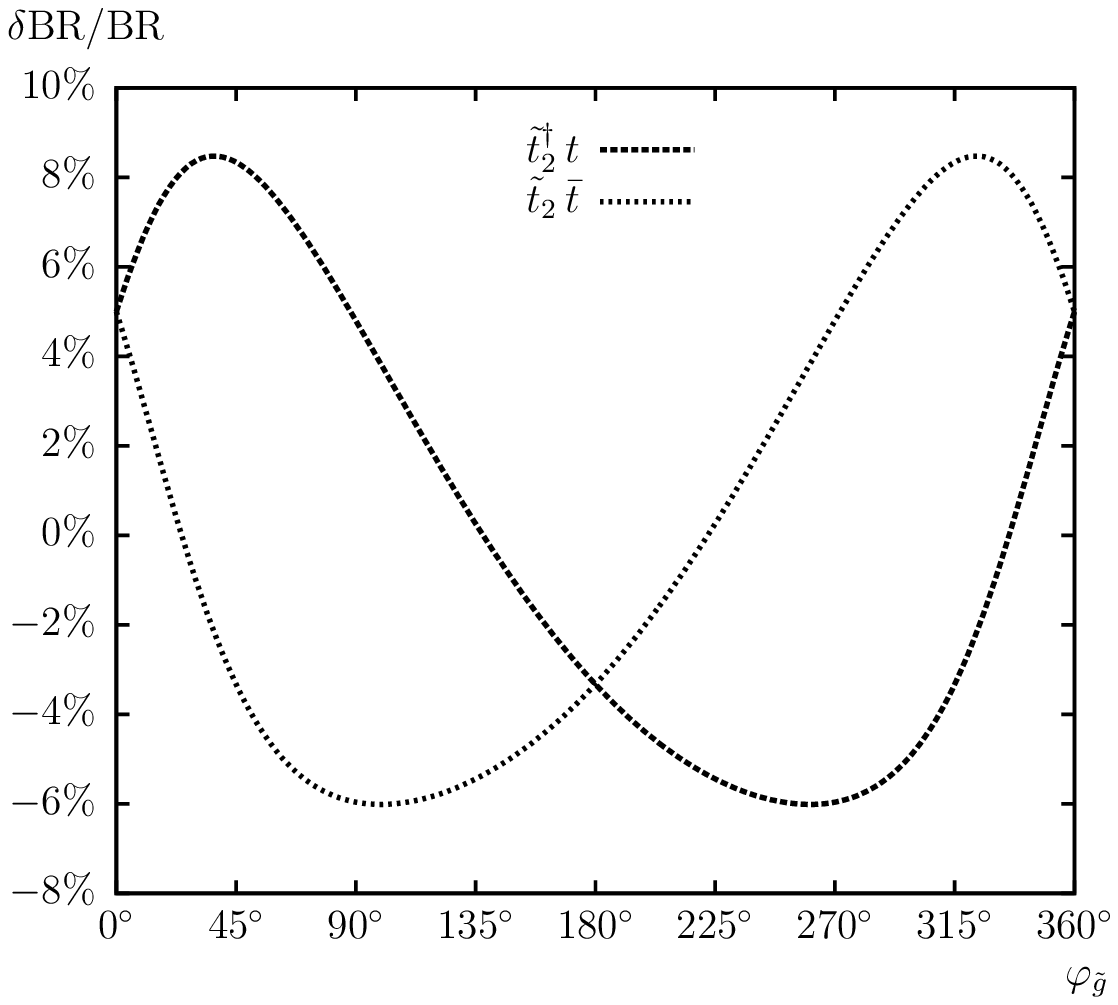}
\end{tabular}
\vspace{2em}
\caption{$\Ga(\decayStz)$.
  Tree-level (``tree'') and full one-loop (``full'') corrected 
  decay widths (including absorptive self-energy contributions) are shown. 
  The parameters are chosen according to \SE\ (see \refta{tab:para}), 
  with $\phigl$ varied.
  The upper left plot shows the decay width, the upper right plot shows 
  the relative size of the corrections. 
  The lower left plot shows the BR, the lower right plot shows 
  the relative size of the BR.
}
\label{fig:PhiM3.glst2t}
\end{center}
\end{figure}

\begin{figure}[htb!]
\begin{center}
\begin{tabular}{c}
\includegraphics[width=0.49\textwidth,height=8.0cm]{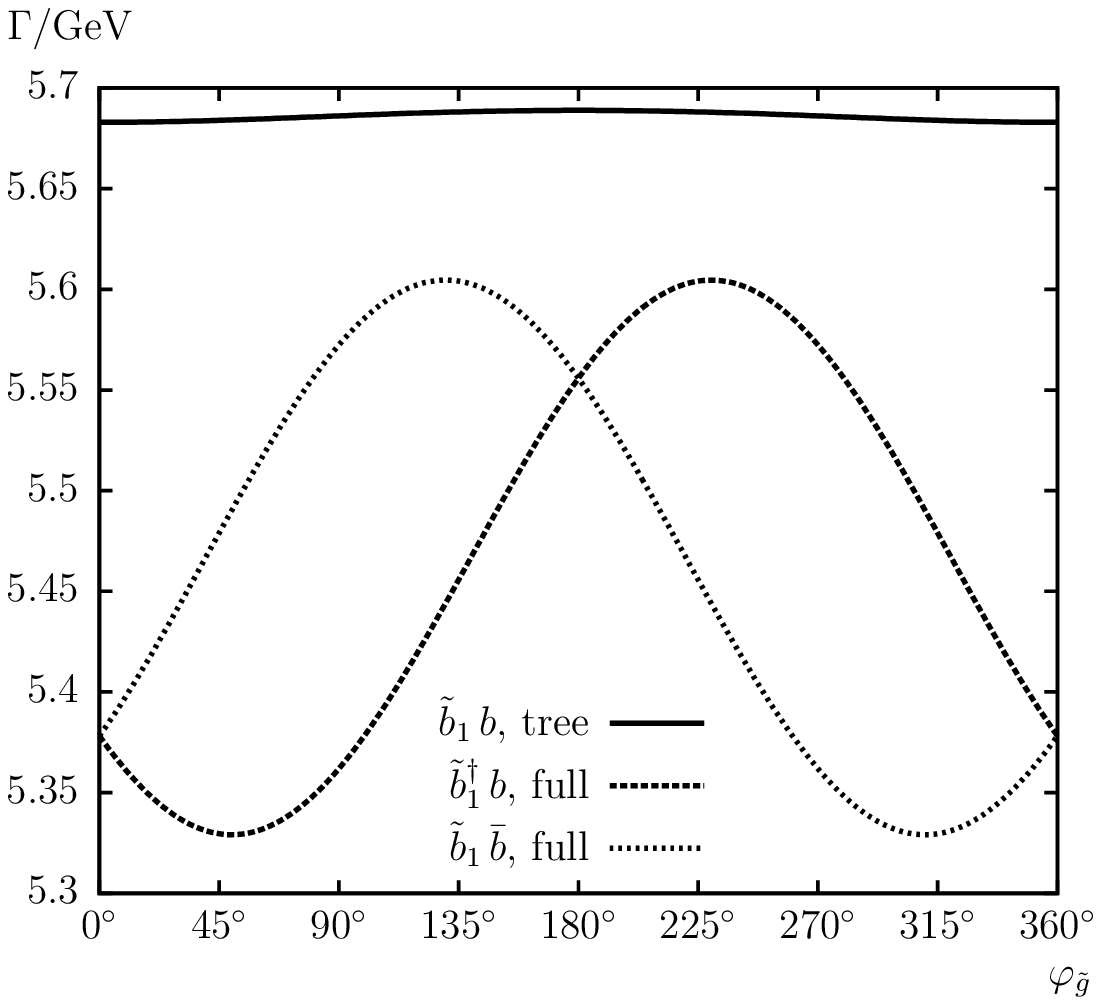}
\hspace{-4mm}
\includegraphics[width=0.49\textwidth,height=8.0cm]{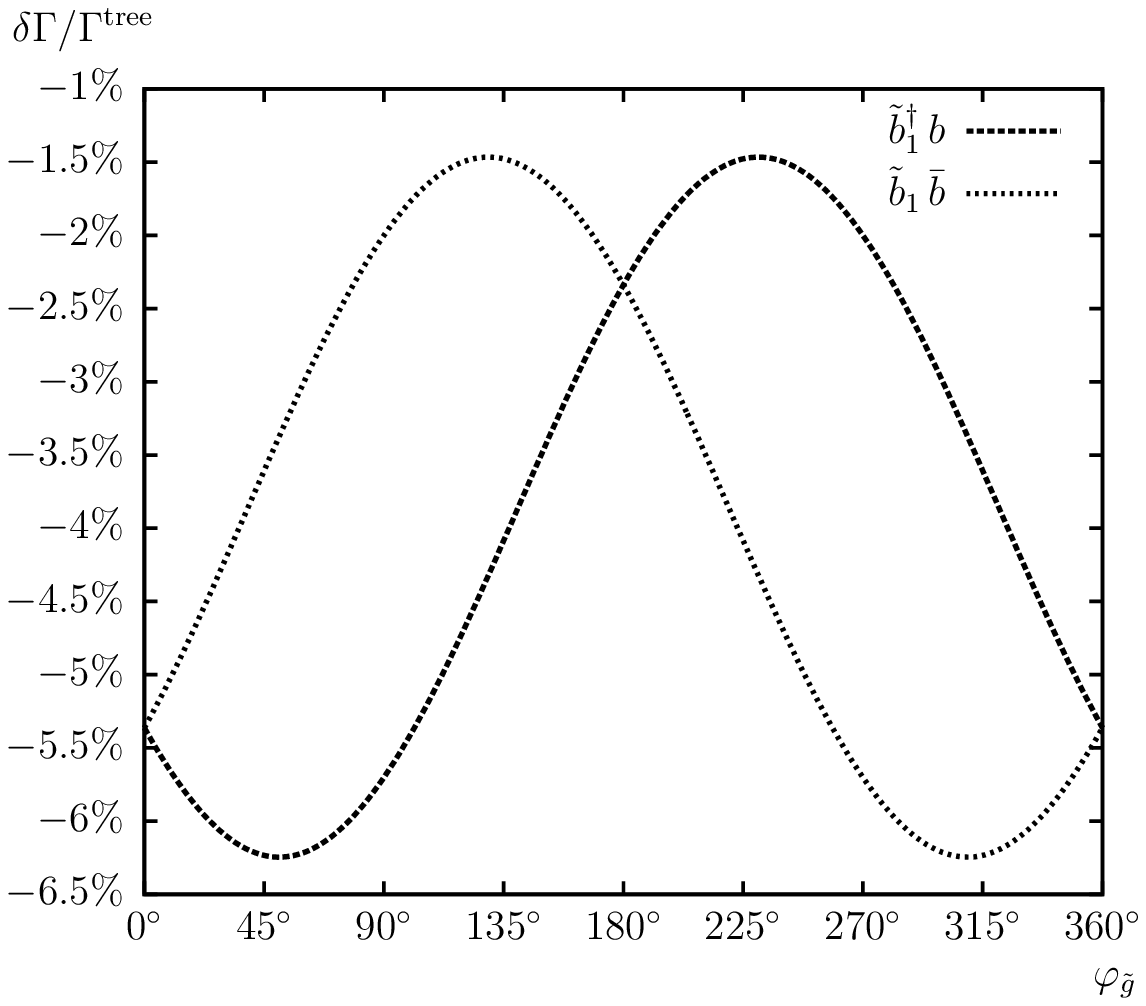}
\\[4em]
\includegraphics[width=0.49\textwidth,height=8.0cm]{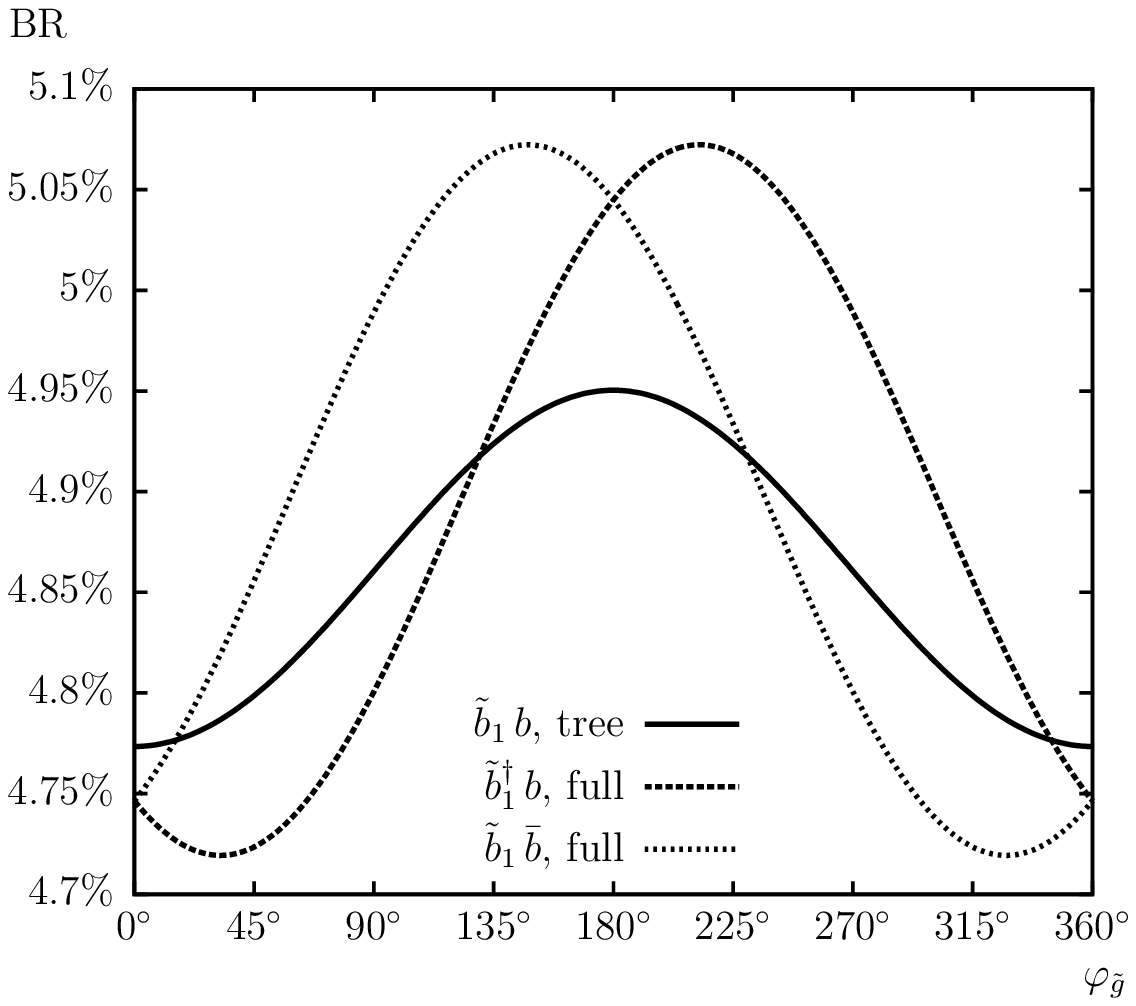}
\hspace{-4mm}
\includegraphics[width=0.49\textwidth,height=8.0cm]{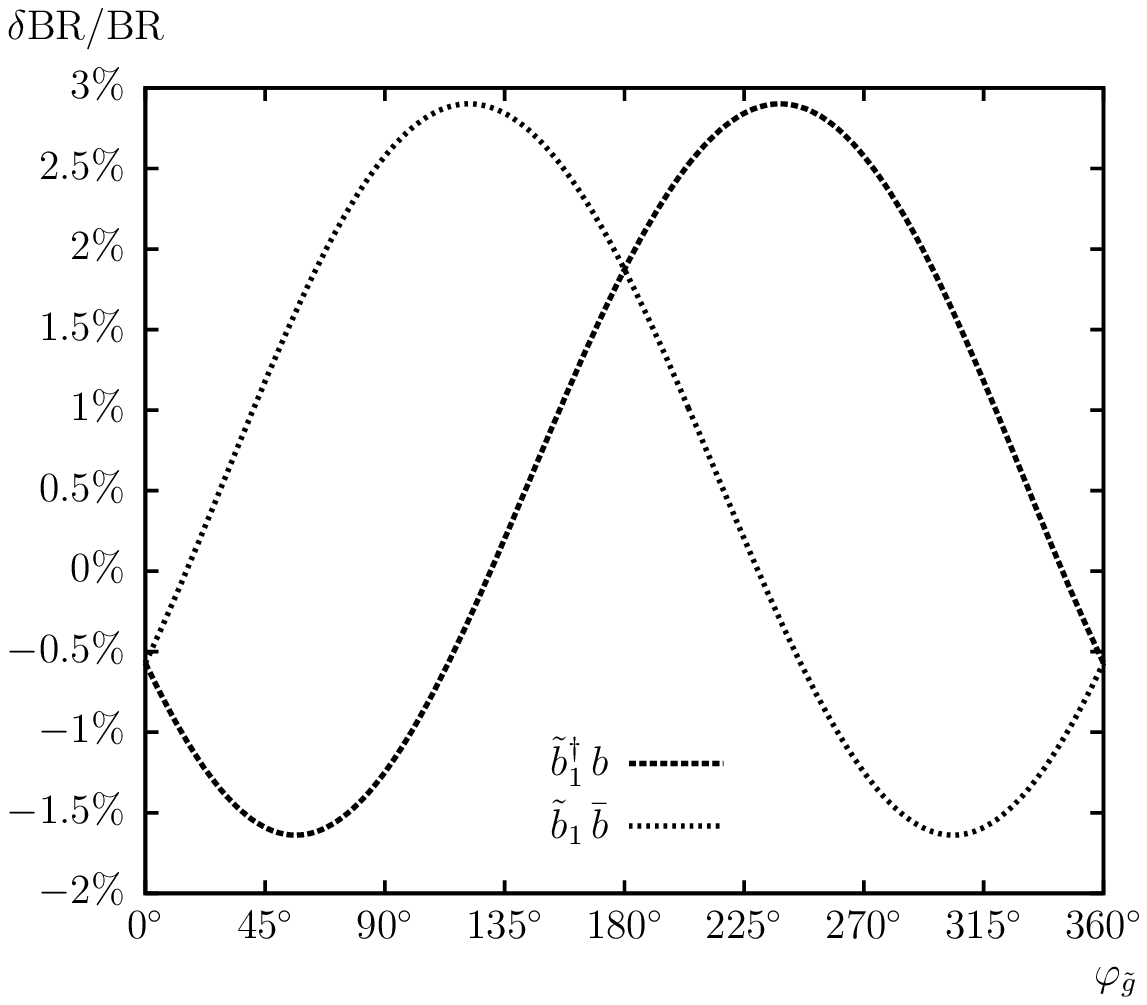}
\end{tabular}
\vspace{2em}
\caption{$\Ga(\decaySbe)$.
  Tree-level (``tree'') and full one-loop (``full'') corrected 
  decay widths (including absorptive self-energy contributions) are shown. 
  The parameters are chosen according to \SE\ (see \refta{tab:para}), 
  with $\phigl$ varied.
  The upper left plot shows the decay width, the upper right plot shows 
  the relative size of the corrections. 
  The lower left plot shows the BR, the lower right plot shows 
  the relative size of the BR.
}
\label{fig:PhiM3.glsb1b}
\end{center}
\end{figure}

\begin{figure}[htb!]
\begin{center}
\begin{tabular}{c}
\includegraphics[width=0.49\textwidth,height=8.0cm]{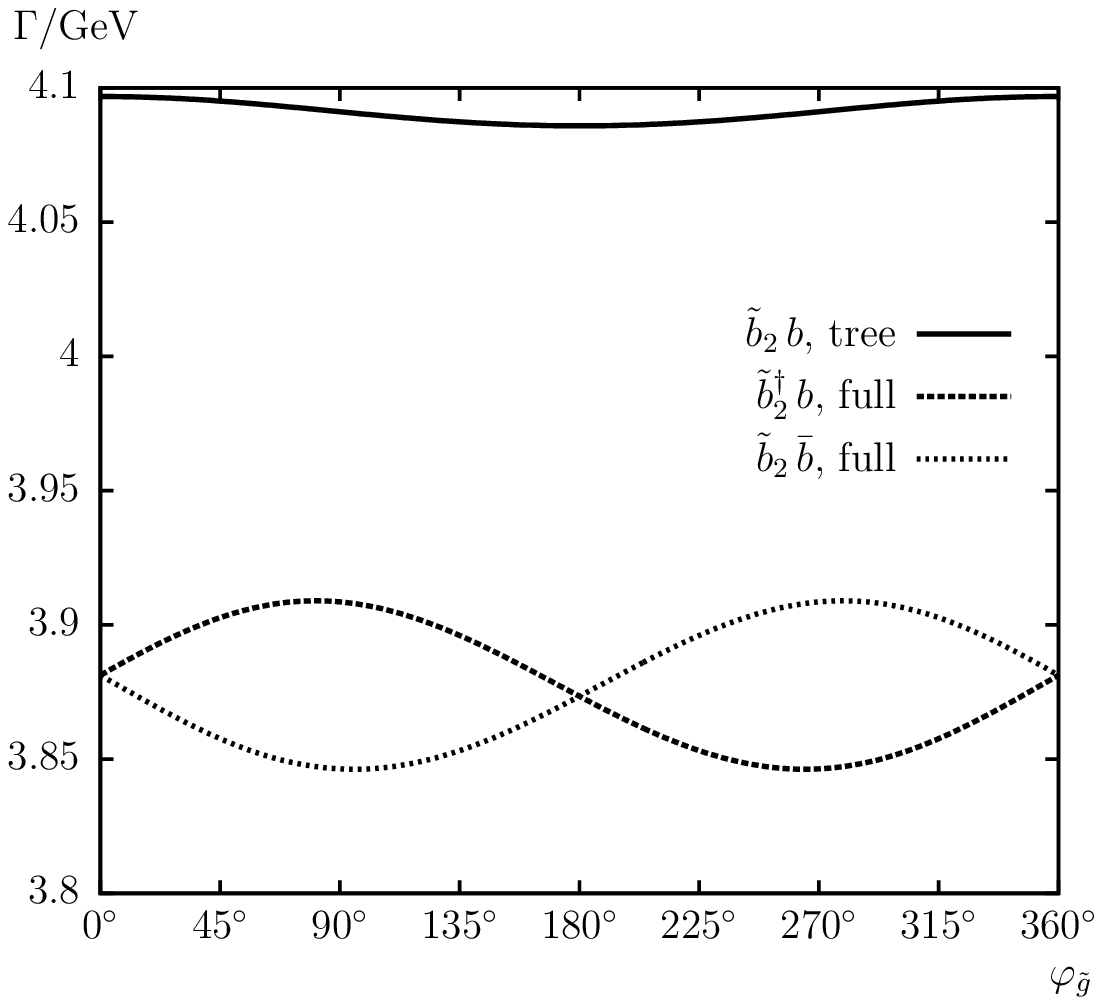}
\hspace{-4mm}
\includegraphics[width=0.49\textwidth,height=8.0cm]{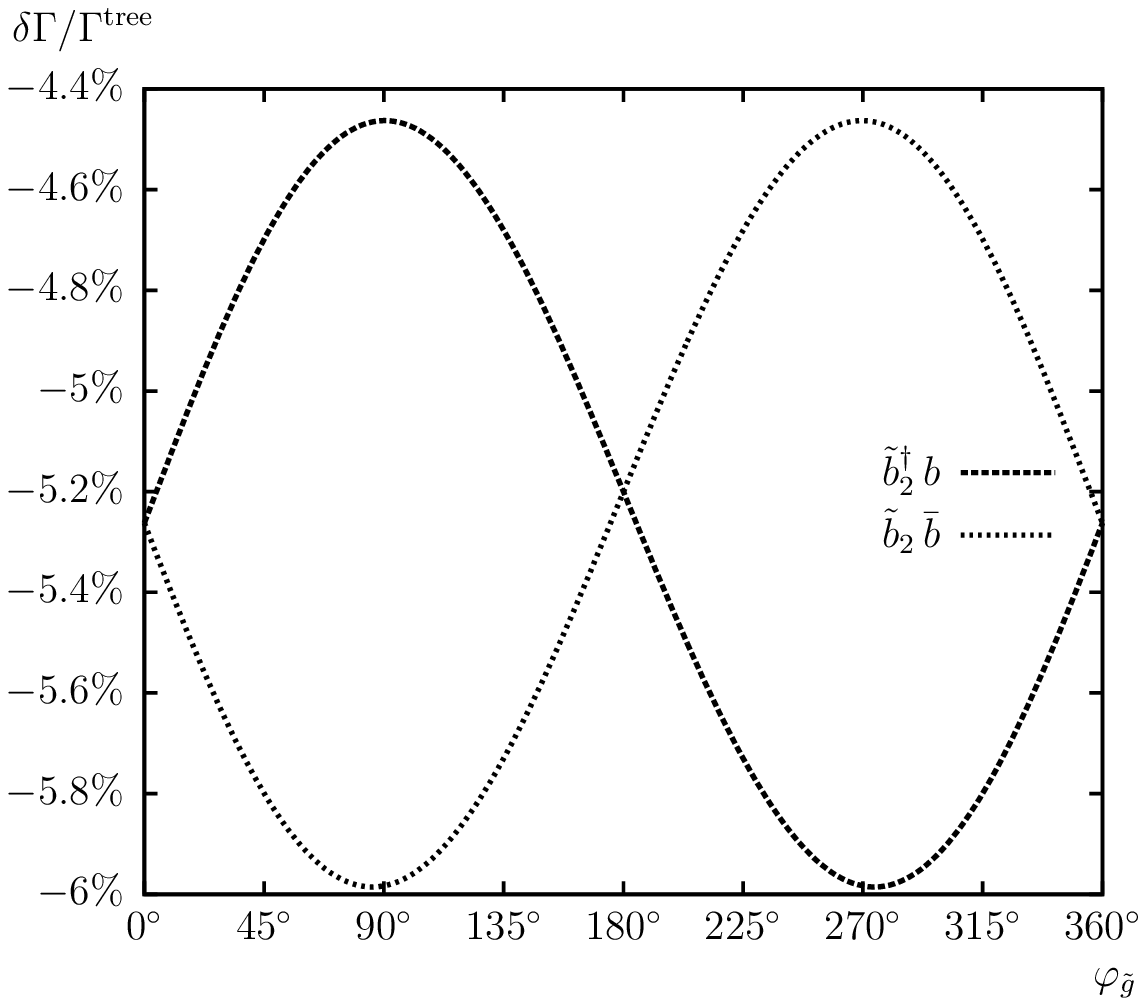}
\\[4em]
\includegraphics[width=0.49\textwidth,height=8.0cm]{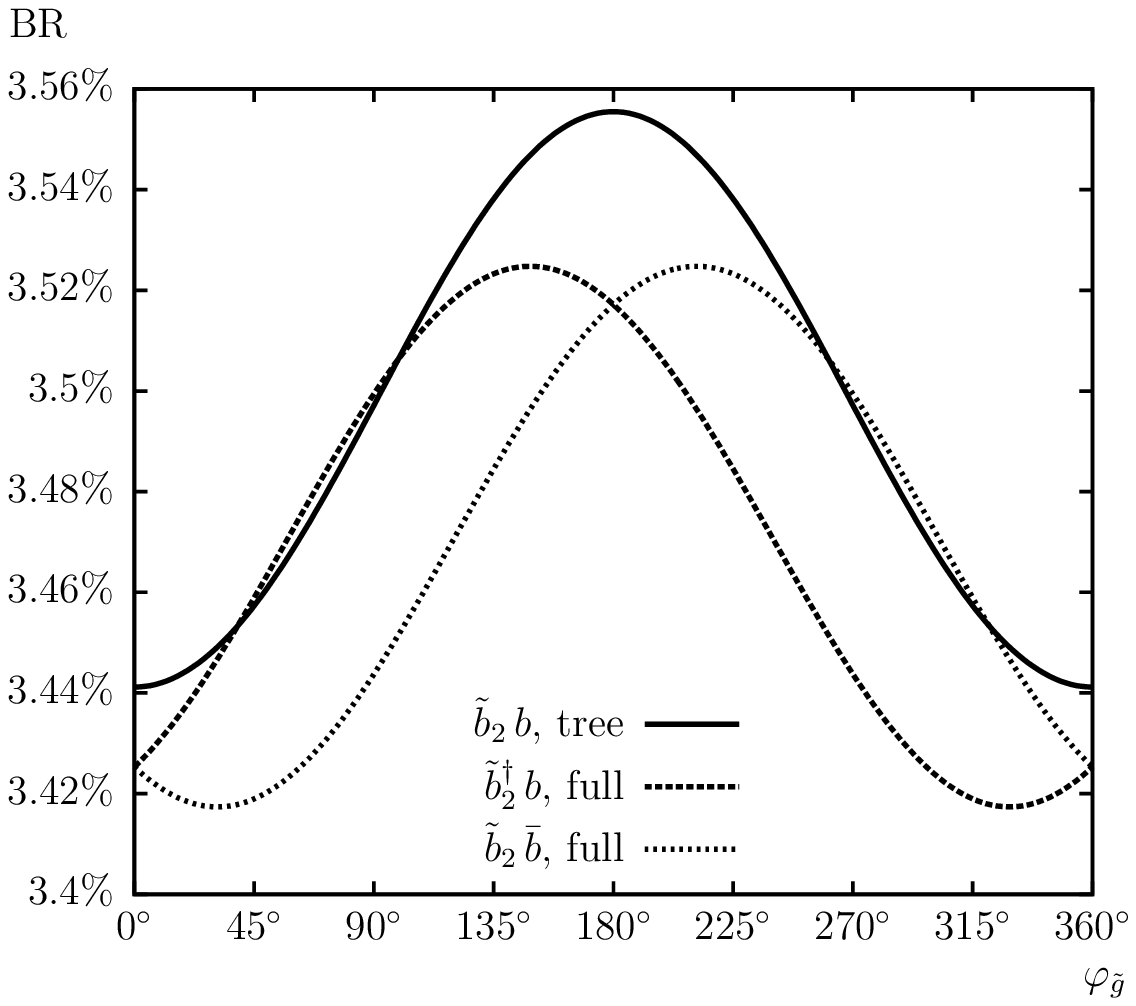}
\hspace{-4mm}
\includegraphics[width=0.49\textwidth,height=8.0cm]{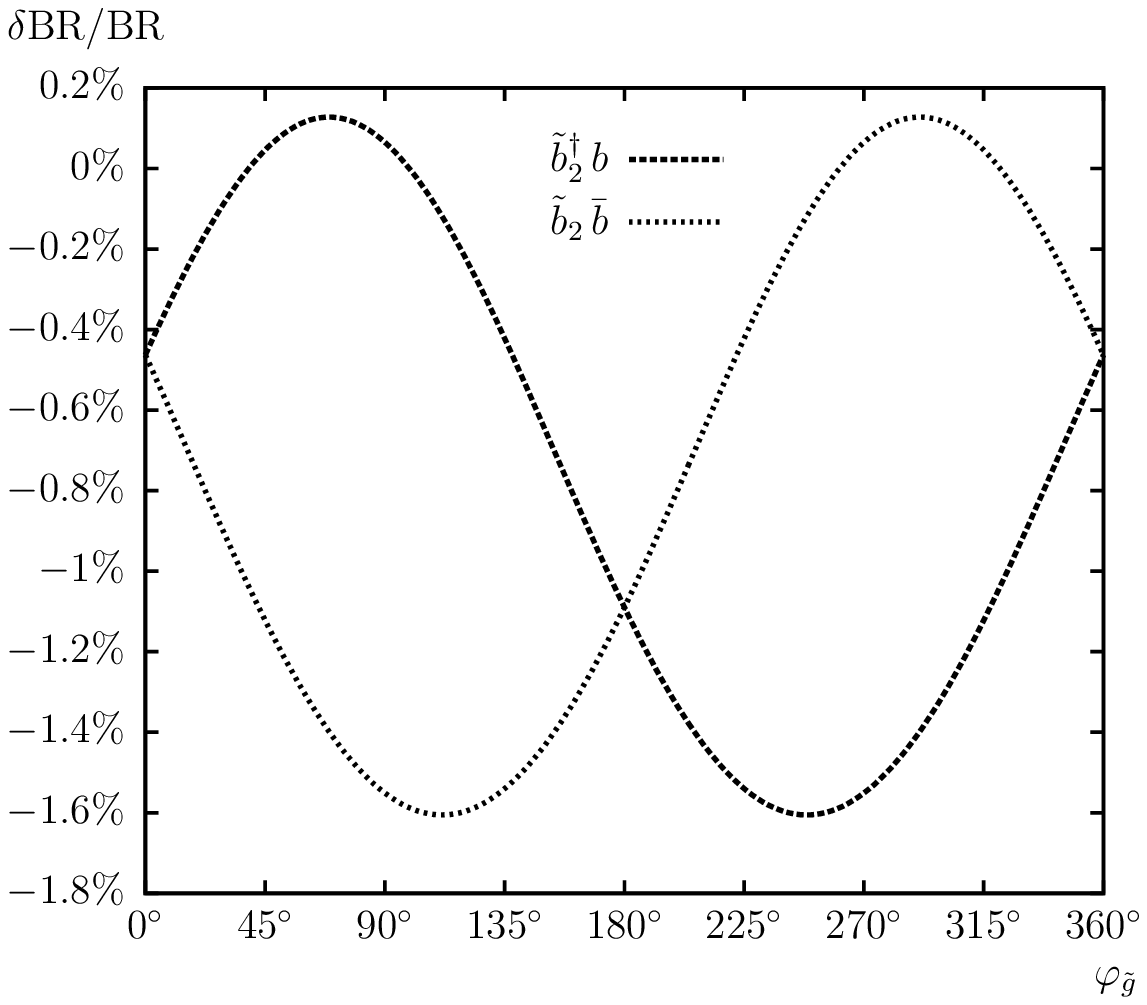}
\end{tabular}
\vspace{2em}
\caption{$\Ga(\decaySbz)$.
  Tree-level (``tree'') and full one-loop (``full'') corrected 
  decay widths (including absorptive self-energy contributions) are shown. 
  The parameters are chosen according to \SE\ (see \refta{tab:para}), 
  with $\phigl$ varied.
  The upper left plot shows the decay width, the upper right plot shows 
  the relative size of the corrections. 
  The lower left plot shows the BR, the lower right plot shows 
  the relative size of the BR.
}
\label{fig:PhiM3.glsb2b}
\end{center}
\end{figure}

\begin{figure}[htb!]
\begin{center}
\begin{tabular}{c}
\includegraphics[width=0.49\textwidth,height=8.0cm]{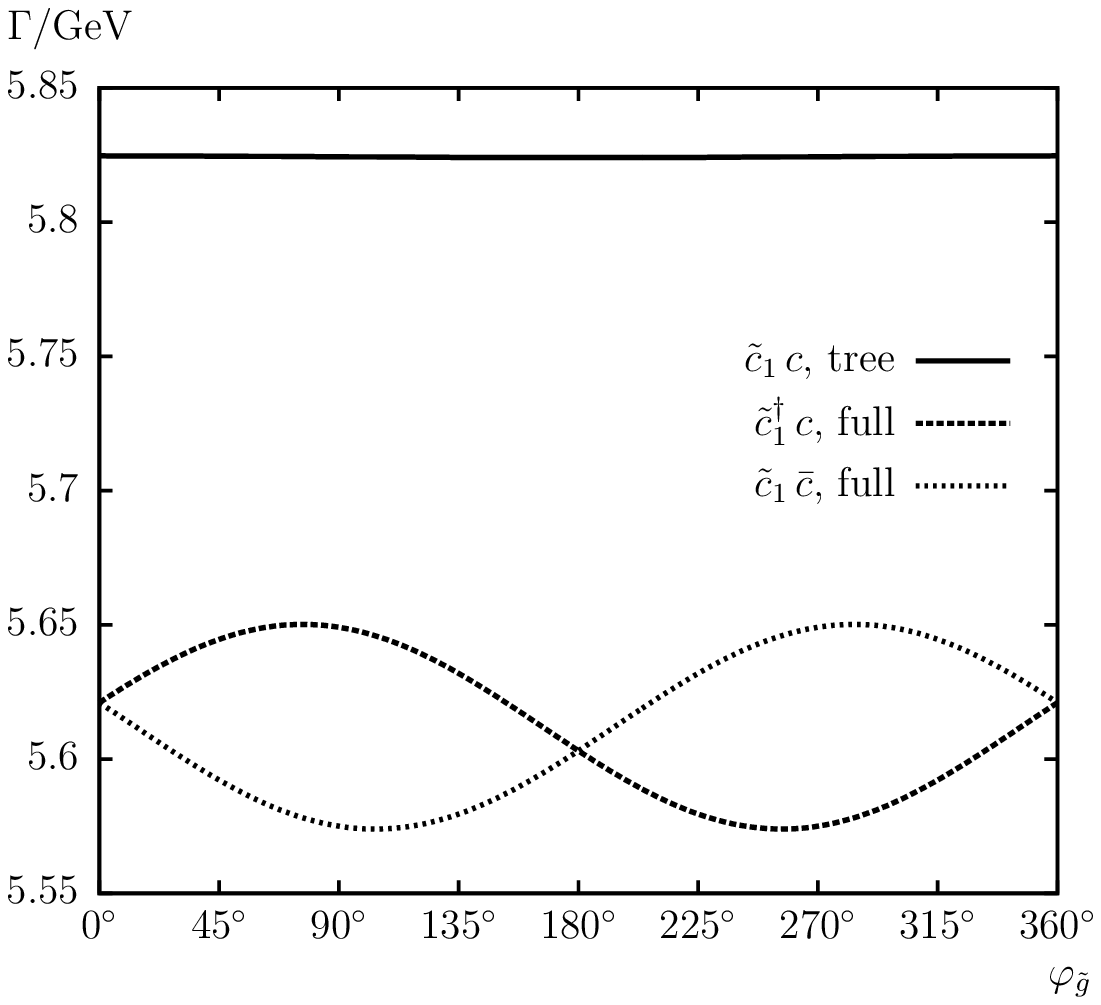}
\hspace{-4mm}
\includegraphics[width=0.49\textwidth,height=8.0cm]{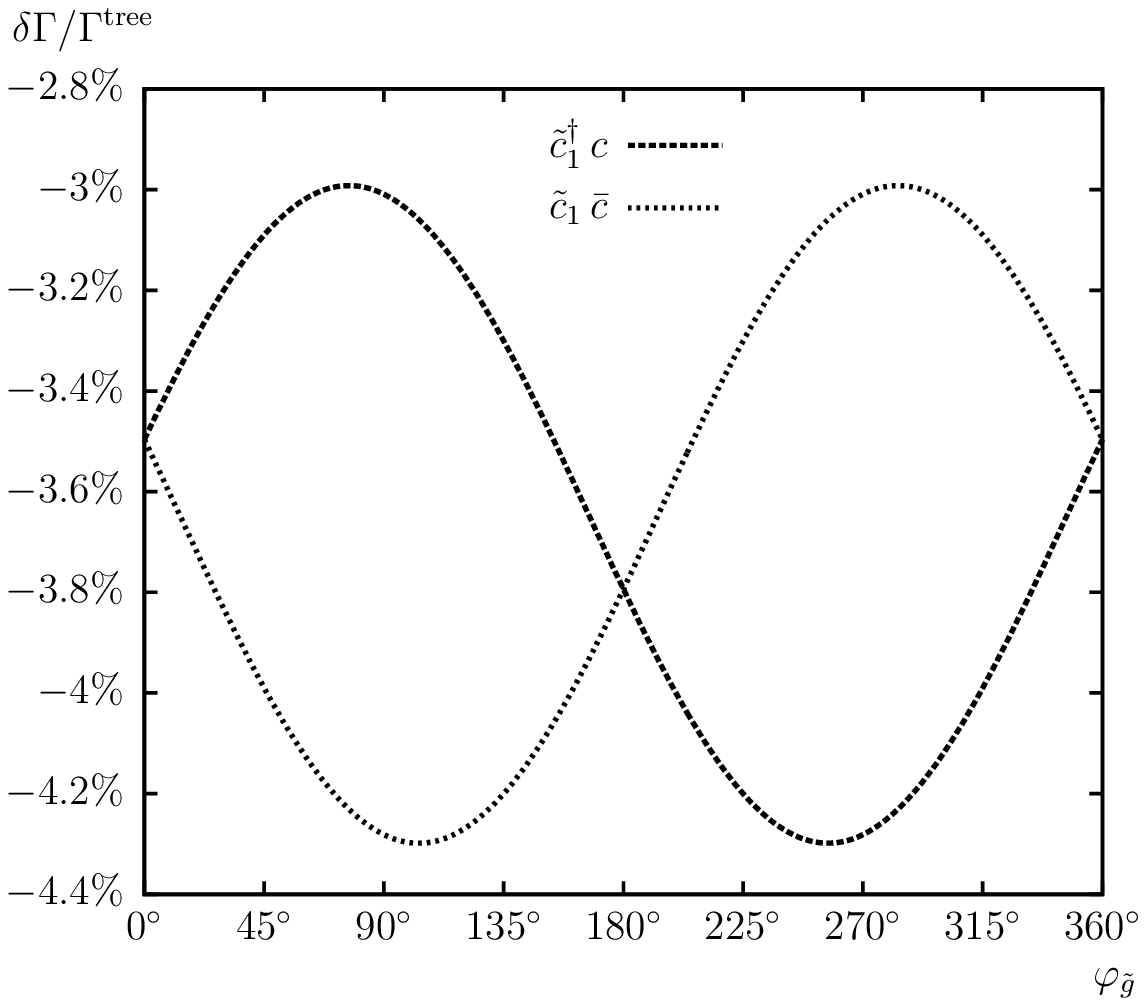}
\\[4em]
\includegraphics[width=0.49\textwidth,height=8.0cm]{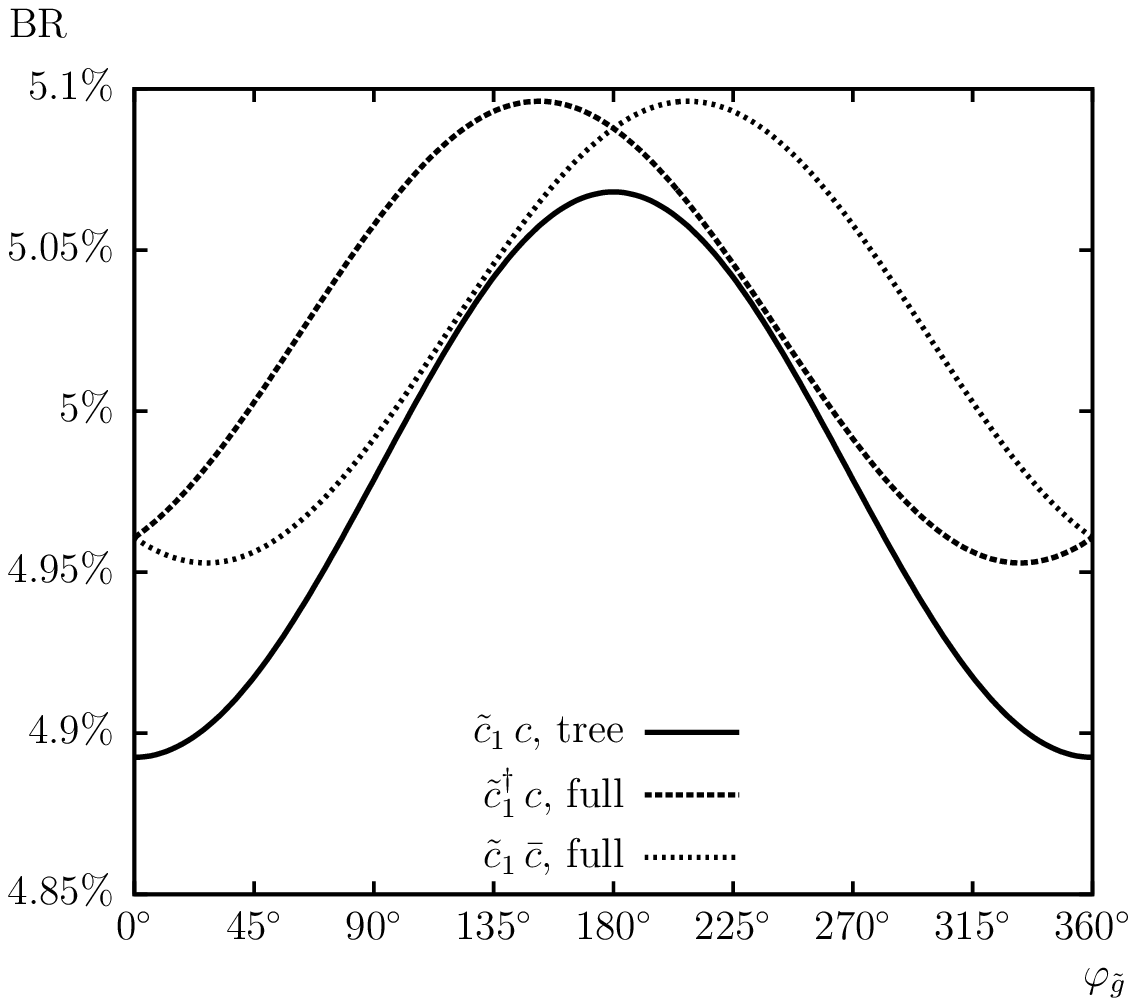}
\hspace{-4mm}
\includegraphics[width=0.49\textwidth,height=8.0cm]{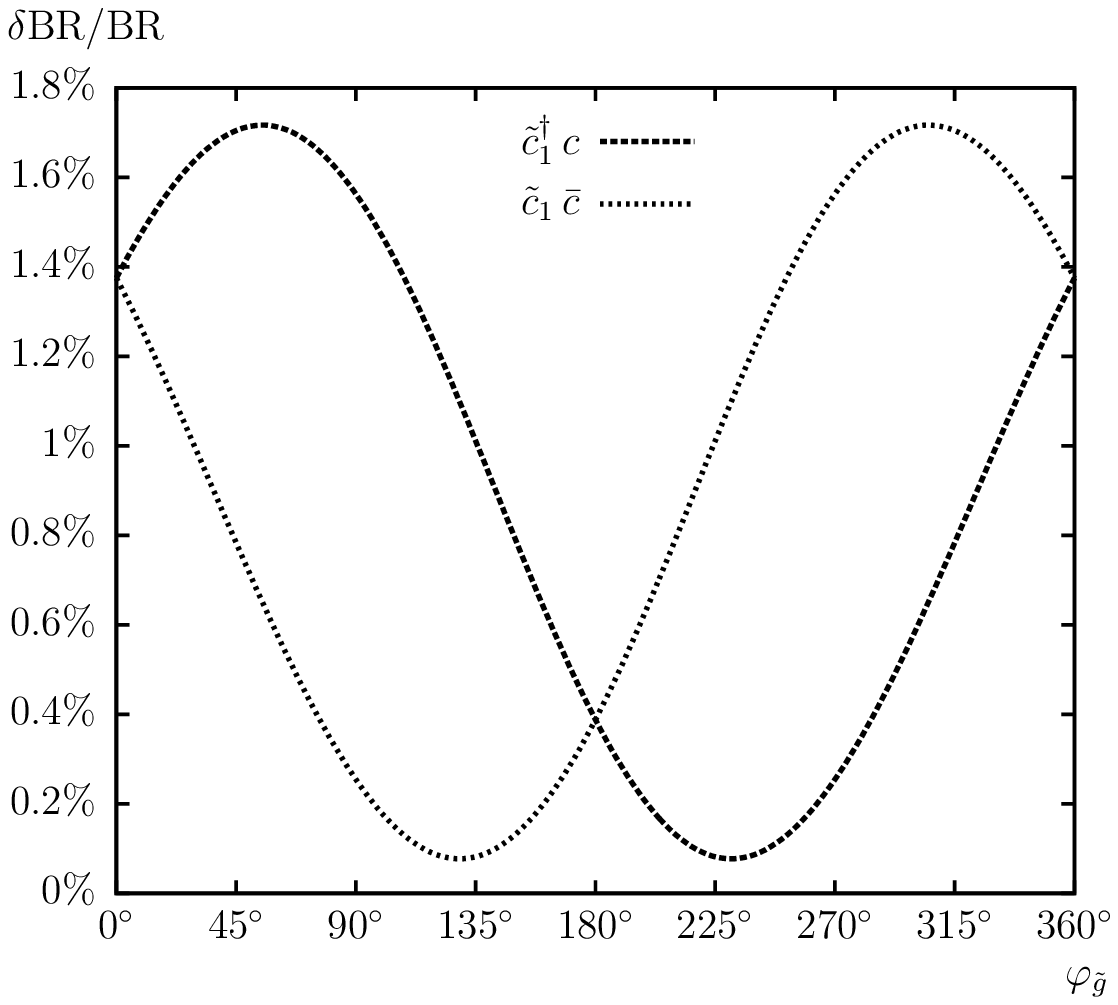}
\end{tabular}
\vspace{2em}
\caption{$\Ga(\decaySce)$.
  Tree-level (``tree'') and full one-loop (``full'') corrected 
  decay widths (including absorptive self-energy contributions) are shown.
  The parameters are chosen according to \SE\ (see \refta{tab:para}), 
  with $\phigl$ varied.
  The upper left plot shows the decay width, the upper right plot shows 
  the relative size of the corrections. 
  The lower left plot shows the BR, the lower right plot shows 
  the relative size of the BR.
}
\label{fig:PhiM3.glsc1c}
\end{center}
\end{figure}

\begin{figure}[htb!]
\begin{center}
\begin{tabular}{c}
\includegraphics[width=0.49\textwidth,height=8.0cm]{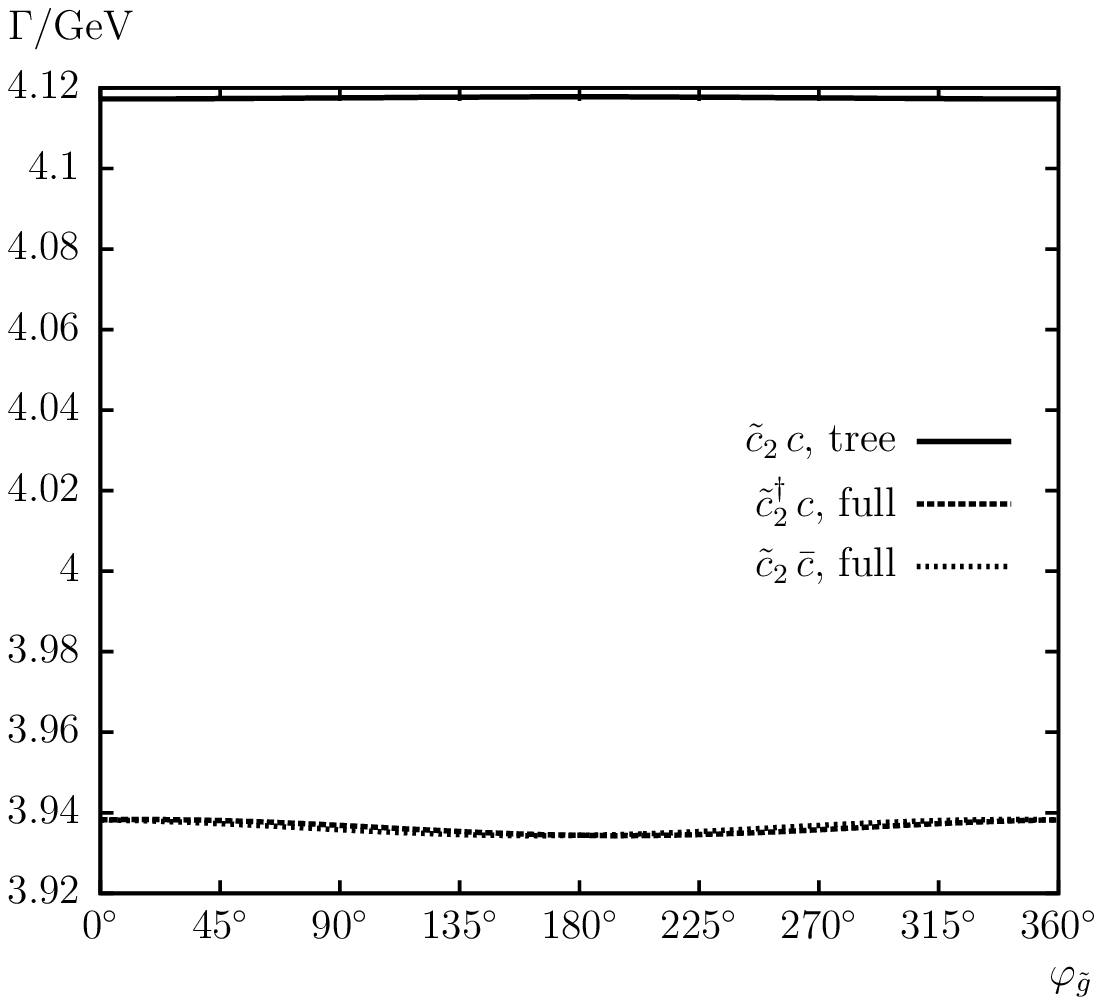}
\hspace{-4mm}
\includegraphics[width=0.49\textwidth,height=8.0cm]{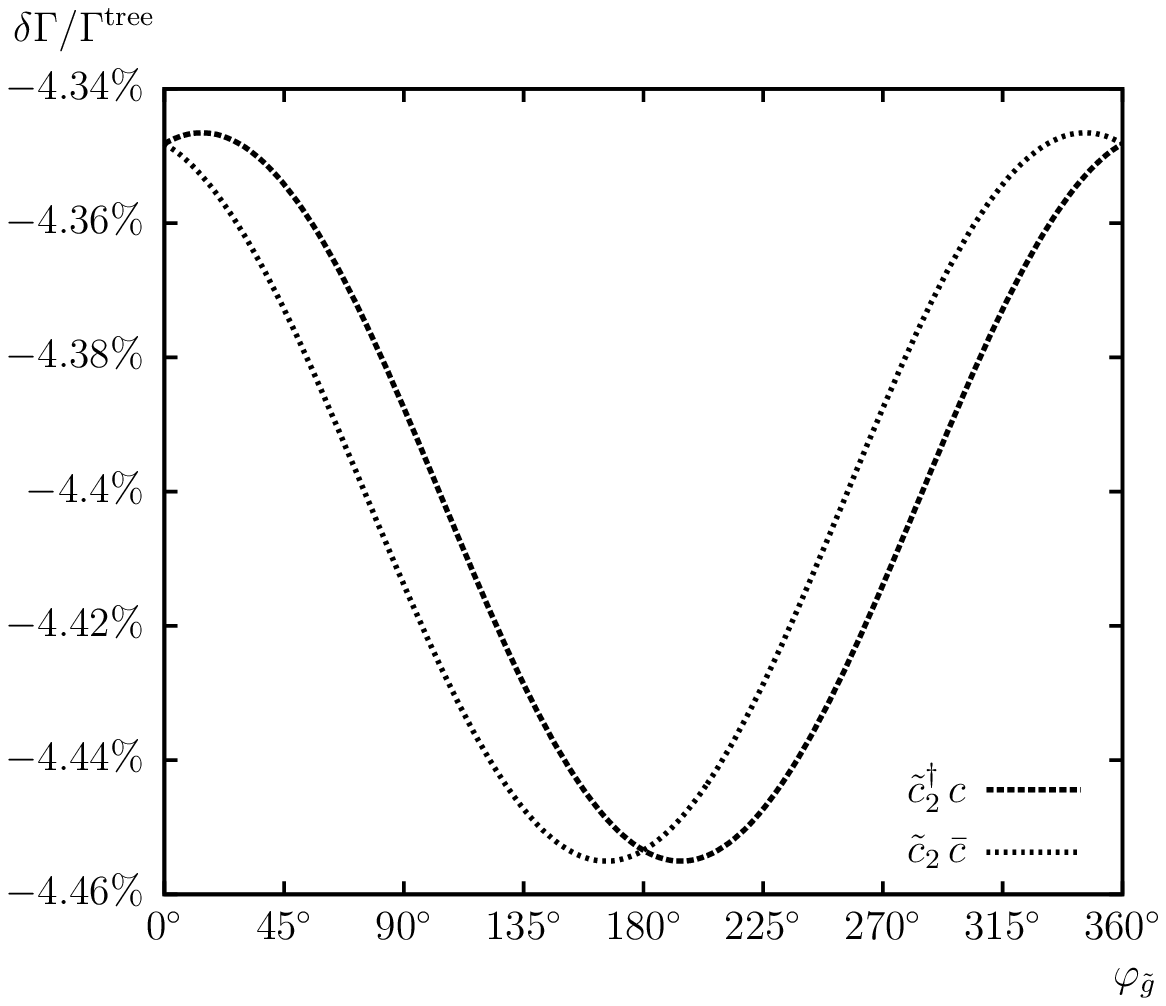}
\\[4em]
\includegraphics[width=0.49\textwidth,height=8.0cm]{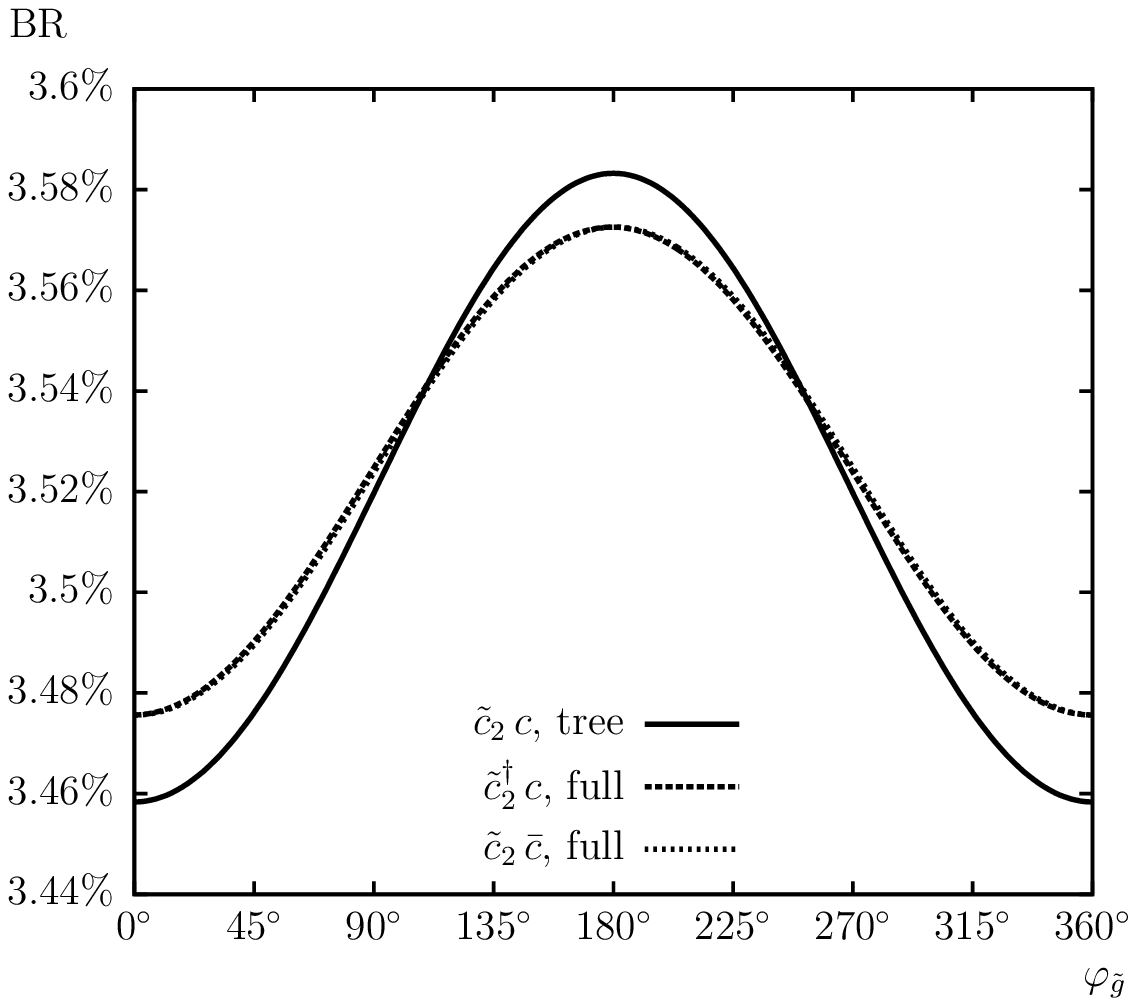}
\hspace{-4mm}
\includegraphics[width=0.49\textwidth,height=8.0cm]{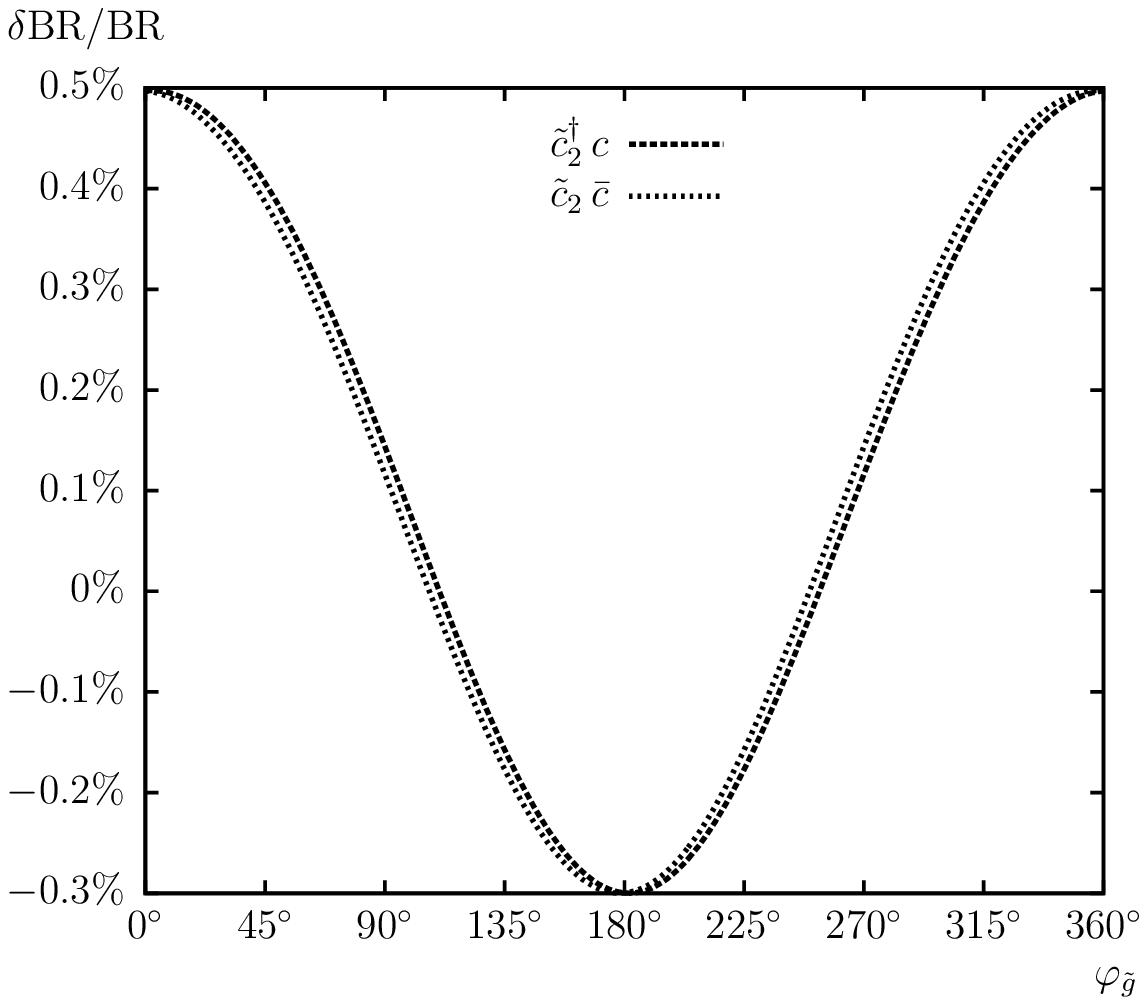}
\end{tabular}
\vspace{2em}
\caption{$\Ga(\decayScz)$.
  Tree-level (``tree'') and full one-loop (``full'') corrected 
  decay widths (including absorptive self-energy contributions) are shown.
  The parameters are chosen according to \SE\ (see \refta{tab:para}), 
  with $\phigl$ varied.
  The upper left plot shows the decay width, the upper right plot shows 
  the relative size of the corrections. 
  The lower left plot shows the BR, the lower right plot shows 
  the relative size of the BR.
}
\label{fig:PhiM3.glsc2c}
\end{center}
\end{figure}

\begin{figure}[htb!]
\begin{center}
\begin{tabular}{c}
\includegraphics[width=0.49\textwidth,height=8.0cm]{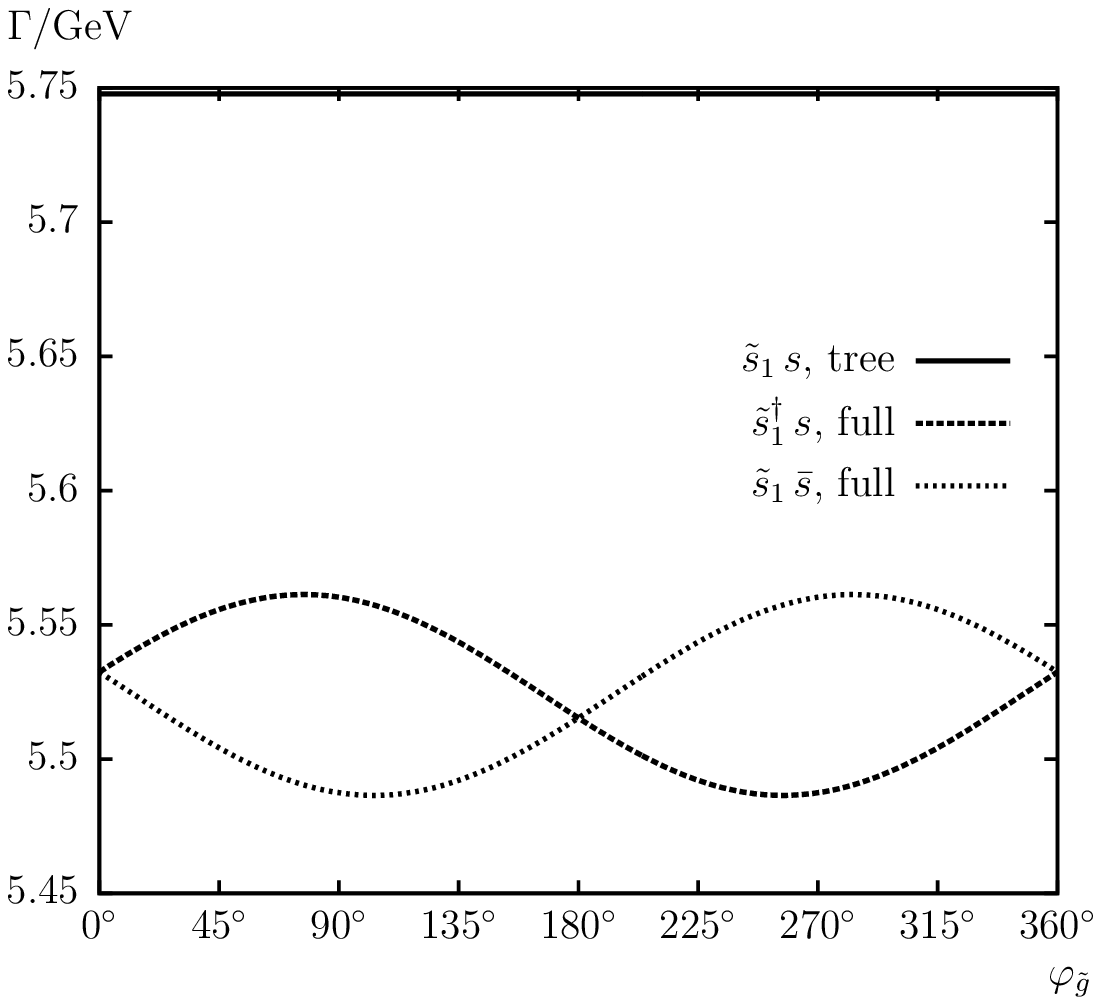}
\hspace{-4mm}
\includegraphics[width=0.49\textwidth,height=8.0cm]{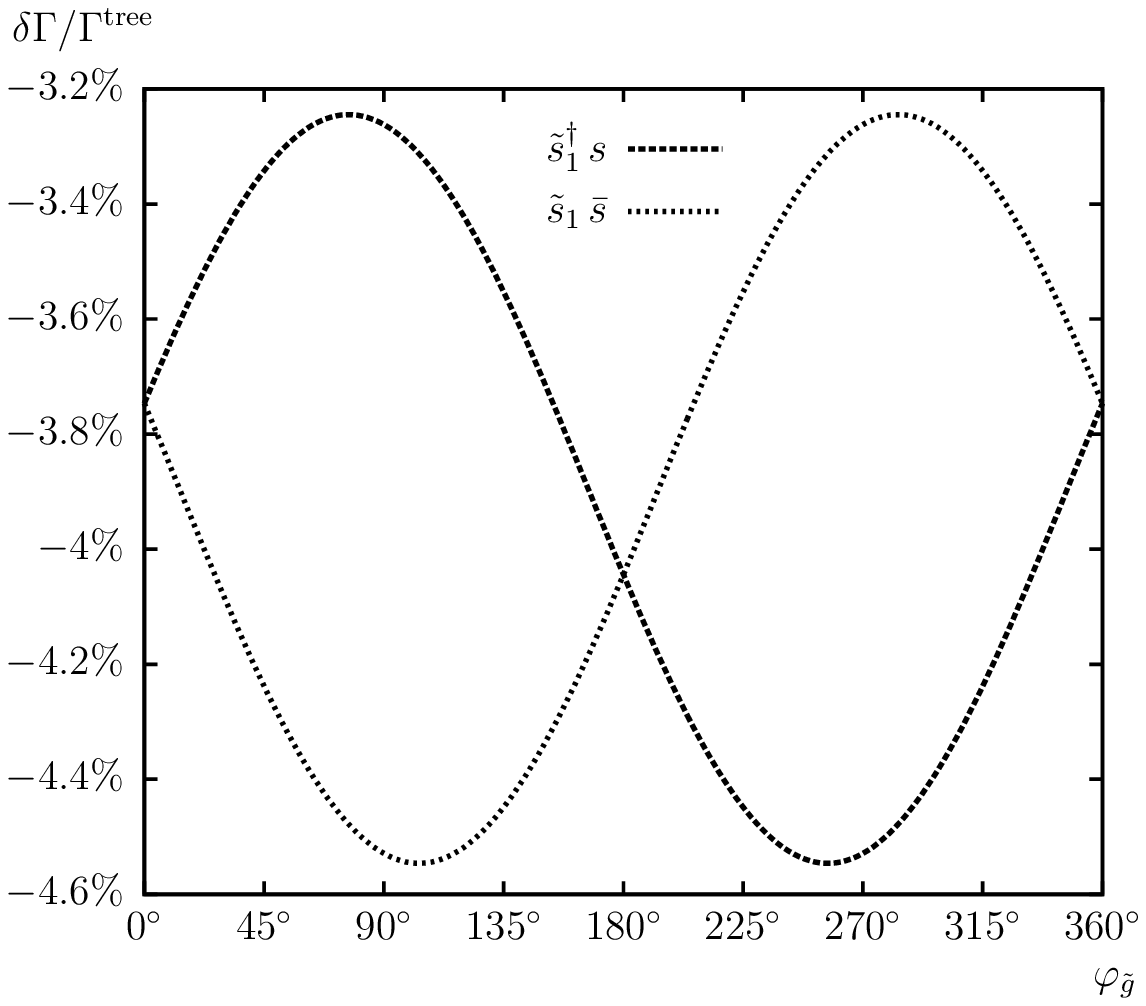}
\\[4em]
\includegraphics[width=0.49\textwidth,height=8.0cm]{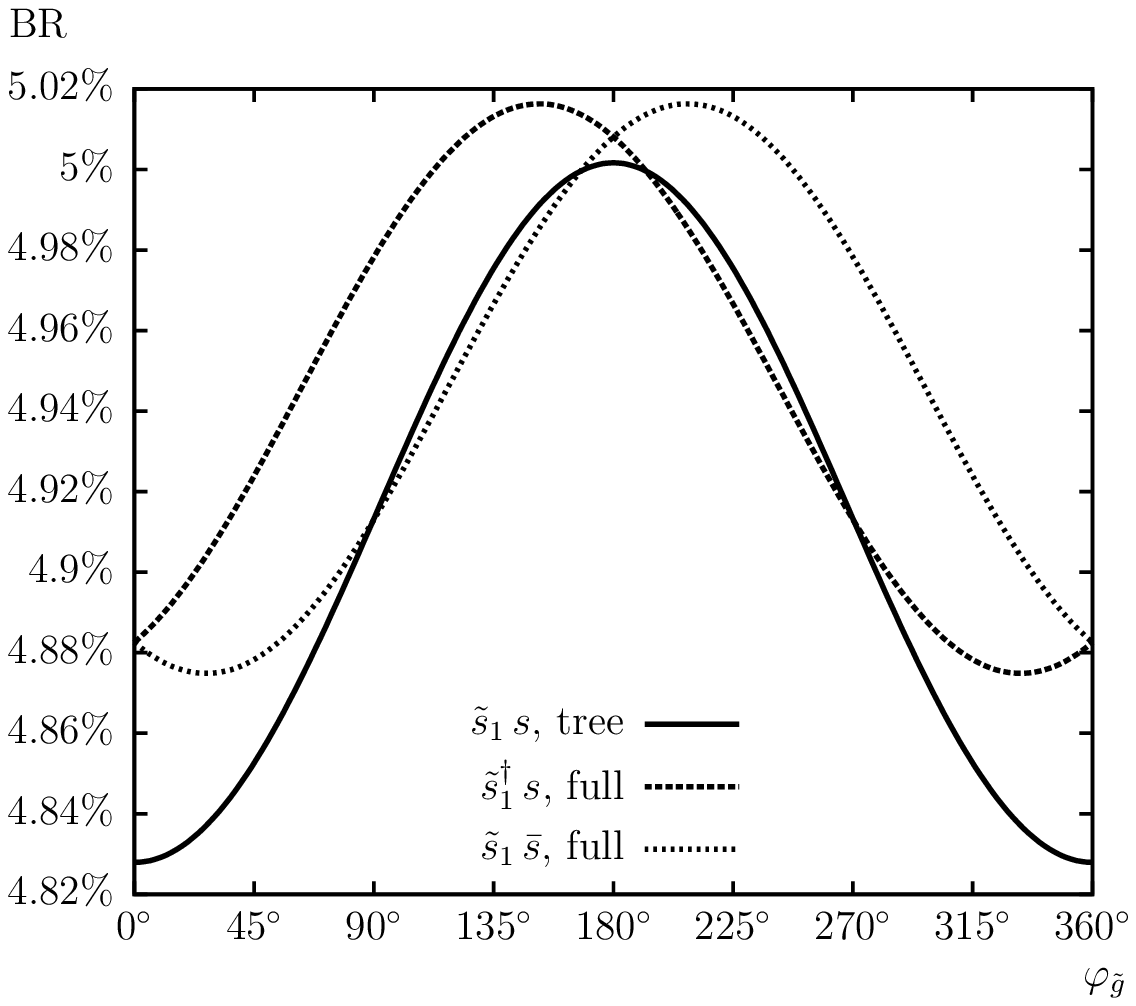}
\hspace{-4mm}
\includegraphics[width=0.49\textwidth,height=8.0cm]{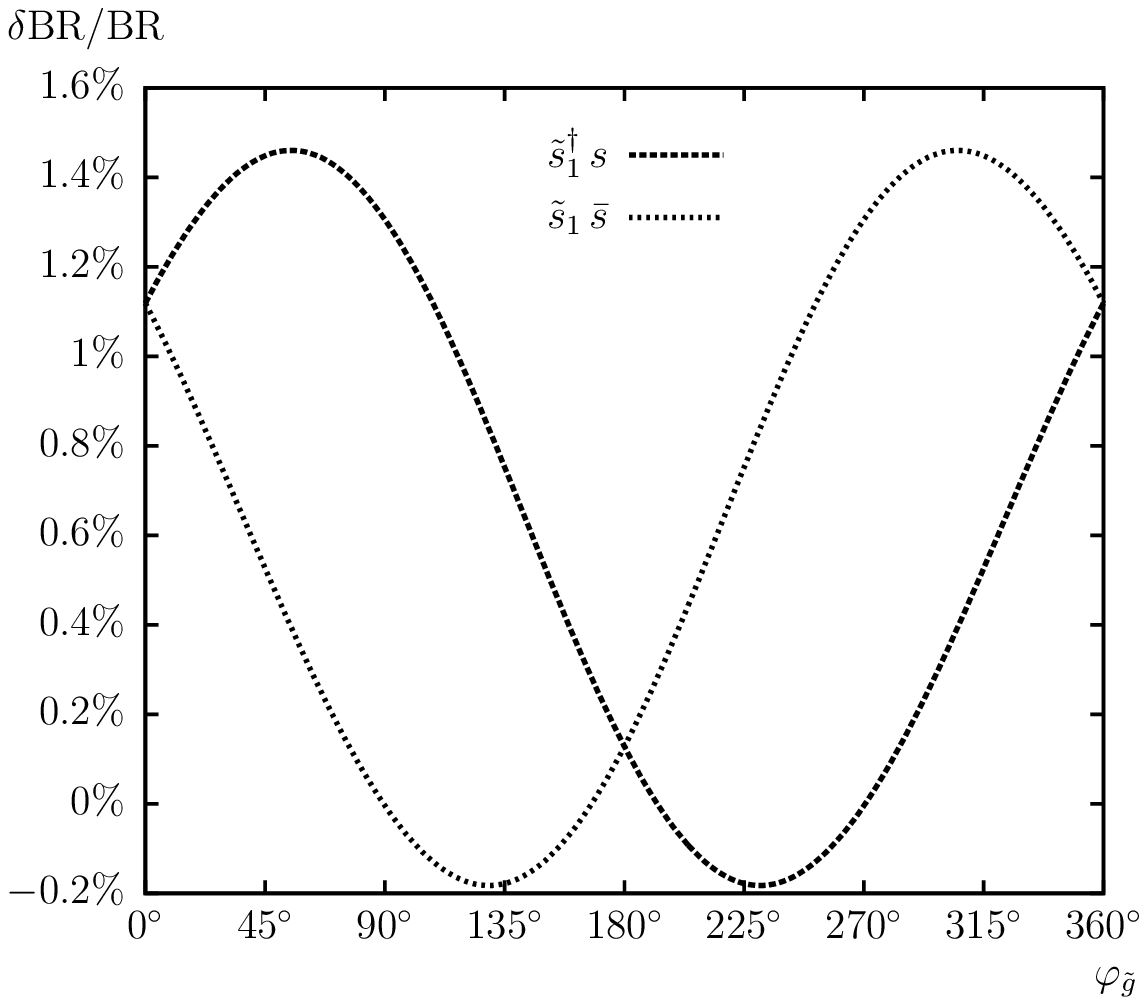}
\end{tabular}
\vspace{2em}
\caption{$\Ga(\decaySse)$.
  Tree-level (``tree'') and full one-loop (``full'') corrected 
  decay widths (including absorptive self-energy contributions) are shown.
  The parameters are chosen according to \SE\ (see \refta{tab:para}), 
  with $\phigl$ varied.
  The upper left plot shows the decay width, the upper right plot shows 
  the relative size of the corrections. 
  The lower left plot shows the BR, the lower right plot shows 
  the relative size of the BR.
}
\label{fig:PhiM3.glss1s}
\end{center}
\end{figure}

\begin{figure}[htb!]
\begin{center}
\begin{tabular}{c}
\includegraphics[width=0.49\textwidth,height=8.0cm]{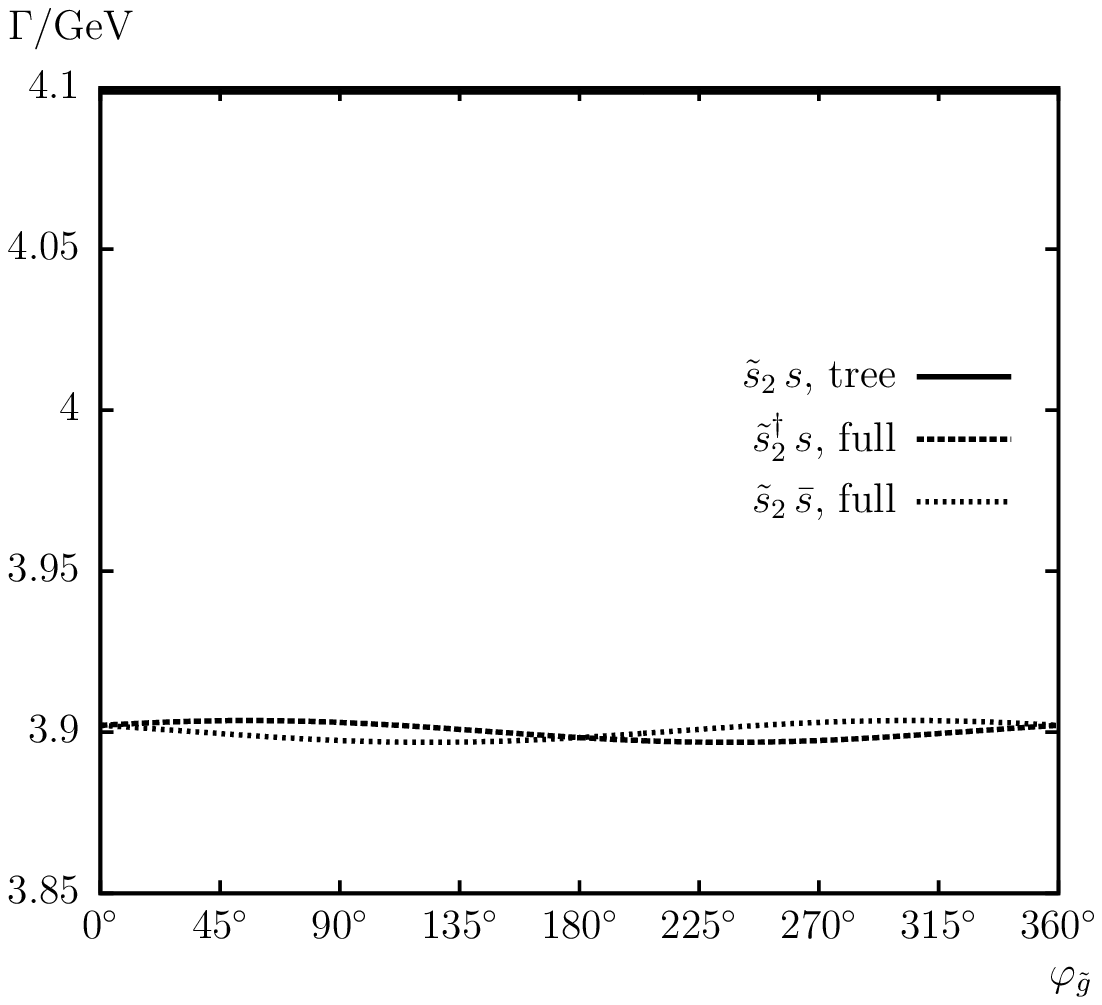}
\hspace{-4mm}
\includegraphics[width=0.49\textwidth,height=8.0cm]{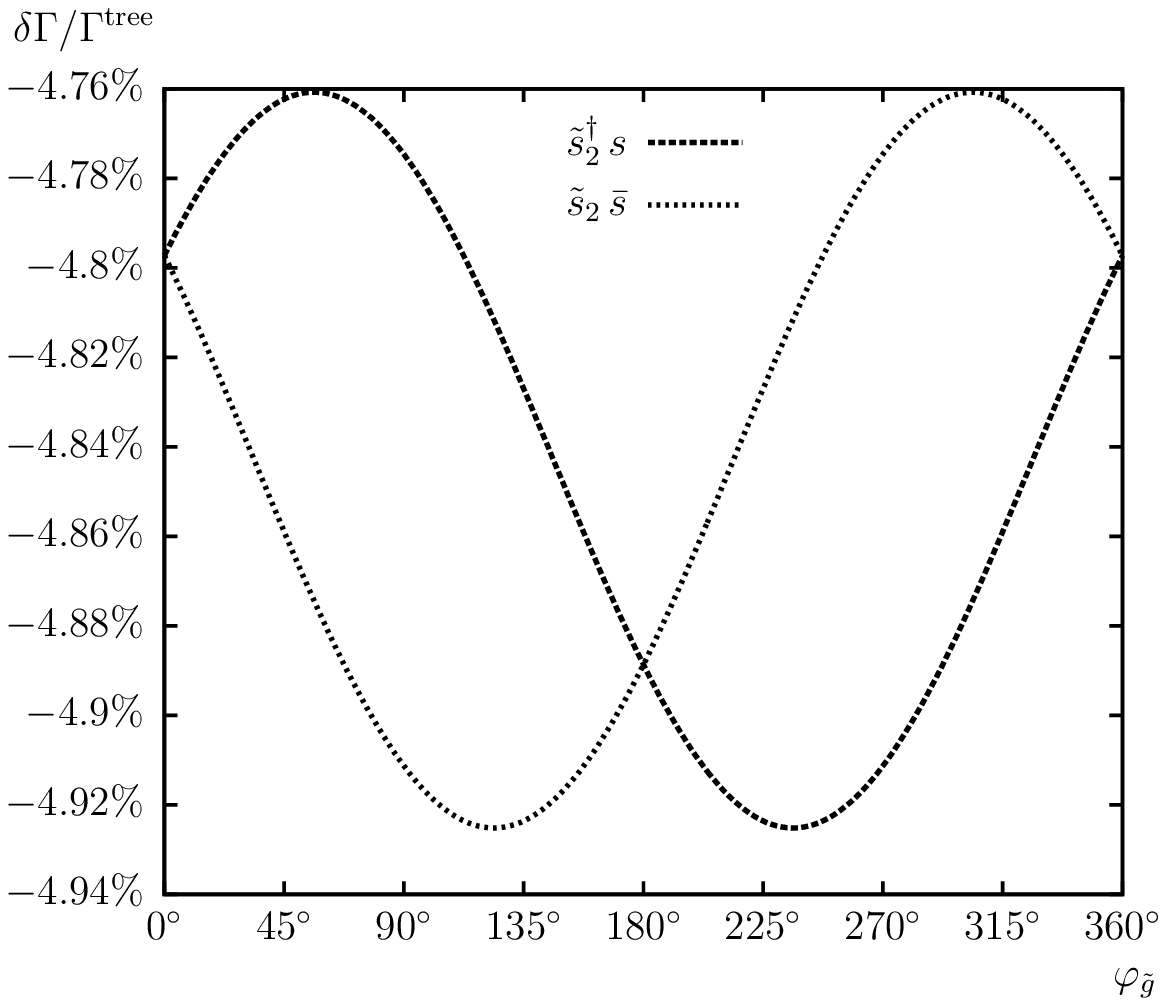}
\\[4em]
\includegraphics[width=0.49\textwidth,height=8.0cm]{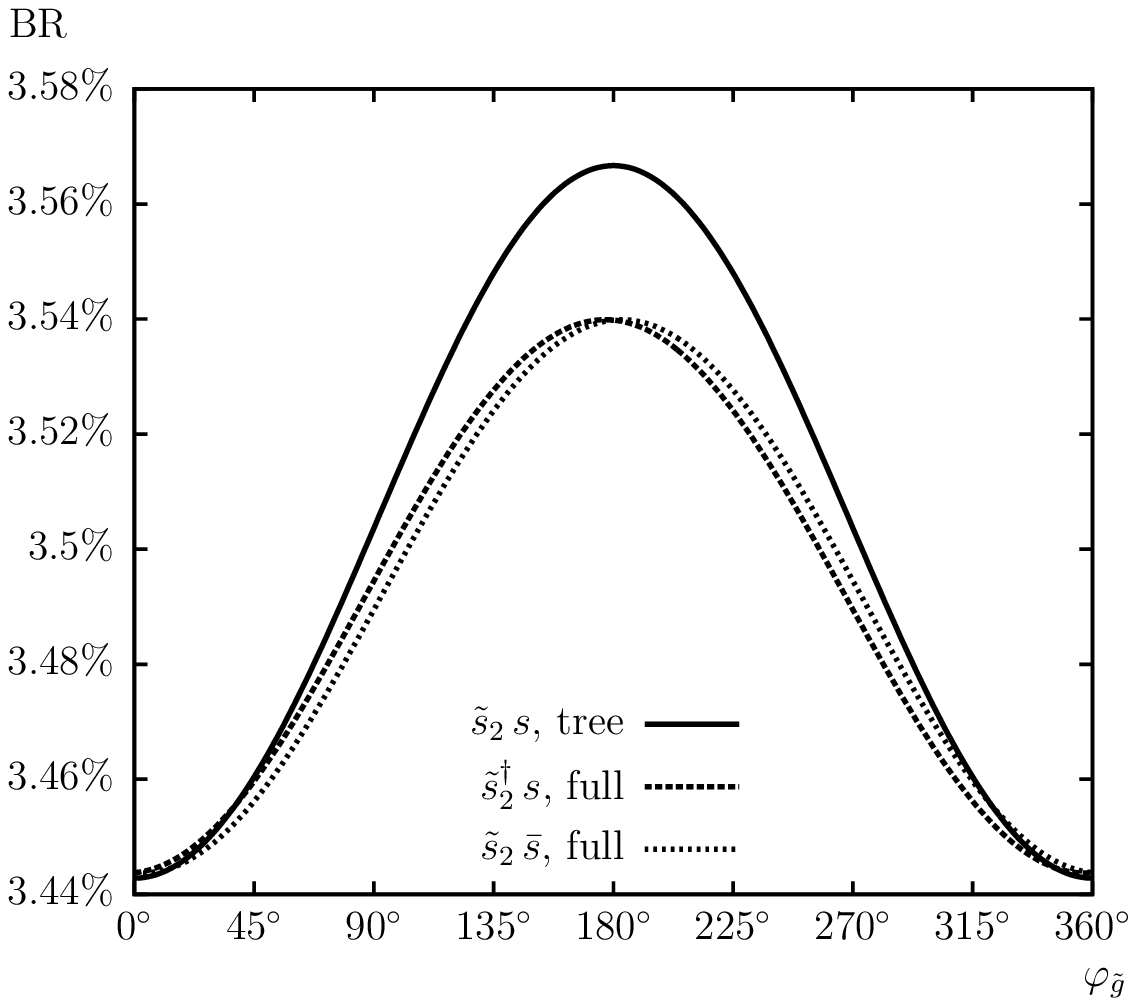}
\hspace{-4mm}
\includegraphics[width=0.49\textwidth,height=8.0cm]{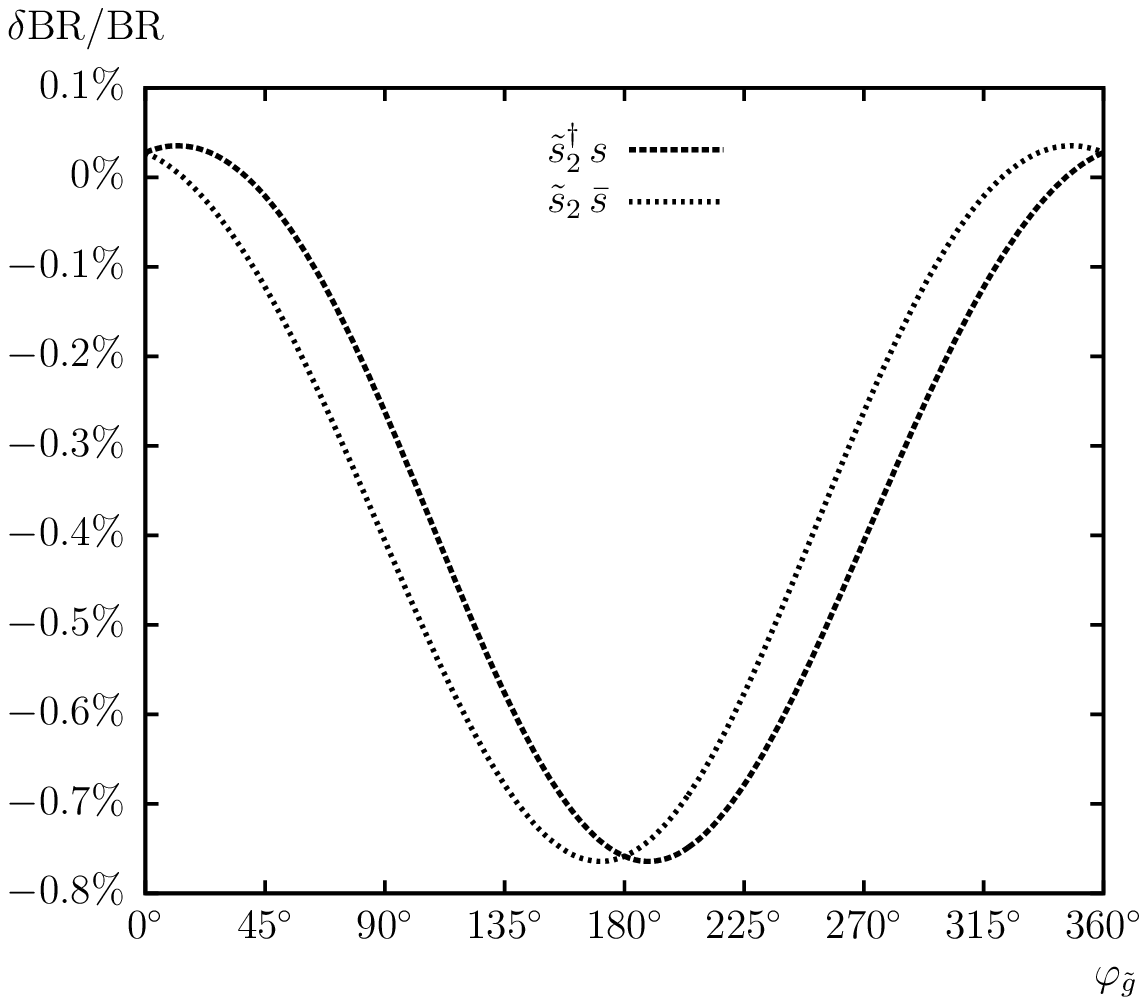}
\end{tabular}
\vspace{2em}
\caption{$\Ga(\decaySsz)$.
  Tree-level (``tree'') and full one-loop (``full'') corrected 
  decay widths (including absorptive self-energy contributions) are shown.
  The parameters are chosen according to \SE\ (see \refta{tab:para}), 
  with $\phigl$ varied.
  The upper left plot shows the decay width, the upper right plot shows 
  the relative size of the corrections. 
  The lower left plot shows the BR, the lower right plot shows 
  the relative size of the BR.
}
\label{fig:PhiM3.glss2s}
\end{center}
\end{figure}

\begin{figure}[htb!]
\begin{center}
\begin{tabular}{c}
\includegraphics[width=0.49\textwidth,height=8.0cm]{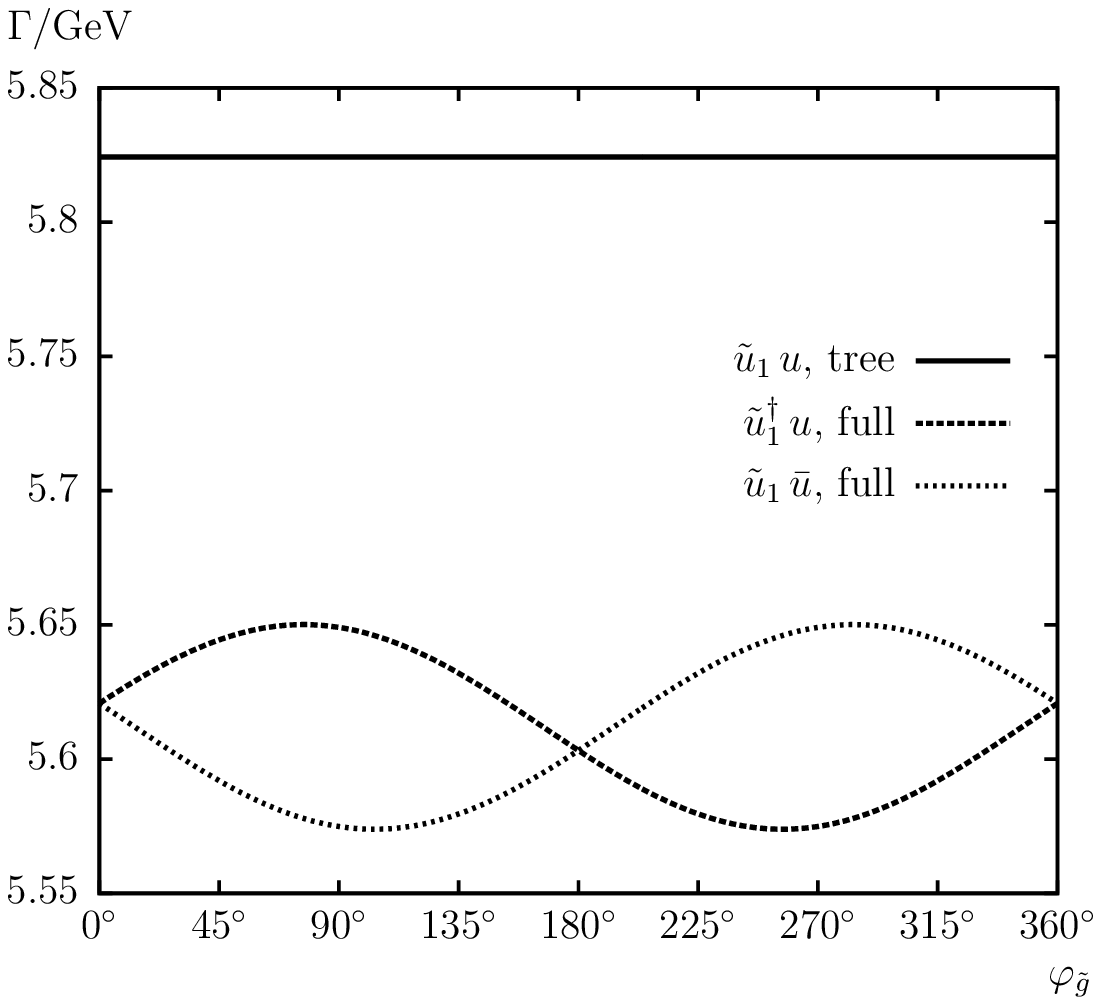}
\hspace{-4mm}
\includegraphics[width=0.49\textwidth,height=8.0cm]{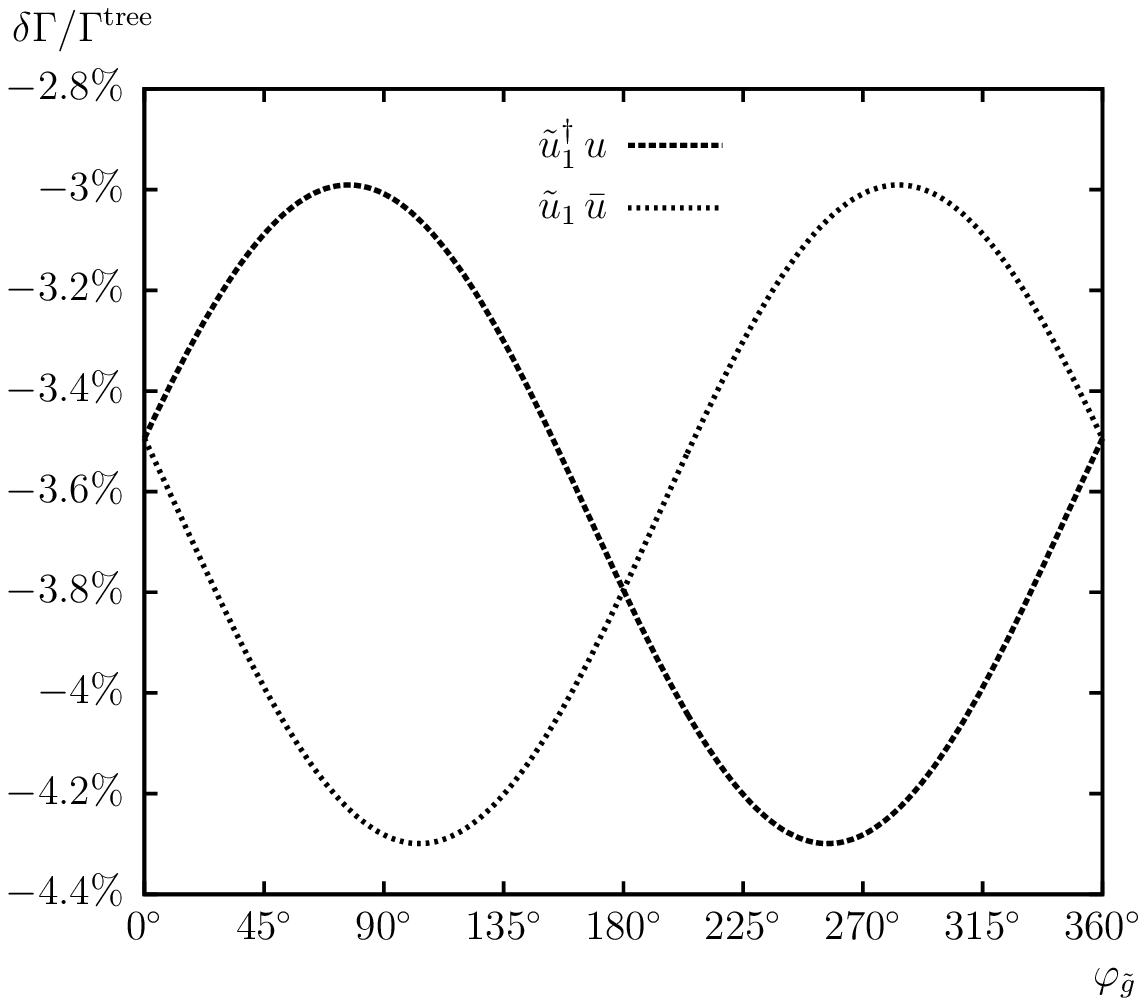}
\\[4em]
\includegraphics[width=0.49\textwidth,height=8.0cm]{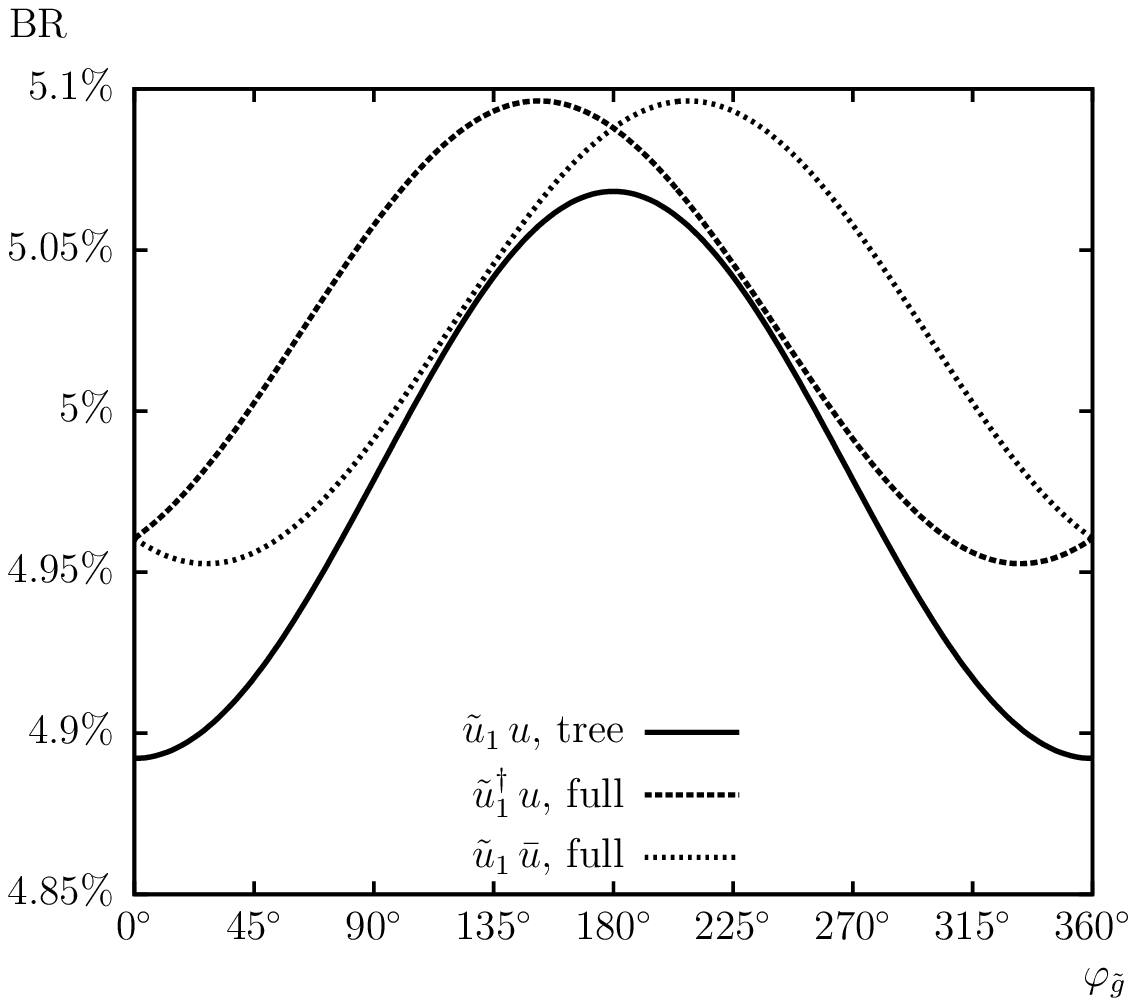}
\hspace{-4mm}
\includegraphics[width=0.49\textwidth,height=8.0cm]{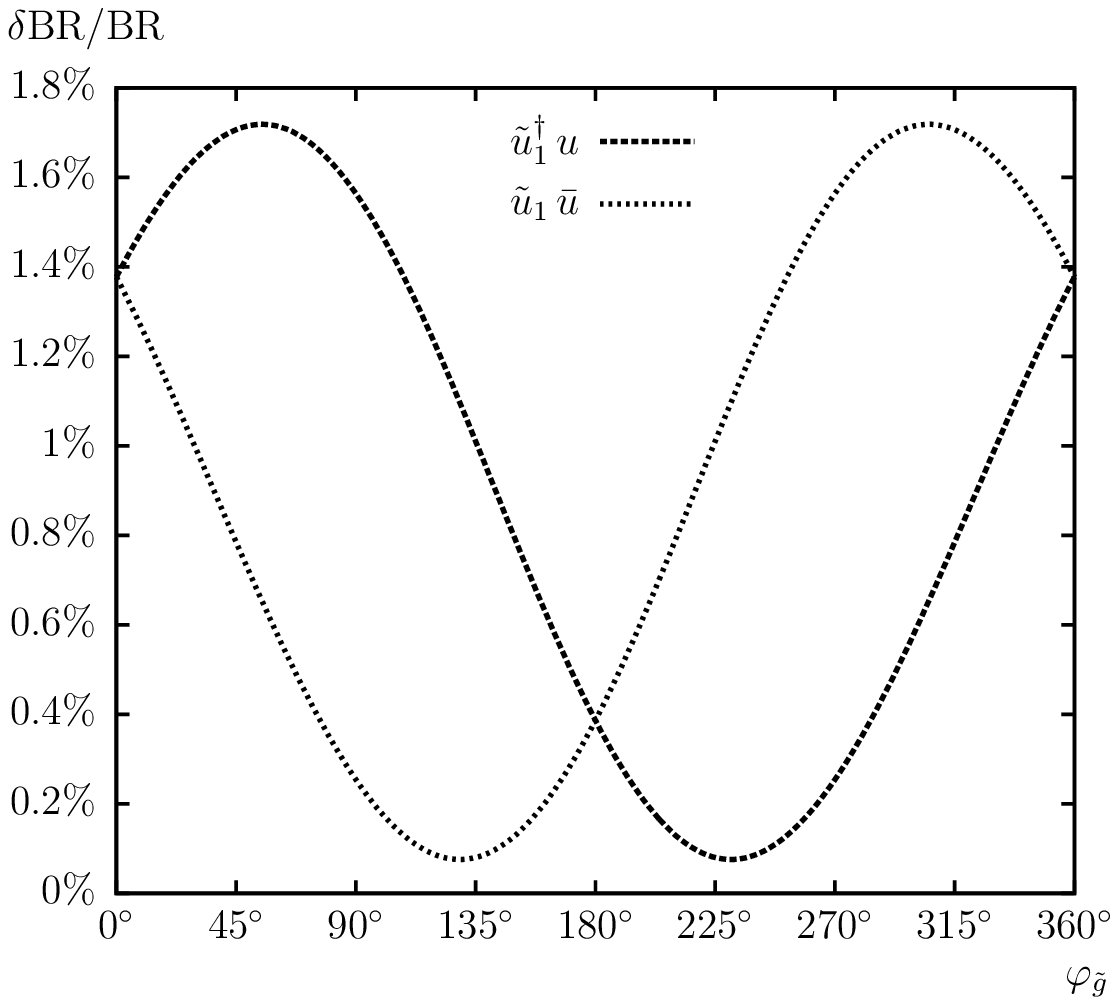}
\end{tabular}
\vspace{2em}
\caption{$\Ga(\decaySue)$.
  Tree-level (``tree'') and full one-loop (``full'') corrected 
  decay widths (including absorptive self-energy contributions) are shown.
  The parameters are chosen according to \SE\ (see \refta{tab:para}), 
  with $\phigl$ varied.
  The upper left plot shows the decay width, the upper right plot shows 
  the relative size of the corrections. 
  The lower left plot shows the BR, the lower right plot shows 
  the relative size of the BR.
}
\label{fig:PhiM3.glsu1u}
\end{center}
\end{figure}

\begin{figure}[htb!]
\begin{center}
\begin{tabular}{c}
\includegraphics[width=0.49\textwidth,height=8.0cm]{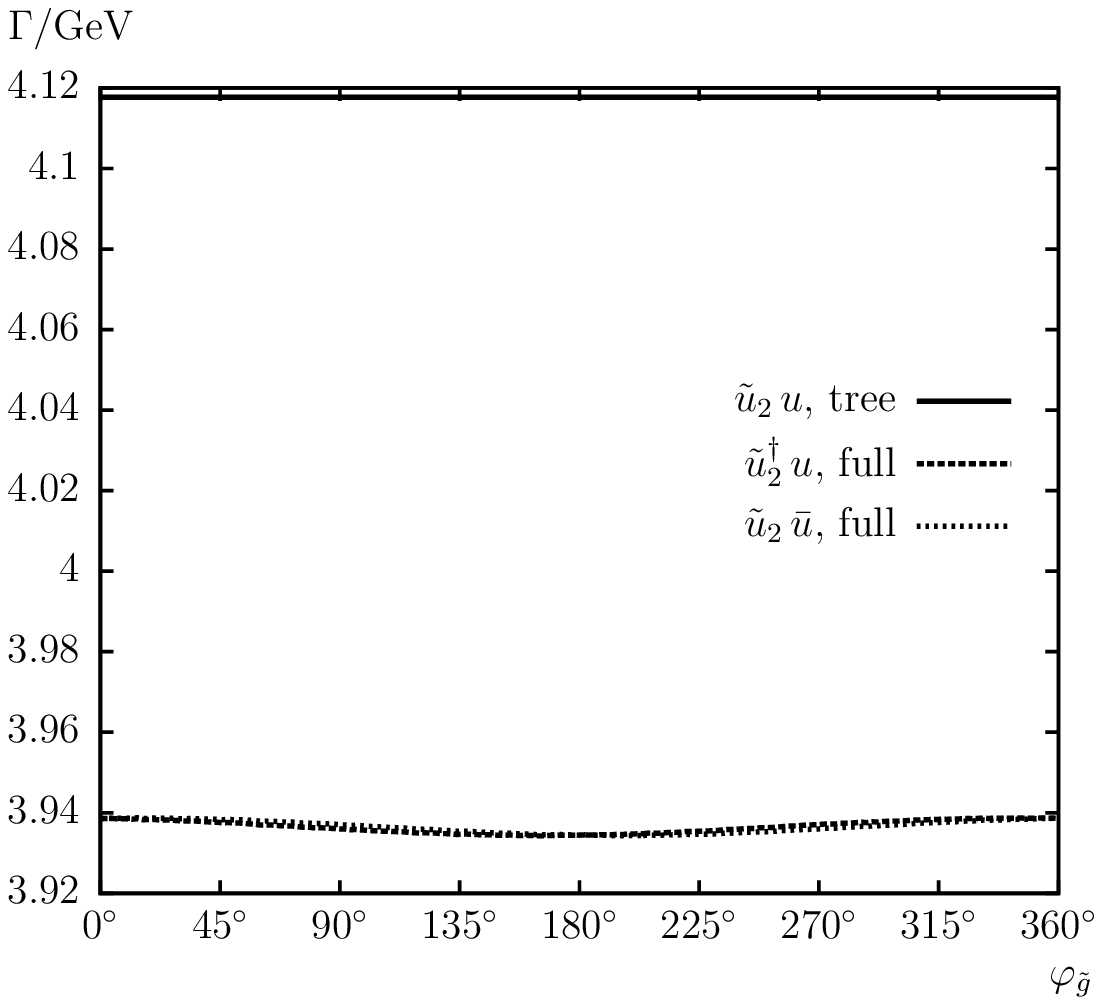}
\hspace{-4mm}
\includegraphics[width=0.49\textwidth,height=8.0cm]{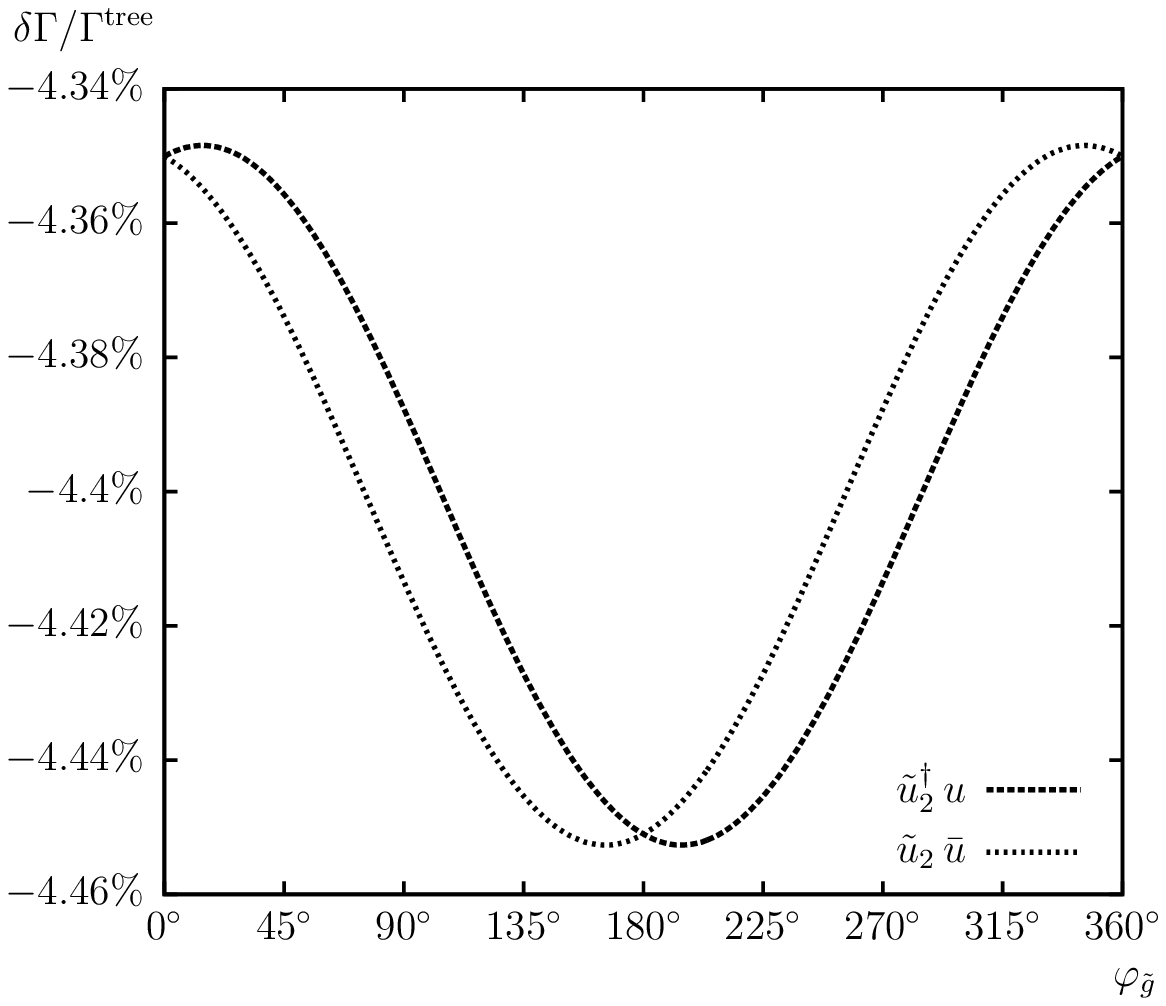}
\\[4em]
\includegraphics[width=0.49\textwidth,height=8.0cm]{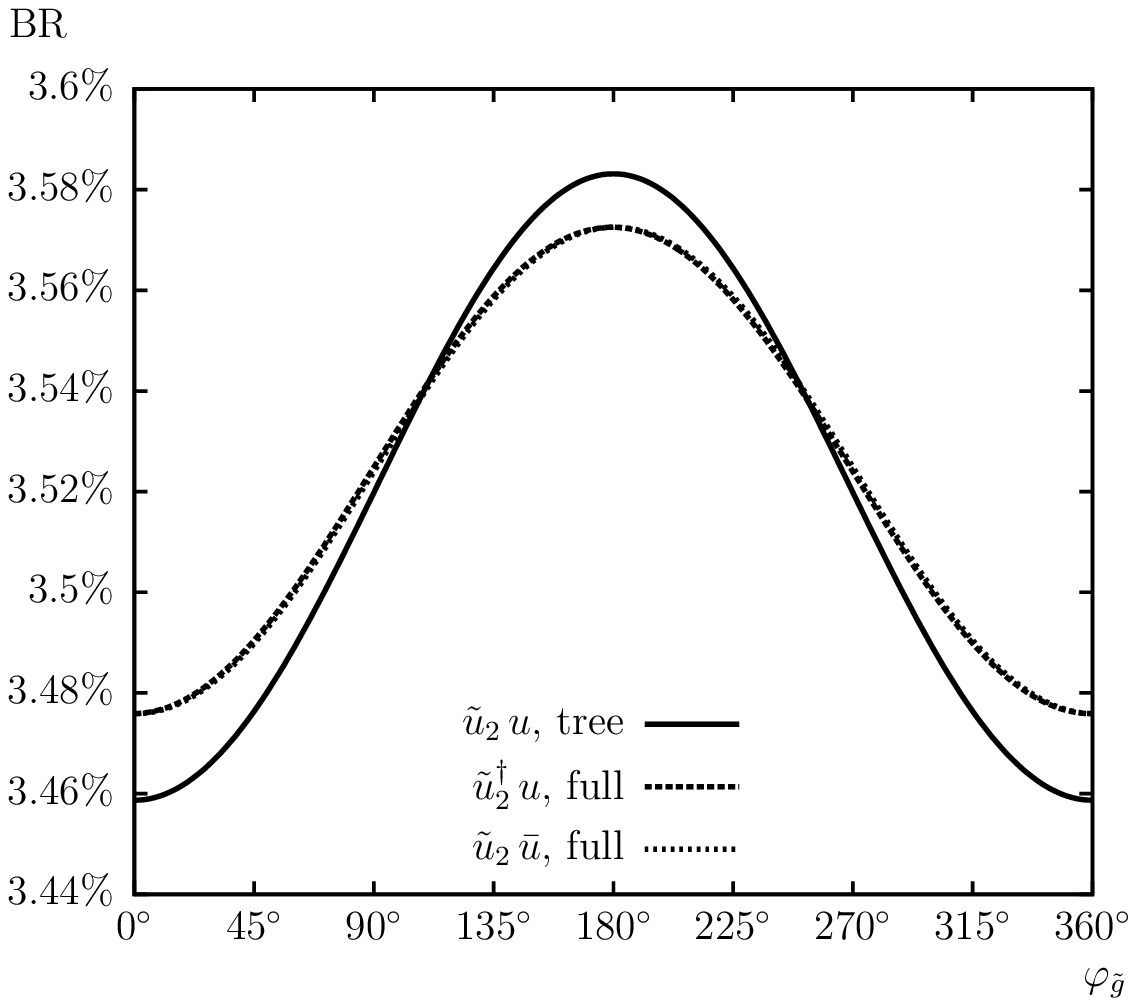}
\hspace{-4mm}
\includegraphics[width=0.49\textwidth,height=8.0cm]{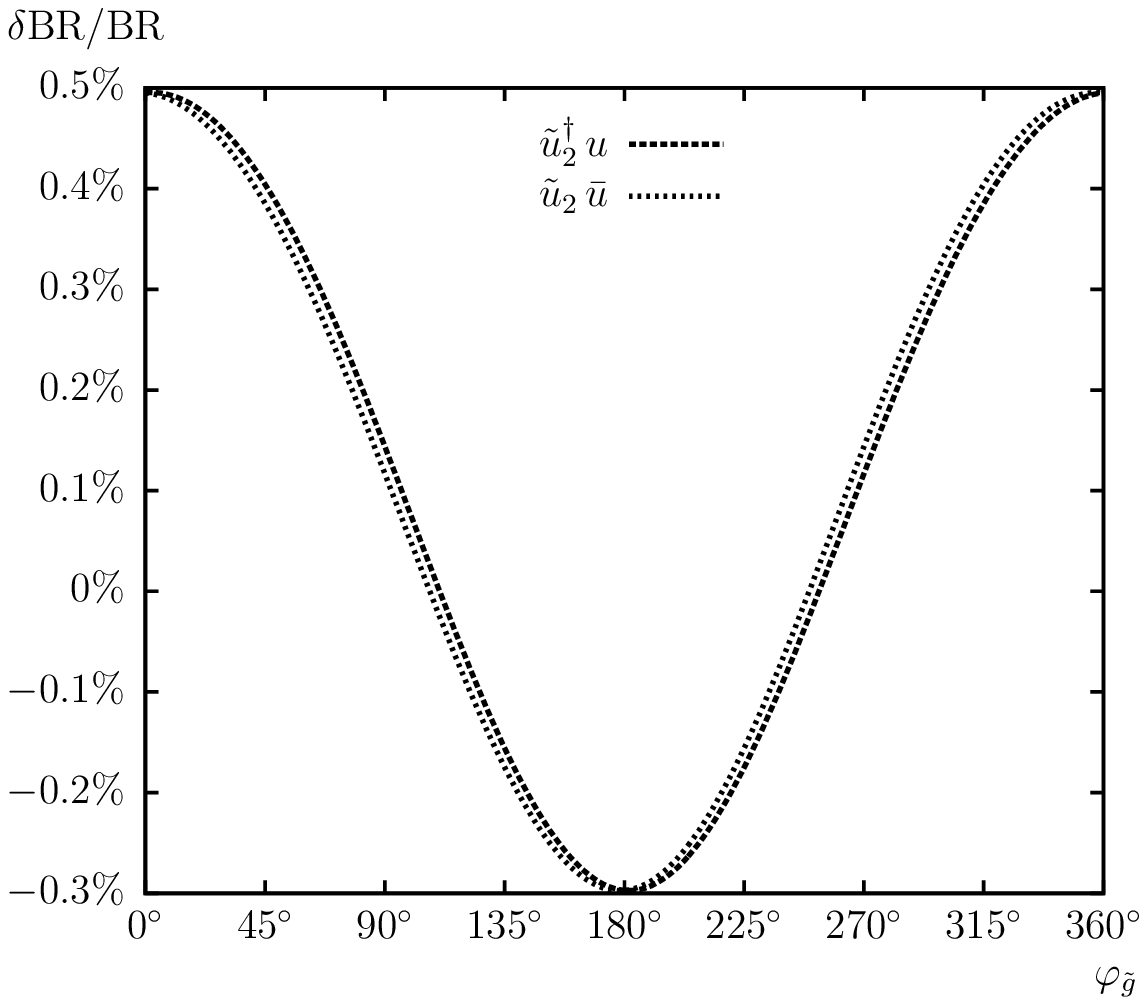}
\end{tabular}
\vspace{2em}
\caption{$\Ga(\decaySuz)$.
  Tree-level (``tree'') and full one-loop (``full'') corrected 
  decay widths (including absorptive self-energy contributions) are shown.
  The parameters are chosen according to \SE\ (see \refta{tab:para}), 
  with $\phigl$ varied.
  The upper left plot shows the decay width, the upper right plot shows 
  the relative size of the corrections. 
  The lower left plot shows the BR, the lower right plot shows 
  the relative size of the BR.
}
\label{fig:PhiM3.glsu2u}
\end{center}
\end{figure}

\begin{figure}[htb!]
\begin{center}
\begin{tabular}{c}
\includegraphics[width=0.49\textwidth,height=8.0cm]{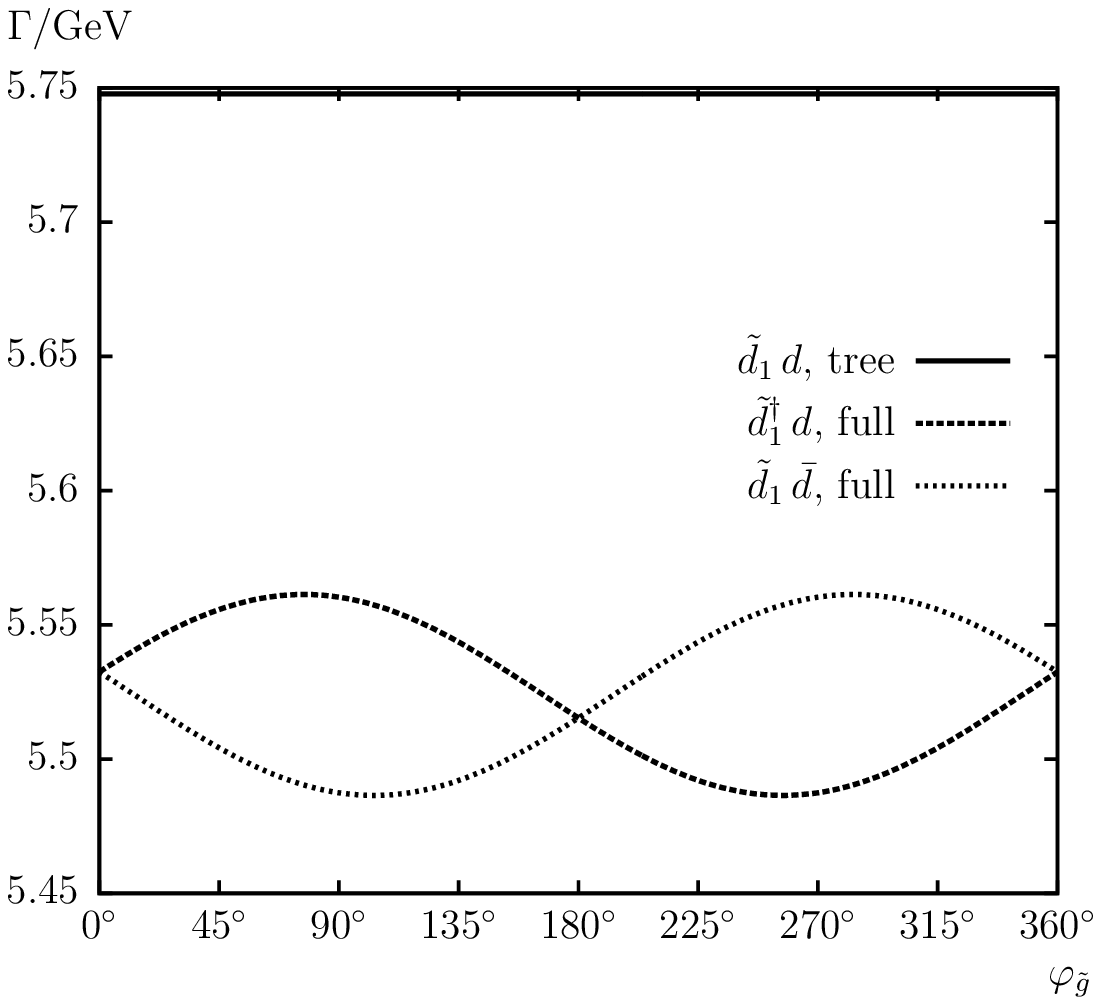}
\hspace{-4mm}
\includegraphics[width=0.49\textwidth,height=8.0cm]{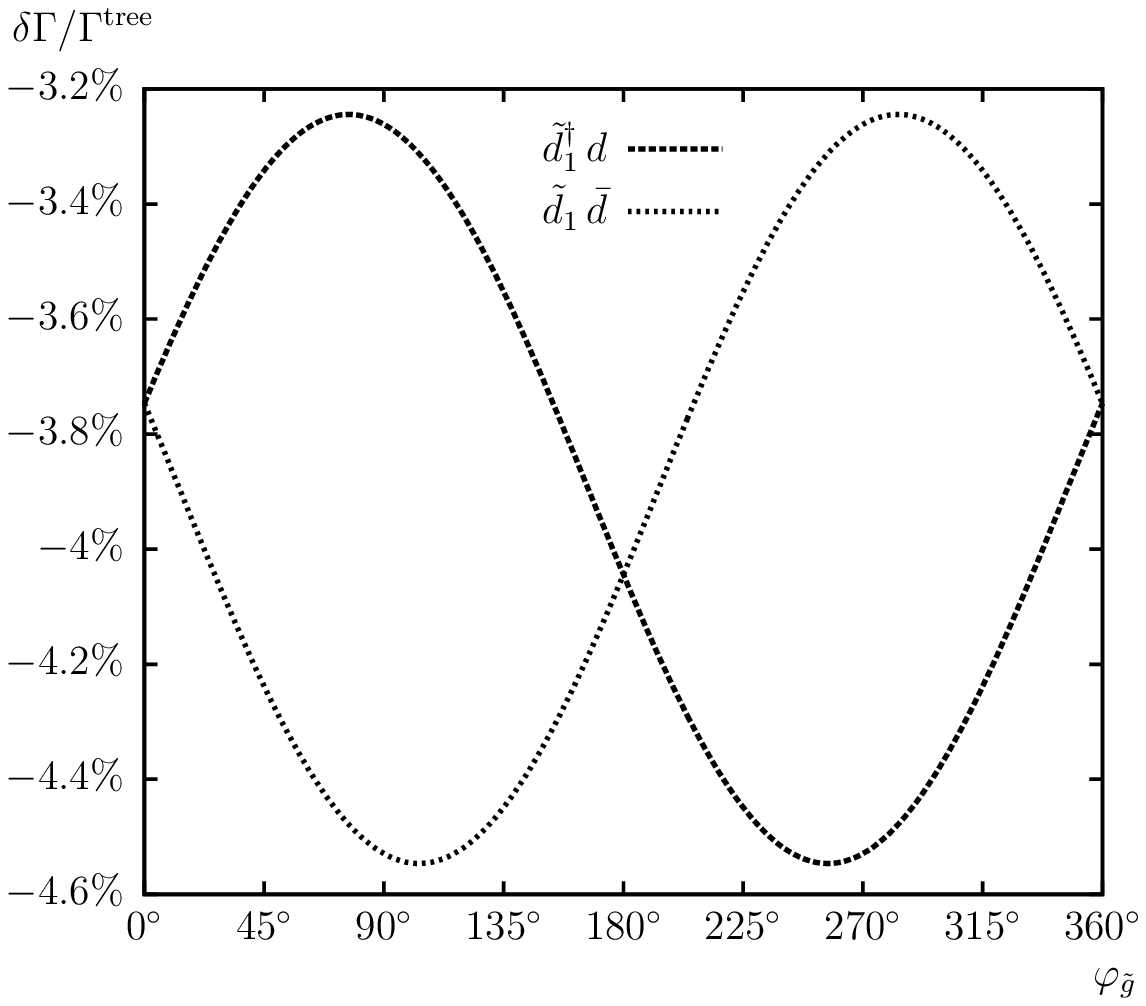}
\\[4em]
\includegraphics[width=0.49\textwidth,height=8.0cm]{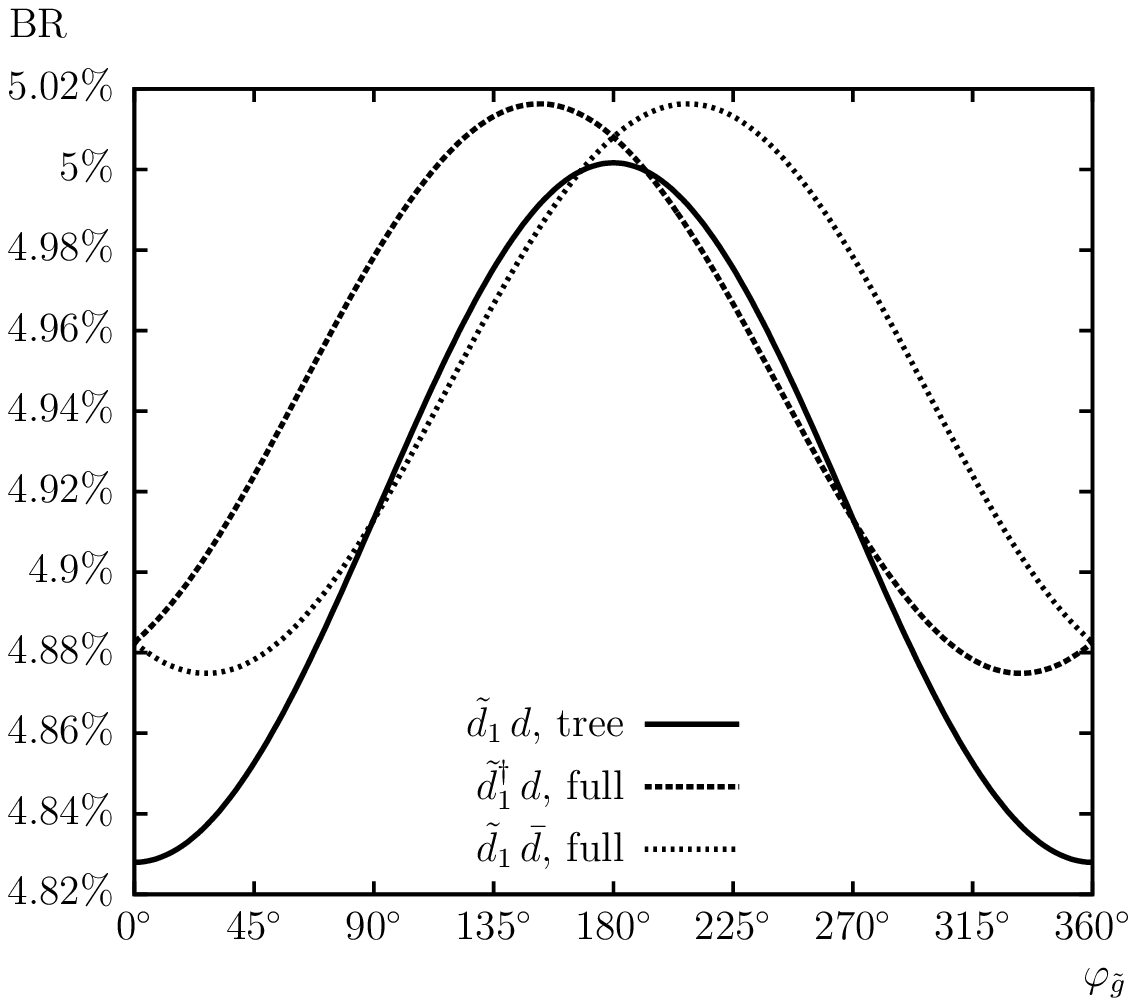}
\hspace{-4mm}
\includegraphics[width=0.49\textwidth,height=8.0cm]{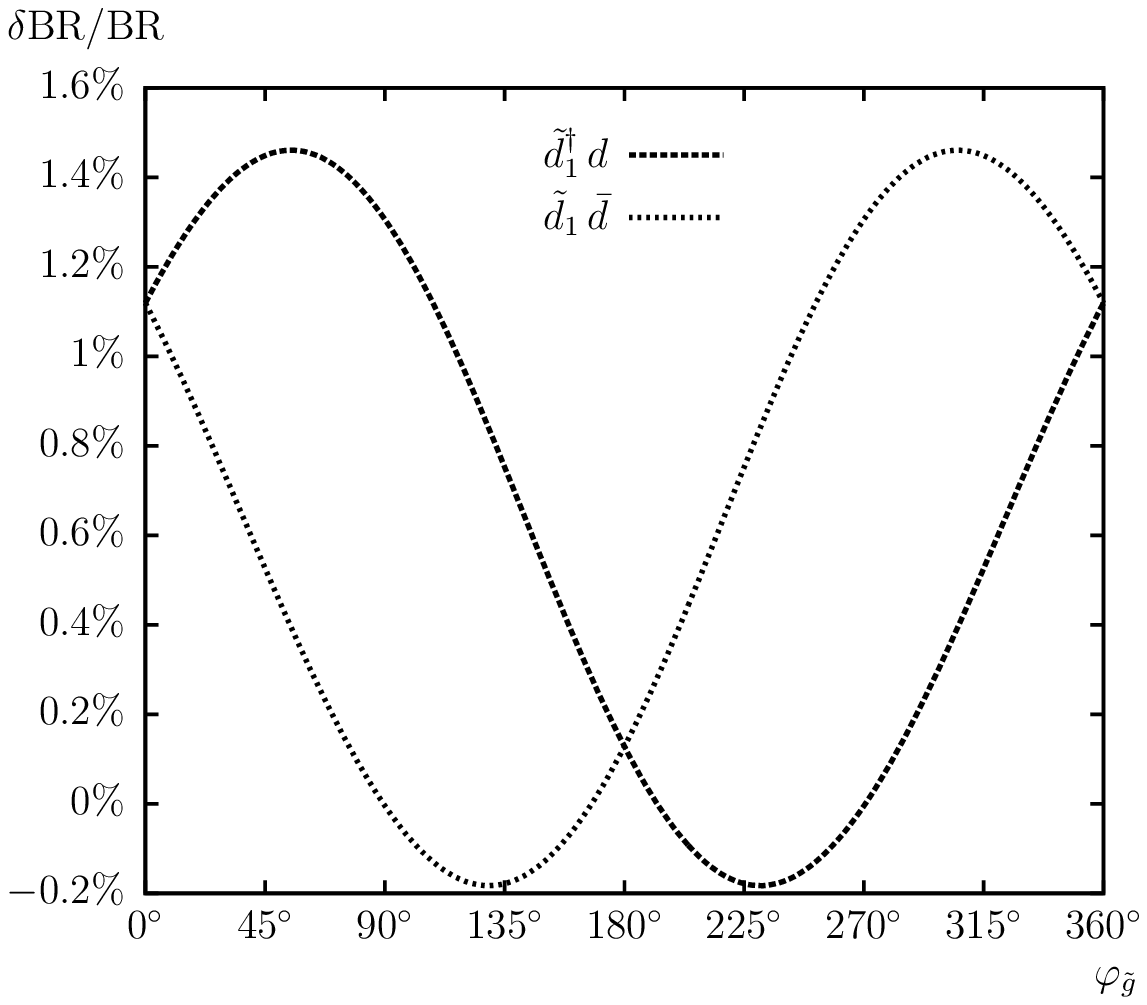}
\end{tabular}
\vspace{2em}
\caption{$\Ga(\decaySde)$.
  Tree-level (``tree'') and full one-loop (``full'') corrected 
  decay widths (including absorptive self-energy contributions) are shown.
  The parameters are chosen according to \SE\ (see \refta{tab:para}), 
  with $\phigl$ varied.
  The upper left plot shows the decay width, the upper right plot shows 
  the relative size of the corrections. 
  The lower left plot shows the BR, the lower right plot shows 
  the relative size of the BR.
}
\label{fig:PhiM3.glsd1d}
\end{center}
\end{figure}

\begin{figure}[htb!]
\begin{center}
\begin{tabular}{c}
\includegraphics[width=0.49\textwidth,height=8.0cm]{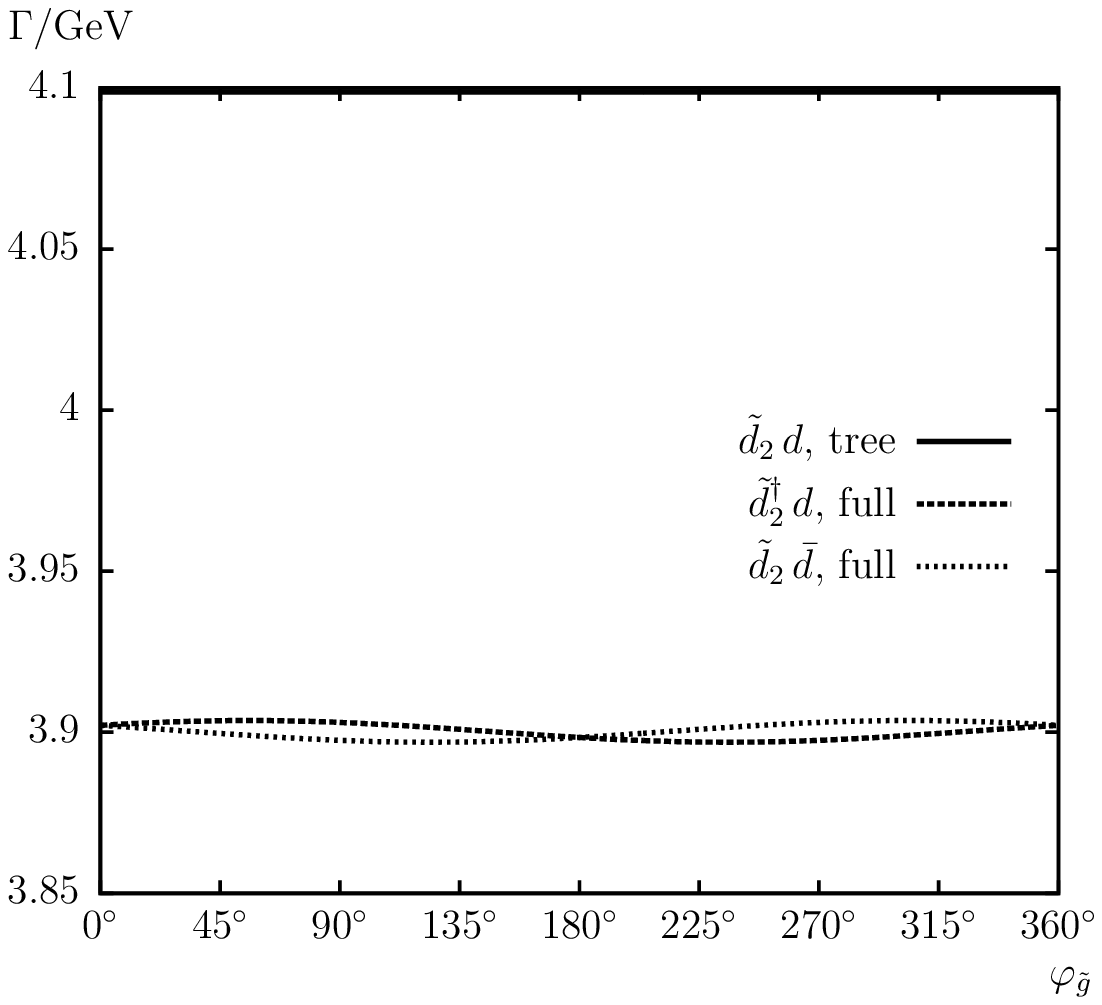}
\hspace{-4mm}
\includegraphics[width=0.49\textwidth,height=8.0cm]{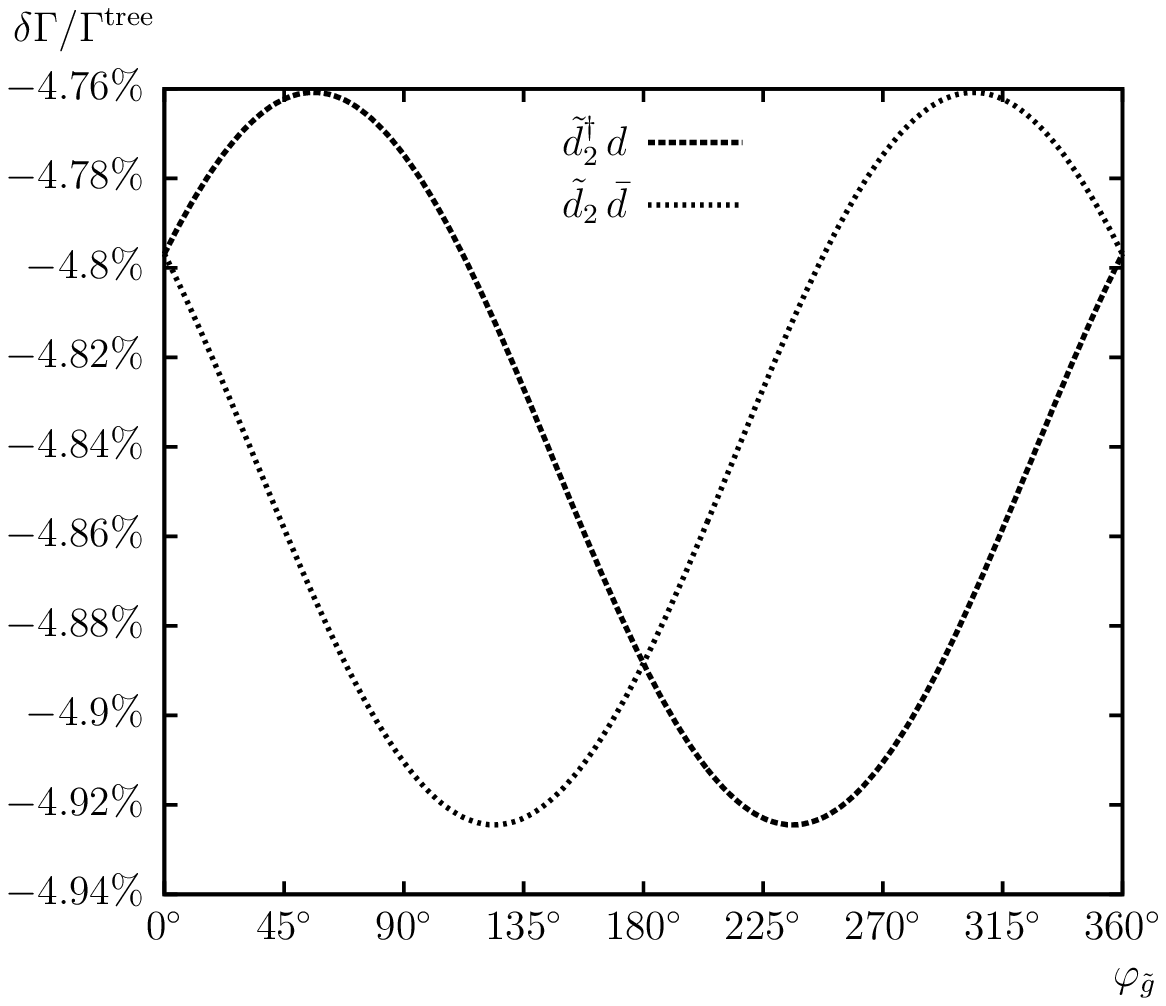}
\\[4em]
\includegraphics[width=0.49\textwidth,height=8.0cm]{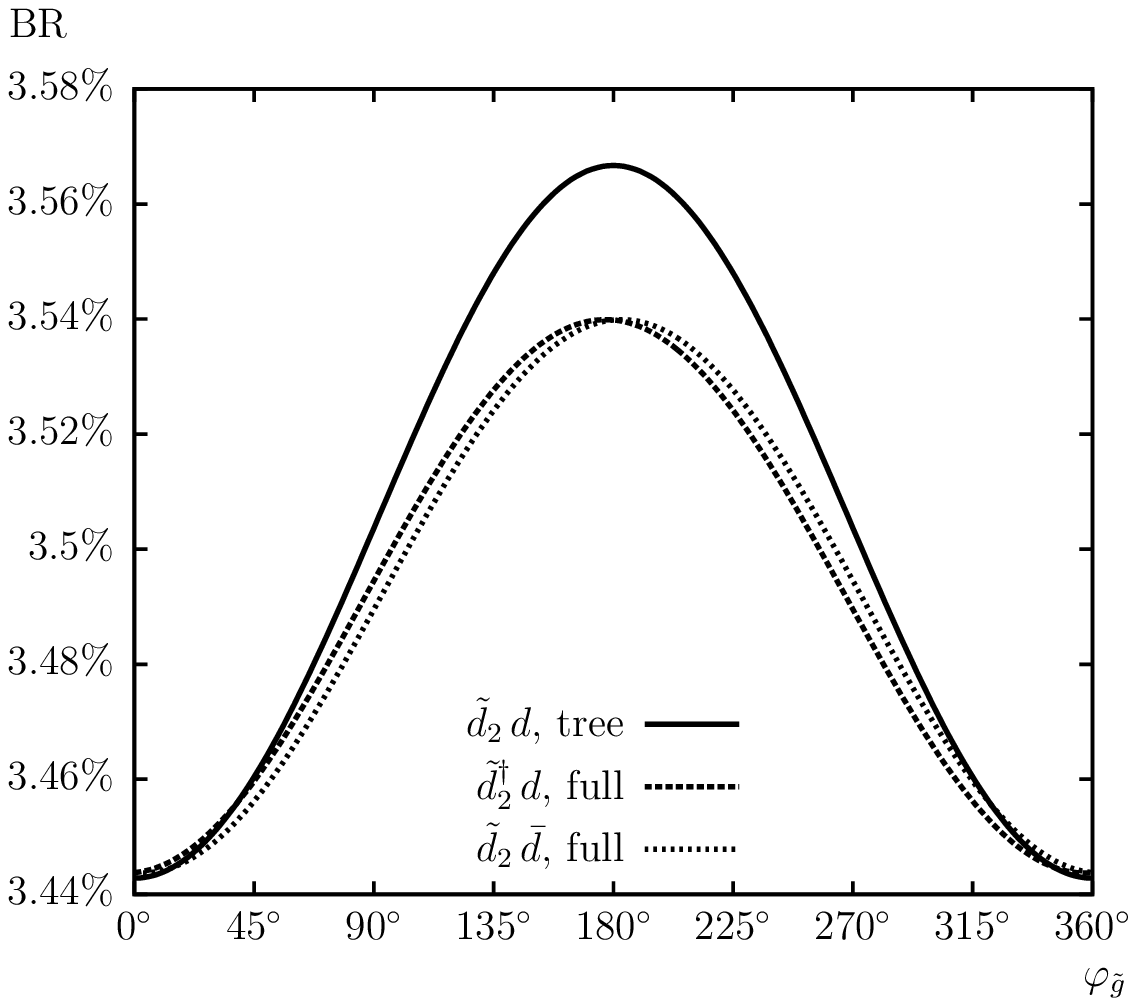}
\hspace{-4mm}
\includegraphics[width=0.49\textwidth,height=8.0cm]{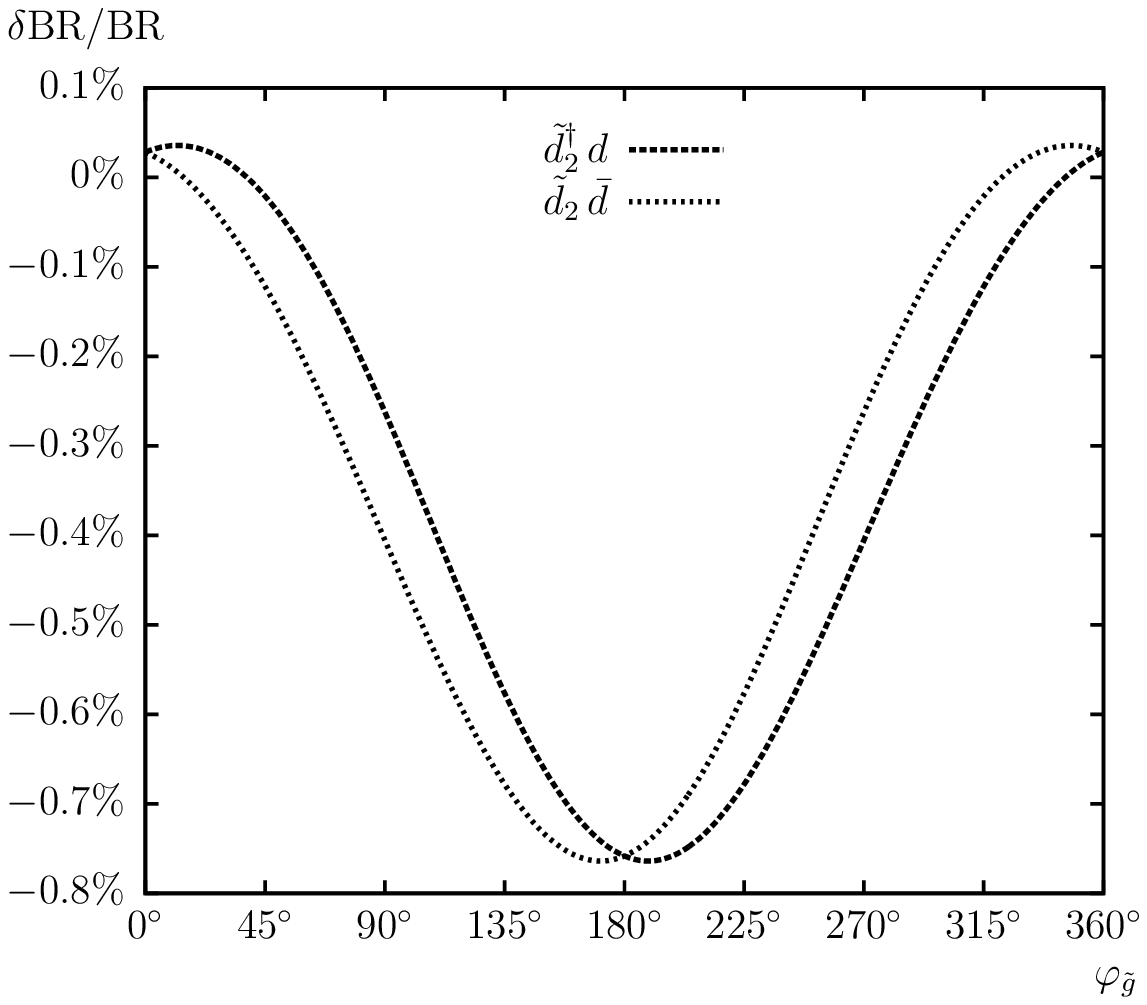}
\end{tabular}
\vspace{2em}
\caption{$\Ga(\decaySdz)$.
  Tree-level (``tree'') and full one-loop (``full'') corrected 
  decay widths (including absorptive self-energy contributions) are shown.
  The parameters are chosen according to \SE\ (see \refta{tab:para}), 
  with $\phigl$ varied.
  The upper left plot shows the decay width, the upper right plot shows 
  the relative size of the corrections. 
  The lower left plot shows the BR, the lower right plot shows 
  the relative size of the BR.
}
\label{fig:PhiM3.glsd2d}
\end{center}
\end{figure}

As before we show the decays in 
\reffi{fig:PhiM3.glst1t} -- \ref{fig:PhiM3.glsd2d}.
The arrangement of the panels is the same as in the previous subsection
(but leaving out the pure SQCD corrections).
According to \refta{tab:para} we choose $\mgl = 1200 \gev$, which is
substantially larger than the squark masses. Accordingly we expect again
that $\br(\decaySqe) + \br(\decaySqz) \sim 1/6$, with possible
deviations in the decays involving scalar tops, where the mass variation
is maximal. 

The results for $\decaySte$ are shown in \reffi{fig:PhiM3.glst1t}, 
where the results are given as a function of $\phigl$.
One can see that both the decay widths as well as the size of the
corrections to it vary substantially with $\phigl$.
While the widths ranges between $\sim 8 \gev$ and $\sim 4 \gev$, 
the corrections vary between $-8.5\%$ and $-2.5\%$ in our numerical scenario
\SE, where also a small difference between $\decayaSte$ and $\decayStae$
can be observed.
The BR's vary correspondingly from $\sim 8.0\%$ to $\sim 4\%$, where
the size of the loop effects in \SE\ goes from $-4\%$ to $\sim +2\%$,
again with a small visible difference between $\decayaSte$ and
$\decayStae$.
The second channel involving scalar tops, $\decayStz$, is shown in
\reffi{fig:PhiM3.glst2t}. In agreement with \reffi{fig:AbsM3.glst2t} the
decay width is substantially smaller than for $\decaySte$, largely due
to the reduced phase space, varying between $\sim 1 \gev$ 
and $\sim 3.5 \gev$. The difference between $\Ga(\decayaStz)$ and
$\Ga(\decayStaz)$ is clearly visible.
The loop corrections range from $+4\%$ to $-10\%$, with a clear symmetry
between $\decayaStz$ and $\decayStaz$.
Correspondingly, the $\br(\decayStz)$ reach values between $\sim 1\%$
and $\sim 3.5\%$ with a variation of $+8\%$ to $-6\%$. The sum of the four
BR's is close to but always a bit smaller than $\sim 1/6$ due to the
non-negligible effects of the scalar top masses and mixings.

For all other decay modes, i.e.\ $\decaySqi$ with $q \neq t$, 
the mass and mixing effects are substantially smaller.%
\footnote{
  Especially the tree-level results depends strongly on the scalar 
  quark mass splitting $m_q X_q$ (see \refeqs{Sfermionmassenmatrix}, 
  (\ref{kappa})) via $U_{\tilde{q}_{ij}} e^{i \phigl/2}$ and is 
  therefore nearly constant with respect to $\phigl$ for $q \ne t$.}
Within \SE\ we find $\Ga(\decaySqe, q \neq t)$ 
(shown in \reffis{fig:PhiM3.glsb1b}, 
\ref{fig:PhiM3.glsc1c}, \ref{fig:PhiM3.glss1s}, 
\ref{fig:PhiM3.glsu1u}, \ref{fig:PhiM3.glsd1d})
at the level of $\sim 5.5 \gev$, where small variations depend on the
squark flavor. 
Also the required symmetry of 
$\Ga(\decayaSqe, q \neq t)|_{\phigl} = \Ga(\decaySqae, q \neq t)|_{-\phigl}$
is clearly visible, and similarly for the BR's and the size of the loop
corrections.
While the size of the one-loop corrections vary between
$-1\%$ and $-6\%$ in the case of $\decaySbe$, they are found to be
within $\sim -3\%$ and $\sim -4.5\%$ for the other flavors.
The corresponding BR's are found at the level of $\sim 5\%$ with less
than $\pm 0.25\%$ of variation. The relative corrections in \SE\ are at
the few per-cent level at most. 

The results for the decay modes to the heavier scalar quarks, 
$\decaySqz, q \neq t$, shown in \reffis{fig:PhiM3.glsb2b}, 
\ref{fig:PhiM3.glsc2c}, \ref{fig:PhiM3.glss2s}, 
\ref{fig:PhiM3.glsu2u}, \ref{fig:PhiM3.glsd2d}, 
are also very similar,
again with a visible difference between $\decayaSqz$ and 
$\decaySqaz, q \neq t$. The widths are 
$\sim 4 \gev$ with a small phase dependence. The size of the
one-loop corrections vary around $-5\%$ for $\decaySbz$, $\sim -4.35\%$
for $\decaySqz, q = c, u$, and $\sim -4.85\%$ 
for $\decaySqz, q = s, d$.
The BR's for all flavors are found at $\sim 3.5\%$ with loop corrections
at the per-cent level. As ``expected'', for all flavors we find
$\br(\decaySqe) + \br(\decaySqz) \sim 1/6$.


\subsection{The one-loop decays \boldmath{$\decayNkg$}}
\label{sec:glneug}

\begin{figure}[ht!]
\begin{center}
\begin{tabular}{c}
\includegraphics[width=0.49\textwidth,height=8.0cm]{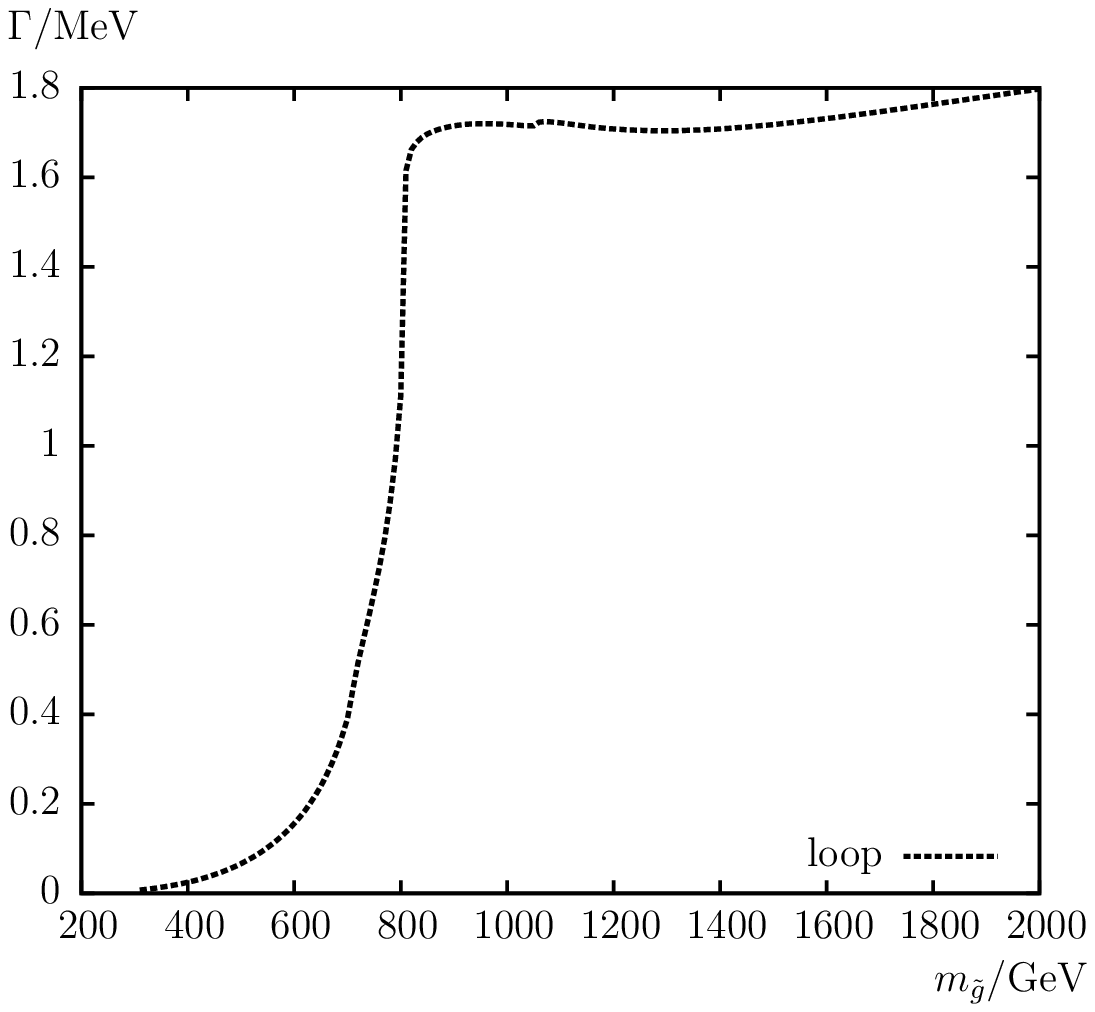}
\hspace{-4mm}
\includegraphics[width=0.49\textwidth,height=8.0cm]{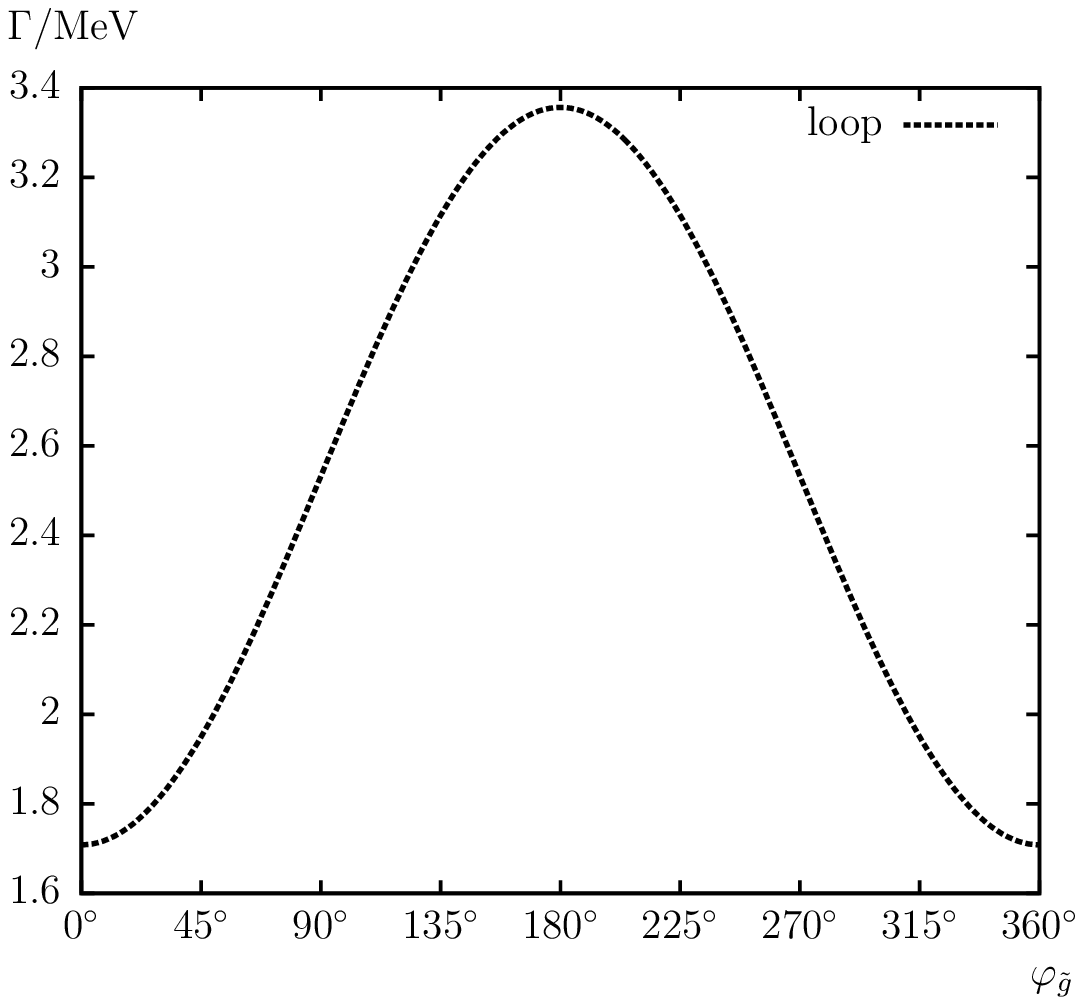}
\hspace{-4mm}
\end{tabular}
\vspace{2em}
\caption{
  $\Ga(\decayNeg)$. One-loop (``loop'') corrected decay widths are 
  shown with the parameters chosen according to \SE\ (see \refta{tab:para}), 
  with $\mgl$ (left plot) and $\phigl$ (right plot) varied.
}
\label{fig:AbsM3.glneu1g}
\end{center}
\end{figure}

\begin{figure}[hb!]
\begin{center}
\begin{tabular}{c}
\includegraphics[width=0.49\textwidth,height=8.0cm]{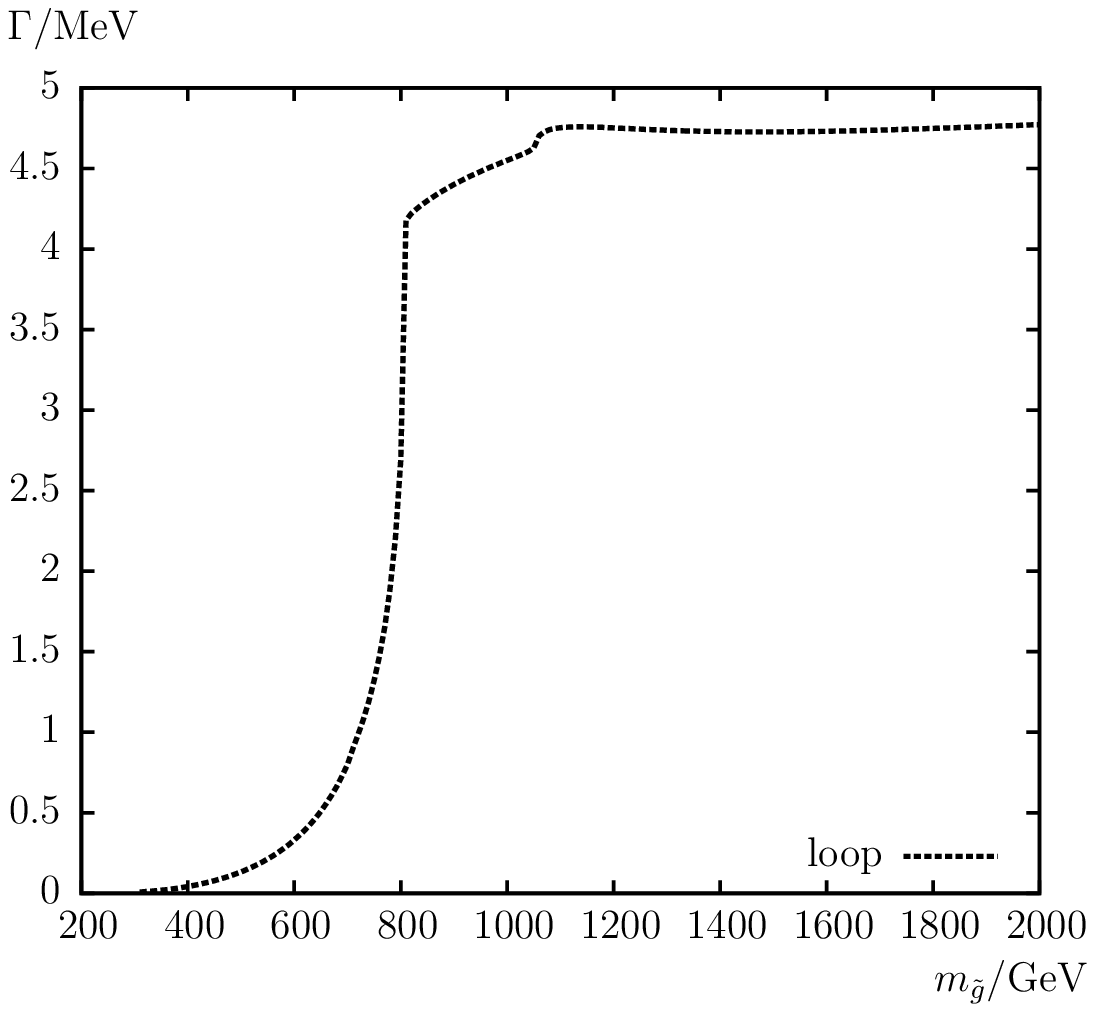}
\hspace{-4mm}
\includegraphics[width=0.49\textwidth,height=8.0cm]{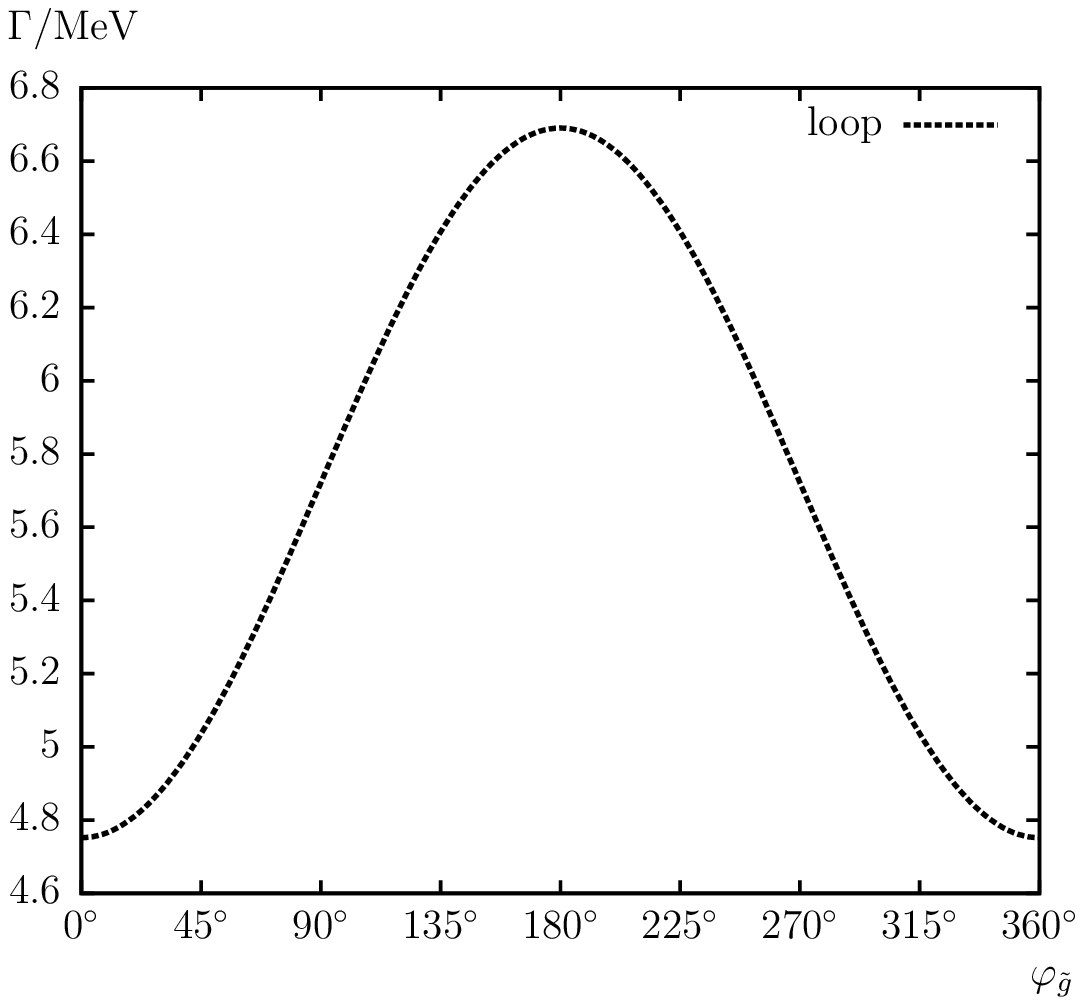}
\end{tabular}
\vspace{2em}
\caption{
  $\Ga(\decayNzg)$. One-loop (``loop'') corrected decay widths are 
  shown with the parameters chosen according to \SE\ (see \refta{tab:para}), 
  with $\mgl$ (left plot) and $\phigl$ (right plot) varied.
}
\label{fig:AbsM3.glneu2g}
\end{center}
\end{figure}

\begin{figure}[ht!]
\begin{center}
\begin{tabular}{c}
\includegraphics[width=0.49\textwidth,height=8.0cm]{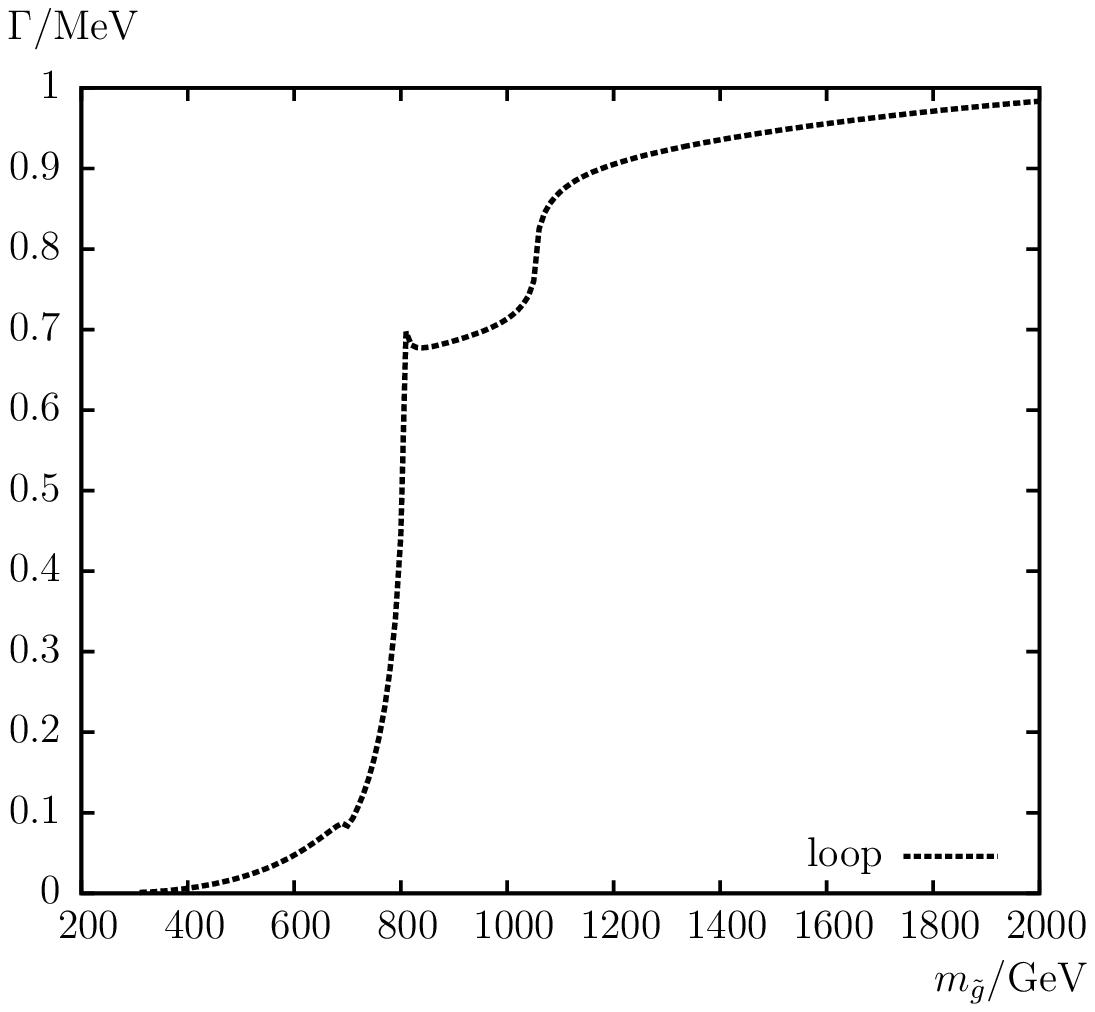}
\hspace{-4mm}
\includegraphics[width=0.49\textwidth,height=8.0cm]{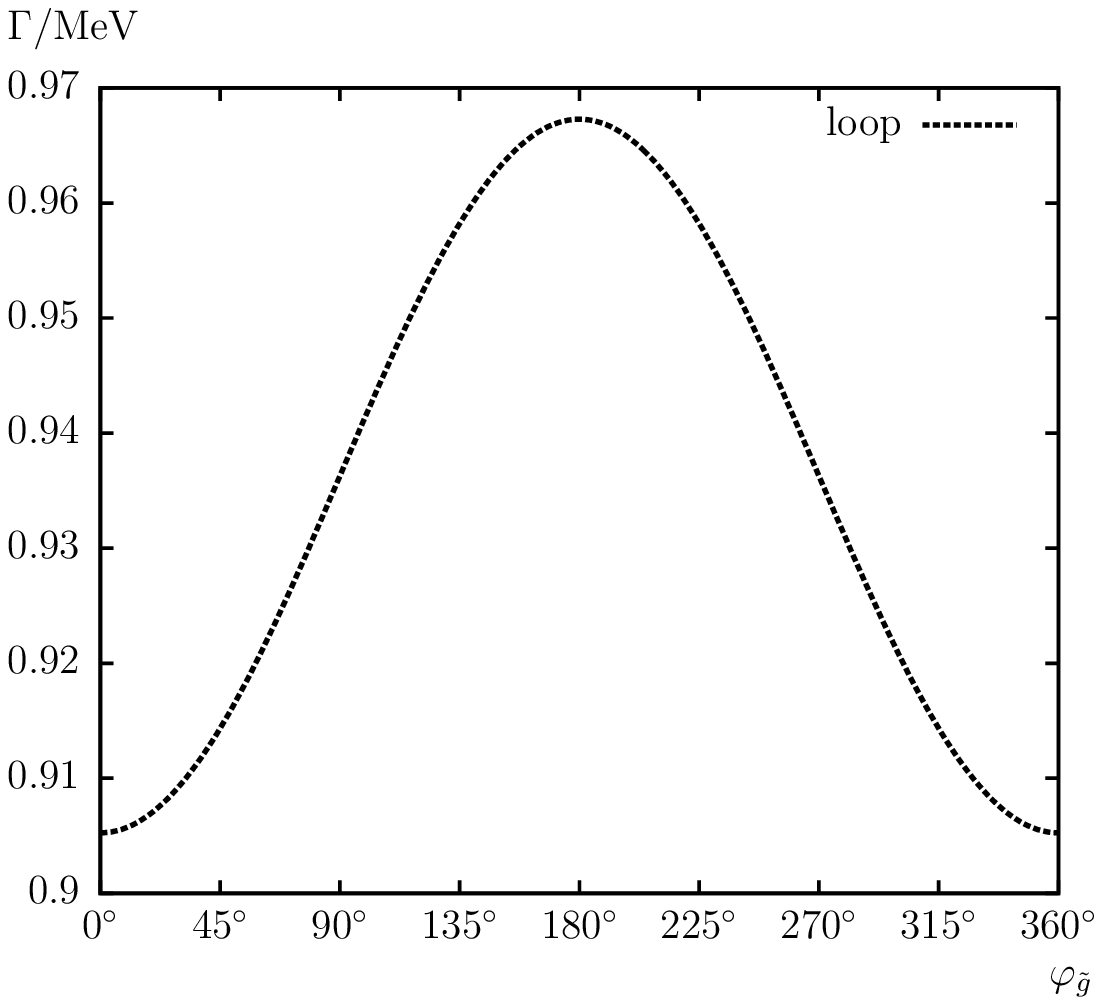}
\hspace{-4mm}
\end{tabular}
\vspace{2em}
\caption{
  $\Ga(\decayNdg)$. One-loop (``loop'') corrected decay widths are 
  shown with the parameters chosen according to \SE\ (see \refta{tab:para}), 
  with $\mgl$ (left plot) and $\phigl$ (right plot) varied.
}
\label{fig:AbsM3.glneu3g}
\end{center}
\end{figure}

\begin{figure}[hb!]
\begin{center}
\begin{tabular}{c}
\includegraphics[width=0.49\textwidth,height=8.0cm]{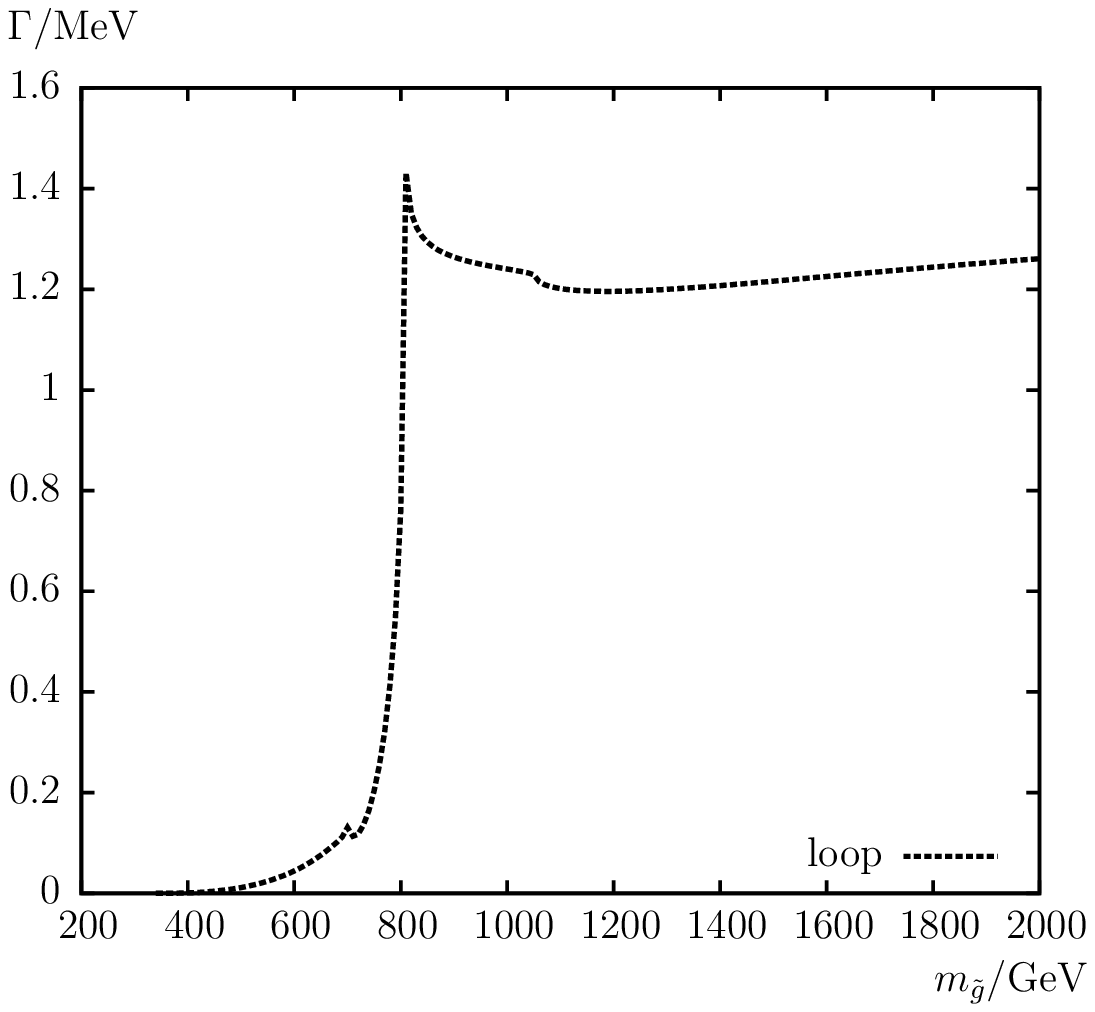}
\hspace{-4mm}
\includegraphics[width=0.49\textwidth,height=8.0cm]{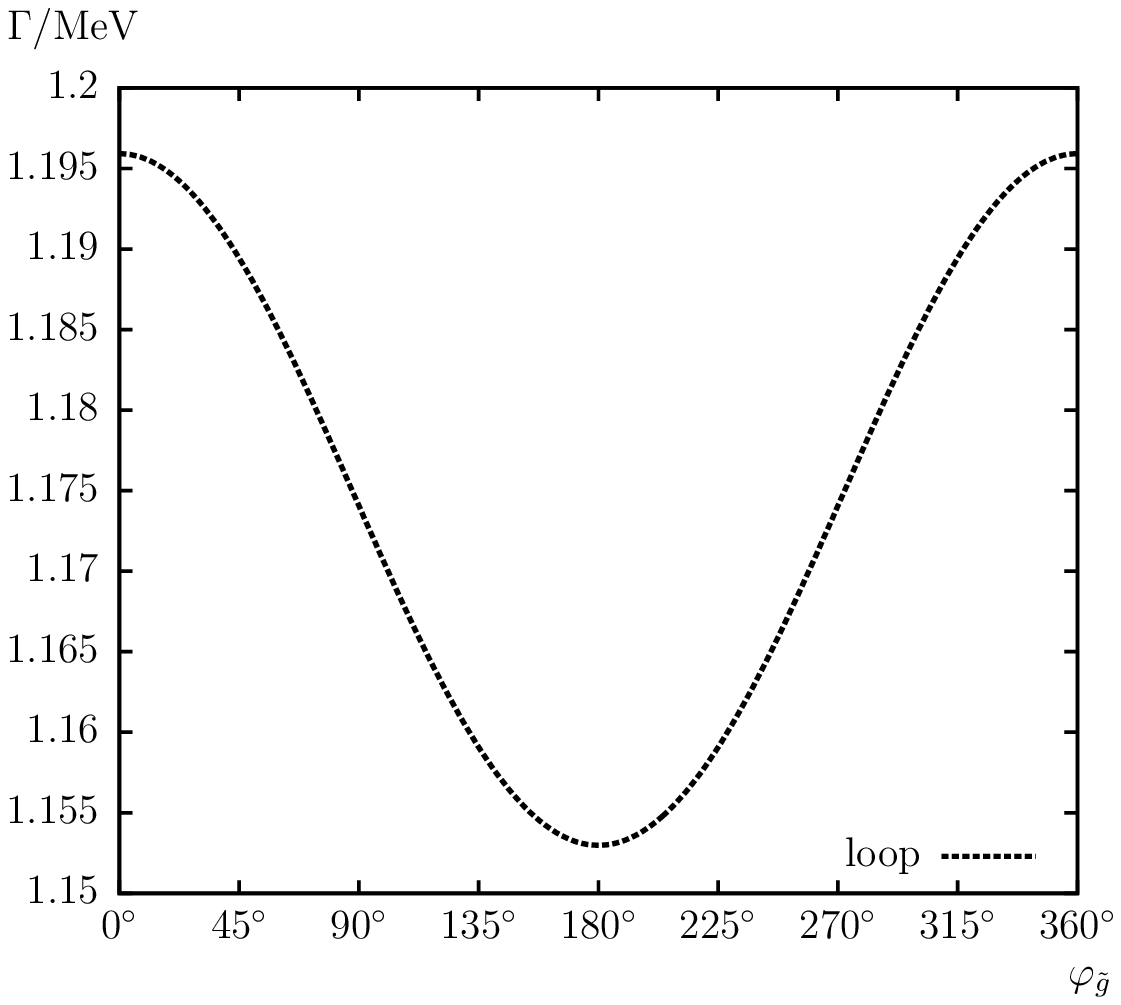}
\end{tabular}
\vspace{2em}
\caption{
  $\Ga(\decayNvg)$. One-loop (``loop'') corrected decay widths are 
  shown with the parameters chosen according to \SE\ (see \refta{tab:para}), 
  with $\mgl$ (left plot) and $\phigl$ (right plot) varied.
}
\label{fig:AbsM3.glneu4g}
\end{center}
\end{figure}

In \reffis{fig:AbsM3.glneu1g} -- \ref{fig:AbsM3.glneu4g}
we present the variation of $\Ga(\decayNkg), (k = 1,2,3,4)$ 
as a function of $\mgl$ and $\phigl$ in \SE, see \refta{tab:para}. 
The structure at $\mgl \approx 1054.6$ in 
\reffi{fig:AbsM3.glneu1g} and \reffi{fig:AbsM3.glneu2g}
is the vertex production threshold $\mgl = \mstz + \mt$.
In \reffi{fig:AbsM3.glneu3g} and \reffi{fig:AbsM3.glneu4g}
the dip at $\mgl \approx 699.2 \gev$ stems from the vertex production
threshold $\mgl = \msce + m_c$,  the peak at $\mgl \approx 809.3 \gev$ 
is the vertex production threshold $\mgl = \mste + \mt$.

With our choice of parameters, see \refta{tab:para}, we find the
neutralino masses at $\sim 151$, $206$, $218$, $338 \gev$, i.e.\ relatively
similar with respect to the larger gluino mass.
The size of the decay widths is very small for all four neutralinos with
respect to the hadronic final states discussed in the previous
subsection. The largest values are reached for $\decayNzg$, going up to 
$\sim 0.007 \gev$. The dependence on $\mgl$ is very similar, as expected
from the large gap between neutralino and gluino mass. We see a
threshold on-set, going up to the maximum values around $800 \gev$. 
Due to the smallness of the decay widths the corresponding branching
ratios are negligible in \SE.%
\footnote{
  They are not included in the calculation of the total decay
  widths shown in \refse{sec:tot}.} 
However, in a situation where the gluino
is lighter than all scalar quarks these loop-induced two-body decay
modes can become relevant and may constitute a significant source for
$\neu{1}$ production at the LHC (which could be used to measure the
properties of the SUSY CDM candidate). In such a situation, of course,
also three particle final states will be relevant, which are, however,
beyond the scope of this paper.

\vspace{1cm}

\vspace{1cm}

\clearpage


\subsection{The total decay width}
\label{sec:tot}

\begin{figure}[htb!]
\begin{center}
\begin{tabular}{c}
\includegraphics[width=0.49\textwidth,height=8.0cm]{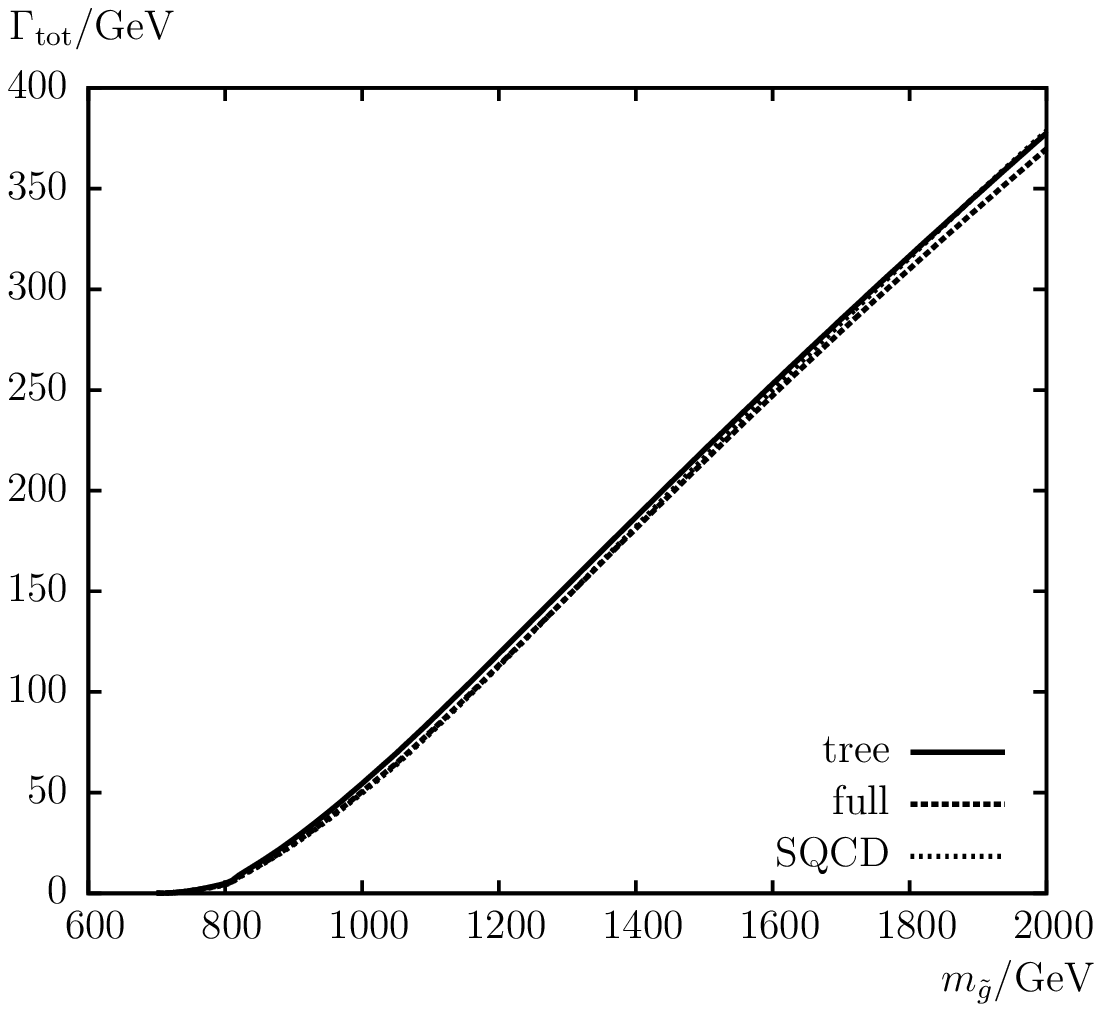}
\hspace{-4mm}
\includegraphics[width=0.49\textwidth,height=8.0cm]{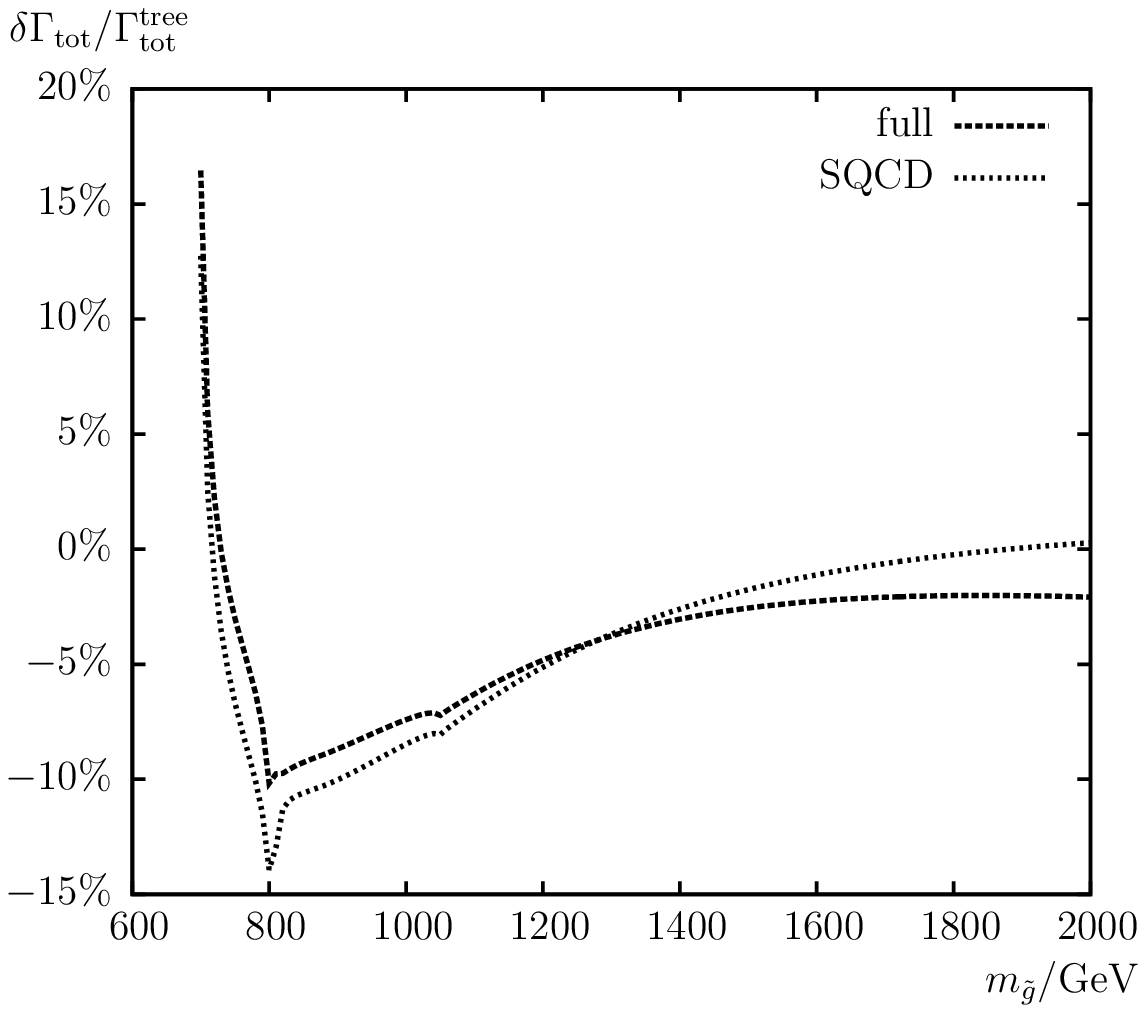}
\\[4em]
\includegraphics[width=0.49\textwidth,height=8.0cm]{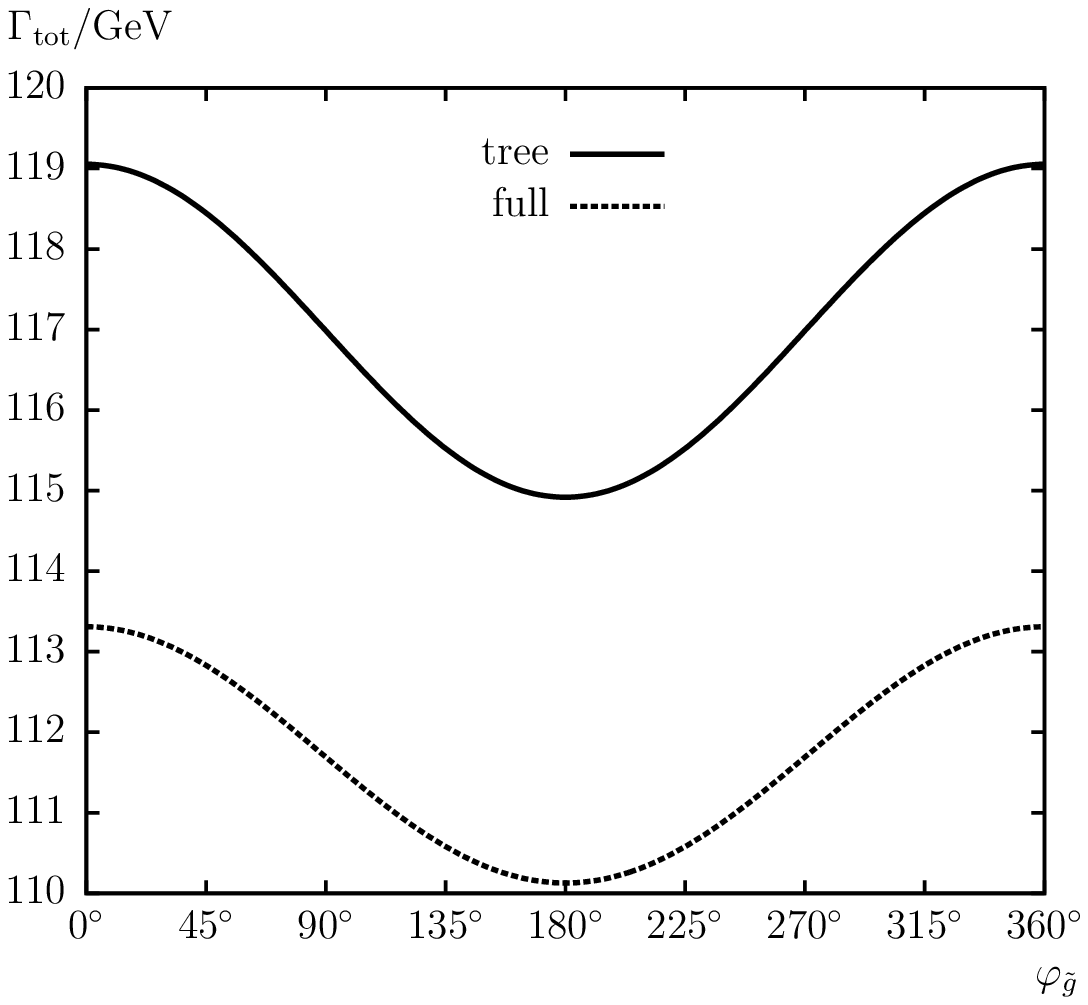}
\hspace{-4mm}
\includegraphics[width=0.49\textwidth,height=8.0cm]{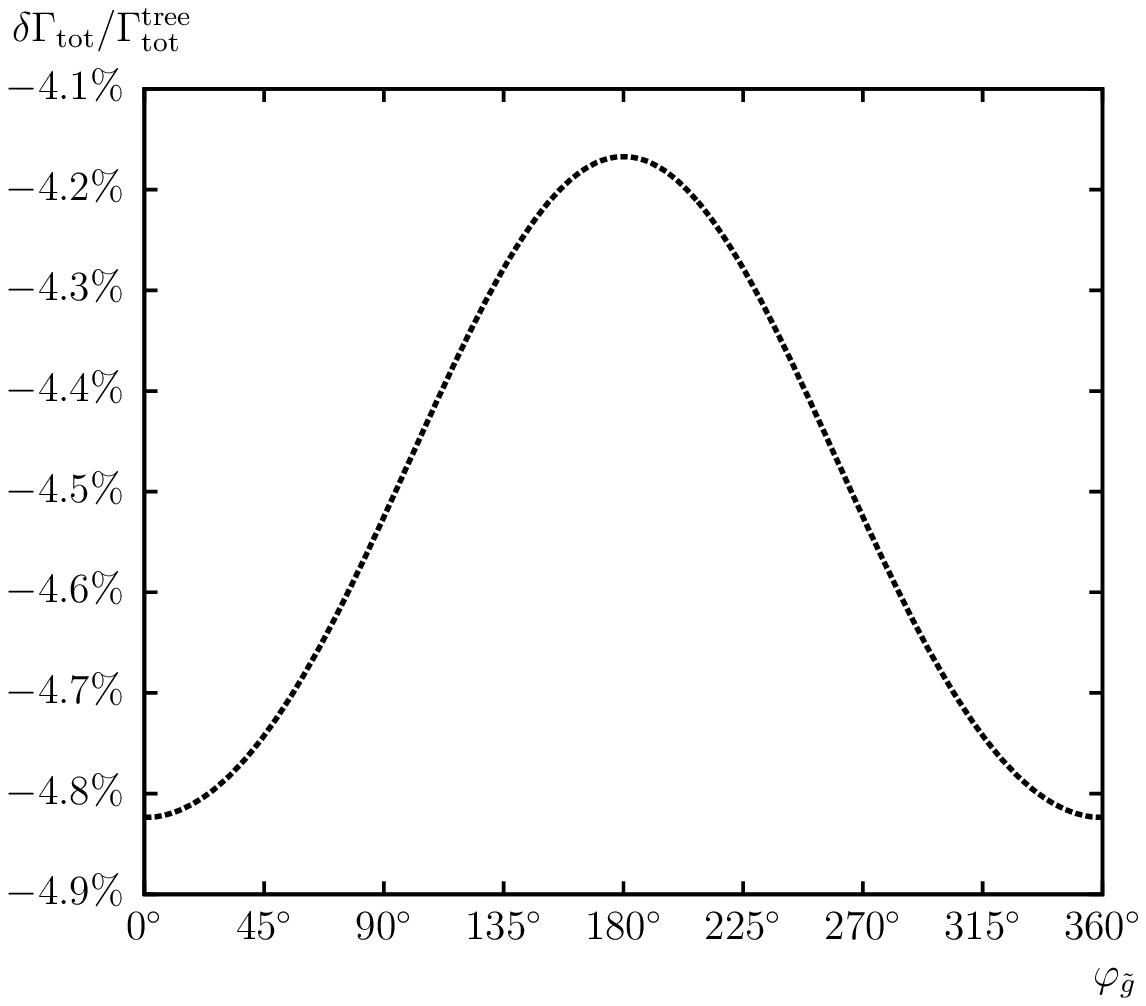}
\end{tabular}
\vspace{2em}
\caption{$\Ga_{\rm tot}$.
  Tree-level (``tree''), full one-loop (``full'') and pure SQCD
  (``SQCD'') corrected decay widths are shown with the parameters chosen 
  according to \SE\ (see \refta{tab:para}).
  The upper left plot shows the total decay width, the upper right plot 
  shows the relative size of the total corrections, with $\mgl$ varied. 
  The lower plots show the same but with $\phigl$ varied (including 
  absorptive self-energy contributions).
}
\label{fig:GammaTot}
\end{center}
\end{figure}

Finally we show the results for the total decay width of the gluino, 
see \refeqs{tot}. 
In \reffi{fig:GammaTot} the upper panels show the absolute and 
relative variation with $\mgl$.
We observe a nearly linear rise from the threshold region on up to
$\mgl = 2 \tev$, where $\Ga_{\rm tot} \approx 370 \gev$ is reached. 
The relative size of the corrections are around $+15\%$ close to the
production threshold, reaches $-10\%$ at $\mgl = 800 \gev$ and levels
out at $\sim -2.5\%$ for large gluino masses. Depending on the choice of
SUSY parameters and the location of (production) thresholds the size of
the one-loop corrections are non-negligible at the LHC. 
We also included here the pure SQCD corrections, which tend to
overestimate the full result by $\sim -5\%$ at low $\mgl$ and go to
zero at large $\mgl$, where the EW corrections are dominating.

The lower panels of \reffi{fig:GammaTot} show the result as a function
of $\phigl$. 
For the nominal value of $\mgl = 1200 \gev$ we find the
loop-corrected total decay width between $\sim 113 \gev$ for 
$\phigl = 0^\circ$ and $\sim 110 \gev$ for $\phigl = 180^\circ$. 
The relative size of the corrections vary between $-4.8\%$ and $-4.2\%$, 
possibly still relevant for LHC measurements.

\section{Conclusions}
\label{sec:conclusions}

We have evaluated all two-body decay widths of the gluino in
the Minimal Supersymmetric Standard Model with complex parameters (cMSSM). 
The decay modes are given in \refeqs{glsqq} -- (\ref{glneg}). 
After the (pair) production of gluinos they are expected to
decay via cascades to lighter SUSY particles and quarks and/or leptons. 
Consequently, the decay of a gluino to a scalar quark and a quark will
one of the most relevant decays in such a cascade. In order to determine
the masses, couplings and parameters of SUSY particles precisely in such
cascades, higher-order corrections are necessary to keep the theory
error on an acceptable level.

Our evaluation is based on a full
one-loop calculation of all decay channels, also including soft and
hard QED and QCD radiation, necessary to derive a reliable prediction 
of any two-body branching ratio. 
We also include the purely loop induced decays $\decayNkg$ ($k = 1,2,3,4$).
These modes as well as possible 
three-body decay modes can become sizable only if all the two-body channels
are kinematically (nearly) closed. We have not investigated these
three-body decay modes, which are, beyond the scope of this paper.

We first reviewed the relevant sectors of the cMSSM, including their
renormalization (where extensive explanations can be found in
\citeres{SbotRen,Stop2decay,LHCxC}). 
We have discussed the calculation of the one-loop diagrams, the
treatment of UV- and IR-divergences (where the latter are canceled 
by the inclusion of real soft QCD and QED radiation), and in 
addition the hard QCD and QED bremsstrahlung.

For the numerical analysis we have chosen one parameter set that allows
simultaneously {\em all} two-body decay modes and respect the current
experimental bounds on Higgs boson and SUSY searches. 
In a first step we have shown the results as a function of $\mgl$. For
the largest value included in our analysis, $\mgl = 2 \tev$ we find
decay widths of $\sim 15 \gev$, and 
$\br(\decaySqe) + \br(\decaySqz) \sim 1/6$. The largest deviations
were found for the decays involving the third generation (s)quarks,
where effects of the squark mixing and corresponding changes in the
masses and couplings are maximal.
The size of the one-loop corrections is found to be largest close to 
the production thresholds and reaches several per-cent for large $\mgl$.  
We have also investigated the quality of the approximation of the
pure SQCD corrections. While for some channels, especially involving
lighter quarks, the SQCD contributions yield a good approximation, in
the channels involving scalar tops and bottoms the EW corrections, which
were evaluated here for the first time, are crucial to yield a reliable
result.

In a second step we have analyzed the dependence of the decay widths and
branching ratios on the phase of the gluino mass parameter, $\phigl$,
for $\mgl = 1200 \gev$ fixed. Again, the results are to a large extent
flavor independent, except for decays involving (scalar) tops. In the
latter case a variation of the BR of up to $14\%$ was found, 
and the size of the loop corrections can yield $\sim 10\%$. For the other
flavors we find as before $\br(\decaySqe) + \br(\decaySqz) \sim 1/6$,
and a smaller variation with $\phigl$. 

The size of the widths of the purely loop-induced decays is
substantially smaller, staying below $0.01 \gev$ for our parameters. 
This could change if the gluino were lighter than all scalar quarks. 
In that case also three-body decays would be relevant, which, however,
were not investigated here.

Furthermore, it has to be kept in mind that we have evaluated the full
one-loop corrections for one set of parameters only, which was chosen
to have many decay modes open, and {\em not} to emphasize the loop
corrections. Consequently, it can easily be imagined that substantially
larger or smaller corrections can be found for other choices of SUSY
parameters. The full one-loop corrections should (eventually) be
taken into account in precision analyses of cascade decays at the LHC.
Nevertheless, the size of the loop corrections found give an idea of the
relevance of the full one-loop corrections. 

Following our analysis it is evident that the full one-loop corrections
are mandatory for a precise prediction of the various branching ratios.
The results for the gluino decays will be implemented into the
Fortran code {\tt FeynHiggs}.


\subsection*{Acknowledgements}

We thank
A.~Denner,
T.~Hahn,
F.~von~der~Pahlen, 
H.~Rzehak
and 
G.~Weiglein
for helpful discussions.
We thank also S.~Liebler 
for his help by the implementation of the photon bremsstrahlung.
The work of S.H.\ was partially supported by CICYT 
(grant FPA 2010--22163-C02-01) and by the
Spanish MICINN's Consolider-Ingenio 2010 Program under grant MultiDark
CSD2009-00064.





\begin{thebibliography}{99} 

\bibitem{mssm} H.P.~Nilles, 
               {\em Phys.\ Rept.} {\bf 110} (1984) 1; \\ 
               H.E.~Haber and G.L.~Kane, 
               {\em Phys.\ Rept.} {\bf 117} (1985) 75; \\  
               R.~Barbieri, 
               {\em Riv.\ Nuovo Cim.} {\bf 11} (1988) 1. 

\bibitem{atlas}
  G.~Aad et al.\ [The ATLAS Collaboration],
  arXiv:0901.0512.

\bibitem{cms}
  G.~Bayatian et al.\ [CMS Collaboration],
  {\em J.\ Phys.} {\bf G 34} (2007) 995.

\bibitem{mc3}
  O.~Buchmueller et al.,
  {\em Eur.\ Phys.\ J.} {\bf C 64} (2009) 391
  [arXiv:0907.5568 [hep-ph]].

\bibitem{glsqq_als} W.~Beenakker, R.~H\"opker and P.~Zerwas,
                    {\em Phys.\ Lett.} {\bf B 378} (1996) 159
                    [arXiv:hep-ph/9602378].

\bibitem{sdecay} M.~M\"uhlleitner, A.~Djouadi and Y.~Mambrini,
                 {\em Comput.\ Phys.\ Commun.} {\bf 168} (2005) 46
                 [arXiv:hep-ph/0311167].

\bibitem{majoranagluinos} S.~Choi, M.~Drees, A.~Freitas and P.~Zerwas,
                          {\em Phys.\ Rev.} {\bf D 78} (2008) 095007
                          [arXiv:0808.2410 [hep-ph]].

\bibitem{gluinopol} M.~Kr\"amer, E.~Popenda, M.~Spira and P.~Zerwas,
                    {\em Phys.\ Rev.} {\bf D 80} (2009) 055002
                    [arXiv:0902.3795 [hep-ph]].

\bibitem{lhc2fc} A.~De Roeck et al.,
                 {\em Eur.\ Phys.\ J.} {\bf C 66} (2010) 525
                 [arXiv:0909.3240 [hep-ph]].

\bibitem{feynhiggs} S.~Heinemeyer, W.~Hollik and G.~Weiglein,
                    {\em Comput. Phys. Commun.} {\bf 124} (2000) 76
                    [arXiv:hep-ph/9812320];
                    see {\tt www.feynhiggs.de} .

\bibitem{mhiggslong} S.~Heinemeyer, W.~Hollik and G.~Weiglein,
                     {\em Eur. Phys. J.} {\bf C 9} (1999) 343
                     [arXiv:hep-ph/9812472].

\bibitem{mhiggsAEC} G.~Degrassi, S.~Heinemeyer, W.~Hollik,
                    P.~Slavich and G.~Weiglein, 
                    {\em Eur. Phys. J.} {\bf C 28} (2003) 133
                    [arXiv:hep-ph/0212020].

\bibitem{mhcMSSMlong} M.~Frank, T.~Hahn, S.~Heinemeyer, W.~Hollik, 
                      R.~Rzehak and G.~Weiglein,
                      {\em JHEP} {\bf 02} (2007) 047
                      [arXiv:hep-ph/0611326].

\bibitem{dissTF} T.~Fritzsche,
                 PhD thesis, Cuvillier Verlag, G\"ottingen 2005,
                 ISBN 3--86537--577--4.

\bibitem{SbotRen} S.~Heinemeyer, H.~Rzehak and C.~Schappacher,
                  {\em Phys.\ Rev.} {\bf D 82} (2010) 075010
                  [arXiv:1007.0689 [hep-ph]];\\
                  PoS C {\bf CHARGED2010} (2010) 039
                  [arXiv:1012.4572 [hep-ph]].

\bibitem{Stop2decay} T.~Fritzsche, S.~Heinemeyer, H.~Rzehak
                     and C.~Schappacher, 
                     arXiv:1111.7289 [hep-ph].

\bibitem{imim} A.~Fowler and G.~Weiglein,
               {\em JHEP} {\bf 1001} (2010) 108
               [arXiv:0909.5165 [hep-ph]].

\bibitem{denner} A.~Denner,
                 {\em Fortsch.\ Phys.} {\bf 41} (1993) 307
                 [arXiv:0709.1075 [hep-ph]].

\bibitem{pdg} K.~Nakamura et al.\ [Particle Data Group],
              {\em J.\ Phys.} {\bf G 37} (2010) 075021.

\bibitem{RunDec} K.~Chetyrkin, J.~K\"uhn and M.~Steinhauser, 
                 {\em Comput. Phys. Commun.} {\bf 133} (2000) 43
                 [arXiv:hep-ph/0004189].

\bibitem{deltab1} R.~Hempfling,
                  {\em Phys. Rev.} {\bf D 49} (1994) 6168;\\
                  L.~Hall, R.~Rattazzi and U.~Sarid,
                  {\em Phys. Rev.} {\bf D 50} (1994) 7048
                  [arXiv:hep-ph/9306309];\\
                  M.~Carena, M.~Olechowski, S.~Pokorski and C.~Wagner,
                  {\em Nucl. Phys.} {\bf B 426} (1994) 269
                  [arXiv:hep-ph/9402253].

\bibitem{deltab2} M.~Carena, D.~Garcia, U.~Nierste and C.~Wagner,
                  {\em Nucl. Phys.} {\bf B 577} (2000) 577
                  [arXiv:hep-ph/9912516].

\bibitem{alsDRbar} R.~Harlander, L.~Mihaila and M.~Steinhauser,
                   {\em Phys.\ Rev.} {\bf D 72} (2005) 095009
                   [arXiv:hep-ph/0509048].

\bibitem{Peccei} R.~Peccei and H.~Quinn,
                 {\em Phys.\ Rev.\ Lett.} {\bf 38} (1977) 1440;
                 {\em Phys.\ Rev.} {\bf  D 16} (1977) 1791.

\bibitem{MSSMcomplphasen} S.~Dimopoulos and S.~Thomas,
                          {\em Nucl.\ Phys.} {\bf  B 465} (1996) 23
                          [arXiv:hep-ph/9510220].

\bibitem{feynarts} J.~K\"ublbeck, M.~B\"ohm and A.~Denner, 
                   {\em Comput. Phys. Commun.} {\bf 60} (1990) 165;\\
                   T.~Hahn, 
                   {\em Comput. Phys. Commun.} {\bf 140} (2001) 418
                   [arXiv:hep-ph/0012260];\\
                   T.~Hahn and C.~Schappacher, 
                   {\em Comput. Phys. Commun.} {\bf 143} (2002) 54
                   [arXiv:hep-ph/0105349].\\
                   The program, the user's guide and the MSSM model files
                   are available via\\ {\tt www.feynarts.de} .

\bibitem{LHCxC} S.~Heinemeyer, F.~v.d.~Pahlen and C.~Schappacher,
                arXiv:1112.0760 [hep-ph].


\bibitem{formcalc} T.~Hahn and M.~P\'erez-Victoria,
                   {\em Comput. Phys. Commun.} {\bf 118} (1999) 153
                   [arXiv:hep-ph/9807565].

\bibitem{cdr} F.~del Aguila, A.~Culatti, R.~Munoz Tapia and M.~Perez-Victoria,
              {\em Nucl. Phys.} {\bf B 537} (1999) 561
              [arXiv:hep-ph/9806451].

\bibitem{dred} W.~Siegel, 
               {\em Phys. Lett.} {\bf B 84} (1979) 193; \\
               D.~Capper, D.~Jones, and P.~van Nieuwenhuizen,
               {\em Nucl. Phys.} {\bf B 167} (1980) 479. 

\bibitem{dredDS} D.~St\"ockinger,
                 {\em JHEP} {\bf 0503} (2005) 076
                 [arXiv:hep-ph/0503129].

\bibitem{dredDS2} W.~Hollik and D.~St\"ockinger,
                  {\em Phys.\ Lett.} {\bf B 634} (2006) 63
                  [arXiv:hep-ph/0509298].

\bibitem{feynarts-mf} The couplings can be found in the files
                      {\tt MSSM.ps.gz} and {\tt MSSMQCD.ps.gz} 
                      as part of the \fa~package~\cite{feynarts}.

\bibitem{LEPHiggsSM} [LEP Higgs working group],
                     {\em Phys. Lett.} {\bf B 565} (2003) 61
                     [arXiv:hep-ex/0306033].

\bibitem{LEPHiggsMSSM} [LEP Higgs working group],
                       {\em Eur.\ Phys.\ J.} {\bf C 47} (2006) 547
                       [arXiv:hep-ex/0602042].

\bibitem{LHCsusy} V.~Khachatryan et al. [CMS Collaboration], 
                  arXiv:1101.1628 [hep-ex];\\
         {\tt http://cdsweb.cern.ch/record/1342547/files/SUS-11-001-pas.pdf};\\
         {\tt http://cdsweb.cern.ch/record/1343076/files/SUS-10-005-pas.pdf};\\
                  G.~Aad et al. [ATLAS Collaboration],
                  arXiv:1102.2357 [hep-ex];
                  arXiv:1102.5290 [hep-ex].

\bibitem{ccb} J.~Frere, D.~Jones and S.~Raby,
              {\em Nucl.\ Phys.} {\bf B 222} (1983) 11;\\
              M.~Claudson, L.~Hall and I.~Hinchliffe,
              {\em Nucl.\ Phys.} {\bf B 228} (1983) 501;\\
              C.~Kounnas, A.~Lahanas, D.~Nanopoulos and M.~Quiros,
              {\em Nucl.\ Phys.} {\bf B 236} (1984) 438;\\
              J.~Gunion, H.~Haber and M.~Sher,
              {\em Nucl.\ Phys.} {\bf B 306} (1988) 1;\\
              J.~Casas, A.~Lleyda and C.~Munoz,
              {\em Nucl.\ Phys.} {\bf B 471} (1996) 3
              [arXiv:hep-ph/9507294];\\
              P.~Langacker and N.~Polonsky,
              {\em Phys.\ Rev.} {\bf D 50} (1994) 2199
              [arXiv:hep-ph/9403306];\\
              A.~Strumia,
              {\em Nucl.\ Phys.} {\bf B 482} (1996) 24
              [arXiv:hep-ph/9604417].

\bibitem{SUSYphases} M.~Dugan, B.~Grinstein and L.~Hall,
                     {\em Nucl.\ Phys.} {\bf B 255} (1985) 413.

\bibitem{EDMDoink} W.~Hollik, J.~Illana, S.~Rigolin and D.~St\"ockinger,
                   {\em Phys. Lett.} {\bf B 416} (1998) 345
                   [arXiv:hep-ph/9707437];
                   {\em Phys. Lett.} {\bf B 425} (1998) 322
                   [arXiv:hep-ph/9711322].

\bibitem{EDMrev2} D.~Demir, O.~Lebedev, K.~Olive, M.~Pospelov and A.~Ritz,
                  {\em Nucl. Phys.} {\bf B 680} (2004) 339
                  [arXiv:hep-ph/0311314].

\bibitem{EDMPilaftsis} D.~Chang, W.~Keung and A.~Pilaftsis,
                       {\em Phys. Rev. Lett.} {\bf 82} (1999) 900
                       [Erratum-ibid.\  {\bf 83} (1999) 3972]
                       [arXiv:hep-ph/9811202];\\
                       A.~Pilaftsis,
                       {\em Phys. Lett.} {\bf B 471} (1999) 174
                       [arXiv:hep-ph/9909485].

\bibitem{EDMRitz} O.~Lebedev, K.~Olive, M.~Pospelov and A.~Ritz,
                  {\em Phys. Rev.} {\bf D 70} (2004) 016003
                  [arXiv:hep-ph/0402023].

\bibitem{EDMrev1} S.~Abel, S.~Khalil and O.~Lebedev,
                  {\em Nucl. Phys.} {\bf B 606} (2001) 151
                  [arXiv:hep-ph/0103320].

\bibitem{EDMrev3} Y.~Li, S.~Profumo and M.~Ramsey-Musolf,
                  {\em JHEP} {\bf 1008} (2010) 062
                  [arXiv:1006.1440 [hep-ph]].


\end{thebibliography}
\end{document}